\documentclass[A4paper,twoside,openright,11pt]{book}

\usepackage[latin1]{inputenc}
\usepackage{amssymb}
\usepackage{latexsym}
\usepackage{mathrsfs}
\usepackage{euscript}
\usepackage{amsmath}
\usepackage{graphicx}
\usepackage{psfrag}
\usepackage{epsfig}
\usepackage{axodraw}
\usepackage[dvips]{color}
\usepackage{fancyhdr}

\renewcommand{\chaptermark}[1]{\markboth{#1}{}}
\renewcommand{\sectionmark}[1]{\markright{\thesection\ #1}}
\renewcommand{\headrulewidth}{0.5pt}

\fancypagestyle{plain}{\fancyhead{}\renewcommand{\headrulewidth}{0pt}}

\newcommand{\nn}{\nonumber}
\newcommand{\ket}[1]{\vert #1 \rangle}
\newcommand{\bra}[1]{\langle #1 \vert}
\newcommand{\braket}[2]{\langle #1 \vert #2 \rangle}
\newcommand{\eq}[1]{\begin{equation} #1 \end{equation}}
\newcommand{\eqa}[1]{\begin{eqnarray} #1 \end{eqnarray}}
\newcommand{\Adir}{\mathcal{A}_{\rm dir}}
\newcommand{\Amix}{\mathcal{A}_{\rm mix}}
\newcommand{\sss}{\scriptscriptstyle}
\newcommand{\scs}{\scriptstyle}
\newcommand{\txs}{\textstyle}
\newcommand{\mo}[1]{\left\vert #1 \right\vert}
\newcommand{\m}[1]{\vert #1 \vert}
\newcommand{\av}[1]{\langle #1 \rangle}
\newcommand{\n}{\noindent}
\newcommand{\q}{q\!\!\!/}
\newcommand{\p}{p\!\!/}
\newcommand{\A}{A\hspace{-0.23cm}/}
\newcommand{\Bs}{B\hspace{-0.25cm}/}
\newcommand{\pa}{\partial\hspace{-0.23cm}/}
\newcommand{\Heff}{{\cal H}_{\rm eff}}
\newcommand{\clearemptydoublepage}{\newpage{\thispagestyle{empty}\cleardoublepage}}

\def\sss{\scriptscriptstyle}

\begin{document}

\thispagestyle{plain}

\ \vspace{3cm}

\begin{center}

{\LARGE \textbf{TOPICS IN HADRONIC B DECAYS}}

\vspace{1.8cm}

{\Large Javier Virto}

\vspace{0.5cm}

Institut de Física d'Altes Energies\\
Universitat Autònoma de Barcelona

\vspace{0.2cm}

\today

\end{center}

\vspace{3.8cm}

\begin{center}
\includegraphics[height=3cm,width=5cm]{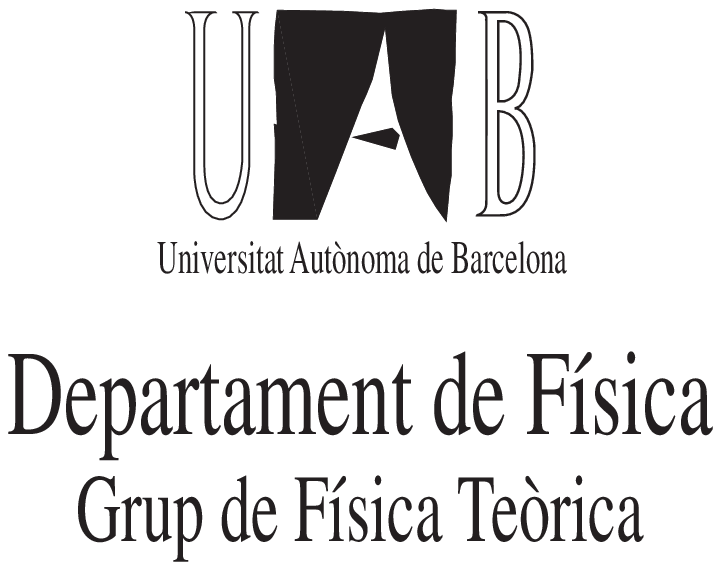}
\end{center}

\vspace{1cm}

\begin{center}
\nn\underline{\ \ \ \ \ \ \ \ \ \ \ \ \ \ \ \ \ \ \ \ \ \ \ \ \ \ \ \ \ \ \ \ \ \ \ \ \ \ \ \ \ \ \ \ \ \ \ \ \ \ \ \ \ \
\ \ \ \ \ \ \ \ \ \ \ \ \ \ \ \ \ \ \ \ \ \ \ \ \ \ \ \ \ \ \ \ \ \ \ \ \ \ \ \ \ \ \ \ \ \ }\\

\vspace{0.3cm}

\parbox{12.5cm}{\footnotesize \emph{Submitted in partial fulfillment of the requirements for the degree of Doctor of Philosophy.}}

\end{center}

\clearemptydoublepage

\thispagestyle{plain}

\pagenumbering{roman}

\tableofcontents

\clearemptydoublepage



\renewcommand{\chaptermark}[1]{\markboth{#1}{}}
\renewcommand{\sectionmark}[1]{\markright{\thesection\ #1}}
\fancyhf{}
\fancyhead[LE,RO]{\it \thepage}
\fancyhead[LO]{\it Preface}
\fancyhead[RE]{\it Preface}
\addtolength{\headheight}{0.5pt}
\fancypagestyle{plain}{\fancyhead{}\renewcommand{\headrulewidth}{0pt}}

\chapter*{Preface}
\addcontentsline{toc}{chapter}{Preface}

\pagenumbering{arabic}

The main reason to study B meson decays is their sensitivity to the flavor structure of nature. Indeed, the fact that the $b$ quark is so heavy makes
B physics a rich source of very different processes, and leads to a very rich phenomenology. An additional consequence of the large mass of the $b$
quark, that makes B decays interesting on the theoretical side is the fact that the always troublesome strong interaction effects can be handled
within a heavy-quark expansion, allowing for theoretical predictions of acceptable accuracy. In fact, one of the differences between B- and D-physics
is that decays of D mesons are theoretically much ``dirtier''. B decays have therefore triggered intensive research on the QCD side, that has
witnessed a huge progress in the last decade.

One of the main points in the B physics program is the search for physics beyond the Standard Model (SM). To that end, a number of dedicated
facilities have been taking data for many years, achieving a long list of new discoveries. Specifically, the B-factories Babar and Belle, and the
hadronic machine at Tevatron, with its experiments CDF and D0, have made this possible. Joint work by theorists and experimentalists has led to the
appearance of several \emph{puzzles} in B decays which could be interpreted to be due to New Physics (NP). However, these may as well disappear as
more accurate measurements are made, or new understanding is accomplished on the theoretical side. The imminent start up of the Large Hadron Collider
(LHC), with its B physics experiment LHCb, and the possibility of a super-B factory will certainly play a very exciting role in the search for
physics beyond the SM, and once NP is found, in understanding its nature.

The motivations that guide us in the search for NP --that so many times have triggered false alarms-- are many. They can be classified in three major
classes. Motivations belonging to the first class are related to observational facts that our theory (the SM) cannot reproduce. The baryon asymmetry
in the universe, as understood by now, requires the three so-called \emph{Sakharov conditions}. One of such conditions is an amount of CP violation
that the SM cannot account for. The solution most probably requires new sources of CP violation. The presence of dark matter and dark energy is also
an observational fact that has not found its solution within the standard theory, and which points towards the existence of new particle content.

Motivations belonging to the second class are related to observations that \emph{can} in principle fit into the theory but which would then seem
extremely \emph{unnatural}. The SM predicts two independent sources of CP violation in strong interactions. One is driven by the presence of
instantons and the other comes from the diagonalization of the quark mass matrix. The measurement of the electric dipole moment of the neutron tells
us that these two --in principle unrelated-- contributions must cancel to one part in $10^{9}$. The classical solution to this strong CP-problem is
to postulate a $U(1)_A$ (Peccei-Quinn) symmetry that dynamically sets this small number to zero. This solution requires the existence of (at least) a
new particle, the axion. The hypothesis of inflation, introduced originally as a solution to the horizon and flatness problems, is pretty much
accepted by the physics community nowadays. However, whatever drives inflation is still unknown, and again the many proposed mechanisms involve
physics beyond the SM. The most striking fine-tuning problem is maybe the one related to the cosmological constant, which states that several
unrelated contributions to the vacuum energy and the bare cosmological constant must cancel to one part in $10^{120}$ in order to agree with
observations, which is preposterous. The Higgs fine-tunning problem arises from the instability of the mass of a scalar particle to radiative
corrections. The difference between the electroweak (EW) and the planck scales requires a fine-tunning of one part in $10^{30}$. Supersymmetry solves
this problem protecting the mass of the scalars, which are not protected by gauge invariance, but the solution might as well be ultimately of
different nature. In any case it should manifest itself as NP at $\sim 1{\rm TeV}$.

Motivations belonging to the third class are not related to problems, but to unanswered issues. The inclusion of gravity in the standard picture is
an issue that has led to the study of extra dimensions and string theory. A thorough theoretical investigation over decades indicates that quantum
gravity can only be merged with the SM together with a significant amount of NP. The SM suffers by itself from its own unanswered issues, more
related to this thesis than the problem of gravity. The SM contains 28 free independent parameters, a feature that goes directly against the quest
for unification that motivated its foundations. An attempt to reduce the number of gauge parameters, through the gauge coupling unification, requires
Supersymmetry and a long list of new particles, introducing even more parameters. From the 28 free parameters in the SM, 22 are directly related to
the flavor sector. The hierarchy of the CKM matrix and the hierarchy in the masses of the quarks and leptons of different families should have an
explanation beyond the SM, and constitutes part of the SM flavor problem. The existence of 3 generations of quarks and leptons is intriguing, and the
existence of a fourth family is neither theoretically nor experimentally excluded. The fact that neutrinos are not massless provides new puzzles
concerning the mixings, hierarchies and flavor violations in the leptonic sector to which any NP would have a definite impact.

So it is clear that the search for new physics is strongly motivated. Past experience has also taught us that purely theoretical arguments are a
powerful tool that leads to actual discovery. The positron was predicted by Dirac as a product of the unification of relativity and quantum
mechanics. The charm quark was postulated in order to provide a GIM mechanism to suppress FCNC's. The third family of quarks was postulated in order
to allow CP violation through the KM mechanism. The weak gauge bosons W, Z were predicted with the right masses as a realization of the GWS theory of
weak interactions. These and other successes are the proof that the theoretical method is in the right track, and that research on the problems
mentioned above will bring discoveries as impressive as old problems did in the past. However, as important as theoretical elucidation might be,
experiment is the only way we get to know the world. Both must be connected through the ever crucial link: phenomenology.

\begin{figure}
\begin{center}
\includegraphics[width=12cm]{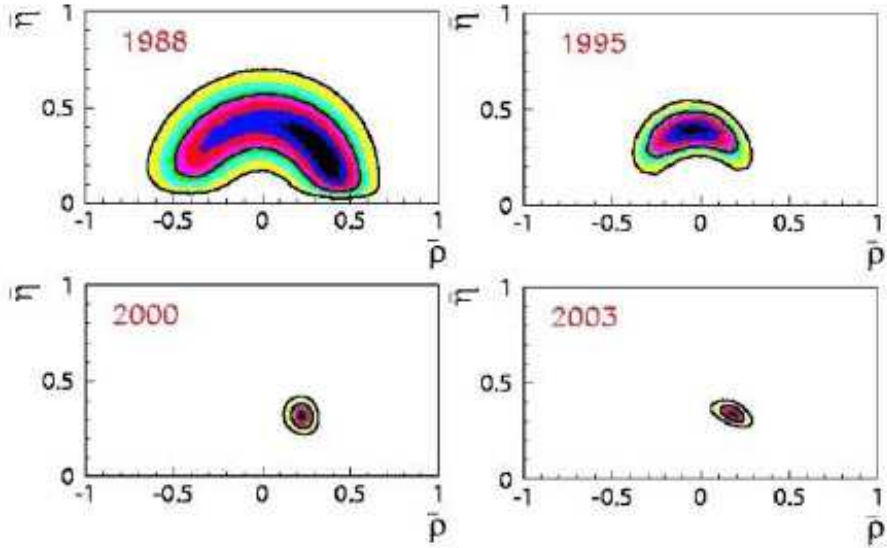}
\end{center}
\vspace{-0.5cm} \caption{\small The progress in the determination of the Unitarity Triangle.} \label{progressCKM}
\end{figure}

The role of flavor physics is central to this enterprise at the phenomenological level. The search for new physics begins with the understanding of
the SM itself, and the precise determination of its parameters. The majority of these parameters are flavor parameters that are not very well
determined in the present, but where impressive progress has been made (see Fig.\ref{progressCKM}). In fact the determination of the flavor
parameters of the SM will soon become of the \emph{precision} type. Flavor parameters can only be determined through flavor physics, and the
determination of the flavor parameters is necessary in order to infer not only quantitative but also qualitative aspects of the physics above the EW
scale, and of the mechanism responsible for the flavor structure that we see at low energies.

The importance of flavor physics cannot be overestimated. If there is really new physics at the TeV scale, it will be most probably detected at the
imminent LHC at CERN within 2 to 5 years from now. However, revealing the true nature of those new particles is a much more uncertain task. The
relevance of the discovery of NP is certainly to find the answers to the problems raised above. Will the discovered new particles provide a clue
about the origin of flavor? Or about the mechanism of electroweak symmetry breaking (EWSB)? Or a sensible reason for the existence of large
hierarchies? As far as we know all these problems may be related, or they may be as well completely disconnected. Will it be clear from the beginning
that we are discovering Supersymmetry, or extra dimensions, or something else? Indeed, the direct search for new particles is an important but not
the only part of the New Physics program. And this is where flavor physics comes at hand. Flavor physics constitutes a powerful arena on which to
investigate \emph{detailed} aspects and properties of the new physics and its possible role on the observed phenomenology. Indeed, flavor physics has
already provided important constraints on the properties of the new physics concerning its CP and flavor violating structure. It has even raised its
own problems. The suppression of flavor-changing neutral currents (FCNC's) is an observational fact that introduces a strong requirement in the
flavor nature of the new physics. Without any mechanism that suppresses FCNC processes, any generic scenarios for NP should be suppressed by a scale
larger than $\sim 10^3{\rm TeV}$. Therefore, any TeV NP must have a very specific flavor structure in order to satisfy the flavor bounds. This is the
NP flavor problem. The non standard CP violating phases that appear, for example, in supersymmetry, are constrained to be of order $10^{-3}$. If
supersymmetry is invoked to tame the hierarchy problem of the Higgs mass, it should provide a reasonable explanation for the smallness of these
phases. Without a mechanism of this sort, such scenarios of supersymmetry are problematic.

\begin{figure}
\begin{center}
\includegraphics[width=10cm,height=7cm]{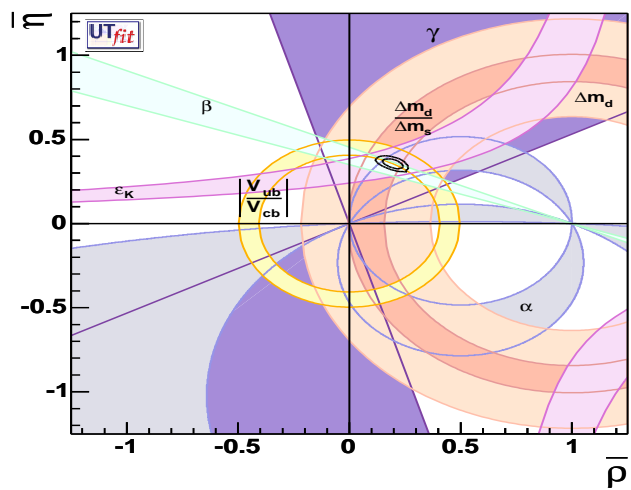}
\end{center}
\vspace{-0.5cm} \caption{\small Fit to the apex of the standard Unitarity Triangle as of October 2007.} \label{UTfit}
\end{figure}

In the Standard Model, the description of CP violation is given by the CKM mechanism. This mechanism is extremely economic, allowing for one
\emph{single} CP violating weak phase. Therefore, within the SM CP is violated ``minimally''. This provides us with an incredibly predictive
framework, which can be tested meticulously by over-constraining the four parameters of the CKM matrix. It is somehow surprising, and certainly
remarkable, that this simple picture is so far in quantitative agreement with all laboratory experiments made up to now. The combination of all
observables that constrain the CKM matrix is usually carried out through a fit to the apex of the Unitarity Triangle (UT). Today, this fit defines a
consistent and very much constrained Unitarity Triangle (see Fig.\ref{UTfit}), giving the following results for the real ($\bar{\rho}$) and imaginary
($\bar{\eta}$) parts of its apex:
\eq{\bar{\rho}=0.164\pm 0.029\ ,\quad \bar{\eta}=0.340\pm 0.017}
The general hope, however, is that more precise measurements and more precise theoretical predictions will at some point reveal some inconsistency.
Once a number of well identified observables have been observed to deviate from the SM expectations, they can be used to study the nature of the New
Physics. If these are flavor observables, their deviations will be a holy grail for the understanding of the flavor structure of the new physics that
might be directly observed at the LHC.

There is a major theoretical difficulty, however, that affects the determination of the SM parameters and the search for New Physics. This difficulty
is related to the non-perturbative nature of QCD at low energies. Without it, the question of whether the SM is in agreement with all the measured
observables would be a strict matter of experimental precision (combined possibly with a perturbative calculation up to a large enough number of
loops).

There are several features of strong interactions that are well known. First, at high energies the hadrons show a ``partonic'' structure, as can be
inferred from deep inelastic scattering experiments, revealing the existence of quarks. This picture of hadrons in terms of ``partons'' unravels one
of the characteristic features of strong interactions, which is \emph{asymptotic freedom}. Second, the mere existence of hadrons means that for some
reason the elementary quarks from which they are composed, are \emph{confined}. This is verified by the fact that quarks are not observed in
isolation, and the fact that colored ``objects'' are not observed either. Although \emph{quark} confinement and \emph{color} confinement are slightly
different concepts, it is clear that evidence favors both. Both features, asymptotic freedom and confinement, can be understood qualitatively in
terms of the scaling of the strong interaction \emph{strength}. In this picture, strong interactions are very strong at long-distances, and they
decrease with increasing energy to become very weak at short distances.

The reasons to believe that QCD is \emph{the} theory of strong interactions are compelling. First, the theory is formulated exactly as it is
understood nowadays that a relativistic quantum theory of particle interactions should be. It is a renormalizable quantum field theory based on a
local gauge principle. Its gauge group is SU(3), where the fact that there are 3 colors is known from $e^+e^-\to\,{\rm hadrons}$. From the non
abelian nature of the gauge group it follows that, at least for a reasonable number of fermion families, the theory is asymptotically free, and that
the coupling constant increases at long distances. Moreover, perturbative computations in QCD are very successful at large energies, where
perturbation theory applies nicely. So, at least in the perturbative regime, QCD shows the qualitative features of strong interactions and reproduces
quantitatively the experimental results.

In the non-perturbative regime, however, things are less clear. First, it has not been proven that QCD implies confinement. Second, it is not known
how to extract the hadronic spectrum from QCD. For example, the Bethe-Salpeter equation, as a dictionary that translates from quarks to hadrons and
back, it is extremely difficult to solve. Despite these formal deficiencies, however, a great progress has been made since the development of QCD in
the 70's, and many qualitative features of strong interactions at low energies can be connected to features of ``QCD-like'' theories. For example,
the large-N limit of QCD is able to give qualitative explanations for the Zweig's rule, the dominance of the leading fock states in the mesons, or
the success of Regge phenomenology. Other successes of QCD itself in the non perturbative regime concern predictions derived for example from QCD
sum-rules or from lattice simulations.

Concerning the strong interactions in B decays, the progress has been driven by the observation that one can perform a perturbative expansion in the
small parameter $\Lambda_{QCD}/m_b$. The first step was the development of an effective theory for mesons containing heavy quarks, called Heavy Quark
Effective Theory (HQET). This theory manifests explicitly the symmetries that arise in the heavy quark limit and allows, for example, to relate
different heavy-heavy form factors to a single (Isgur-Wise) function. For inclusive B decays, the large value of the $b$ mass allows to use the so
called Heavy Quark Expansion (HQE), which predicts, for example, that the inclusive decay of a B meson is dominated by the partonic decay of the b
quark alone. In the case of exclusive decays, the establishment of a power counting in terms of $\Lambda_{QCD}/m_b$ led to the development of the QCD
factorization approach that has been used extensively to make predictions for all two body hadronic, and radiative B decays. Finally, a consistent
effective field theory of exclusive and inclusive heavy meson decays at large momentum transfer has been formulated under the name of Soft Collinear
Effective Theory (SCET). However, much has to be done before we are able to give theoretical predictions for exclusive hadronic
decays with uncertainties at the percent level.\\

In this thesis I present some topics related to my work in non-leptonic decays of $B$ mesons. It is divided it two major parts. The first part is an
overview of the basic matters that constitute the background on which the original work is based. Such a general overview is not an easy task in this
field, since there are many topics involved, some of them very well established, and some of them not so much. Therefore I might have been too
extensive on very well known areas, maybe short in topics that require more explanation, I may have been repetitive on some parts and on the contrary
have skipped completely things that some people will miss. However, I believe that the whole text is quite self-contained, and that this first part
constitutes a rather complete link to the second part. I have divided this part in three chapters: first I discuss the weak effective hamiltonian in
$B$ physics, both in the context of the SM and for the study of NP. Then I present the concept of factorization for the computation of matrix
elements, and also the use of flavor symmetries as an alternative to direct computations. Finally I present the theory of CP violation in meson
decays. All these three topics are necessary ingredients in the work presented in the second part.

The second part of this thesis contains a series of applications of the theory presented in the first part. It is based on some papers that have been
published within the last three years. These chapters are intimately related, but constitute different ``topics'' in the field; this is what
motivates the title of this thesis. The central issue is the study of $B_s\to KK$ modes within and beyond the SM, but it requires an interconnection
with the whole grand compact field of flavor physics in general, and in particular of $B$ physics.
I hope the concepts are written clearly and the structure of this thesis makes it easy to read, and easy to pose questions, and comments.\\

Finally, I would like to acknowledge my collaborators, David London, Seungwon Baek, Sebastien Descotes-Genon and Rafel Escribano for nice
collaborations and thousands of mails of enlightening content. I am also indebted to the theory groups at Dortmund and Rome for the hospitality
during my stays there, where also some part of this work was done. Many thanks also to Felix Schwab, who so kindly made a thorough reading of the
first part of this thesis and contributed with intelligent comments and suggestions. I am particularly grateful to Guido Martinelli, Bernardo Adeva,
Sebastien Descotes, Santi Peris, Joan Soto, Gudrun Hiller and Ramon Miquel for reading the thesis, and participating as members of the committee.
Some of them, specially Guido, contributed with important suggestions, improving further the quality of this writing. Very special thanks go to my
thesis advisor, Joaquim Matias, with whom I've had endless conversations about physics and life since April 2004, and has gone through painful
revisions of this text. I can hardly imagine a better tutor. Personal acknowledgements have been kept personal.

%


\part{Fundamentals}

\chapter{Effective Hamiltonians for B physics}

\renewcommand{\chaptermark}[1]{\markboth{#1}{}}
\renewcommand{\sectionmark}[1]{\markright{\thesection\ #1}}
\fancyhf{}
\fancyhead[LE,RO]{\it \thepage}
\fancyhead[LO]{\it \rightmark}
\fancyhead[RE]{\it \leftmark}
\addtolength{\headheight}{0.5pt}
\fancypagestyle{plain}{\fancyhead{}\renewcommand{\headrulewidth}{0pt}}

\label{RenormGroup}

B physics describes \emph{weak} decays of \emph{B mesons}. This means that it studies physical processes which involve simultaneously --at least--
three different energy scales. First, the fact that they are \emph{weak} decays implies that they involve the scale of weak interactions, given by
the mass of the $W$ boson, $M_W$. Second, since the energy of the process is that of the decaying meson, a second scale involved is the mass of the B
meson $m_B$. Third, the fact that we are dealing with \emph{mesons} implies that the physics of strong interactions of bound states is also
important. The scale introduced in this case is the hadronic scale $\Lambda_{QCD}$. Moreover, since we are looking for new physics, it is reasonable
to allow for the existence of at least another energy scale, $\Lambda_{NP}$, the scale at which the SM breaks down as an effective theory.

The mere existence of several energy scales does not by itself recall for the use of effective hamiltonians. The utility of an effective theory
arises when two different physical phenomena are mostly independent of one another because they operate at completely different scales. For example,
the vibration of the atoms in a macroscopic object is too fast to affect at all the movement of the object, and the movement of the object is too
slow to affect at all the vibration of its atoms, so both physical processes can be studied independently. Weak decays of B mesons are most suitably
studied within an effective theory approach because of the large separation between the energy scales involved:
\eq{\Lambda_{QCD}\ll m_B \ll M_W,\ \Lambda_{NP}}
In fact, $\Lambda_{QCD}\sim 0.2-1\,{\rm GeV}$, $m_B\sim 5\,{\rm GeV}$, $M_W\sim 100\,{\rm GeV}$ and $\Lambda_{NP} > {\rm few\ TeV}$. The large scale
of weak interactions and New Physics with respect to the mass of the B meson motivates the use of the weak effective Hamiltonian, which is introduced
in this chapter. The low scale of the strong interaction inside hadrons with respect to the mass of the B meson (or the b quark in this case) is what
motivates the use of Heavy Quark Effective Theory (HQET). A classical exhaustive reference for HQET is the review by Neubert \cite{Neubert:1993mb}.
Other scales that appear in exclusive decays of B mesons are related to the collinear and hard-collinear degrees of freedom, that introduce a scale
of order $m_b\Lambda_{QCD}$ which motivates the use of Soft Collinear Effective Theory (SCET). The theoretical and phenomenological importance of
this recent development calls at least for a set of references linked to its formulation
\cite{Bauer:2000ew,Bauer:2000yr,Bauer:2001yt,Beneke:2002ph,Chay:2002vy,Hill:2002vw}.

\section{The Weak Effective Hamiltonian}

\label{WEH}

The study of hadronic weak decays is rooted to the concept of the \emph{effective weak Hamiltonian}. It describes weak interactions at low energy,
relevant for energies below $M_W$, and it has the structure of an expansion of local effective operators (OPE):
\eq{\mathcal{H}_{\rm eff}=\frac{G_F}{\sqrt{2}}\sum_i C_i(\mu)\,Q_i \label{HeffOPE}}
It can be regarded as a generalization of the Fermi theory to include all quarks and leptons and the electroweak and strong interactions described by
the SM. Furthermore, it constitutes a suitable framework for the inclusion of physics beyond the SM. The effective Hamiltonian is defined so that the
amplitude is given by
\eq{A(i\to f)=\bra{f}\mathcal{H}_{\rm eff}\ket{i}=\frac{G_F}{\sqrt{2}}\sum_i C_i(\mu)\bra{f}Q_i(\mu)\ket{i}\label{A(i->f)}}

When dealing with such an effective description, there are two ways to proceed. The first one is to treat the coefficients in the OPE as unknown
phenomenological parameters, to be measured or constrained through experiment. Once these are measured, the effective theory can be tested by
checking its predictions over different observables. This approach was the one followed in Fermi's theory, and it is relevant when the fundamental
theory is unknown. Yet the SM itself is often seen as an effective theory in this context. Alternatively, if the underlying theory is known (or
assumed) then these coefficients can be calculated in terms of the fundamental parameters.

When following the later approach, one must compute the amplitudes in the full theory. One will, in general, encounter the typical divergencies that
can be absorbed through conventional renormalization, and that introduces a renormalization scale $\mu$. When dealing with the effective theory, the
inclusion of QCD effects also introduces divergencies. Some of them can be reduced through field renormalization; however, the resulting expressions
are still divergent. Hence, one is forced to introduce an \emph{operator renormalization}, to remove these divergencies. This process generically
\emph{mixes} different operators in the OPE, in such a way that new operators can arise that were not present at tree level (without QCD
corrections). It is therefore important to work with a basis of operators which is ``closed'' under renormalization.

There are several very important issues related with the appearance of a renormalization scale in the OPE. First, the full amplitude cannot depend on
the arbitrary scale $\mu$. As a result, the $\mu$-dependence of the coefficients $C_i(\mu)$ has to cancel the $\mu$-dependence of the matrix elements
$\av{Q_i(\mu)}$. This cancellation involves in general several terms in the expansion in (\ref{A(i->f)}). Since this scale can be chosen freely, it
is a matter of choice what exactly belongs to $C_i(\mu)$ and what to $\av{Q_i(\mu)}$. In general, by running the value of $\mu$ one assigns to these
two quantities different energy ranges, such that $C_i(\mu)$ contains \emph{short distance} effects above $\mu$ and $\av{Q_i(\mu)}$ contains the
\emph{long distance} non-perturbative contributions with energies below $\mu$.

A second comment on the renormalization scale has to do with the perturbative regime of QCD and the suitable scale to ``compute'' the matrix elements
of the operators. As the coefficients are calculated perturbatively, they should be computed at a scale in which the QCD coupling constant is small.
This means a scale $\mu_W\sim \mathcal{O}(M_W)$. However, the scale at which the matrix elements can be factorized in a meaningful way is much lower,
for instance $\mu_f\sim \mathcal{O}(m_b,m_c,1\,{\rm GeV})$ for $B$, $D$, and $K$ decays respectively. The problem that arises here is that the large
difference of the two scales involved spoils perturbation theory through large logarithms. Fortunately, the Renormalization Group (RG) provides a
tool to recover the validity of the perturbative series by a resummation to all orders of large logarithms.

In this chapter we will review schematically the specific details that allow to realize these ideas in a quantitative way. A thorough but pedagogical
treatment of these matters can be found, for instance, in \cite{hep-ph/9512380,Buras:1998raa}. They will play a central role in the forthcoming
chapters.

\subsection{Operator Product Expansion in Weak decays}

\label{OPE}

\begin{figure}
\begin{center}
\psfrag{u}{$u_\alpha$} \psfrag{d}{$d_\alpha$} \psfrag{ub}{$u'_\beta$} \psfrag{db}{$d'_\beta$} \psfrag{f}{\Large{$\longrightarrow$}} \psfrag{W}{$W$}
\psfrag{Q}{$Q_2$} \psfrag{a}{$(a)$} \psfrag{b}{$(b)$}
\includegraphics[height=3.5cm]{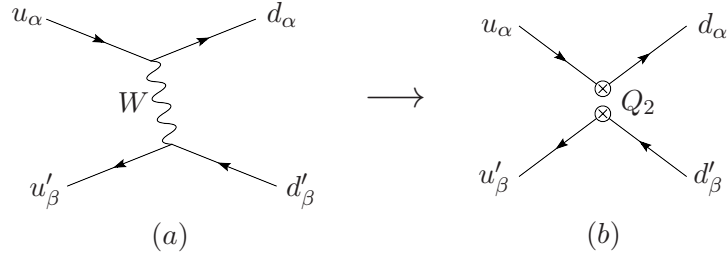}
\end{center}
\vspace{-0.4cm} \caption{\small $(a)$ Leading order contribution to $u\to u'\bar{d}d'$ in the full theory. (b) Contribution in the effective theory.
Integrating out the $W$ boson in the full theory generates the local operator $Q_2=(\bar{d}u)_{\sss V-A}(\bar{u'}d')_{\sss V-A}$ in the effective
theory. Here, $u$ and $u'$ denote $u$-type quarks and $d$, $d'$ denote $d$-type quarks.} \label{EffVert}
\end{figure}

The basic idea that motivates the use of \emph{low energy} effective theories is that there is a lower limit on the distances that can be resolved
through a process of given energy. This means that short-distance effects below this limit $\sim 1/k$, with $k$ an external momentum, can be treated
as local. In particular, one can ``integrate out'' heavy modes, in such a way that non-local interactions mediated by heavy particles are reduced to
local interactions. A simple example of this process is shown in Figure \ref{EffVert}, where the $W$ boson is integrated out to give a local
four-quark operator, $Q_2$.

To be specific, we show how this works in the case of weak charged current interactions. The relevant part of the weak interaction Lagrangian density
is
\eq{\mathcal{L}_{cc}=-\frac{1}{2}(\partial_\mu W_\nu^+\!\!-\partial_\nu W_\mu^+)(\partial^\mu W^{-\nu}\!\!-\partial^\nu W^{-\mu}) +M_W^2 W_\mu^+
W^{-\mu}\!+\frac{g_2}{2\sqrt{2}}(J_\mu^+W^{+\mu}\!+J_\mu^-W^{-\mu}) \label{Lcc}}
which, integrating the $W$ fields in the generating functional, leads to a non-local action of the form
\eq{S_{cc}=-\frac{g_2^2}{8}\int d^4x d^4y J_\mu^-(x)\Delta^{\mu\nu}(x,y)J_\nu^+(y)\label{Scc}}
where $\Delta^{\mu\nu}$ is the $W$ propagator. In the unitary gauge,
\eq{\Delta^{\mu\nu}(x,y)=\int \frac{d^4k}{(2\pi)^4}\frac{(g^{\mu\nu}-k^\mu k^\nu/M_W^2)}{M_W^2-k^2}\ e^{-ik(x-y)} \approx
\frac{g^{\mu\nu}}{M_W^2}\delta^{(4)}(x-y)+\mathcal{O}\,(1/M_W^4)}
The expansion in powers of $1/M_W^2$ is meaningful for energies well below $M_W^2$, so this step identifies the low energy expansion. Inserting this
propagator in (\ref{Scc}) and reading off the lagrangian density we have, keeping only $\mathcal{O}\,(1/M_W^2)$ terms,
\eq{\mathcal{L}_{\rm eff}=-\frac{G_F}{\sqrt{2}}J_\mu^-J^{+\mu}(x)= -\frac{G_F}{\sqrt{2}}\sum_{u,u',d,d'}V^*_{ud}V_{u'd'}\,(\bar{d}u)_{\sss
V-A}(\bar{u'}d')_{\sss V-A}}
where $G_F=g_2^2/8M_W^2$. This lagrangian no longer contains the $W$ degrees of freedom, as they have been integrated out. It has the structure
(\ref{HeffOPE}) of a set of local four quark operators, with coefficients which contain information on the short-distance physics, mainly the mass of
the $W$ boson. This expansion is called an \emph{Operator Product Expansion} (OPE), and the coefficients of the operators are called \emph{Wilson
coefficients}. Often, the top quark appears in loops and must be integrated out too, introducing in the Wilson coefficients a $m_t$ dependence. In
Chapter \ref{SUSYcontributions} we shall be integrating out supersymmetric particles, and the coefficients in the OPE will contain the masses of
these particles.

This is an example of how to calculate the Wilson coefficients in (\ref{HeffOPE}) from the underlying theory. For theories beyond the SM one shall
begin with a more general Lagrangian than (\ref{Lcc}), which may in general introduce a different set of local operators, and whose coefficients will
contain new parameters, couplings and masses.

\begin{figure}
\begin{center}
\psfrag{Q}{$Q_2$} \psfrag{+}{$+$} \psfrag{+ps}{$+\cdots$} \psfrag{q}{\footnotesize $Q_2$} \psfrag{=}{$=$} \psfrag{dq1}{\footnotesize $\delta Q_1$}
\psfrag{dq2}{\footnotesize $\delta Q_2$} \psfrag{C}{\hspace{-0.5cm} $=(\bar{d}u)_{\sss V-A}(\bar{u'}d')_{\sss V-A}\quad$ at $s=-\mu^2$}
\includegraphics[height=4cm,width=14cm]{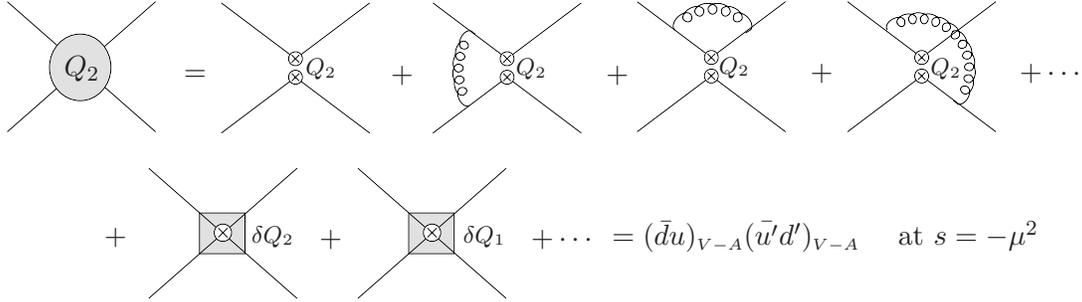}
\end{center}
\vspace{-0.4cm} \caption{\small Renormalization of the local operator $Q_2$. The vertex corrections such as the third diagram cancel the
contributions from the field-strength renormalization. The counterterms in the second line are thus defined as to cancel the divergencies from the
other diagrams, and to satisfy the renormalization condition introducing the scale $\mu$. Note that it is necessary to include the counterterm of the
operator $Q_1$. This is a consequence of operator mixing.} \label{LocOps}
\end{figure}

Up to some point, however, we must introduce QCD corrections to go beyond tree level. In the effective theory, this corrections will renormalize the
local operators (see Figure \ref{LocOps}). Since these operators have dimension greater than four, the field renormalization does not cancel all the
divergencies, and an additional \emph{operator renormalization} is necessary. Furthermore, the inclusion of QCD corrections generates new operators
with new color structure (or new flavor structure due to Fierz rearranging, if preferred). This \emph{operator mixing} is responsible for the
renormalization constant being a matrix,
\eq{Q_i^{(0)}=Z_{ij}Q_j}
The important fact about the operator renormalization is that it does not depend on the full theory; once we have renormalized a complete set of
operators to find $Z_{ij}$, this will be used for the effective theory of any fundamental theory that generates this set of operators.

Since the renormalization scale dependence introduced in this step will play a central role, it is rather more convenient to introduce the
\emph{anomalous dimension matrix} of the operators, defined as
\eq{\hat{\gamma}=\hat{Z}^{-1}\frac{d}{d\ln{\mu}}\hat{Z} \label{ano-malo}}
Now we give the basic guidelines to calculate the leading order anomalous dimension matrix for a general set of local operators.

\subsection{Calculation of the Anomalous Dimension Matrix}

In order to compute the $i$'th row of the matrix $Z_{ij}$ to leading order we must compute the one-loop QCD corrections to the operator $Q_i$, as
shown in Figure \ref{LocOps}. For the sake of generality (focusing on four-quark operators for simplicity), we denote this operator as
\eq{Q_i=(\bar{d}_\alpha \Gamma_1 V_1^{\alpha\beta} u_\beta)\otimes(\bar{u'}_\gamma \Gamma_2 V_2^{\gamma\delta} d'_\delta)}
where $V_1^{\alpha\beta}\otimes V_2^{\gamma\delta}$ refers to the color structure, and $\Gamma_1\otimes\Gamma_2$ to the Dirac structure
\cite{Buras:1998raa}. We must calculate the last three diagrams in the fist row of Figure \ref{LocOps} (together with the symmetric ones not shown)
with the insertion of this general operator. We denote these (pairs of) diagrams by $\mathcal{D}_a$, $\mathcal{D}_b$ and $\mathcal{D}_c$. Because we
are interested in computing the $Z$ constants, we just need the divergent parts of these diagrams. A straightforward calculation (in dimensional
regularization) gives
\eqa{ &&\mathcal{D}_a=\frac{\alpha_s}{4\pi}\frac{1}{4}\frac{1}{\epsilon}\, \Big( \mathcal{C}_a^{(1)}\Gamma_1\gamma_\rho\gamma_\mu \otimes
\Gamma_2\gamma^\rho\gamma^\mu +
\mathcal{C}_a^{(2)}\gamma_\mu\gamma_\rho\Gamma_1 \otimes \gamma^\mu\gamma^\rho\Gamma_2 \Big)\nn\\
&&\mathcal{D}_b=\frac{\alpha_s}{4\pi}\frac{1}{4}\frac{1}{\epsilon}\, \Big( \mathcal{C}_b^{(1)}\gamma_\mu\gamma_\rho\Gamma_1\gamma^\rho\gamma^\mu
\otimes \Gamma_2 +
\mathcal{C}_b^{(2)}\Gamma_1 \otimes \gamma_\mu\gamma_\rho\Gamma_2\gamma^\rho\gamma^\mu \Big)\nn\\
&&\mathcal{D}_c=\frac{\alpha_s}{4\pi}\frac{1}{4}\frac{1}{\epsilon}\, \Big( \mathcal{C}_a^{(1)}\Gamma_1\gamma_\rho\gamma_\mu \otimes
\gamma^\mu\gamma^\rho\Gamma_2 + \mathcal{C}_a^{(2)}\gamma_\mu\gamma_\rho\Gamma_1 \otimes \Gamma_2\gamma^\rho\gamma^\mu \Big) \label{Di}}
where the color factors are given by
\eqa{
\mathcal{C}_a^{(1)}=V_1T^a\otimes V_2T^a && \quad\mathcal{C}_a^{(2)}=T^aV_1\otimes T^aV_2 \nn\\
\mathcal{C}_b^{(1)}=T^aV_1T^a\otimes V_2 && \quad\mathcal{C}_b^{(2)}=V_1\otimes T^aV_2T^a \nn\\
\mathcal{C}_c^{(1)}=V_1T^a\otimes T^aV_2 && \quad\mathcal{C}_c^{(2)}=T^aV_1\otimes V_2T^a }

The operators appearing in (\ref{Di}) can be transformed into the operators $\av{Q_j}$ if the set closes under renormalization. The sum of these
divergent parts of the diagrams give the counterterms of the operators plus the counterterm corresponding to the renormalization of the quark fields,
\eq{\sum_i\mathcal{D}_i=(Z_q^{-2}-1)\,\av{Q_i}+(Z_{ij}-\delta_{ij})\,\av{Q_j} \label{Zij-1}}
Since the dependence of $\hat{Z}$ on $\mu$ comes only from $\alpha_s$, the leading order anomalous dimension matrix (\ref{ano-malo}) is
\eq{\hat{\gamma}^{(0)}\frac{\alpha_s}{4\pi}=-2(Z_{ij}-\delta_{ij})}
with $(Z_{ij}-\delta_{ij})$ obtained from (\ref{Zij-1}) and (\ref{Di}).

\begin{table}
\begin{center}
\begin{tabular}{|cc|cccc|cccc|cc|}
\hline $-2$ & 6 & $ -2/9 $ &  $ 2/3 $ &  $-{2/9} $ & ${2/3}$ & $ 0 $ & $ 0 $ & $ 0 $ & $ 0 $ & $ {416/81} $ & $ {70/27} $
\\
6 & $-2$ & 0 & 0 & 0 & 0 & 0 & 0 & 0 & 0 & 0 & 3 \\
\hline $0 $ & $ 0 $ & $ -{22/9} $ & $ {22/3} $ & $ -{4/9} $ & $ {4/3} $ & $
 0 $ & $ 0 $ & $ 0 $ & $ 0 $ & $ -{464/81} $ & $ {545/27} $ \\
$ 0 $ & $ 0 $ & $ {44/9} $ & $ {4/3} $ & $ -{10/9} $ & $ {10/3} $ & $
 0 $ & $ 0 $ & $ 0 $ & $ 0 $ & $ {136/81} $ & $ {512/27} $ \\
$0 $ & $ 0 $ & $ 0 $ & $ 0 $ & $ 2 $ & $ -6 $ & $
 0 $ & $ 0 $ & $ 0 $ & $ 0 $ & $ {32/9} $ & $ -{59/3}$ \\
$0 $ & $ 0 $ & $ -{10/9} $ & $ {10/3} $ & $ -{10/9} $ & $ -{38/3} $ & $
 0 $ & $ 0 $ & $ 0 $ & $ 0 $ & $ -{296/81} $ & $ -{703/27} $
\\
\hline $ 0 $ & $ 0 $ & $ 0 $ & $ 0 $ & $ 0 $ & $ 0 $ & $ 2 $ & $ -6 $ & $ 0 $ & $
0 $& $ -{16/9} $ & $ {5/6} $ \\
$ 0 $ & $ 0 $ & $ -{1/9} $ & $ {1/3} $ & $ -{1/9} $ & $ {1/3} $ & $ 0 $ & $ -16 $ & $ 0 $ & $ 0 $ & $ -{1196/81} $ &
$ -{11/54}$  \\
$0 $ & $ 0 $&$ {2}{9} $&$ -{2/3} $&$ {2/9} $&$ -{2/3}
$&$ 0 $&$ 0 $&$ -2 $&$ 6 $&$ {232/81} $&$ -{59/54} $ \\
$0 $&$ 0 $&$ -{1/9} $&$ {1/3} $&$ -{1/9} $&$ {1/3}
$&$ 0 $&$ 0 $&$ 6 $&$ -2 $&$ {1180/81} $&$ -{46/27} $ \\
\hline $0 $&$ 0 $&$ 0 $&$ 0 $&$ 0 $&$ 0 $&$ 0 $&$ 0 $&$ 0 $&$ 0 $&$ {32/3} $
&$ 0 $ \\
$0 $&$ 0 $&$ 0 $&$ 0 $&$ 0 $&$ 0 $&$ 0 $&$ 0 $&$ 0 $&$ 0 $
&$ -{32/9} $&$ {28/3} $ \\
\hline
\end{tabular}
\end{center}
\vspace{-0.5cm} \caption{\small The anomalous dimension matrix at leading logarithms for the set of twelve operators $O_{1,\ldots,10}$,
$O_{7\gamma}$, $O_{8g}$ defined in eq. (\ref{OpsSM}) \cite{Buchalla:2005us}.} \label{tableADM}
\end{table}

For example, the leading order anomalous dimension matrix for the set of operators (\ref{OpsSM}) is given in Table \ref{tableADM}. The computation of
the anomalous dimension matrix beyond leading order is a much more formidable task, especially when, besides QCD corrections, the desired precision
requires electromagnetic or even electroweak corrections.

\subsection{Renormalization Group Evolution}

Up to this point, let us recall the basic procedure discussed so far. The basic steps in the calculation of amplitudes in weak decays are the
following:

\begin{itemize}

\item Calculation of the amplitudes within the full theory. This is done at a high scale ($\mu_W$) suitable for a perturbative treatment. In this
step one can see which are the operators generated at this scale, and write the relevant OPE for this theory.

\item Renormalization of the effective operators and computation of the anomalous dimensions. Once this is done, the full theory can be
matched into the effective theory to find the Wilson coefficients. These will be free from divergencies, but will contain logarithms of the form
$\log{(M_W^2/\mu_W^2)}$. This $\mu$ dependence is cancelled by the logs of the form $\log{(\mu_W^2/s)}$ in the matrix elements $\av{Q_i}$.

\item Calculation of the matrix elements. As we will see, this has to be done at a low scale, and thus the $\mu$ scale has to
run down from $M_W$ to the lower scale, which for the five flavour effective theory will be $m_b$.

\end{itemize}

The last step introduces a subtlety. The renormalization scale is arbitrary and therefore one could think that after computing the Wilson
coefficients it can be set to $m_b$ without further care. However, this introduces the large logarithms $\log{(M_W^2/m_b^2)}$, which appear in the
perturbative expansion as $[\alpha_s \log{(M_W^2/\mu^2)}]^n$, $\alpha_s [\alpha_s \log{(M_W^2/\mu^2)}]^n$, etc., and spoil the convergence of the
perturbative series. Fortunately, the Renormalization Group allows to sum the terms $[\alpha_s \log{(M_W^2/\mu^2)}]^n$ to all orders in perturbation
theory at leading order, the terms $\alpha_s [\alpha_s \log{(M_W^2/\mu^2)}]^n$ at next-to-leading order, and so on, yielding what is called the
\emph{RG improved perturbation theory}.

Consider the necessary $\mu$-independence of the amplitudes, in the form of a RG equation,
\eq{\frac{d}{d\ln{\mu}}\Big(\sum_i C_i(\mu)\av{Q_i}\Big)=0\label{RGE}}
Recalling the $\mu$-independence of the bare matrix elements $\av{Q_i}^{(0)}$ and the definition (\ref{ano-malo}), equation (\ref{RGE}) can be
written as a RG equation for the Wilson coefficients,
\eq{\frac{d}{d\ln{\mu}}\vec{C}(\mu)=\hat{\gamma}^T(\mu)\vec{C}(\mu)\label{RGE2}}
where for simplicity we use vector notation. This differential equation can be solved rather trivially at leading order, to give
\eq{\vec{C}(\mu)=\left[ \frac{\alpha_s(M_W)}{\alpha_s(\mu)} \right]^\frac{\scriptstyle \hat{\gamma}^{T(0)}}{\scriptstyle 2\beta_0} \vec{C}(M_W)
\label{C(mu)}}
where we have used the definition of the $\beta$-function at leading order, given by $d\alpha_s/d\ln{\mu}=-2\beta_0(\alpha_s^2/4\pi)$. In general,
beyond leading order, one can still write
\eq{\vec{C}(\mu)=U_5(\mu,M_W)\,\vec{C}(M_W) \label{Cevol}}
but the $\mu$-evolution matrix $U_5(\mu,M_W)$ takes a more complicated form than in (\ref{C(mu)}). Furthermore, when running the renormalization
scale below the threshold of $m_b$, one should start integrating out the $b$ quark to go to the four-flavor effective theory, and the charm quark to
go beyond $m_c$, and so on.

By looking at the solution (\ref{C(mu)}) of the RG equation, one can immediately see that an order $\alpha_s^1$ matching of the Wilson Coefficients
provides in fact terms of all orders in $\alpha_s$. A more careful inspection shows that the terms are of the type $[\alpha_s \log{\mu}]^n$ with $n$
any positive power. It is, however, far from evident that higher loop computations do not contribute terms of this type. This is indeed the case, and
that is why after a one loop matching a \emph{resummation} of all the terms $[\alpha_s \log{\mu}]^n$ is possible. Here I give a recursive proof.

Write the Wilson Coefficients as an analytic expansion in powers of $\alpha_s$ and $\ell\equiv \log{\mu}$,
\eq{\vec{C}(\mu)=\sum_{i,j}\vec{a}_{ij}\Big(\frac{\alpha_s}{4\pi}\Big)^i\ell^j \label{expC}}
with arbitrary coefficients $\vec{a}_{ij}$. Consider also the perturbative expansion for the beta function and the anomalous dimension matrix,
\eqa{
\frac{d\alpha_s}{d\ell}&=&-8\pi\Big(\frac{\alpha_s}{4\pi}\Big)^2\sum_{i}\beta_i\Big(\frac{\alpha_s}{4\pi}\Big)^i\\
\hat{\gamma}^T&=&\Big(\frac{\alpha_s}{4\pi}\Big)\sum_{i}\hat{\gamma}^{T(i)}\Big(\frac{\alpha_s}{4\pi}\Big)^i }
It is convenient to let $i,j\in \mathbb{Z}$ and take $\vec{a}_{ij}=\beta_i=\hat{\gamma}^{T(i)}=0$ for $i,j<0$. This takes into account that
$(d\alpha_s/d\ell)$ appears first at second order in $\alpha_s$ and the anomalous dimension at first order. Now, because the Wilson Coefficients
depend on $\ell$ explicitly and implicitly through $\alpha_s$, one usually writes the RG equation as
\eq{\Big( \frac{\partial}{\partial\ell}+ \frac{d\alpha_s}{d\ell}\frac{\partial}{\partial\alpha_s}\Big)\vec{C}=\hat{\gamma}^T\vec{C}}
Now, by plugging into this equation the expansions for $(d\alpha_s/d\ell)$, $\hat{\gamma}^T$ and $\vec{C}$ in powers of $\alpha_s$ and $\ell$, and
requiring equality at each order, one arrives to the following recursive relation for the coefficients $\vec{a}_{ij}$,
\eq{(j+1)\,\vec{a}_{k,j+1}-8\pi\sum_{i\ge 0}(i+1)\,\vec{a}_{i+1,j}\,\beta_{k-i-2} =\sum_{i\ge 0}\vec{a}_{i,j}\,\hat{\gamma}^{T(k-i-1)}\qquad
\forall\,j,k \label{recursion}}
From this recursive relation one can extract all the coefficients $\vec{a}_{ij}$ from initial conditions $\vec{a}_{n0}$ (that is, the Wilson
coefficients at the matching scale). The important thing is to see what coefficients $\beta_i$ and $\hat{\gamma}^{T(i)}$ are required for each
$\vec{a}_{ij}$. All the necessary results to understand how these coefficients are related can be proved by induction from (\ref{recursion}). For
example, the first result is that
\eq{\vec{a}_{ij}=0\qquad \forall\,j>i}
which tells us that the power of the logarithms cannot exceed the order in $\alpha_s$. The next thing one can prove is that
\eq{\vec{a}_{kk}=\frac{1}{k}\,\Big( \hat{\gamma}^{T(0)}+8\pi(k-1)\beta_0 \Big)\,\vec{a}_{k-1,k-1}}
This is a very important result. It tells us that all the terms of the form $[\alpha_s \log{\mu}]^n$ in the Wilson Coefficients are known from the
matching condition $\vec{C}(M_W)$ and the leading order coefficients $\hat{\gamma}^{T(0)}$ and $\beta_0$. This is exactly what allows for a
\emph{resummation of leading logarithms} (LL). It is essential for the convergence of the series in (\ref{expC}) when the difference between the two
scales $M_W$ and $\mu$ makes $[\alpha_s\log{(M_W/\mu)}]$ a number of order one.

Next to leading logarithms (NLL) can also be resummed in a similar fashion. For example, all the terms of the type $\alpha_s[\alpha_s \log{\mu}]^n$
are known once one has computed the beta function and the anomalous dimension matrix at next to leading order. The corresponding recursion relation
from which this fact follows is
\eq{\vec{a}_{k,k-1}=\frac{1}{k-1}\Bigg[ \Big(\hat{\gamma}^{T(1)}+8\pi(k-2)\beta_1 \Big)\,\vec{a}_{k-2,k-2} + \Big(
\hat{\gamma}^{T(0)}+8\pi(k-1)\beta_0 \Big)\,\vec{a}_{k-1,k-2} \Bigg]}
and thus requires also a one-loop matching of the Wilson coefficients (notice that besides $\vec{a}_{00}$ also $\vec{a}_{10}$ is needed).

One can proceed in this way proving more complicated recursion relations. In general, for a resummation of terms of order $\alpha_s^n[\alpha_s
\log{\mu}]^k$ for all $k$ one needs:
\begin{itemize}

\item $\vec{a}_{00}$, $\vec{a}_{10}$, \dots, $\vec{a}_{n0}$ (n-loop matching of the Wilson Coefficients).

\item $\hat{\gamma}^{T(0)}$, $\hat{\gamma}^{T(1)}$, \dots, $\hat{\gamma}^{T(n)}$ ((n+1)-loop anomalous dimension matrices).

\item $\beta_0$, $\beta_1$, \dots, $\beta_n$ (beta function at n+1 loops).

\end{itemize}
Of course, in practice the resummations are not done recursively. One just solves the RG equation differentially as in (\ref{C(mu)}). In my opinion,
however, the recursive proof is more transparent and allows to keep control on the required orders in perturbation theory.

\section{The Weak Effective Hamiltonian in the SM}

\label{SMHeff}

In the SM, only a reduced set of operators are generated in the weak effective Hamiltonian. Here we show only the effective Hamiltonian relevant for
$|\Delta B|=1$ and $|\Delta S|=0,1$ processes. In this case, the operators are usually classified as current-current operators, QCD-penguin
operators, electroweak penguin operators, and electromagnetic and chromomagnetic operators. The Wilson coefficients always contain the same products
of CKM elements, so these are factored out and written explicitly in the effective Hamiltonian, denoted by $\lambda_p^{\sss (D)}\equiv
V_{pb}V^*_{pD}$. Here $D=d$ for $\Delta S=0$ and $D=s$ for $\Delta S=1$ decays. The effective Hamiltonian is given by
\eq{ {\cal H}_{\rm eff} = \frac{G_F}{\sqrt2} \sum_{p=u,c} \!
   \lambda_p^{\sss (D)} \bigg( C_1\,Q_1^p + C_2\,Q_2^p
   + \!\sum_{i=3,\dots, 10}\! C_i\,Q_i + C_{7\gamma}\,Q_{7\gamma}
   + C_{8g}\,Q_{8g} \bigg) + \mbox{h.c.} \,,
\label{HeffSM}}
where $Q_{1,2}^p$ are the left-handed current-current operators, $Q_{3\dots6}$ and $Q_{7\dots 10}$ are the QCD and electroweak penguin operators and
$Q_{7\gamma}$ and $Q_{8g}$ are the electromagnetic and chromomagnetic dipole operators. Their explicit form reads
\eqa{ Q_1^p &=& (\bar p b)_{V-A} (\bar D p)_{V-A} \,,
    \hspace{2.5cm}
    Q^p_2 = (\bar p_i b_j)_{V-A} (\bar D_j p_i)_{V-A} \,, \nonumber\\
   Q_3 &=& (\bar D b)_{V-A} \sum{}_{\!q}\,(\bar q q)_{V-A} \,,
    \hspace{1.7cm}
    Q_4 = (\bar D_i b_j)_{V-A} \sum{}_{\!q}\,(\bar q_j q_i)_{V-A} \,,
    \nonumber\\
   Q_5 &=& (\bar D b)_{V-A} \sum{}_{\!q}\,(\bar q q)_{V+A} \,,
    \hspace{1.7cm}
    Q_6 = (\bar D_i b_j)_{V-A} \sum{}_{\!q}\,(\bar q_j q_i)_{V+A} \,,
    \nonumber\\
   Q_7 &=& (\bar D b)_{V-A} \sum{}_{\!q}\,{\textstyle\frac32} e_q
    (\bar q q)_{V+A} \,, \hspace{1.11cm}
    Q_8 = (\bar D_i b_j)_{V-A} \sum{}_{\!q}\,{\textstyle\frac32} e_q
    (\bar q_j q_i)_{V+A} \,, \nonumber \\
   Q_9 &=& (\bar D b)_{V-A} \sum{}_{\!q}\,{\textstyle\frac32} e_q
    (\bar q q)_{V-A} \,, \hspace{0.98cm}
    Q_{10} = (\bar D_i b_j)_{V-A} \sum{}_{\!q}\,{\textstyle\frac32} e_q
    (\bar q_j q_i)_{V-A} \,, \nonumber\\
   Q_{7\gamma} &=& \frac{-e}{8\pi^2}\,m_b\,
    \bar D\sigma_{\mu\nu}(1+\gamma_5) F^{\mu\nu} b \,,
    \hspace{0.81cm}
   Q_{8g} = \frac{-g_s}{8\pi^2}\,m_b\,
    \bar D\sigma_{\mu\nu}(1+\gamma_5) G^{\mu\nu} b \,,
\label{OpsSM}}
where $(\bar{q}_1 q_2)_{V\pm A}=\bar{q}_1 \gamma_\mu(1\pm\gamma_5)q_2$ and the sum runs over all active quark flavours in the effective theory, i.e.
$q=u,d,s,c,b$. If no colour index $i,j$ is given, the two operators are assumed to be in a colour singlet state.

The Wilson coefficients at NLO are given in Table \ref{tableWC}, at six different scales of interest (we will neglect electroweak penguins). These
will be used in latter chapters.

\begin{table}
\begin{center}
\begin{tabular}{|l|cccccccc|}
\hline
                             &   $C_1$  &  $C_2$  &  $C_3$  &  $C_4$  &  $C_5$ &  $C_6$  &  $C_{7\gamma}^{\rm eff}$  &  $C_{8g}^{\rm eff}$ \\
\hline
$\mu=2m_b$                   &   1.045  & -0.113  &  0.009  & -0.025  &  0.007 & -0.027  &       -0.281              &      -0.136 \\
$\mu=m_b$                    &   1.082  & -0.191  &  0.013  & -0.036  &  0.009 & -0.042  &       -0.318              &      -0.151 \\
$\mu=m_b/2$                  &   1.139  & -0.296  &  0.021  & -0.051  &  0.010 & -0.066  &       -0.364              &      -0.169 \\
$\mu=\sqrt{2m_b \Lambda_h}$  &   1.141  & -0.230  &  0.021  & -0.051  &  0.011 & -0.067  &       -0.366              &      -0.170 \\
$\mu=\sqrt{m_b \Lambda_h}$   &   1.184  & -0.371  &  0.027  & -0.062  &  0.011 & -0.086  &       -0.395              &      -0.182 \\
$\mu=\sqrt{m_b \Lambda_h/2}$ &   1.243  & -0.464  &  0.035  & -0.076  &  0.010 & -0.115  &       -0.432              &      -0.198 \\
\hline
\end{tabular}
\end{center}
\vspace{-0.5cm} \caption{\small NLO Wilson coefficients of the operators $O_{1,\dots,6}$, $O_{7\gamma}$ and $O_{8g}$ at six different scales of
interest. The scale $\Lambda_h=0.5\,{\rm GeV}$.} \label{tableWC}
\end{table}

\section{The Weak Effective Hamiltonian beyond the SM}

\label{BSMHeff}

In the presence of NP, many more operators can be generated other than those in (\ref{OpsSM}). If NP exists at a scale $\Lambda_{NP}$, the exchange
of these new particles between the low energy SM fields will in general generate standard and non-standard operators, with Wilson coefficients
suppressed by the new physics scale $\Lambda_{NP}$:
\eq{{\cal H}_{\rm eff}^{\rm \sss NP}=\sum_{D=5}^\infty \sum_i \frac{c_i^{[D]}}{\Lambda_{NP}^{D-4}} \mathcal{O}_i^{[D]}}
Indeed, the effective theory approach is the most general and model independent way of dealing with NP at low energies. It relies on the following
assumptions:
\begin{itemize}
\item The low energy degrees of freedom are the SM particles with masses below $M_W$,
so these fields are the only building blocks in the operators in the OPE.
\item All the operators must be gauge invariant, the gauge group being that
of the SM. Other global symmetries might be imposed in order to avoid proton decay and other phenomenological requirements.
\item A prescription is taken to cut the infinite set of operators. This prescription consists in dropping the operators with dimension
higher than a given dimension $d$. Operators of dimension $D$ are suppressed by $\Lambda_{NP}^{4-D}$, so if the scale $\Lambda_{NP}$ is relatively
high, the ``cutting'' dimension $d$ can be low, leaving a reasonable number of operators in the Hamiltonian. In the case of B decays this means to
keep only operators up to dimension six.
\end{itemize}
Following this procedure, however, one finds that the most general effective Hamiltonian contains too many operators. And since the Wilson
coefficients are just phenomenological unknown parameters of the effective theory, the theory has too many unknowns. The general set of operators was
derived in \cite{Buchmuller:1985jz}. The concept of a general effective Hamiltonian will be used in Chapter \ref{Bdecays} to write a general
parametrization of the NP amplitudes for $\bar{b} \to \bar{s} q \bar{q}$ transitions, where the issue of the large number of operators is
circumvented with an argument concerning the NP strong phases.

A general structure of the new physics is, as mentioned before, difficult to reconcile with a relatively low value for $\Lambda_{NP}$. Flavor physics
constrains severely many NP contributions to flavor violating operators and CP violating phases in the NP Wilson coefficients. This leads to
effective scenarios of new physics that reduce considerably the complexity of the general effective Hamiltonian and fulfill the experimental bounds
almost virtually. An example is given by the minimal flavor violation (MFV) hypothesis, which assures that the flavor violation in the effective
theory comes entirely from the CKM matrix \cite{Buras:2003jf}. While there are several non-equivalent definitions of MFV in the literature, they all
share this property.

This formalism is also useful to study model independently concrete scenarios of new physics. For example, there are several models that describe the
origin of supersymmetry breaking. While these models might be wrong, low energy supersymmetry might be right, and the study of supersymmetry at low
energies should not rely on the specific scenario chosen, for example, at the Planck scale. For a derivation of a low energy supersymmetry effective
Hamiltonian see \cite{Hall:1990ac}.

For model dependent studies of specific new physics, the use of the weak effective Hamiltonian is much the same as in the SM. The only difference is
the extended set of operators that is generated, and of course that the matching conditions lead to different Wilson coefficients. In Chapter
\ref{SUSYcontributions} we will match gluino-squark box and penguin contributions to $\bar{b} \to \bar{s} q \bar{q}$ transitions into the NP weak
effective Hamiltonian.


\chapter{Hadronic Matrix Elements}

\label{HME}

The concept of the Operator Product Expansion, discussed in the previous chapter in the context of the weak effective Hamiltonian, is a powerful
tool. It allows for a \emph{factorization} of long and short distance physics; the short distance contained in the Wilson Coefficients and the long
distance contained in the matrix elements of the operators. The Wilson Coefficients are calculable within perturbation theory at a high energy scale
(e.g $M_W$), and are process-independent. Moreover, this large scale can be dragged down to a phenomenologically sensible scale (e.g $m_b$) by means
of the RG resummation of large logarithms of the form $\log (M_W/m_b)$ that would spoil the perturbative series.

In this way, a general process-independent effective Hamiltonian is derived, which includes QCD (and possibly electroweak) corrections to a given
order (LL, NLL, NNLL,\dots) from gluons of virtuality above the scale $\mu$. The effective Hamiltonian depends on $\mu$ through the Wilson
Coefficients. All this is process independent, and the model dependence comes solely from the matching conditions of the WC (this can very well
account for the fact that many operators might have been dropped from the beginning: one just includes \emph{all} the operators, but puts to zero the
WC's of those operators that are not wanted).

In order to compute the amplitudes, this effective Hamiltonian must be sandwiched between the initial and final states (\emph{cf.} (\ref{A(i->f)})),
\eq{A(i\to f)=\bra{f}\mathcal{H}_{\rm eff}\ket{i}=\frac{G_F}{\sqrt{2}}\sum_i C_i(\mu)\av{Q_i(\mu)}.}
The last step of the process is then the computation of the \emph{Hadronic Matrix Elements} of the operators $\av{Q_i(\mu)}$, between the initial and
the final states.

The scale dependence of the matrix elements is conceptually subtle but enlightening. The operators, by themselves, do not have any scale dependence.
Neither do the initial or final states. However, it is clear that the matrix elements must have a scale dependence that cancels the scale dependence
of the WC's. (The same is also true for the \emph{scheme} dependence, which leads to a criticism to some generalized factorization approaches as will
be discussed later.) The scale dependence of the matrix elements can be partially understood through the following physical reasoning. Since the
effective Hamiltonian is process independent, the matrix elements are precisely what distinguishes between different final states (lets say the
initial state is always a B meson). But the final state is not characterized only by the quark fields of which is composed. Indeed, one can always
rearrange the quarks in the final state to get different \emph{hadronic} final states. This rearrangement is due to strong interaction rescattering.
The point is that QCD corrections with virtualities above the scale $\mu$ (\emph{hard} gluons) are accounted for in the Wilson Coefficients, which
are process independent, so that these corrections have no impact on the final state hadronization\footnote{The values of the WC definitely tell what
are the most favorable final states, but that is a feature of the structure of the interactions themselves, not of the hadronization process.}.
Therefore, the final state hadronization takes place through interchange of \emph{soft} gluons (QCD corrections with virtualities below $\mu$), which
must then be contained inside the hadronic matrix elements.

Up to now, however, there is no fully consistent way to compute hadronic matrix elements from QCD (apart from, arguably, lattice methods). This is a
longstanding problem that goes back to the foundation of QCD and that partially motivates this thesis.

The concept of \emph{factorization} of matrix elements, as a second step after the OPE, has proven a powerful tool to face this problem. The idea is
to reduce the matrix elements to products of form factors and decay constants, which are process independent quantities. However, even in the cases
where factorization can be strictly justified, it just solves the problem partially, since a systematic way of computing form factors and decay
constants is also necessary. This is a less pathological problem, though, since their process-independent nature allows to extract them from data.
Also, because they are intrinsically simpler objects, it is possible to extract them from lattice simulations, which is up to now the most pure
``QCD-based'' technique for non-perturbative computations.

Factorization is strict in the case of leptonic and semi-leptonic B decays. The simplest ones --concerning QCD corrections-- are the leptonic decays,
which are those with only leptons in the final state. Because leptons do not interact strongly, QCD effects take place only within the B meson in the
initial state. Therefore, when computing the matrix elements, factorization is strictly valid, and all the long-distance strong interaction effects
are contained in
\eq{\bra{0}\,\bar{b}\,\Gamma\, q\,\ket{B_q}\sim f_B }
which defines the \emph{decay constant} $f_B$.

Semi-leptonic decays are those with leptons and hadrons in the final state. Strong interactions take place inside the initial and final hadrons, in
the process of hadronization, and between them during the decay. But these corrections affect only one vertex of the weak current (since the other
vertex is leptonic), and thus factorization is still exact. The non-perturbative physics is parameterized now in terms of a \emph{form factor},
\eq{\bra{M_q}\,\bar{b}\,\Gamma\, q\,\ket{B}\sim F^{B\to M}}
The case of non-leptonic decays is more complicated. Since the final state is composed purely of hadrons, factorizable and non-factorizable effects
take place. Therefore, factorization is no longer justified. The observation that, surprisingly, strict (\emph{naive}) factorization works reasonably
well also for non-leptonic decays, led originally to what has become an intensive line of research. Now we understand the reason, since we have
factorization theorems that tell about the \emph{limitations} of naive factorization.

An alternative approach to deal with non-leptonic B decays beyond factorization is to take advantage of approximate \emph{symmetries} of QCD. Flavor
symmetries arise in the limit in which the masses of the quarks are equal, and it turns out that for the three light quarks $u,d,s$ this is a good
approximation. This introduces a $SU(3)_{\sss V}$ symmetry group under which the QCD lagrangian is approximately invariant. Symmetries always allow
to establish relations between matrix elements, according to the Wigner-Eckart theorem. Therefore, up to symmetry breaking corrections, $SU(3)$
relations between amplitudes of different processes hold, and allow to make predictions \emph{without} the need of computing matrix elements. Of
course, how big the symmetry breaking corrections are is a dynamical question.

In Section \ref{factorization} we introduce more formally the idea of factorization for two body non-leptonic B decays, and give the relevant
formulae for the evaluation of amplitudes within the framework of QCD-Factorization. In Section \ref{symmetries} we explain the main features of the
use of flavor symmetries in hadronic B decays. Both subjects will be important in the development of the following chapters.

\section{Factorization}

\label{factorization}

\subsection{Beyond naive factorization}

\label{preliminaries}

Consider a decay of a $\bar{B}_q$ meson into two mesons $M_1$ and $M_2$. We would like to calculate the matrix element
\eq{\bra{M_1  M_2}O_i\ket{\bar{B}_q} \label{<P2P|0|P1>}\ .}
The meson $M_2$ can be generated directly by a quark current containing the appropriate flavor and Lorentz quantum numbers, say $\bar{q}_1\gamma_\mu
\gamma_5 q_2$ for a pseudoscalar. A dimension-six operator $\mathcal{O}_i$ containing this particular current will contribute to the decay through a
factorized product of two current matrix elements if the other current has the correct quantum numbers to generate the transition $\bar{B}_q\to M_1$.
Let's say this current is $\bar{q}'\gamma_\mu b$. So factorization amounts to the following simplification,
\eq{\bra{M_1 M_2}(\bar{q}_1\gamma_\mu \gamma_5 q_2)(\bar{q}'\gamma_\mu b)\ket{\bar{B}_q} \stackrel{\rm \sss Fact}{--\longrightarrow}
\bra{M_2}\bar{q}_1\gamma_\mu \gamma_5 q_2\ket{0}\bra{M_1}\bar{q}'\gamma_\mu b\ket{\bar{B}_q} \sim f_{M_2} \cdot F^{\bar{B}_q\to M_1} \label{fact} }
If this simplification is justified, then one can express the matrix elements such as (\ref{<P2P|0|P1>}) in terms of products of a decay constant and
a form factor. This oversimplified presentation gives an idea of the concept of factorization. The strongest version of this process consists simply
to promote (\ref{fact}) to an equality, and it has been called \emph{naive factorization}.

In the weak effective Hamiltonian, one in general will have to deal with pairs of operators of the following type
\eqa{
O_i&\!\!\!\!=&\!\!\!(\bar{q}_\alpha b_\alpha)_{\Gamma_1} (\bar{q}_{1\,\beta} q_{2\,\beta})_{\Gamma_2}\ , \nn \\
O_j&\!\!\!\!=&\!\!\!(\bar{q}_\alpha b_\beta)_{\Gamma_1} (\bar{q}_{1\,\beta} q_{2\,\alpha})_{\Gamma_2}\ . \label{opsnonfact} }
where $\alpha$, $\beta$ are color indices and $\Gamma_1$, $\Gamma_2$ are Lorentz structures. These operators can be Fierz rearranged and written as
\eqa{
O'_i&\!\!\!\!=&\!\!\!(\bar{q}_{1\,\beta} b_\alpha)_{\Gamma'_1} (\bar{q}_{\alpha} q_{2\,\beta})_{\Gamma'_2}\ , \nn \\
O'_j&\!\!\!\!=&\!\!\!(\bar{q}_{1\,\alpha} b_\alpha)_{\Gamma'_1} (\bar{q}_\beta q_{2\,\beta})_{\Gamma'_2}\ . \label{opsfact} }
where in general the Lorentz structures $\Gamma'_1$, $\Gamma'_2$ are different from $\Gamma_1$, $\Gamma_2$. Now, depending on the flavor structure of
the decay, it is possible that the operators can be factorized in the form $O_i$, in the reordered form $O'_j$, in both forms or in none of them. In
the case that both forms contribute, both have to be taken into account since they correspond to different quantum mechanical paths (different ways
to reorder the quarks to form the final mesons). The factorized matrix elements are defined as
\eqa{ \av{O_i}_F &\equiv& \bra{M_1}(\bar{q} b)_{\Gamma_1}\ket{\bar{B}_q}\bra{M_2}(\bar{q}_{1} q_{2})_{\Gamma_2}\ket{0} +
\bra{M_2}(\bar{q} b)_{\Gamma_1}\ket{\bar{B}_q}\bra{M_1}(\bar{q}_{1} q_{2})_{\Gamma_2}\ket{0}\ ,\nn\\[2pt]
\av{O'_j}_F &\equiv& \bra{M_1}(\bar{q}_1 b)_{\Gamma'_1}\ket{\bar{B}_q}\bra{M_2}(\bar{q} q_{2})_{\Gamma'_2}\ket{0} + \bra{M_2}(\bar{q}_1
b)_{\Gamma'_1}\ket{\bar{B}_q}\bra{M_1}(\bar{q} q_{2})_{\Gamma'_2}\ket{0}\ . }

In order to introduce factorization it is useful to work in the singlet-octet basis for the operators. Using the identity for the $SU(3)$ generators
\eq{ T^a_{\alpha\beta}T^a_{\gamma\sigma}=\frac{1}{2}\Big[\delta_{\alpha\sigma}\delta_{\beta\gamma}- \frac{1}{N}
\delta_{\alpha\beta}\delta_{\gamma\sigma}\Big] \label{SU3}}
one can write
\eqa{
C_i\,O_i+C_j\,{O_j}&=&\Big( C_i + \frac{1}{N} C_j \Big)\,{O_i} + 2 C_j \,{O_i^8} \nn \\
C_i\,{O'_i}+C_j\,{O'_j}&=&\Big( C_j + \frac{1}{N} C_i \Big)\,{O'_j} + 2 C_j \,{O^{'8}_j} }
where $O_i^8=(\bar{q} T^a b)_{\Gamma_1} (\bar{q}_{1} T^a q_{2})_{\Gamma_2}$, and similar for $O^{'8}_j$. It's clear that in order for the octet
operators to contribute through a factorized matrix element, strong interactions must somehow change the color structure, so it is reasonable to
think that these operators contribute only at order $\alpha_s$. Be that as it may, in naive factorization, by definition, the matrix elements of
octet operators are set to zero. At the end, in naive factorization (NF), the contribution to the amplitude of this pair of operators is given by
\eq{C_i(\mu)\av{O_i}+C_j(\mu)\av{O_j} = a_i(\mu) \av{O_i}_F + a_j(\mu) \av{O'_j}_F \qquad\quad {\rm (NF)} \label{NF}}
where
\eq{a_i(\mu)= C_i(\mu) + \frac{1}{N} C_j(\mu)\ ,\quad a_j(\mu)=C_j(\mu) + \frac{1}{N} C_i(\mu)\ .}

The hadronic matrix elements of the currents that result from factorization do not show any renormalization scale dependence that can cancel the
scale dependence in the Wilson coefficients. Therefore, naive factorization cannot be correct. However, one may hope that a particular
\emph{factorization scale} $\mu_f$ exists, at which the Wilson coefficients are evaluated, for which this is a good approximation. As we will see,
there is a major inconvenient to this point.

The attempts to give a formulation for factorization that didn't have the problem of the renormalization scale dependence led to different
\emph{generalized factorization} approaches. We first present the formulation by Neubert and Stech \cite{Neubert:1997uc}.

The structure of factorization can be made \emph{exact} by introducing process dependent non-perturbative parameters $\varepsilon_1(\mu)$ and
$\varepsilon_8(\mu)$ that parametrize the non-factorizable contributions. For example, in the case of a decay with a flavor structure such that
$\av{O'_j}_F=0$, they are defined as
\eq{\varepsilon_1(\mu)\equiv \frac{\av{O_i}}{\av{O_i}_F}-1\ ,\quad \varepsilon_8(\mu)\equiv 2\frac{\av{O^8_i}}{\av{O_i}_F}\ ,}
so that $C_i\av{O_i}+C_j\av{O_j}=[(C_i+C_j/N)(1+\varepsilon_1)+\varepsilon_8 C_j]\av{O_i}_F$ is \emph{exact}. Moreover, naive factorization arises in
the limit $\varepsilon_1,\varepsilon_8\to 0$. The formula (\ref{NF}) is then generalized to
\eq{C_i(\mu)\av{O_i}+C_j(\mu)\av{O_j} = a_i^{\rm eff} \av{O_i}_F + a_j^{\rm eff} \av{O'_j}_F \qquad\quad {\rm (GF)} \label{GF}}
where the effective parameters $a_{i,j}^{\rm eff}$ are given by
\eqa{
a_i^{\rm eff}=\Big( C_i(\mu) + \frac{1}{N} C_j(\mu) \Big)\big(1+\varepsilon_1(\mu)\big)+\varepsilon_8(\mu) C_j(\mu)\nn\\[2pt]
a_j^{\rm eff}=\Big( C_j(\mu) + \frac{1}{N} C_i(\mu) \Big)\big(1+\varepsilon'_1(\mu)\big)+\varepsilon'_8(\mu) C_i(\mu) }

Since the factorized matrix elements do not have any scale dependence, all the $\mu$-dependence is contained inside the Wilson coefficients and the
parameters $\epsilon_i(\mu)$. As the full amplitude is $\mu$-independent, the hadronic parameters $\epsilon_i(\mu)$ restore the correct
$\mu$-dependence of the matrix elements, which is lost in the naive factorization approximation. This observation is enough to extract the
$\mu$-dependence of the hadronic parameters by means of the RG equations. The condition for this amplitude to be independent of the scale is that,
for any particular scale $\mu_0$,
\eq{\Big(\! C_i(\mu) + \frac{1}{N} C_j(\mu) \!\Big)\big(1+\varepsilon_1(\mu)\big)+\varepsilon_8(\mu) C_j(\mu)= \Big(\! C_i(\mu_0) + \frac{1}{N}
C_j(\mu_0) \!\Big)\big(1+\varepsilon_1(\mu_0)\big)+\varepsilon_8(\mu_0) C_j(\mu_0) }
holds. Now, using the RG equation (\ref{Cevol}) we can write the Wilson coefficients at the $\mu$ scale in terms of those evaluated at $\mu_0$.
Taking into account that the evolution matrix $U(\mu_0,\mu)$ cannot depend on the values of the Wilson coefficients, we find
\eqa{
1+\varepsilon_1(\mu)&\!\!\equiv&\!\!\Big(\kappa_{ii}+\frac{1}{N}\,\kappa_{ji}\Big)(1+\varepsilon_1(\mu_0))+\kappa_{ji}\,\varepsilon_8(\mu_0) \nn \\
\varepsilon_8(\mu)&\!\!\equiv&\!\!\Big(\kappa_{ij}-\frac{1}{N}\,(\kappa_{ii}-\kappa_{jj})-\frac{1}{N^2}\,\kappa_{ji}\Big)(1+\varepsilon_1(\mu_0))
+(\kappa_{jj}-\frac{1}{N}\,\kappa_{ji})\,\epsilon_8(\mu_0) \qquad \label{RGeps} }
where $\kappa=U(\mu_0,\mu)$. The main idea in generalized factorization is to extract from data the non-factorizable parameters $\varepsilon_1(m_b)$,
$\varepsilon_8(m_b)$ and then run these parameters according to (\ref{RGeps}) to find the factorization scale $\mu_f$ for which they vanish.

There is, however, a major drawback to the generalized factorization procedure as presented in \cite{Neubert:1997uc}. While the scale dependence is
properly taken into account, at next-to-leading order in the renormalization group improved perturbation theory the Wilson coefficients depend also
on the renormalization \emph{scheme}. This scheme dependence can only be compensated by non-factorizable scheme dependent contributions in the matrix
elements. It has been proven \cite{Buras:1998us} that for any chosen scale $\mu_f=\mathcal{O}(m_b)$ it is possible to find a renormalization scheme
for which the parameters $\varepsilon_{1,8}(\mu_f)$ vanish simultaneously. Therefore, the factorization scale is scheme dependent and can have no
physical meaning as such.

A different approach to generalized factorization that doesn't suffer from the scheme dependence problem is the one discussed in
\cite{Cheng:1994zx,Cheng:1998uy,Ali:1997nh,Ali:1998eb}. The idea is to calculate in perturbation theory the matrix elements of the operators between
the quark states. In this way one can extract the scale and scheme dependence of the hadronic matrix elements. Combining these scale and scheme
dependent contributions with the Wilson coefficients $C_i(\mu)$, one obtains effective coefficients $C_i^{\rm eff}$ that are scale and scheme
independent. Then one can write
\eq{C_i\av{O_i}+C_j\av{O_j}=C_i^{\rm eff}\av{O_i}^{\rm tree}+C_j^{\rm eff}\av{O_j}^{\rm tree}}
where $\av{O_{i,j}}^{\rm tree}$ denote tree level matrix elements. Once this is done, factorization can be applied to the tree level matrix elements
without the problem of the scale and scheme dependence. The result can be cast in the form of eq.(\ref{GF}), with $a_{i,j}^{\rm eff}$ calculable up
to a phenomenological parameter (an effective number of colors) that should in principle give information on the pattern of non-factorizable
contributions.

This approach, however, has also its own drawbacks. As noted in \cite{Buras:1998us}, the effective Wilson coefficients are in general gauge
dependent, and also depend on an infrared regulator. These dependencies originate in the perturbative evaluation of the scheme dependent finite
contributions to the matrix elements, that are necessary to cancel the scheme dependence in the Wilson coefficients.

Some papers have been written attempting to solve the problems that arise in generalized factorization approaches (see for example
\cite{Cheng:1999gs}). Here we will focus on the so called `BBNS' approach or QCD-factorization (QCDF), presented initially in \cite{Beneke:1999br} in
the context of $B\to \pi\pi$, and extended later to general $B_q\to M_1M_2$ decays
\cite{Beneke:2000ry,Beneke:2001ev,Du:2001ns,Du:2002up,hep-ph/0308039,hep-ph/0612290}. An overview is postponed until Sections \ref{QCDFth} and
\ref{QCDF}.

We finish this section writing a general formula for the amplitudes in SM within any generalized factorization approach in terms of the coefficients
$a_i^{\rm eff}$. The SM operators in eq.(\ref{OpsSM}) can be rearranged to fit the form of eq.(\ref{GF}),
\eqa{
   Q_1^p &=& (\bar p b)_{V-A} (\bar D p)_{V-A} \,,
    \hspace{2.5cm}
    Q^{'p}_2 = (\bar D b)_{V-A} (\bar p p)_{V-A} \,, \nonumber\\
   Q_3 &=& \sum{}_{\!q}\,(\bar D b)_{V-A} (\bar q q)_{V-A} \,,
    \hspace{1.7cm}
    Q'_4 = \sum{}_{\!q}\,(\bar q b)_{V-A} (\bar D q)_{V-A} \,,
    \nonumber\\
   Q_5 &=& \sum{}_{\!q}\,(\bar D b)_{V-A} (\bar q q)_{V+A} \,,
    \hspace{1.7cm}
    Q'_6 = -2\sum{}_{\!q}\,(\bar q b)_{S-P} (\bar D q)_{S+P} \,,
    \nonumber\\
   Q_7 &=& \sum{}_{\!q}\,{\textstyle\frac32} e_q\, (\bar D b)_{V-A} (\bar q q)_{V+A} \,,
    \hspace{1.11cm}
    Q'_8 = -\sum{}_{\!q}\,3 e_q\, (\bar q b)_{S-P} (\bar D q)_{S+P} \,,
    \nonumber \\
   Q_9 &=& \sum{}_{\!q}\,{\textstyle\frac32} e_q\, (\bar D b)_{V-A} (\bar q q)_{V-A} \,,
    \hspace{0.98cm}
    Q'_{10} = \sum{}_{\!q}\,{\textstyle\frac32} e_q\, (\bar q b)_{V-A} (\bar D q)_{V-A} \,,
\label{OpsSMfierz}}
Notice that the difference between the contributions from operators $Q_3$ and $Q_5$ in their factorized form is just a minus sign if the $\bar qq$
meson has odd parity. The same is true for $Q_7$ and $Q_9$. The difference between the contributions from operators $Q'_4$ and $Q'_6$ is, besides the
factor $-2$, a minus sign if the $\bar Dq$ meson has odd parity and a chiral factor $r_\chi$ from the scalar dirac structure of the currents (see
below). Then, the amplitude for a $\bar{B}_{q_s}\to M_1M_2$ decay is given by
\eq{\bra{M_1M_2}\Heff \ket{\bar{B}}=\sum_{p=u,c}\lambda_p^{\sss (D)*}\bra{M_1M_2}T_A^p \ket{\bar{B}}.}
with the transition operator $T_A^p$ given by
\eqa{ T_A^p & = & a_1^{\rm eff} A([\bar q_s p][\bar p D]) + a_2^{\rm eff} A([\bar q_s D][\bar p p])
          + (a_3^{\rm eff}\mp a_5^{\rm eff}) \sum_q A([\bar q_s D][\bar q q])\nn\\
      &   & + (a_4^{\rm eff}\pm r_\chi a_6^{\rm eff}) \sum_q A([\bar q_s q][\bar q D])
            + (a_9^{\rm eff}\mp a_7^{\rm eff}) \sum_q{\txs \frac32}\,e_q\,A([\bar q_s D][\bar q q])\nn\\
      &   & + (a_{10}^{\rm eff}\pm r_\chi a_8^{\rm eff}) \sum_q{\txs \frac32}\,e_q\,A([\bar q_s q][\bar q D])\ .
\label{GFformula}}
The upper signs correspond to the case when the second meson has odd parity and the lower signs when it's even. The matrix elements of the operators
$A([\cdots][\cdots])$ are non-zero only if the flavor of the mesons match the quarks inside the $[\cdots]$ in either order, and will be expressed in
terms of form factors and decay constants in the following section. In naive factorization the coefficients $a_i^{\rm eff}$ are given, for $i$ odd,
by
\eq{a_i^{\rm NF}= C_i + \frac{1}{N} C_{i+1}\ ,\quad a_{i+1}^{\rm NF}=C_{i+1} + \frac{1}{N} C_i\ .}
As a final comment, just note that contributions from weak annihilation or hard interaction with the spectator quark are not taken into account in
this formulation. Naively they are expected to be small, but at some point they can have an impact in phenomenology. The inclusion of these
contributions is an issue in the QCDF approach.

\subsection{Form factors, decay constants and meson distribution amplitudes}

\label{FFsDCsMDAs}

In this section we give the definitions for the decay constants and form factors of pseudoscalar and vector mesons, and give the expressions for the
matrix elements of the operators $A([\cdots][\cdots])$ that appear in the factorized amplitudes. We also give the definitions of the meson
distribution amplitudes that appear in the hard scattering kernels in QCDF, and present their representation in terms of Gegenbauer polynomials .

The decay constant $f_P$ of a pseudoscalar meson $P$ with 4-momentum $q$ is defined as
\eq{\bra{P(q)}\bar q \gamma_\mu\gamma_5 q'\ket{0}\equiv -i f_P q_{\mu}}
For a vector meson $V(q,\varepsilon^*)$ with 4-momentum $q$ and polarization vector $\varepsilon^*_\mu$, the longitudinal ($f_V$) and transverse
($f_V^\bot$) decay constants are defined as
\eq{\bra{V(q,\varepsilon^*)}\bar q \gamma_\mu q'\ket{0}\equiv -i f_V m_V \varepsilon^*_\mu \ ,\qquad \bra{V(q,\varepsilon^*)}\bar q \sigma_{\mu\nu}
q'\ket{0}\equiv -i f_V^\bot (q_\mu \varepsilon^*_\nu-q_\nu \varepsilon^*_\mu)\ .}
Using the Dirac equation for the quark fields, the following identity follows
\eq{\partial_\mu (\bar{q}\gamma^\mu\gamma_5 q')=(\partial_\mu \bar{q})\gamma^\mu\gamma_5 q'-\bar{q}\gamma_5 \gamma^\mu(\partial_\mu q')
=i(m_q+m_{q'})\ \bar{q}\gamma_5 q'\ .}
Therefore, we have $\ (m_q+m_{q'})\bra{P(q)}\bar{q}\gamma_5 q'\ket{0}=-i\partial_\mu(-if_P q^\mu)=i f_P m_P^2$, and
\eq{\bra{P(q)}\bar{q}\gamma_5 q'\ket{0}=\frac{i f_P m_P^2}{m_q+m_{q'}}\ .\label{pseudoscalarDC}}
The scalar matrix element $\bra{V(q,\varepsilon^*)}\bar{q} q'\ket{0}$ is zero because it can only depend on $q\cdot \varepsilon^*$, which is zero as
can be easily seen going to the rest frame of the vector meson.

For the form factors we use the conventions in \cite{Beneke:2000wa}. For a $\bar B \to P$ transition, the form factors $F_0$ and $F_+$ are defined as
\eq{\bra{P(p')}\bar q \gamma^\mu b \ket{\bar B(p)}=\left[ (p+p')^\mu-\frac{m_B^2-m_P^2}{q^2}\ q^\mu \right] F_+(q^2) + \frac{m_B^2-m_P^2}{q^2}\ q^\mu
F_0(q^2)\ ,\label{axialFF}}
where $q^\mu=(p-p')^\mu$ is the momentum transfer. The identity
\eq{\partial_\mu (\bar{q}\gamma^\mu b)=(\partial_\mu \bar{q})\gamma^\mu b+\bar{q} \gamma^\mu(\partial_\mu b) =i(m_{q}-m_{b})\ \bar{q}\,b\ ,}
and the fact that the $F_+$ term in (\ref{axialFF}) vanishes when contracted with $q^\mu$, implies that for the scalar current
\eq{\bra{P(p')}\bar{q}\,b \ket{B(p)}=\frac{m_B^2-m_P^2}{m_{b}-m_{q}}\ F_0(q^2) \label{scalarFF}}
For a $\bar B \to V$ transition, we will need the form factors $A_0$, $A_1$, $A_2$ and $V$ defined as
\eqa{ \bra{V(p',\varepsilon^*)}\bar q \gamma^\mu b \ket{\bar B(p)} &
        = & \frac{2iV(q^2)}{m_B+m_V}\epsilon^{\mu\nu\rho\sigma}\varepsilon^*_\nu p'_\rho p_\sigma \nn \\
\bra{V(p',\varepsilon^*)}\bar q \gamma^\mu \gamma_5 b \ket{\bar B(p)} &
        = & 2m_V A_0(q^2)\frac{\varepsilon^*\cdot q}{q^2}q^\mu
          + (m_B+m_V)A_1(q^2)\left[\varepsilon^{*\mu} - \frac{\varepsilon^*\cdot q}{q^2}q^\mu\right] \nn \\
       &  & - A_2(q^2)\frac{\varepsilon^*\cdot q}{m_B+m_V} \left[ (p+p')^\mu-\frac{m_B^2-m_P^2}{q^2}\ q^\mu \right]
}

Now we are ready to write the matrix elements of the operators $A([\dots][\dots])$ that appear in the formula for the amplitudes. We define
\eq{A_{\sss M_1M_2}\equiv {\frac{G_F}{\sqrt{2}}}\,\bra{M_1}(\bar{q} b)_{V-A}\ket{\bar{B}_q}\bra{M_2}(\bar{q}_1 q_2)_{V-A}\ket{0}}
Using the definitions given above for the decay constants and form factors and neglecting terms of $\mathcal{O}(m_M^2/m_B^2)$, we have
\eqa{
A_{\sss P_1P_2} & = & i\txs{\frac{G_F}{\sqrt{2}}}\,m_B^2 f_{\sss P_2} F_0^{\sss B\to P_1}   \nn\\
A_{\sss P_1V_2} & = & -2i\txs{\frac{G_F}{\sqrt{2}}}\,m_V\,\varepsilon^*_{\sss V_2}\cdot p_B\, f_{\sss V_2} F_+^{\sss B\to P_1}  \nn\\
A_{\sss V_1P_2} & = & -2i\txs{\frac{G_F}{\sqrt{2}}}\,m_V\,\varepsilon^*_{\sss V_1}\cdot p_B\, f_{\sss P_2} A_0^{\sss B\to V_1}  \nn\\
A_{\sss V_1V_2}^0 & = & i\txs{\frac{G_F}{\sqrt{2}}}\,m_B^2 f_{\sss V_2} A_0^{\sss B\to V_1} \label{Afactors} }
with all the form factors evaluated at $q^2\simeq 0$. The case for $B\to VV$ with transversally polarized mesons can be found in
\cite{hep-ph/0405134,hep-ph/0612290}; we omit it because we will be more concerned about longitudinal polarizations. Now, we finally have
\eq{\bra{M_1M_2}a_i^{\rm eff}A([\dots][\dots])\ket{\bar{B}_{q_s}}=
    c_1\,a_i^{\rm eff}(M_1M_2)\,A_{\sss M_1M_2}+c_2\,a_i^{\rm eff}(M_2M_1)\,A_{\sss M_2M_1}\ ,}
where $c_i=0,1,\pm1/\sqrt{2}$ are obvious constants related to the flavor composition of the mesons. The chiral factors $r_\chi$, that account for
the difference between $\scs(S-P)\otimes (S+P)$ and $\scs (V-A)\otimes (V+A)$ matrix elements are given by
\eq{r_\chi^P(\mu)=\frac{2m_P^2}{m_b(\mu)(m_{q_1}+m_{q_2})(\mu)}\ ,\quad r_\chi^V(\mu)=\frac{2m_V}{m_b(\mu)}\frac{f_V^\bot(\mu)}{f_V}\ , \label{rxi}}
neglecting light quark masses with respect to $m_b$. At leading order $r_\chi^V$ should be put to zero, since the scalar current gives no
contribution to the vector decay constant.

For the meson light-cone distribution amplitudes (LCDA's), the definitions used are the ones in \cite{Braun:1988qv,Braun:1989iv}. We will need only
the two particle leading twist (twist-2) LCDA's $\Phi_P$ and $\Phi_V$, and the twist-3 LCDA's $\Phi_p$ and $\Phi_v$ for pseudoscalar and
longitudinally polarized vector mesons.

The definitions for the leading-twist light-cone distribution amplitudes are
\eqa{ \bra{P(q)} [\bar{q}(y)\gamma_\mu\gamma_5 q'(x)] \ket{0}\big|_{(x-y)^2=0} & = &
   -if_P q_\mu \int_0^1 du e^{i(\bar{u}qx+uqy)}\Phi_P(u,\mu) \nn \\
\bra{V_\|(q)} [\bar{q}(y)\gamma_\mu q'(x)] \ket{0}\big|_{(x-y)^2=0} & = &
   -if_V q_\mu \int_0^1 du e^{i(\bar{u}qx+uqy)}\Phi_V(u,\mu)
}
with $\bar{u}\equiv 1-u$, and the brackets meaning that the fields at $x$ and $y$ are connected by a Wilson line that makes the matrix element gauge
invariant. They are conventionaly represented by a Gegenbauer expansion,
\eq{\Phi_M(u,\mu)=6u\bar{u}\left[ 1+\sum_{n=1}^\infty \alpha_n^M(\mu) C_n^{(3/2)}(u-\bar{u}) \right] \label{GegenbauerExp}}
where $\alpha_n^M(\mu)$ are the Gegenbauer moments, and $C_n^{(3/2)}(x)$ are the Gegenbauer polynomials. This expansion is usually truncated after
$n=2$. The relevant Gegenbauer polynomials are $C_1^{(3/2)}(x)=3x$ and $C_2^{(3/2)}(x)=3(5x^2-1)/2$.

The definitions for the twist-3 light-cone distribution amplitudes are
\eqa{ \bra{P(q)} [\bar{q}(y)\gamma_5 q'(x)] \ket{0}\big|_{(x-y)^2=0} &\!\! = \!\!\!&
   -if_P \mu_P \int_0^1 du e^{i(\bar{u}qx+uqy)}\Phi_p(u,\mu)  \\
\bra{V_\|(q)} [\bar{q}(y)\sigma_{\mu\nu} q'(x)] \ket{0}\big|_{(x-y)^2=0} &\!\! = \!\!\!&
   -if_V^\bot m_V \int_0^1 du e^{i(\bar{u}qx+uqy)}(u-\bar{u})\frac{q_\mu z_\nu - q_\nu z_\mu}{q\cdot z}\Phi_v(u,\mu) \nn
}
with $z=y-x$, and $\mu_P=m_P/(m_q+m_{q'})$. When three-particle amplitudes are neglected, $\Phi_p(x)=1$ and
\eq{\Phi_v(u,\mu)=3\left[ (u-\bar{u})+\sum_{n=1}^\infty \alpha_{n,\bot}^V(\mu)P_{n+1}(u-\bar{u})\right]\ . \label{3twistV}}
Here $P_n(x)$ are Legendre polynomials. At second order, $P_2(x)=(3x^2-1)/2$ and $P_3(x)=(5x^3-3x)/2$. It should be emphasized that twist-3
distribution amplitudes only contribute formally at order $1/m_b$ to the decay amplitudes. However, they can appear in terms that are chirally
enhanced by the factors $r_\chi$, so they must be taken into account.

An example of how meson distribution amplitudes arise in the computation of matrix elements in QCDF is the following. The outgoing meson in the decay
has momentum q. At leading order, the meson can be described by it's leading fock state: it is composed of a quark with momentum $\bar{u} q$ and an
antiquark with momentum $uq$, with $u+\bar{u}=1$. Then, when computing, for example, a vertex correction to the matrix element, one will come up with
a term of the form
\eq{\bar{u}_{\alpha a}(uq)\Gamma(u,\dots)_{\alpha\beta,ab,\dots}v_{\beta b}(\bar{u}q)}
sandwiched between hadronic states. Let's say for definiteness that the meson is pseudoscalar. Then, the prescription is that this term between the
vacuum and the meson state, must be changed to
\eq{\frac{if_P}{4N}\int_0^1 du \Phi_P(u)(\q\gamma_5)_{\beta\alpha}\Gamma(u,\dots)_{\alpha\beta,aa,\dots}}
This prescription is a manifestation of factorization, and requires the proper disentanglement of long- and short-distance contributions. The fact
that this disentanglement occurs in the heavy b-quark limit is discussed in the following section.

The B meson distribution amplitudes arise in the computation of contributions where the spectator quark suffers a hard-scattering interaction (hard
spectator-scattering terms). This is due to the fact that a hard interaction can resolve the inner structure of the B meson, probing the momentum
distribution of the b and spectator quarks.

At leading power in $1/m_b$ the B meson is described by two scalar wave functions. In the case in which the transverse momentum of the spectator
quark can be neglected in the hard-scattering amplitude \cite{Beneke:2000ry}, these scalar wave functions can be defined through the following
decomposition of the B meson LCDA,
\eq{\bra{0} [\bar{q}_\alpha (z) b_\beta(0) ] \ket{\bar{B}(q)}\Big|_{z_{+,\bot}=0}= -\frac{if_B}{4}[(\p+m_b)\gamma_5]_{\beta\gamma}\int_0^1 d\xi
e^{-i\xi p_+ z_-}[\Phi_{B1}(\xi)+n\!\!\!/_-\Phi_{B2}(\xi)]_{\gamma\alpha} }
where $n_=(1,0,0,-1)$ and the subscripts ($+$, $-$, $\bot$) refer to the usual light-cone decomposition of 4-vectors. Specifically, at the leading
order the hard spectator scattering contribution depends only on the first LCDA $\Phi_{B1}$. This dependence is of the form
\eq{\int_0^1 d\xi \frac{\Phi_B(\xi)}{\xi}\equiv \frac{m_B}{\lambda_B} \label{lambdaB}}
which defines a new hadronic parameter $\lambda_B$ of order $\Lambda_{QCD}$ \cite{Beneke:1999br}.

\subsection{Factorization in the heavy quark limit}

\label{QCDFth}

Taking advantage of the fact that the b quark mass $m_b$ is large compared to the hadronic scale $\Lambda_{QCD}$, it is possible to tackle the
problem of hadronic matrix elements in two-body B decays in a systematic manner. Indeed, once a consistent power counting in terms of
$\Lambda_{QCD}/m_b$ has been established, it can be proven that factorization is a formal prediction of QCD in the heavy quark limit
\cite{Bauer:2001cu}. For this strong statement to be true, however, some comments must be added. The proof is valid for a certain type of decays, and
the mass of the charm quark $m_c$ might have to be included in the power counting as a formally large parameter, as well as the energy of the ejected
meson, which should therefore be light. Nevertheless, the importance of such a result is clearly monumental.

The physical picture is always the same. The most clear example to illustrate this picture is the case of a B meson that decays into a heavy and a
light meson, being the heavy meson who pics up the spectator quark. The b quark in the B meson decays into a set of very energetic partons. The charm
and the spectator quarks form the final heavy meson with no difficulty: since the charm is quite heavy, its velocity will not be large. On the other
hand, the two light quarks formed in the weak vertex will be very energetic, so that if they are going to form a meson they must be highly collinear
and in a color-singlet configuration. This compact color-singlet object will leave the interaction region without interacting with the degrees of
freedom that hadronize into the heavy meson, since the soft interactions decouple. This is the usual \emph{color transparency} argument
\cite{Bjorken:1988kk} due to Bjorken.

The fast color-singlet pair of quarks will then hadronize with a probability given by the leading-twist light-cone distribution amplitude of the
light meson $\Phi_M(u)$, depending on the momentum fraction of each quark. The transition from the B meson to the heavy meson is parameterized by a
standard form factor. This is how the color transparency argument links to the concept of factorization. This physical picture was made quantitative
with the development of the QCDF approach \cite{Beneke:1999br,Beneke:2000ry}.

\begin{figure}
\begin{center}
\includegraphics[width=14cm,height=1.7cm]{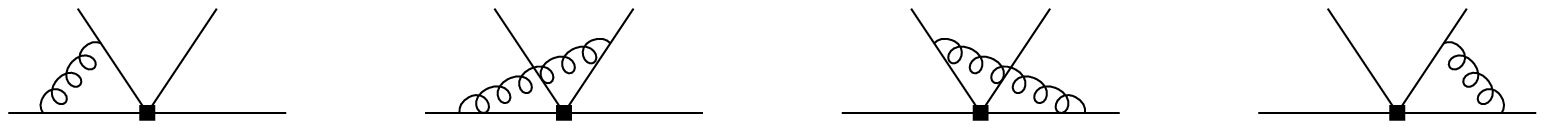}
\end{center}
\vspace{-0.5cm} \caption{\small Order $\alpha_s$ vertex corrections.} \vspace{-0.2cm} \label{fig_vertex}
\end{figure}

The mathematical formulation of this picture is powerful because it allows to compute corrections in a systematic way. For example, it is not
required for the ejected pair of collinear quarks to be in a color-singlet configuration in order to form a meson. Indeed, a hard gluon exchange with
the B - heavy meson system can put this pair in a color-singlet state before hadronization. The important point is that this correction is calculable
in perturbation theory (see Fig.\ref{fig_vertex}). One must prove, however, that this escaping object doesn't interact softly with the interaction
region even if it is not in a color-singlet state. This is the sort of things that must be checked in order to prove factorization.

\begin{figure}
\parbox[t]{8cm}{
\begin{center}\includegraphics[width=7cm,height=1.7cm]{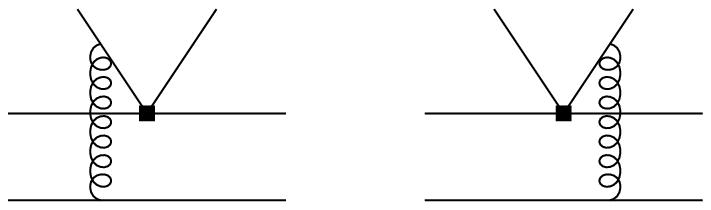}\end{center}
\vspace{-0.5cm} \caption{\label{fig_hardspec}\small Order $\alpha_s$ hard spectator interactions.}}
\parbox[t]{7cm}{
\begin{center}\includegraphics[width=6cm,height=1.7cm]{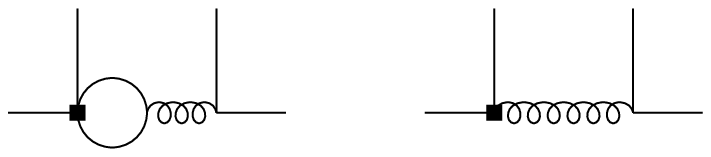}\end{center}
\vspace{-0.5cm} \caption{\label{fig_penguin}\small Penguin contractions.}}
\end{figure}

The case of two light mesons in the final state is more complicated. The light meson that picks up the spectator quark, receives from the weak vertex
a very energetic light quark. Therefore, since the spectator quark is slow, the meson is created in a very asymmetric configuration. The probability
of hadronization is then given by the meson distribution amplitude near its end point, which leads to a suppression of order
$(\Lambda_{QCD}/m_b)^{3/2}$. Hence, one should take into account the competing contribution from a process in which the spectator quark suffers a
hard interaction. If this interaction is the exchange of a hard gluon with the b quark or with the other quark with who is forming the light meson,
this is just a contribution to the heavy-to-light form factor. If the hard gluon is exchanged with the pair of quarks that form the ejected meson,
then this is again a calculable correction, called ``hard-spectator interaction'' (see Fig.\ref{fig_hardspec}).

Other calculable corrections to naive factorization are given by penguin contractions (see Fig.\ref{fig_penguin}), with the insertion of a penguin
operator, or the chromomagnetic dipole operator. These contributions introduce complex phases in the hard-scattering kernel which account for
perturbative strong-interaction phases, due to the rescattering phase of the penguin loop. This is basically the familiar Bander-Silverman-Soni
mechanism for strong phases \cite{Bander:1979px}. However, this is not the only source of strong phases since the vertex corrections in
Fig.\ref{fig_vertex} also generate imaginary parts. These are correctly included in the QCDF approach.

\begin{figure}
\begin{center}
\includegraphics[width=14cm,height=1.7cm]{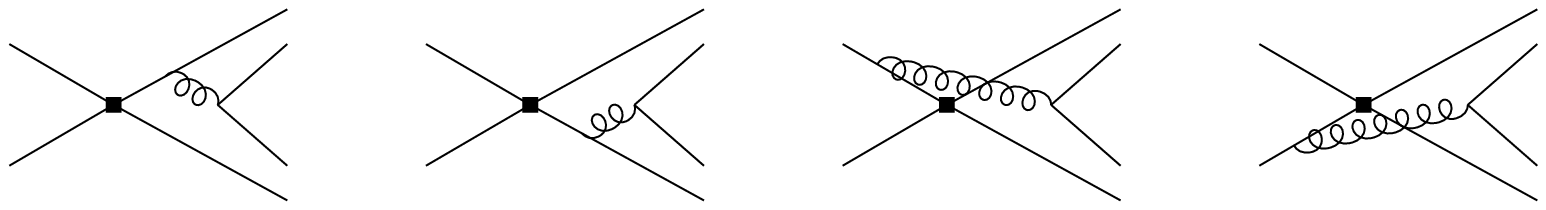}
\end{center}
\vspace{-0.5cm} \caption{\small Weak annihilation contributions.} \vspace{-0.2cm} \label{fig_annh}
\end{figure}

There is another set of contributions to the decay amplitudes that are completely missing in the generalized factorization formula (\ref{GFformula}).
They consist of the processes in which the two quarks inside the B meson annihilate to form the final state partons (see Fig.\ref{fig_annh}). These
contributions are formally suppressed in the heavy quark limit. However, some of the annihilation topologies related to the corresponding twist-3
light meson LCDA's are chirally enhanced by the factors $r_\chi$, and for realistic b-quark masses the suppression $\Lambda_{QCD}/m_b$ might not be
sufficient to neglect these terms. Unfortunately, logarithmic end-point divergencies do not cancel properly within these terms, leading to the
appearance of non-factorizable contributions. Some power suppressed terms contributing to hard-spectator graphs also suffer from this disease. These
corrections constitute one of the weak points of QCDF at the phenomenological level. We shall come back to this issue later.

The final formula for the amplitudes in QCDF is then exactly the same one as Eq.(\ref{GFformula}), except that the coefficients $a_i^{\rm eff}$ are
not phenomenological parameters anymore, but are completely calculable in perturbation theory. Moreover, one should add the hard-spectator and
annihilation contributions that do not fit in the scheme of Eq.(\ref{GFformula}). The final formulae for the amplitudes at order $\alpha_s$ will be
given in the following section. An important observation is that naive factorization arises formally from this picture as a prediction of QCD at
leading order in $\alpha_s$ and $\Lambda_{QCD}/m_b$. This justifies mathematically the phenomenological successes and limitations of the naive
factorization approach. Moreover, beyond leading order in $\alpha_s$ the scale and scheme dependences are correctly cancelled between the Wilson
coefficients and the matrix elements. A proof of factorization in QCDF has however been only given at order $\alpha_s^2$ for heavy-light final states
\cite{Beneke:2000ry}, and the all order proof is not straightforward. Towards this end a more systematic formalism based on an effective lagrangian
(SCET) is more useful \cite{Bauer:2001cu}.

The significant property of the factorization approach in phenomenology is the fact that the non perturbative quantities that appear in the formulae
for the amplitudes are much simpler objects than the original hadronic matrix elements. These simpler objects are either related to universal
properties of a single meson state, as is the case for the light-cone distribution amplitudes, or describe the $B\to$ meson transition matrix element
of a local current, parameterized by a form factor. These are objects that appear in many different decay amplitudes, and can therefore be extracted
from data and used to make predictions on other modes. Moreover, one can also use QCD sum-rule techniques to study them, or extract them from the
lattice. Factorization is a consequence of the fact that in the heavy quark limit only hard interactions between the `$B\to$ meson' and the
`ejected-meson' systems are relevant.

\subsection{$B\to M_1M_2$ amplitudes in QCD Factorization}

\label{QCDF}

Here we collect the relevant formulae for the $B\to M_1 M_2$ amplitudes in QCD Factorization according to \cite{hep-ph/0308039,hep-ph/0612290}. The
amplitudes are given as matrix elements of the transition operators $\mathcal{T}_A^p$ and $\mathcal{T}_B^p$,
\eq{\bra{M_1M_2}\Heff \ket{\bar{B}}=\sum_{p=u,c}\lambda_p^{\sss (D)*}\bra{M_1M_2}\mathcal{T}_A^p+\mathcal{T}_B^p \ket{\bar{B}}. \label{ampQCDF}}
The transition operator $\mathcal{T}_A^p$ contains the contribution from the leading order tree diagrams, and the next-to-leading order vertex
corrections, penguin and hard spectator-scattering. It can be written as
\eqa{
   {\cal T}_A^p
   &=& \delta_{pu}\,\alpha_1(M_1 M_2)\,A([\bar q_s u][\bar u D]) + \delta_{pu}\,\alpha_2(M_1 M_2)\,A([\bar q_s D][\bar u u])\nonumber\\[0.2cm]
   &+&\alpha_3^p(M_1 M_2)\,\sum_q A([\bar q_s D][\bar q q])+ \alpha_4^p(M_1 M_2)\,\sum_q A([\bar q_s q][\bar q D]) \nonumber\\[-0.3cm]
   &+&\alpha_{3,\rm EW}^p(M_1 M_2)\,\sum_q\frac32\,e_q\,A([\bar q_s D][\bar q q])+ \alpha_{4,\rm EW}^p(M_1 M_2)\,\sum_q\frac32\,e_q\,
    A([\bar q_s q][\bar q D]) \,,\nonumber\\[-0.3cm]
   &&\label{Aops}
}
where the operators $A([\cdots][\cdots])$ and their matrix elements are the same as introduced above. Notice that this is basically the same formula
(\ref{GFformula}). The sums extend over $q=u,d,s$ and $\bar{q}_s$ is the spectator antiquark. The $\alpha_i^p$ coefficients are given by
\eqa{
\alpha_1(M_1 M_2) &=& a_1(M_1 M_2) \,, \nonumber\\
   \alpha_2(M_1 M_2) &=& a_2(M_1 M_2) \,, \nonumber\\
   \alpha_3^p(M_1 M_2) &=& \left\{
    \begin{array}{cl}
     a_3^p(M_1 M_2) - a_5^p(M_1 M_2) \,; & \quad \mbox{if~} M_1 M_2=PP, \,VP \,, \\
     a_3^p(M_1 M_2) + a_5^p(M_1 M_2) \,; & \quad \mbox{if~} M_1 M_2=PV, \,VV \,,
    \end{array}\right. \nonumber\\
   \alpha_4^p(M_1 M_2) &=& \left\{
    \begin{array}{cl}
     a_4^p(M_1 M_2) + r_{\chi}^{M_2}\,a_6^p(M_1 M_2) \,;  & \quad \mbox{if~} M_1 M_2=PP, \,PV \,, \\[0.1cm]
     a_4^p(M_1 M_2) - r_{\chi}^{M_2}\,a_6^p(M_1 M_2) \,;  & \quad \mbox{if~} M_1 M_2=VP, \,VV \,,
    \end{array}\right.\nonumber\\
   \alpha_{3,\rm EW}^p(M_1 M_2) &=& \left\{
    \begin{array}{cl}
     a_9^p(M_1 M_2) - a_7^p(M_1 M_2) \,;  & \quad \mbox{if~} M_1 M_2=PP, \,VP \,, \\[0.1cm]
     a_9^p(M_1 M_2) + a_7^p(M_1 M_2) \,;  & \quad \mbox{if~} M_1 M_2=PV, \,VV \,,
    \end{array}\right. \\
   \alpha_{4,\rm EW}^p(M_1 M_2) &=& \left\{
    \begin{array}{cl}
     a_{10}^p(M_1 M_2) + r_{\chi}^{M_2}\,a_8^p(M_1 M_2) \,;  & \quad \mbox{if~} M_1 M_2=PP, \,PV \,, \\[0.1cm]
     a_{10}^p(M_1 M_2) - r_{\chi}^{M_2}\,a_8^p(M_1 M_2) \,;  & \quad \mbox{if~} M_1 M_2=VP, \,VV \,.
     \end{array}\right.\nonumber
}
where, at next-to-leading order in $\alpha_s$, the coefficients $a_i^p$ have the following general form
\eqa{ a_i^p(M_1 M_2) &=& \left( C_i + \frac{C_{i\pm 1}}{N_c} \right)
   N_i(M_2) \nonumber\\
  && + \,\frac{C_{i\pm 1}}{N_c}\,\frac{C_F\alpha_s}{4\pi}
   \left[ V_i(M_2) + \frac{4\pi^2}{N_c}\,H_i(M_1 M_2) \right]
   + P_i^p(M_2) \,.
\label{aQCDF}}
The coefficient $N_i(M_2)$ is the leading order term, $V_i(M_2)$ contains the vertex terms, the $P_i^p$ are the penguin terms and the $H_i(M_1M_2)$
are the hard spectator terms. The expressions in terms of meson distribution amplitudes and the various hard-scattering functions can be found in
\cite{hep-ph/0308039,hep-ph/0612290}.

The transition operator $\mathcal{T}_B^p$ contains the contributions from weak annihilation and is written as
\eqa{ {\cal T}_B^p
   &=& \delta_{pu}\,b_1(M_1 M_2)\,\sum_{q'}
    B([\bar u q'][\bar q' u][\bar D b])
    + \delta_{pu}\,b_2(M_1 M_2)\,\sum_{q'}
    B([\bar u q'][\bar q' D][\bar u b]) \nonumber\\
   &\!\!+&\hspace{-0.4cm} b_3^p(M_1 M_2)\,\sum_{q,q'} B([\bar q q'][\bar q' D][\bar q b])
    + b_4^p(M_1 M_2)\,\sum_{q,q'} B([\bar q q'][\bar q' q][\bar D b])
    \nonumber\\[-0.1cm]
   &\!\!\!+&\hspace{-0.4cm} b_{3,\rm EW}^p(M_1 M_2)\,\sum_{q,q'} \frac{3}{2}\,e_q\,
    B([\bar q q'][\bar q' D][\bar q b])
    + b_{4,\rm EW}^p(M_1 M_2)\,\sum_{q,q'} \frac{3}{2}\,e_q\,
    B([\bar q q'][\bar q' q][\bar D b]) \nonumber\\
   &\!\!\!+&\hspace{-0.4cm} \delta_{pu}\,b_{S1}(M_1 M_2)\,\sum_{q'}
    B([\bar u u][\bar q'\!q'][\bar D b])
    + \delta_{pu}\,b_{S2}(M_1 M_2)\,\sum_{q'}
    B([\bar u D][\bar q'\!q'][\bar u b]) \nonumber\\
   &\!\!\!+&\hspace{-0.4cm} b_{S3}^p(M_1 M_2)\,\sum_{q,q'}
    B([\bar q D][\bar q'\!q'][\bar q b])
    + b_{S4}^p(M_1 M_2)\,\sum_{q,q'}
    B([\bar q q][\bar q'\!q'][\bar D b]) \nonumber\\[-0.1cm]
   &\!\!\!+&\hspace{-0.4cm} b_{S3,\rm EW}^p(M_1 M_2)\,\sum_{q,q'} \frac{3}{2}\,e_q\,
    B([\bar q D][\bar q'\!q'][\bar q b])
    + b_{S4,\rm EW}^p(M_1 M_2)\,\sum_{q,q'} \frac{3}{2}\,e_q\,
    B([\bar q q][\bar q'\!q'][\bar D b]) \,, \nonumber\\
    &&
\label{Bops}}
where the matrix elements of the operators $B([\cdots][\cdots][\cdots])$ are given by
\eq{
   \bra{M_1 M_2}B([\ldots][\ldots]][\ldots])\ket{\bar B_q}\equiv c \,B_{\sss M_1 M_2} \,, \quad \mbox{with} \quad
   B_{\sss M_1 M_2} = \pm i\,\frac{G_F}{\sqrt{2}}\,f_{B_q} f_{M_1} f_{M_2} \,,
}
whenever the quark flavors of the three brackets match those of $M_1$, $M_2$, and $\bar B_q$. The constant $c$ is the same as for the $A$ operators.
The upper sign applies when both mesons are pseudoscalar and the lower sign otherwise.

The coefficients with the subscript `S' contribute only to final states containing flavor-singlet mesons or neutral vector mesons. The expressions
for the $b_i$ coefficients can be found in \cite{hep-ph/0308039,hep-ph/0612290}.

Up to this point it useful to look closely at the contributions that are dangerous in QCDF, mainly the hard-spectator and the annihilation terms with
twist-3 LCDA insertion. The hard-spectator terms receive a contribution of this form:
\eq{H(M_1M_2)\supset \frac{B_{\sss M_1M_2}}{A_{\sss M_1M_2}}\frac{m_B}{\lambda_B}\int_0^1 \!dx\, \int_0^1\! dy \ r_\chi^{\sss M_1}
\frac{\Phi_{M_2}(x)\Phi_{m_1}(y)}{x\bar{y}}\ \propto \int_0^1 \frac{dy}{\bar{y}}\Phi_{m_1}(y) }
where $\lambda_B$ is defined in (\ref{lambdaB}). This contribution is formally suppressed in the heavy quark limit, but the presence of the chiral
factor $r_\chi^{\sss M_1}$ makes it phenomenologically important. Moreover, as can be seen by plugging in $\Phi_p(y)=1$ or $\Phi_v(y)$ as given in
(\ref{3twistV}), there is a divergence coming from the region $y\sim 1$. This infrared divergence is a manifestation of the fact that factorization
does not hold in this case.

It is customary to extract this divergence defining a parameter $X_H^{M_1}\equiv \int_0^1 dy/\bar{y}$ such that
\eq{\int_0^1 \frac{dy}{\bar{y}}\Phi_{m_1}(y)=\Phi_{m_1}(1)X_H^{M_1}+\int_0^1\frac{dy}{\bar{y}}\left[ \Phi_{m_1}(y)-\Phi_{m_1}(1) \right]}
Because this divergence is associated with a soft interaction of the ejected meson with the spectator quark, the divergence arises specifically from
the region $\bar{y}\sim \Lambda_{QCD}/m_b$, and therefore one expects that $X_H^{M}\sim \log(m_b/\Lambda_{QCD})$ with an a priori arbitrary strong
phase. The choice for the values of $X_H^M$ introduces an unavoidable model dependence in the predictions.

Consider now the chirally enhanced contributions to the annihilation terms with 3-twist LCDA insertions. For illustration we take the coefficient
$b_1$ in Eq.(\ref{Bops}), which contains a term
\eq{b_1(M_1M_2)\supset \frac{C_F}{N^2}C_1\pi\alpha_s \int_0^1\!dx\int_0^1\!dy\ r_\chi^{M_1}r_\chi^{M_2}\Phi_{m_2}(x)\Phi_{m_1}(y) \frac{2}{\bar{x}
y}\ .}
This contribution is again infrared divergent from the regions $\,y,\bar{x}\sim \Lambda_{QCD}/m_b\,$. As before, the divergencies are parameterized
by the quantities $X_A^{M_1}$ and $X_A^{M_2}$ that introduce an unavoidable model dependence. The rest of the coefficients in Eq.(\ref{Bops}) are
also affected by these divergencies, and signal the breakdown of factorization.

A way to get around these inconveniences at the phenomenological level was proposed in \cite{hep-ph/0603239} and used later in
\cite{hep-ph/0610109,hep-ph/0611280,arXiv:0705.0477,arXiv:0707.2046}. The basic idea is that the so called \emph{GIM penguin} is in many cases free
from these divergencies when computed in QCDF, since it is dominated by short-distance physics. This quantity (that we call $\Delta$), is given
generically by
\eq{\Delta_{\sss M_1M_2}=A_{\sss M_1M_2} \frac{C_F \alpha_s}{4\pi N}C_1 \big[ \bar{G}(m_c^2/m_b^2)-\bar{G}(0) \big]}
where $\bar{G}(x)$ are penguin functions. Using this infrared safe theoretical quantity together with data and some SU(3) relations, predictions can
be made safer and more precise. This will be discussed in Chapters \ref{SymFac1} and \ref{SymFac2}.

\section{Symmetries}

\label{symmetries}

It should be clear by now that the strong interaction dynamics inside the matrix elements is a very big deal. An impressive progress has been
observed in the past decade concerning the idea of factorization in B physics, but direct computations of amplitudes from QCD still suffer from two
diseases. First, the computations involve a great amount of work, combining tedious perturbative calculations, lattice simulations and other
technicalities. Second, the theoretical validity of such approaches at the phenomenological level is threatened by corrections to the heavy quark
limit and by potentially non-factorizable long-distance leading effects, like the case of the long-distance charm loop issue (``charming penguins'')
\cite{Ciuchini:1997hb,Ciuchini:2001gv,Bauer:2004tj,Beneke:2004bn,Jain:2007dy}.

Fortunately there is a complementary approach to B-decay phenomenology that relies on the approximate flavor symmetries of QCD, and incorporates
naturally all the hadronic information that cannot be extracted from first principles. The drawback is of course the uncontrollable symmetry breaking
effects, and the fact that only certain \emph{relations} between observables can be obtained. However, these relations have been very useful in order
to extract CKM parameters from experimental data on non-leptonic decays. In this section we review schematically how this flavor symmetry relations
are obtained.

Consider the lagrangian of massless QCD with $n_f$ quark flavors
\eq{\mathcal{L}_{QCD}=-\frac{1}{4}F^{a\,\mu\nu}F_{\mu\nu}^a + i\,\sum_{k=1}^{n_f}\, \bar{q}_k\, D\hspace{-0.25cm}/\ q_k}
This lagrangian has global symmetry $SU(n_f)_V \otimes SU(n_f)_A \otimes U(1)_V \otimes U(1)_A$. For example, for $n_f=2$ (u \& d quarks), this
symmetry is expected to hold with good precision due to the smallness of $m_u$ and $m_d$. The first term $SU(2)_V$ is isospin, and it is seen to be a
good symmetry. The second term $SU(2)_A$ is not observed as a symmetry of nature, because this symmetry is spontaneously broken by the quark
condensate $\av{\bar{q}q}$, giving rise to goldstone bosons that we observe as pions. The third term $U(1)_V$ is nothing but baryon number
conservation, which is obviously observed. The last term $U(1)_A$ led originally to the so called $U(1)$-problem \cite{Weinberg:1975ui}, but later it
was found that in fact this symmetry is broken by instantons \cite{Hooft:1976up}. So it is clear that, at least for two flavors, the approximate
global symmetry is well understood and under control. In the case of isospin, the symmetry breaking effects are expected at the level of a few
percent.

For $n_f=3$ flavors, the flavor SU(3) symmetry is not so exact as isospin, since the strange quark mass is high compared with $m_{u,d}$. However, the
main feature of a SU(3) flavor symmetry is phenomenologically realized: we observe the $(\pi,K,\eta)$ pseudoscalar octet, and the power of group
theory can be applied to SU(3) in the same way as for isospin. The predictions, nonetheless, will be affected by symmetries breaking uncertainties of
$\sim 20\%$, or maybe higher \cite{Khodjamirian:2003xk}.

\subsection{The Wigner-Eckart Theorem}

The Wigner-Eckart theorem is the mathematical prescription to extract the maximal advantage from a symmetry group in a quantum mechanical system.
Although the Wigner-Eckart theorem can be formulated with total generality, we just quote here the theorem for $SU(2)$, as used in elementary quantum
mechanics when adding angular momenta. In the next section we will use this theorem to write a simple flavor symmetry relation between $B\to \pi K$
amplitudes.

Given a symmetry group of the system under consideration, any operator can in general be decomposed in terms of \emph{irreducible tensor operators},
defined as operators that transform as tensors of a given rank under the symmetry group. In the case of SU(2), an irreducible rank-$k$ operator is a
$(2k+1)$-tuple of operators $Q^{(k)}_q$ ($q=-k,-k+1,\dots,k$), satisfying the proper commutation relations,
\eqa{
\big[ J_{\pm},Q^{(k)}_q \big] & = & \pm \sqrt{(k\mp q)(k\pm q+1)}\ Q^{(k)}_{q\pm 1} \nn \\
\big[ J_z,Q^{(k)}_q \big] & = & q\ Q^{(k)}_q \label{ComRel} }
where $J_{\pm}$ are the ladder operators and $J_z$ is the Cartan operator of SU(2). The Wigner-Eckart theorem states that the matrix element
$\bra{\alpha';j',m'}Q^{(k)}_q\ket{\alpha;j,m}$ depends on the quantum numbers $m$, $m'$ and $q$ only through Clebsch-Gordan coefficients,
specifically
\eq{\bra{\alpha';j',m'}Q^{(k)}_q\ket{\alpha;j,m}=C(j',m'|j,k;m,q)\,\langle \alpha';j'\| Q^{k} \| \alpha;j \rangle\ ,}
where $C(j',m'|j,k;m,q)$ is the corresponding Clebsch-Gordan coefficient and the object $\langle \alpha';j'\| Q^{k} \| \alpha;j \rangle$ is a so
called \emph{reduced matrix element}.

The power of this result is that, although the reduced matrix element cannot be computed from symmetry principles alone, it only depends on the
representation on which the objets live in, and therefore one can establish relations between transitions that connect states in the same
representations.

\subsection{A simple example: $B\to \pi K$ Amplitudes}

As a simple example of how the Wigner-Eckart theorem applies to flavor symmetries, we derive the famous $B\to \pi K$ amplitude relation
\cite{Gronau:1991dq,Nir:1991cu,Lipkin:1991st}.

We want to extract the isospin structure of the matrix elements
\eq{\bra{\pi K}\Heff\ket{B}\ ,}
where $\Heff$ is the weak effective Hamiltonian, given in Eqs.(\ref{HeffSM}),(\ref{OpsSM}) for the SM case. First we decompose the initial and final
states in the isospin basis. $(B^0,B^+)$ and $(K^0,K^+)$ are isospin doublets, and $(\pi^-,\pi^0,\pi^+)$ is an isospin triplet:
\eqa{
& \ket{B^0}=\ket{1/2,-1/2}\ ,\ \ket{B^+}=\ket{1/2,1/2}\ ,&\nn\\[5pt]
&\ket{K^0}=\ket{1/2,-1/2}\ ,\ \ket{K^+}=\ket{1/2,1/2}\ , & \nn\\[5pt]
& \ket{\pi^-}=\ket{1,-1}\ ,\ \ket{\pi^0}=\ket{1,0}\ ,\ \ket{\pi^+}=\ket{1,1}& }
The isospin decomposition of the $\ket{\pi K}$ states is given by the $(1\otimes\,1/2)$ SU(2) Clebsch-Gordan coefficients,
\eqa{ \ket{\pi^+ K^0} = \sqrt{\frac{1}{3}}\ \left| \frac{3}{2}\,,\frac{1}{2} \right\rangle +
                      \sqrt{\frac{2}{3}}\ \left| \frac{1}{2}\,,\frac{1}{2} \right\rangle &\!\!\!,&
\ket{\pi^- K^+} = \sqrt{\frac{1}{3}}\ \left| \frac{3}{2},-\frac{1}{2} \right\rangle -
                      \sqrt{\frac{2}{3}}\ \left| \frac{1}{2},-\frac{1}{2} \right\rangle \qquad\nn\\
\ket{\pi^0 K^+} = \sqrt{\frac{2}{3}}\ \left| \frac{3}{2}\,,\frac{1}{2} \right\rangle -
                      \sqrt{\frac{1}{3}}\ \left| \frac{1}{2}\,,\frac{1}{2} \right\rangle &\!\!\!,&
\ket{\pi^0 K^0} = \sqrt{\frac{2}{3}}\ \left| \frac{3}{2},-\frac{1}{2} \right\rangle +
                      \sqrt{\frac{1}{3}}\ \left| \frac{1}{2},-\frac{1}{2} \right\rangle
\label{IsoDec}}

The next step is to decompose the effective Hamiltonian in terms of irreducible $SU(2)$ tensor operators. But this is easy: the current-current
operators $Q_{1,2}^u\sim \bar{u}b\bar{s}u$ have components $\Delta I=0$ ad $\Delta I=1$, and so do the electroweak penguin operators
$Q_{7,\dots,10}\sim \bar{D}b\sum e_q \bar{q}q$; the current-current operators $Q_{1,2}^c\sim \bar{c}b\bar{s}c$ and the QCD-penguin operators
$Q_{3,\dots,6}\sim \bar{s}b\sum \bar{q}q$ only have $\Delta I=0$ components. Therefore, the effective Hamiltonian can be decomposed as the sum of
triplet and singlet tensor operators:
\eq{\Heff=\Heff^{(1)}\oplus \Heff^{(0)}}

Now the matrix elements can be expressed in terms of reduced matrix elements. First we have $\bra{3/2,\pm 1/2}\Heff^{(0)}\ket{1/2,\pm 1/2}=0$ due to
the commutation relations (\ref{ComRel}). For the rest, the Wigner-Eckart theorem implies that
\eqa{ \left\langle \frac{3}{2}\,,\pm\frac{1}{2}\,\right|\Heff^{(1)}\left| \frac{1}{2}\,,\pm\frac{1}{2} \right\rangle & \!\!\!=\!\! &
C\left(\frac{3}{2},\pm\frac{1}{2}\,\Big|\,\frac{1}{2},1;\pm\frac{1}{2},0\right) \left\langle
\frac{3}{2}\right\|\!\Heff^{(1)}\!\left\|\frac{1}{2}\right\rangle =
\sqrt{\frac{2}{3}}\,\left\langle \frac{3}{2}\right\|\!\Heff^{(1)}\!\left\|\frac{1}{2}\right\rangle \nn\\[7pt]
\left\langle \frac{1}{2}\,,\pm\frac{1}{2}\,\right|\Heff^{(1)}\left| \frac{1}{2}\,,\pm\frac{1}{2} \right\rangle & \!\!\!=\!\! &
C\left(\frac{1}{2},\pm\frac{1}{2}\,\Big|\,\frac{1}{2},1;\pm\frac{1}{2},0\right) \left\langle
\frac{1}{2}\right\|\!\Heff^{(1)}\!\left\|\frac{1}{2}\right\rangle =
\mp\sqrt{\frac{1}{3}}\,\left\langle \frac{1}{2}\right\|\!\Heff^{(1)}\!\left\|\frac{1}{2}\right\rangle \nn\\[7pt]
\left\langle \frac{1}{2}\,,\pm\frac{1}{2}\,\right|\Heff^{(0)}\left| \frac{1}{2}\,,\pm\frac{1}{2} \right\rangle & \!\!\!=\!\! & \left\langle
\frac{1}{2}\right\|\!\Heff^{(0)}\!\left\|\frac{1}{2}\right\rangle \label{RedMat}}
It is conventional to define the invariant isospin amplitudes $A_{3/2}$, $A_{1/2}$ and $B_{1/2}$ ,
\eq{A_{3/2}\equiv \frac{\sqrt{2}}{3}\left\langle \frac{3}{2}\right\|\!\Heff^{(1)}\!\left\|\frac{1}{2}\right\rangle\ ,\ A_{1/2}\equiv
-\frac{\sqrt{2}}{3}\left\langle \frac{1}{2}\right\|\!\Heff^{(1)}\!\left\|\frac{1}{2}\right\rangle\ ,\ B_{1/2}\equiv \sqrt{\frac{2}{3}}\left\langle
\frac{3}{2}\right\|\!\Heff^{(0)}\!\left\|\frac{1}{2}\right\rangle\ , \nn }
so that combining Eqs.(\ref{IsoDec}) and (\ref{RedMat}) the amplitudes can be written as
\eqa{
A(B^+\to \pi^+ K^0)&=&B_{1/2}+A_{1/2}+A_{3/2}\ , \nn\\
-\sqrt{2}A(B^+\to \pi^0 K^+)&=&B_{1/2}+A_{1/2}-2A_{3/2}\ , \nn\\
-A(B^0\to \pi^- K^+)&=&B_{1/2}-A_{1/2}-A_{3/2}\ , \nn\\
\sqrt{2}A(B^0\to \pi^0 K^0)&=&B_{1/2}-A_{1/2}+2A_{3/2}\ . \label{BpiKamps}}
From this representation of the amplitudes in terms of reduced matrix elements, the promised $B\to \pi K$ amplitude relation follows:
\eq{A(B^+\!\to \pi^+ K^0)-A(B^0\!\to \pi^- K^+)+\sqrt{2}A(B^+\!\to \pi^0 K^+)-\sqrt{2}A(B^0\!\to \pi^0 K^0)=0 \label{BpiK}}

\subsection{Discussion}

Following the same procedure as for $B \to \pi K$, it is possible to write other isospin relations. For example, for $B\to \pi\pi\,$, isospin implies
\eq{A(B^0\to \pi^+\pi^-)-\sqrt{2}A(B^+\to \pi^+\pi^0)+A(B^0\to \pi^0\pi^0)=0\ . \label{Bpipi}}
Relations arising from SU(3) flavor symmetry are richer, since they mix many other decays that are not just connected by an isospin transformation.
The complexity of the group SU(3) with respect to the simplicity of SU(2) is the main reason to avoid a detailed discussion of the technical aspects
of the SU(3) analysis here, and the important idea is clear from the example above. The SU(3) Clebsch-Gordan coefficients were collected for the
representations of interest by de Swart \cite{Swart:1963gc}, and the complete SU(3) decomposition of the non-leptonic decay amplitudes of the triplet
$(B^+_u,B^0_d,B^0_s)$ into pseudoscalar mesons was given by Grinstein and Lebed \cite{Grinstein:1996us}.

However, the general decomposition of SU(3) amplitudes introduces too many SU(3) invariant amplitudes, so that in order to extract amplitude
relations it is often necessary to neglect some of them. But the relative sizes of the reduced matrix elements is a dynamical question, so this
process requires a dynamical assumption. Because the language of invariant amplitudes is often obscure in this sense, it is often convenient to use a
\emph{diagrammatic} approach in which the invariant amplitudes are related to flavor-flow topologies (see for example \cite{Gronau:1994rj}). The
amplitudes can then be written in terms of $T$ (tree), $P$ (penguin), $C$ (color suppressed), $E$ (exchange), $A$ (annihilation), $PA$ (penguin
annihilation) and electroweak penguins $P_{EW}$ and $P_{EW}^C$. Including $b\to s$ transitions doubles the number of parameters, which are usually
denoted with primes. Then one can argue that the electroweak penguins are small, and also neglect the contributions $E$, $A$ and $PA$ on the basis of
helicity or $f_B/m_B$-type suppressions \cite{Gronau:1995hn}. It is, however, far from clear whether these assumptions can be made safely.

An analysis along these lines allows to find flavor symmetry relations that do not rely on any dynamical assumption, like for example
\cite{Gronau:2006eb,Escribano:2007mq}
\eqa{
&A(B^0\to K^0\pi^0)-\sqrt{3}A(B^0\to K^0\eta_8)=0\ ,& \nn\\[1ex]
&A(B^+\to K^+\pi^0)+\sqrt{2}A(B^+\to K^0\pi^+)-\sqrt{3}A(B^+\to K^+\eta_8)=0\ ,& \\[1ex]
&A(B^0\to K^+\pi^-)+A(B^+\to K^0\pi^+)-\sqrt{6}A(B^+\to K^+\eta_8)+\sqrt{6}A(B^0\to K^0\eta_8)=0\ .&\nn }
From these relations the isospin $B\to \pi K$ relation (\ref{BpiK}) follows.

Up to this point it not obvious what is the usefulness of these amplitude relations. Indeed, a powerful aspect of flavor symmetry relations is to be
able to write relations between \emph{observables} and not just amplitudes. These relations can be then tested experimentally to probe the size of
the SU(3) breaking. Since they are specially sensitive to, for instance, non standard isospin violating contributions, they provide a window for New
Physics (see for example \cite{hep-ph/9909297} for a discussion concerning $B\to \pi K$).

But obtaining relations between observables is not as easy as for the amplitudes. Observables are constructed, roughly, by squaring amplitudes.
Therefore, the observables depend on the magnitudes of the invariant amplitudes and also on the \emph{strong phases}, since there is interference
between different invariant amplitudes. For example, in the case of Eq.(\ref{BpiKamps}), there are two independent CKM parameters entering the decay
amplitudes, introducing two weak phases. Each invariant amplitude receives contributions from both weak phases. Up to a global phase, one ends up
with five independent strong phases and six independent real amplitudes, to be fitted with four branching ratios and four CP asymmetries
\cite{Neubert:1998re}. This example shows how a model independent \emph{amplitude} relation is not sufficient to obtain a model independent
\emph{observable} relation.

Despite the apparent negativity of the discussion above, flavor symmetries have been used in phenomenology with the outcome of very interesting
analyses and results, helping to achieve an understanding on B decays that would have not been possible with dynamical approaches alone
\cite{Neubert:1998re,Savage:1989ub,Fleischer:1997um,Buras:1997cv,Grossman:1997gr,Neubert:1998pt,Fleischer:1999pa,Grossman:2003qp,Matias:2001ch}
\footnote{To name a few.}. Flavor symmetry relations will be used extensively throughout this thesis, specially for $B\to KK$ modes.

%


\chapter{CP Violation in Meson Decays}

\label{CPV}

The fact that nature is not invariant under parity transformations (P) was an early surprise. It was of general common sense to think that one could
never tell whether a physical process was being observed through a mirror or directly. Indeed, electromagnetism and strong interactions conserve
parity, and in the 1950's experiments in particle physics were mainly probing the recently developed theory of QED, and the strong interactions of
the zoo of hadrons that was just being discovered. The exception was the experiments on $\beta$-decay, that was at that time described by Fermi's
theory. This theory had just been proposed to be universal (e.g, neutron decay and muon decay would have the same coupling constant), and this was
the main issue in weak interactions.

However, in 1955 the $\tau$-$\theta$ puzzle raised some confusion. The $\theta^+$ would decay in to a CP-even two pion state, and the $\tau^+$ into a
CP-odd three pion state. Parity invariance necessarily implied that these were two different particles with opposite parity, but the puzzling point
was that they were otherwise \emph{identical}; same charge, same mass, same width. This led to Lee and Yang to point out that while P invariance had
been tested in strong and electromagnetic interactions, it had never been verified in weak interactions, and they proposed several experiments. This
proposal was taken seriously by Wu's group, who in 1957 provided first evidence of Parity violation studying the angular distribution of nuclear
$\beta$-decay with polarized ${\rm Co}^{60}$ \cite{Wu:1957my}. Their discovery was simultaneously confirmed by Lederman and collaborators with a
measurement of the electron asymmetry in pion decay \cite{Garwin:1957hc}. The discovery of P violation immediately solved the $\tau$-$\theta$ puzzle,
a crucial brick in the physics of kaons.

In parallel, the theory of strangeness arose as an explanation to the kaon-hyperon problem. These particles were produced very easily \emph{in
pairs}, but they decayed very slowly. Gell-Mann and Nishijima proposed the existence of a quantum number (strangeness) that was only violated by weak
interactions; pair production could then be a strong process, but single decay had to be weak. This was the foundation of flavor physics. An
immediate consequence was the existence of two different neutral kaons with opposite strangeness. The separate identity of these two kaons was
established very soon by Gell-Mann and Pais in 1955 \cite{GellMann:1955jx}, a discovery that gave birth to the physics of neutral meson mixing.

The oscillation of neutral mesons quickly opened up a chapter in the history of particle physics. The measurement of the oscillation period of
neutral kaons through time dependent semileptonic decays allowed the extraction of the tiny mass splitting between the two physical kaons, $\Delta
M_K/M_K\sim 10^{-15}$. Later, studies of regeneration of short-lived kaons in matter revealed an excess of regenerated kaons, a result that had an
unexpected origin. In 1964 an experiment by Christenson and collaborators \cite{Christenson:1964fg} established that both the short-lived and the
long-lived components of the neutral kaon system were decaying into a CP even $\pi^+\pi^-$ mode. The clear interpretation of this discovery was the
violation of CP invariance.

Today we have much evidence of CP violating phenomena in kaon and $B_d$ meson decays \cite{Yao:2006px}. Measurements of CP asymmetries in $B_s$
decays are still compatible with zero at the $2\sigma$ level, but since the experimental prospects are good at Tevatron, and theoretically these
asymmetries can be as large as for $B_d$ decays, it is safe to say that it will not take long. CP violation has neither been observed in $D$ meson
decays, which is very suppressed in the SM. However, at the LHC the sensitivity could reach the 1 per mil for $D$ meson CP asymmetries, making CP
violation in the $D$ sector a powerful probe of New Physics.

The SM description of CP violation, the CKM mechanism \cite{Kobayashi:1973fv}, is successful and economic, involving just one CP violating phase. The
fits to the Unitarity Triangle are getting tight \cite{UTfit,CKMfitter}, and all data is consistent (within experimental and theoretical errors) with
a single value for its apex. However, the main reason to believe that there \emph{must} be sources of CP violation beyond the SM, is the issue of the
baryon asymmetry in the universe.

The measurement of baryon asymmetry has recently become accurate (about 5\% error) due to the measurements of the fluctuations of the cosmic
microwave radiation background. The order of magnitude of the baryon-to-photon ratio is $(n_B-n_{\bar{B}})/n_\gamma\sim 10^{-10}$. The fact that this
asymmetry could have been an initial condition at the big bang is excluded, since any baryon asymmetry present in the early universe would have been
diluted by inflation to completely negligible levels. So it is very much accepted that the baryon asymmetry must have a dynamical origin.

The necessary conditions for the dynamical generation of a baryon asymmetry in the universe were studied by Andrei Sakharov in 1967
\cite{Sakharov:1967dj}. There are three main conditions that have to be fulfilled. First, there must exist baryon number violating interactions (this
is clear). Second, there must be a departure from thermal equilibrium during a sufficiently long period of time. This is because the baryon number
violating interactions must occur in the \emph{forward} direction predominantly, so there must be an arrow of time. Third, there must be CP
violation, since CP invariance would equal the rates of CP-conjugated baryon number violating interactions, thus giving a zero net contribution.

Now, baryon number is broken in the SM by sphalerons at high temperature. The electroweak phase transition, occurring at a temperature $T\sim
100\,{\rm GeV}$, could provide the necessary departure from thermal equilibrium (this is called \emph{electroweak baryogenesis}). It turns out that
this phase transition is not strong enough to produce and maintain the observed baryon asymmetry, but let's assume it were. Then, the baryon
asymmetry can be estimated to be $(n_B-n_{\bar{B}})/n_\gamma\sim 10^{-2}\cdot J/T^{12}$, where $J$ is the Jarlskog determinant (a measure of CP
violation in the SM). For $T\sim 100\,{\rm GeV}$, then $(n_B-n_{\bar{B}})/n_\gamma\sim 10^{-23}$, so clearly the CP violation in the SM is far too
small to account for the observed baryon asymmetry. On the other hand, almost any extension of the SM provides new sources of CP violation, so the
search for physics beyond the SM is much in contact with the study of CP violation.

In general, CP violation is the cause that a process and it's CP-conjugate don't share the same rights. This means that it is possible to distinguish
\emph{objectively} between CP-conjugated events\footnote{For example, it is possible to define the electric charge in an absolute sense: positive
charge is the charge of the lepton more often produced in the semileptonic $K_L$ decays.}. In particular, a difference between the decay rate of a
given process, $A\to B$, and that of the CP transformed, $\bar{A}\to \bar{B}$, will signal CP violation. This is generally quantified by a so called
\emph{CP asymmetry}, defined as
\eq{A_{CP}(A\to B)\equiv \frac{\Gamma(A\to B)-\Gamma(\bar{A}\to \bar{B})}{\Gamma(A\to B)+\Gamma(\bar{A}\to \bar{B})}}

The deviation of any CP asymmetry from zero is a measure of CP violation. In this chapter we will review the general features of CP violation in
meson decays and the different types of CP asymmetries that one may encounter.

\section{Mixing of neutral mesons}

\label{Mixing}

The phenomenon of neutral meson mixing is a consequence of flavor violation. Therefore, the mixing is a \emph{weak} process, and since the mesons
themselves are regarded as asymptotic states of the strong Hamiltonian, it is sensible to study the mixing in terms of mesonic degrees of freedom.

The quantum mechanical idea for this phenomenon is very simple. The weak Hamiltonian is not invariant under flavor rotations. This means that flavor
eigenstates deviate from mass eigenstates and mix under time evolution. This deviation, as already mentioned, is due to weak effects, and this allows
to study the oscillations in a perturbative fashion.

Experimentally, oscillations have been observed in all the low-lying neutral meson systems. The oscillation of neutral kaons was observed for the
first time in 1961 \cite{Kmixing}. The oscillations of $B_d$ mesons was observed in 1987 \cite{Albajar:1986it,Albrecht:1987dr}. The observation of
$B_s$ and $D$ meson oscillations has taken a much longer time, but they have been observed finally, very recently
\cite{Abulencia:2006mq,Aubert:2007wf}.

In this section we present a general description of particle-antiparticle oscillations in a quantum mechanical approach. We will see in later
chapters that the physics of meson oscillations is strongly related to the physics of CP violation.

\subsection{The effective Hamiltonian}

\label{effhamil}

Let $P^0$ be a neutral meson carrying an internal additive quantum number $F\ne 0$. For our purposes $F$ is a flavor quantum number. Charge
conjugation will change the sign of this quantum number, so the C-conjugate state $\bar{P}^0$ is a \emph{different} neutral meson.

Now, let's say that $F$ is not conserved, that is, the interaction Hamiltonian contains (weak) breaking terms with $\Delta F\ne 0$. Transitions
$P^0\leftrightarrow \bar{P}^0$ are then possible as a one-step process (through $H_W^{\sss \Delta F=2}$) or through an intermediate state (at second
order in $H_W^{\sss \Delta F=1}$).  To be specific, we write the full Hamiltonian as
\eq{H=H_0+H_W^{\sss \Delta F=1}+H_W^{\sss \Delta F=2},}
where $H_0$ is the flavor invariant (strong) Hamiltonian. The states $\ket{P^0}$ and $\ket{\bar{P}^0}$ are then defined as eigenstates of $H_0$. As
mentioned above, that it actually makes sense to speak about $P^0$ and $\bar{P}^0$ is due to the fact that $H_W$ is weak. On the other hand, the
presence of the terms $H_W$ will produce $P^0-\bar{P}^0$ mixing.

Consider now an initial state which is a mixture of a $P^0$ and a $\bar{P}^0$,
\eq{ \ket{\psi(t=0)}=c\,\ket{P^0}+\bar{c}\,\ket{\bar{P}^0}. \label{InitPsi}}
At $t>0$ this state will evolve in two different ways. First, transitions between $P^0$ and $\bar{P}^0$ will give rise to unitary rotations in the
$c-\bar{c}$ space (oscillations). Second, the fact that these mesons can decay into lighter particles will allow the initial state to evolve outside
the two meson system:
\eq{\ket{\psi(t)}=c(t)\ket{P^0}+\bar{c}(t)\ket{\bar{P}^0}+\sum_n c_n(t)\ket{n}}
where $\ket{n}=\ket{\pi\pi},\,\ket{3\pi},\,\ket{\pi l\bar{\nu}_l},...$ represents any state of any number of particles which does not violate
symmetry requirement --mainly decay modes of the original mesons. We would like to study the time evolution of a general initial state
(\ref{InitPsi}) \emph{inside} the 2 dimensional $P^0-\bar{P}^0$ subspace. To that end we split the full Hamiltonian into a part $H_0\equiv H_{\Delta
F=0}$ and a part $H_W\equiv H_{\Delta F\ne0}$, $H=H_0+H_W$, and we follow a time-dependent perturbation theory formalism treating $H_W$ as a small
correction.

In the interaction picture, the state vectors and operators are defined as follows
\eqa{
\ket{\psi(t)}_I&=&e^{iH_0t}\ket{\psi(t)}_S\\
O_I(t)&=&e^{iH_0t}O_S\,e^{-iH_0t} }
The time evolution in this picture defines the evolution operator $U_I(t)$ through
\eq{ \ket{\psi(t)}_I=U_I(t)\ket{\psi(0)}, }
so that the evolution of the coefficients $c(t)$ and $\bar{c}(t)$ is given \emph{exactly} by
\eq{ \left\{
\begin{array}{l}
c(t)=\bra{P^0}U_I(t)\ket{\psi(0)}=c\,\bra{P^0}U_I(t)\ket{P^0}+\bar{c}\,\bra{P^0}U_I(t)\ket{\bar{P}^0}\\
\bar{c}(t)=\bra{\bar{P}^0}U_I(t)\ket{\psi(0)}=c\,\bra{\bar{P}^0}U_I(t)\ket{P^0}+\bar{c}\,\bra{\bar{P}^0}U_I(t)\ket{\bar{P}^0}
\end{array}
\right. \label{ccbar} }

The unitary operator $U_I(t)$ verifies the Schrödinger equation
\eq{ i\partial_tU_I(t)=H_W(t)\cdot U_I(t), \label{SEIP} }
where $H_W(t)\equiv (H_W)_I$. The solution of (\ref{SEIP}) with initial condition $U_I(0)=1$ can be computed perturbatively
\eq{ U_I(t)=1-i\int_0^t \!dt' H_W(t')-\int_0^t \!dt'\int_0^{t''} \!dt' H_W(t')H_W(t'')+ \cdots \label{UI} }

We are interested only in the time evolution of the \emph{projection} of the full state $\ket{\psi(t)}$ on the two-dimensional subspace of the
$P^0-\bar{P}^0$ system. This evolution is NOT unitary, and effectively is described by a Schrödinger-like equation:
\eq{ i\partial_t \left(
\begin{array}{c}
\!\!c(t)\!\!\\
\!\!\bar{c}(t)\!\!
\end{array}
\right) = \mathcal{H}_{\rm eff} \left(
\begin{array}{c}
\!\!c(t)\!\!\\
\!\!\bar{c}(t)\!\!
\end{array}
\right) \equiv \left(
\begin{array}{cc}
\bra{P^0}\hat{\mathcal{H}}_{\rm eff}\ket{P^0} & \bra{P^0}\hat{\mathcal{H}}_{\rm eff}\ket{\bar{P}^0} \\
\bra{\bar{P}^0}\hat{\mathcal{H}}_{\rm eff}\ket{P^0} & \bra{\bar{P}^0}\hat{\mathcal{H}}_{\rm eff}\ket{\bar{P}^0}
\end{array}
\right) \left(
\begin{array}{c}
\!\!c(t)\!\!\\
\!\!\bar{c}(t)\!\!
\end{array}
\right) \label{effSE} }
This is an effective description valid at time scales much larger than the typical strong interaction scale, and it's called the
\emph{Wigner-Weisskopf approximation} \cite{Weisskopf:1930au}. In particular, the effective Hamiltonian $\mathcal{H}_{\rm eff}$ will not be
hermitian.

By introducing (\ref{ccbar}) in (\ref{effSE}), and using (\ref{UI}), we find an equation for $\mathcal{H}_{\rm eff}$:
\eqa{
\bra{a}H_W(t)\ket{b}-i\bra{a}H_W(t)\int_0^t dt'H_W(t')\ket{b}+\cdots\hspace{4cm} \nonumber\\
=\bra{a}\mathcal{H}_{\rm eff}\ket{b}-i\bra{a}\sum_m \mathcal{H}_{\rm eff} \ket{m}\bra{m}\int_0^t dt'H_W(t')\ket{b}+\cdots }
where $a,b,m\in \{P^0,\bar{P}^0\}$. We can solve this equation order by order in $H_W$. For our purpose here it will be enough to stop at second
order:
\eq{ \mathcal{H}_{\rm eff}=H_W+\sum_n H_W\ket{n}\bra{n}H_W \left( \mathcal{P}\frac{1}{m_P-E_n+i\epsilon}-i\pi\delta(m_P-E_n)\right)+\cdots
\label{Heff(HW)} }
Here, $\mathcal{P}$ stands for the principal part prescription, and the sum runs over all intermediate states $n$. $m_P$ and $E_n$ are the energies
in the center of mass frame defined as $H_0\ket{P^0}=m_P\ket{P^0}$, $H_0\ket{\bar{P}^0}=m_P\ket{\bar{P}^0}$ and $H_0\ket{n}=E_n\ket{n}$.

Up to this point we introduce two \emph{hermitian} operators $M$ and $\Gamma$ defined as
\eqa{ M &\equiv& \frac{1}{2}(\mathcal{H}_{\rm eff}+\mathcal{H}_{\rm
eff}^\dagger)\nonumber\\
\Gamma &\equiv& i(\mathcal{H}_{\rm eff}-\mathcal{H}_{\rm eff}^\dagger). \label{M,Gamma} }
By inversion of (\ref{M,Gamma}) we can write the effective Hamiltonian in terms of $M$ and $\Gamma$,
\eq{ \mathcal{H}_{\rm eff}=M-\frac{i}{2}\Gamma=\left(
\begin{array}{cc}
M_{11}-\frac{i}{2}\Gamma_{11} & M_{12}-\frac{i}{2}\Gamma_{12}\\[2pt]
M_{21}-\frac{i}{2}\Gamma_{21} & M_{22}-\frac{i}{2}\Gamma_{22}
\end{array} \right) \label{M-iGamma}
}
and the explicit matrix elements for $M$ and $\Gamma$ can be read off from equations (\ref{Heff(HW)}) and (\ref{M,Gamma}),
\eqa{
M_{11}&=&m_P+\sum_n \mathcal{P}\frac{\m{\bra{n}H_W^{\scriptscriptstyle\Delta F=1}\ket{P^0}}^2}{m_P-E_n}\nonumber \\
M_{22}&=&m_P+\sum_n \mathcal{P}\frac{\m{\bra{n}H_W^{\scriptscriptstyle\Delta F=1}\ket{\bar{P}^0}}^2}{m_P-E_n}\nonumber \\
M_{12}&=&M_{21}^*=\bra{P^0}H_W^{\scriptscriptstyle\Delta F=2}\ket{\bar{P}^0}+
\sum_n \mathcal{P}\frac{\bra{P^0}H_W^{\scriptscriptstyle\Delta F=1}\ket{n}\bra{n}H_W^{\scriptscriptstyle\Delta F=1}\ket{\bar{P}^0}}{m_P-E_n}\nonumber \\
\Gamma_{11}&=&2\pi \sum_n \delta(m_P-E_n)\m{\bra{n}H_W^{\scriptscriptstyle\Delta F=1}\ket{P^0}}^2 \nonumber\\
\Gamma_{22}&=&2\pi \sum_n \delta(m_P-E_n)\m{\bra{n}H_W^{\scriptscriptstyle\Delta F=1}\ket{\bar{P}^0}}^2 \nonumber\\
\Gamma_{12}&=&\Gamma_{21}^*=2\pi \sum_n \delta(m_P-E_n)\bra{P^0}H_W^{\scriptscriptstyle\Delta F=1}\ket{n}\bra{n}H_W^{\scriptscriptstyle\Delta F=1}
\ket{\bar{P}^0} \label{M's y Gamma's} }

Assuming CPT as a symmetry of $H_W$ leads to the following relations
\eq{ M_{11}=M_{22}\equiv M_0\ ,\quad \Gamma_{11}=\Gamma_{22}\equiv \Gamma_0.\label{CPT} }

\subsection{Mass eigenstates: Diagonalizing the effective Hamiltonian}

The mesons $P^0$ and $\bar{P}^0$ are charge-conjugated states, and hence they differ only by the sign of their internal additive quantum numbers.
They are eigenstates of $H_0$, which conserves such quantum numbers, but not of $H_W$. Therefore these are not physical states (or mass eigenstates)
with the corresponding consequence of mixing and decay. There is a feature, however, that decouples the process of \emph{mixing} with that of
\emph{decay}, and allows to treat them separately: oscillations are unitary in the two dimensional subspace, but decay is not. A new basis
$\{P_L,P_H\}$ which diagonalizes $\mathcal{H}_{\rm eff}$ defines two fields that do not oscillate, just decay. Of course the question arises of
whether $M$ and $\Gamma$ can be diagonalized simultaneously. In general they cannot, so no unitary transformation can diagonalize $\mathcal{H}_{\rm
eff}$, and the states $\{P_L,P_H\}$ will not be orthogonal. The formalism to deal with this issue was developed by T.D. Lee and L. Wolfenstein in
1965 \cite{Lee:1965hi}.

To simplify the notation, we parameterize the effective Hamiltonian in terms of three complex numbers, $A$, $B$ and $r$:
\eq{\mathcal{H}_{\rm eff}=\left(
\begin{array}{cc}
A & B/r\\
r B & A \label{Heff(A,B,r)}
\end{array}
\right) }
where CPT has been assumed through (\ref{CPT}). By comparing (\ref{Heff(A,B,r)}) with (\ref{M-iGamma}) we see that \footnote{In the standard
literature (see for example \cite{Bigi:2000yz,Harrison:1998yr}), what we call $r$ is denoted by $(q/p)$, with
\eq{ p=\frac{1}{\sqrt{1+\m{r}^2}}\ \ , \quad q=\frac{r}{\sqrt{1+\m{r}^2}}.\nonumber }}
\eqa{
A&=&M_0-\frac{i}{2}\Gamma_0\\
B&=&\sqrt{(M_{12}-\frac{i}{2}\Gamma_{12})(M_{12}^*-\frac{i}{2}\Gamma_{12}^*)} \label{B}\\
r&=&\sqrt{\frac{M_{12}^*-\frac{i}{2}\Gamma_{12}^*}{M_{12}-\frac{i}{2}\Gamma_{12}} \label{r}} }

The eigenvalues and eigenstates are then
\eqa{
\mathcal{H}_{\rm eff}\ket{P_L}&=&(A+B)\ket{P_L}\equiv \big(M_L-\frac{i}{2}\Gamma_L\big)\ket{P_L}\nonumber\\
\mathcal{H}_{\rm eff}\ket{P_H}&=&(A-B)\ket{P_H}\equiv \big(M_H-\frac{i}{2}\Gamma_H\big)\ket{P_H} \label{massEigenvls} }
\eqa{
\ket{P_L}&=&\frac{1}{\sqrt{1+\m{r}^2}}\big(\ket{P^0}+r\ket{\bar{P}^0}\big)\nonumber\\
\ket{P_H}&=&\frac{1}{\sqrt{1+\m{r}^2}}\big(\ket{P^0}-r\ket{\bar{P}^0}\big) \label{massEigens} }

As mentioned above, the mass eigenstates $\{P_L,P_H\}$ are not necessarily orthogonal. This can be easily seen by inspection of (\ref{massEigens}),
since the non-hermiticity of $\mathcal{H}_{\rm eff}$ allows $\m{r}\ne 1$. However, the amount of non-orthogonality is constrained by the exclusive
widths of $P_L$ and $P_H$. The proof goes as follows. Consider the braket
\eq{\bra{P_H}\Gamma\ket{P_L}=\bra{P_H} i(\mathcal{H}_{\rm eff}-\mathcal{H}_{\rm eff}^\dagger) \ket{P_L}=\Big[
\frac{1}{2}(\Gamma_L+\Gamma_H)+i(M_L-M_H)\Big]\braket{P_H}{P_L} }
This quantity can be calculated from (\ref{M's y Gamma's}),
\eq{\bra{P_H}\Gamma\ket{P_L}=2\pi\sum_f\delta(m_P-E_f)\bra{f}H\ket{P_H}^*\bra{f}H\ket{P_L}\,.}
We denote $\bra{f}H\ket{P}\equiv \Gamma^f$. Then by means of the Schwartz inequality, $\mo{\bra{P_H}\Gamma\ket{P_L}}^2\le\sum_f\Gamma_L^f\Gamma_H^f$,
we find that
\eq{\m{\braket{P_H}{P_L}}\le \sqrt{\frac{\sum_f 4\Gamma_L^f\Gamma_H^f}{(\Gamma_L+\Gamma_H)^2+4(M_L-M_H)^2}}\ \ . \label{Bell-Steinberger} }

The equation (\ref{Bell-Steinberger}) is called the \emph{Bell-Steinberger inequality}. Note that if $H_W^{\sss\Delta F=1}=0$, then the states do not
decay, only mix (if $H_W^{\sss\Delta F=2}\ne 0$, of course). Therefore $\Gamma^f=0$, but still $M_L\ne M_H$, so $\braket{P_H}{P_L}=0$. This would
correspond to a conventional unitary mixing.

\subsection{Time evolution}

We denote by $\ket{P(t)}$ the state at time $t$ that at $t=0$ was a pure $\ket{P}$. The time evolution of the flavor states $P^0,\bar{P^0}$ is
complicated because of the mixing, in the sense that the states $\ket{P^0(t)}$ and $\ket{\bar{P}^0(t)}$ will be superpositions of $P^0$ and
$\bar{P^0}$ at $t>0$. However, the time evolution of the mass eigenstates is quite simple since they diagonalize the effective Hamiltonian. From
(\ref{effSE}) and (\ref{massEigenvls}) we find the usual time evolution for decaying stationary states:
\eq{\ket{P_{L,H}(t)}=e^{-iM_{L,H}t}e^{-\frac{1}{2}\Gamma_{L,H}t}\ket{P_{L,H}} }
By inversion of (\ref{massEigens}) we can find the time evolution of the states $P^0,\bar{P^0}$,
\eqa{
\ket{P^0(t)}&=&g_+(t)\ket{P^0}+rg_-(t)\ket{\bar{P}^0}\nonumber\\
\ket{\bar{P}^0(t)}&=&g_+(t)\ket{\bar{P}^0}+\frac{1}{r}g_-(t)\ket{P^0} \label{P(t)} }
where, to avoid cumbersome expressions, we have defined
\eq{ g_{\pm}(t)=\frac{1}{2} e^{-iM_Lt}e^{-\frac{1}{2}\Gamma_Lt}\big(1\pm e^{-i\Delta\!M\,t}e^{\frac{1}{2}\Delta\Gamma\,t}\big) \label{g(t)}}
and\footnote{It's not clear yet from the construction if $\Delta M$ and $\Delta\Gamma$ are positive or negative. It is a matter of the choice made in
(\ref{massEigenvls}) as to which mass eigenstate should we call $\ket{P_H}$ or $\ket{P_L}$. We can take by convention to call the heaviest one with
the subscript ``H'' -- not necessarily as in (\ref{massEigenvls})-- so that $\Delta M > 0$. But once this convention is taken, it becomes an
empirical question whether $\Delta\Gamma$ is positive or negative.}
\eq{\Delta M\equiv M_H-M_L,\quad \Delta\Gamma\equiv\Gamma_L-\Gamma_H. \label{DeltaM,DeltaGamma}}

We are interested in the time evolution of observables such as branching ratios, decay rates, and CP asymmetries. We define $A_f$ and $\bar{A}_f$ as
the amplitudes for the decay of $P^0$ and $\bar{P}^0$ into a final state {\it f} , i.e.,
\eq{A_f\equiv \bra{f}H_W^{\scriptscriptstyle\Delta F=1}\ket{P^0}\ ,\quad \bar{A}_f\equiv \bra{f}H_W^{\scriptscriptstyle\Delta F=1}\ket{\bar{P}^0}
\label{A(f)}}

The decay rates are proportional to the square of the time dependent decay amplitudes, the proportionality factor given by a phase space factor
$f_{\sss \textrm{PS}}$ (see the Appendix at the end of this chapter). From (\ref{Gamma=fps*A}), (\ref{P(t)}) and (\ref{A(f)}) we find the following
master equations for the time dependent decay rates,
\eqa{ \Gamma(P^0(t)\!\rightarrow\!f)&\!\!\!\!=&\!\!\!\!\frac{1}{4}\,f_{\sss \textrm{PS}}\, e^{-\Gamma_Lt}\,\m{A_f}^2\Big[ K_+(t) +
K_-(t)\,\m{\lambda_f}^2 +
2\textrm{Re}\big( L(t)\,\lambda_f \big)\Big] \label{master1}\\
\Gamma(\bar{P}^0(t)\!\rightarrow\!f)&\!\!\!\!=&\!\!\!\!\frac{1}{4}\,f_{\sss \textrm{PS}}\, e^{-\Gamma_Lt}\,\m{A_f}^2\mo{\frac{1}{r}}^2\Big[
K_+(t)\,\m{\lambda_f}^2 + K_-(t) + 2\textrm{Re}\big( L(t)\,\lambda_f^* \big)\Big]\label{master2} }
where
\eqa{
K_{\pm}(t)&\equiv&1+e^{\Delta\Gamma t}\pm 2e^{\frac{1}{2}\Delta\Gamma t}\cos{\Delta\! Mt}\nonumber\\
L(t)&\equiv&1-e^{\Delta\Gamma t}+ 2ie^{\frac{1}{2}\Delta\Gamma t}\sin{\Delta\! Mt} }
and
\eq{\lambda_f\equiv\,r\,\frac{\bar{A}_f}{A_f}\label{lambda_f}}

\subsection{CP Violation in the neutral meson system} \label{CPV in the neutral meson system}

Up to now all the discussion about the neutral meson system did not make any reference to CP symmetry. All the previous results are independent of
whether CP is or isn't a good symmetry of the full Hamiltonian. However, it turns out that some of the quantities defined so far are closely related
to the amount of CP violation. Therefore, the study of CP violation is very useful when extracting physical content to the formalism above. On the
other hand, the study of such systems provides a lot of information about the nature of CP violation, to the extent that CP violation can be
quantified even from CP conserving processes.

The role of CP-violation in the mixing of neutral mesons can be identified by answering the following question: How are the results above modified
(or simplified) if we impose CP as a symmetry? The answer is twofold:
\begin{itemize}
\item We must identify the mass eigenstates (\ref{massEigens}) with the CP eigenstates. The action of CP on the flavor eigenstates
is\footnote{The phase $\xi_P$ is convention dependent and hence unphysical. Because the convention in the election of such phase is related to other
conventions in the definition of other quantities used in the text, it is useful to keep track of it for the moment.}
\eq{CP\ket{P^0}=e^{i\xi_P}\ket{\bar{P}^0}\,,\quad CP\ket{\bar{P}^0}=e^{-i\xi_P}\ket{P^0}\label{CP|P>}}

so the CP eigenstates (defined as $CP\ket{P_\pm}=\pm P_\pm$) are
\eqa{
\ket{P_+}&=&\frac{1}{\sqrt{2}}\big(\ket{P^0}+e^{i\xi_P}\ket{\bar{P}^0}\big)\nonumber\\
\ket{P_-}&=&\frac{1}{\sqrt{2}}\big(\ket{P^0}-e^{i\xi_P}\ket{\bar{P}^0}\big) \label{CPEigens} }

Comparing (\ref{massEigens}) with (\ref{CPEigens}) we see that in the limit of CP invariance, $r=e^{i\xi_P}$. This is a convention dependent
quantity, so it is not quite interesting. However, its modulus
\eq{\m{r}\stackrel{\sss CP}{\longrightarrow}1 \label{rcp}}

it's an observable, and its deviation from 1 measures CP violation in the mixing of neutral mesons.

\item The amplitudes $A_f$ and $\bar{A}_{\bar{f}}$ defined in (\ref{A(f)}) are equal under CP invariance up to
an unphysical phase:
\eq{ \bar{A}_{\bar{f}}=\bra{\bar{f}}H\ket{\bar{P}^0}=\bra{\bar{f}}CP^{\dagger}(CP H CP^{\dagger})CP\ket{\bar{P}^0} \stackrel{\sss
CP}{\longrightarrow}\bra{f}e^{i\xi_f}He^{-i\xi_P}\ket{P^0}=e^{i(\xi_f-\xi_P)}A_f \label{Abar=A}}

where the phase $\xi_f$ is the analogue to (\ref{CP|P>}) in the definition of $\ket{\bar{f}}$. The ratio of the moduli of these amplitudes is
independent of these phase conventions and its difference from unity is a measure of CP violation in the decay process,
\eq{\mo{\bar{A}_{\bar{f}}/A_f}\stackrel{\sss CP}{\longrightarrow}1 \label{Abar/A}}

In the absence of mixing, as in the case of charged meson decays, this is the only source of CP violation.\\
If $f_{\rm CP}$ is a CP eigenstate, i.e. $CP\ket{f_{\rm CP}}=\eta_f\ket{f_{\rm CP}}$ with $\eta_f=\pm 1$, then
\eq{ \bar{A}_{\bar{f}_{\rm CP}}=\bra{\bar{f}_{\rm CP}}H\ket{\bar{P}^0}=\eta_f e^{i\xi_f}\bra{f_{\rm CP}}H\ket{\bar{P}^0}= \eta_f
e^{i\xi_f}\bar{A}_{f_{\rm CP}} \label{A(fCP)}}

and comparing (\ref{Abar=A}) and (\ref{A(fCP)}) we find that
\eq{ \frac{\bar{A}_{f_{\rm CP}}}{A_{f_{\rm CP}}}\stackrel{\sss CP}{\longrightarrow}\eta_f\,e^{-i\xi_P} }

This ratio is convention dependent, as contains the phase $\xi_P$. But this dependence is cancelled by $r$ inside the quantity $\lambda_f$ defined in
(\ref{lambda_f})
\eq{\lambda_{f_{\rm CP}}\stackrel{\sss CP}{\longrightarrow}\eta_{f} \label{lambda_fcp}}

This is a physical quantity, whose modulus and phase (or equivalently its modulus and its imaginary part) signal CP violation in the decay and mixing
processes.

\end{itemize}

The results (\ref{rcp}), (\ref{Abar/A}) and (\ref{lambda_fcp}) contain most of the qualitative information on how CP is violated in meson decays.

\section{Classification of CP-violating effects}

Now that we have seen how CP manifests itself in physical quantities, we are ready to make a classification of CP-violating effects. Fortunately, the
neutral meson system is rich in CP-violating phenomenology, so mostly everything was said and done in section \ref{CPV in the neutral meson system}.
There are three main independent ways in which CP is broken in meson decays:

\subsection*{$\star$ {\bf \emph{CP Violation in decay: }} $\m{\bar{A}_{\bar{f}}/A_f}\ne 1$}

Consider the situation in which no oscillations occur, i.e.,
\eq{B=0 \quad {\rm or} \quad \Delta M=\Delta \Gamma =0}
Then, following from the master equations (\ref{master1}) and (\ref{master2}), we see that
\eqa{
\Gamma(P\rightarrow f)&=&f_{\sss \textrm{PS}}\,e^{-\Gamma_Lt}\,\m{A_f}^2 \nonumber \\
\Gamma(\bar{P}\rightarrow \bar{f})&=&f_{\sss \textrm{PS}}\,e^{-\Gamma_Lt}\,\m{\bar{A}_{\bar{f}}}^2 }

A difference in the CP conjugated decay rates $\Gamma(P\rightarrow f)\ne \Gamma(\bar{P}\rightarrow \bar{f})$ is a signal of CP violation, and arises
if
\eq{\mo{\bar{A}_{\bar{f}}/A_f}\ne 1}
as advanced in (\ref{Abar/A}). This type of CP violation is the only possible one in charged meson decays, where mixing effects are absent, and hence
CP violation in decay is best measured in these modes. The relevant asymmetry is then
\eq{\mathcal{A}_{\rm CP}(P^\pm\rightarrow f^\pm)\equiv\frac{\Gamma(P^+\rightarrow f^+)-\Gamma(P^-\rightarrow f^-)} {\Gamma(P^+\rightarrow
f^+)+\Gamma(P^-\rightarrow f^-)}=\frac{1-\m{\bar{A}_{f^-}/A_{f^+}}^2}{1+\m{\bar{A}_{f^-}/A_{f^+}}^2}}

\subsection*{$\star$ {\bf \emph{CP Violation in mixing: }} $\m{r}\ne 1$}

Consider a final state $f$ that can only come from $P^0$, not from $\bar{P}^0$, and also the CP analogue, that is
\eq{A_{\bar{f}}=\bar{A}_f=0}
These are called \emph{flavor-specific} decays. Then, the transitions $P^0(t)\to \bar{f}$ and $\bar{P}^0(t)\to f$ are only possible because of
mixing; for example $P^0(t)\to \bar{f}=P^0\stackrel{t}{\to}\bar{P}^0\to \bar{f}$. These transitions are called ``wrong-sign'' decays.  The wrong-sign
decay rates, according to (\ref{master1}) and (\ref{master2}), are given by
\eqa{\Gamma(P^0(t)\rightarrow \bar{f})&=&\frac{1}{4}\,f_{\sss \textrm{PS}}\,
e^{-\Gamma_Lt} K_-(t)\,\m{r}^2\m{\bar{A}_{\bar{f}}}^2\ , \nonumber \\
\Gamma(\bar{P}^0(t)\rightarrow f)&=&\frac{1}{4}\,f_{\sss \textrm{PS}}\, e^{-\Gamma_Lt}\,K_-(t)\mo{\frac{1}{r}}^2\!\m{A_f}^2\ . }

Consider also that there is no CP violation in decay, such that $|A_f|=|\bar{A}_{\bar{f}}|$. The standard example of this situation is the case of
charged-current semileptonic neutral meson decays. These are decays
 \mbox{$P^0,\bar{P}^0\rightarrow l^\pm X$}, which in the Standard Model verify $\m{A_{l^+X}}=\m{\bar{A}_{l^-X}}$ and
 $A_{l^-X}=\bar{A}_{l^+X}=0\ $ to lowest order in $G_F$.
The CP asymmetry in this case is given by
\eq{\mathcal{A}_{\rm SL}\equiv\frac{\Gamma(P^0(t)\rightarrow \bar{f})-\Gamma(\bar{P}^0(t)\rightarrow f)} {\Gamma(P^0(t)\rightarrow
\bar{f})+\Gamma(\bar{P}^0(t)\rightarrow f)}=\frac{1-\m{r}^4}{1+\m{r}^4}}
which is nonzero whenever
\eq{\m{r}\ne 1}
in agreement with (\ref{rcp}). Note that in this case the CP asymmetry between time-dependent decay rates is actually time-independent. These CP
asymmetries are suited for the extraction of the mixing parameter $|r|$.

\subsection*{$\star$ {\bf \emph{CPV in the interference between mixing and decay: }} $Im(\lambda_f)\ne 0$}

In the case of flavor non-specific decays, that is when a final state $f$ can be reached by both $P^0$ and $\bar{P}^0$, two amplitudes interfere in
the process:
\eq{A(P^0(t)\rightarrow f)\sim A(P^0\rightarrow f)+A(P^0\rightarrow \bar{P}^0\rightarrow f)}

Even if neither the decay itself nor the mixing introduce CP violation, the interference between these two decay channels can produce a nonzero CP
asymmetry. The most transparent case arises in the situation in which the final state is a CP eigenstate $f_{\rm CP}$. In the absence of CP violation
in mixing and decay, $\m{\lambda_{f_{\rm CP}}}=1$ holds. Still a deviation from (\ref{lambda_fcp}) can arise from the phase of $\lambda_{f_{\rm
CP}}$. In order to fix notation we define the \emph{mixing angle} $\phi_{\sss M}$ and the \emph{decay angle} $\phi_{\sss D}$ as\footnote{This
definition for $\phi_{\sss M}$ corresponds to $\phi_d=2\beta$ in $B^{\sss 0}_d-\bar{B}^{\sss 0}_d$ mixing, as will become clear later.}
\eq{\phi_{\sss M}\equiv \arg{(r^*)}\ ,\qquad\phi_{\sss D}\equiv \arg{(A_f/\bar{A}_f)}\label{phi_M,phi_D}}
so that
\eq{\arg{(\lambda_{f_{\rm CP}})}=-(\phi_{\sss M}\!\!+\!\!\phi_{\sss D})}
In this situation, eqs. (\ref{master1}) and (\ref{master2}) imply
\eqa{\Gamma(P^0(t)\rightarrow f_{\rm CP})&=&\frac{1}{2}\,f_{\sss \textrm{PS}}\,
e^{-\Gamma_Lt} \m{A_{f_{\rm CP}}}^2 \Big( 1+e^{\Delta \Gamma t}+{\rm Re}\big(L(t)\,e^{-i(\phi_{\sss M}+\phi_{\sss D})}\big) \Big) \nonumber \\
\Gamma(\bar{P}^0(t)\rightarrow \bar{f}_{\rm CP})&=&\frac{1}{2}\,f_{\sss \textrm{PS}}\, e^{-\Gamma_Lt} \m{A_{f_{\rm CP}}}^2 \Big( 1+e^{\Delta \Gamma
t}+{\rm Re}\big(L(t)\,e^{i(\phi_{\sss M}+\phi_{\sss D})}\big) \Big) }
The CP asymmetry is then
\eq{\mathcal{A}_{\rm CP}(t)\equiv \frac{\Gamma(P^0(t)\rightarrow f_{\rm CP})-\Gamma(\bar{P}^0(t)\rightarrow \bar{f}_{\rm CP})}
{\Gamma(P^0(t)\rightarrow f_{\rm CP})+\Gamma(\bar{P}^0(t)\rightarrow \bar{f}_{\rm CP})}= \frac{\sin{(\phi_{\sss M}\!\!+\!\!\phi_{\sss D})}\sin{\Delta
M t}}{\cosh{(\frac{1}{2}\Delta \Gamma t)}- \sinh{(\frac{1}{2}\Delta \Gamma t)}\cos{(\phi_{\sss M}\!\!+\!\!\phi_{\sss D})}}}
This time-dependent CP asymmetry is nonzero only in the presence of $P^0-\bar{P}^0$ oscillations generating $\Delta M\ne 0$ and whenever
\eq{{\rm Im}(\lambda_f)\ne 0}

\section{Strong and Weak phases}

\label{StrongAndWeakPhases}

Consider a decay process and its CP conjugate. If CP is not conserved, the two amplitudes don't need to be correlated and can be completely different
in modulus and phase. Two arbitrary complex numbers $A_f$ and $\bar{A}_{\bar{f}}$ can always be decomposed in the following way:
\eqa{
A_f&=&\m{a_1}e^{i\delta_1}e^{i\phi_1}+\m{a_2}e^{i\delta_2}e^{i\phi_2}\nonumber\\
\bar{A}_{\bar{f}}&=&\m{a_1}e^{i\delta_1}e^{-i\phi_1}+\m{a_2}e^{i\delta_2}e^{-i\phi_2} \label{CPphaseDeco} }

So the amount of CP violation can be encoded inside the so called CP violating phases $\phi_i$, which change sign under CP. These phases arise from
complex parameters in the Lagrangian that contribute to the amplitudes, since the couplings in the Lagrangian appear in complex conjugate form in CP
conjugate amplitudes (see for instance \cite{Harrison:1998yr}). In the Standard Model complex parameters appear only in the weak sector -- through
Yukawa couplings--, so these CP violating phases are due to weak interaction and therefore are usually called \emph{weak phases}.

The CP conserving phases $(\delta_1\!-\!\delta_2\ne 0)$ are necessary (together with $\phi_1\!-\!\phi_2\ne 0$) to account for a difference in modulus
between the two CP conjugate amplitudes. They arise even if the Lagrangian is real and generally through a process know as \emph{rescattering}. This
is a contribution from a possible final state interaction, with on-shell intermediate states. In the Standard Model the main source of CP conserving
phases is the strong interaction, so these phases are commonly referred to as \emph{strong phases}.

Obviously, the weak and strong phases by themselves are not convention independent, since they depend (at least) on the choice made for the phases
$\xi_{P,f}$. However, the relative strong and weak phases between the different terms in the amplitudes are physical.

The decomposition (\ref{CPphaseDeco}) is particularly explicit in the SM. As can be seen from the $\Delta F=1$ SM effective Hamiltonian in eq.
(\ref{HeffSM}), the $\Delta F=1$ amplitudes are the sum of a term proportional to $\lambda_u^{\sss (D)}$ and a term proportional to $\lambda_c^{\sss
(D)}$. This is a consequence of CKM unitarity as will become clear in Chapter \ref{Bdecays}. The factors $\lambda_{u,c}^{\sss (D)}$ contain all the
CKM information, so the CP violating phases are all contained in these factors. On the other hand, the matrix elements of the operators will contain
strong interaction phases, as explained in Chapter \ref{HME}. At the end, the two CP conjugated amplitudes can be written in full generality as
\eq{A=\lambda_u^{\sss (D)*}\,T + \lambda_c^{\sss (D)*}\,P\ ,\qquad \bar{A}=\lambda_u^{\sss (D)}\,T + \lambda_c^{\sss (D)}\,P\ , \label{CKMdeco}}
where $T=|T|e^{i\delta_T}$ and $P=|P|e^{i\delta_P}$ are complex hadronic parameters, called ``tree'' and ``penguin'', and $\delta_{T,P}$ are strong
phases. This representation of the amplitudes is exactly of the form of eq. (\ref{CPphaseDeco}). We shall go very much into detail with this
decomposition in the following chapters.

\section{CP Violation in B decays}

\label{sectionCPB}

The discussion about CP violation in the previous sections is quite general and describes within the same framework all mesons of different families,
such as $K$'s, $D$'s, $B$'s, etc. However, the variety of values that nature has chosen for their lifetimes, masses and oscillation patterns results
in very different phenomenologies. The phenomenology of $B$ decays is particularly interesting and contains a number of characteristic features
\cite{Bigi:2000yz}:

\begin{itemize}
\item Nature has provided us with two neutral $B$ meson systems, with a good measure of similarities and differences. The
mass differences for both systems are \cite{HFAG,Abulencia:2006ze}
\eqa{
&&\Delta M_d=0.507\pm 0.004\ {\rm ps}^{-1} \label{DeltaMd_exp}\\
&&\Delta M_s=17.77\pm 0.12\ {\rm ps}^{-1} \label{DeltaMs_exp} }

\item CP violation in mixing can be ignored completely in present analyses of $B$ decays, both in the $B_d$ and $B_s$ systems.
This statement is based on recent measurements of the CP violating parameter $r$ \cite{HFAG}:
\eqa{
&&\m{r}_d=1.0033\pm 0.0017 \label{rd_exp}\\
&&\m{r}_s=0.9998\pm 0.0046 \label{rs_exp} }

\item The width difference in the $B_d$ system is found to be small. While an initial beam of $K^{\sss 0}$ and $\bar{K}^{\sss 0}$ mesons
is transformed with time into a practically pure $K_L$ beam, this does not happen with $B_d/\bar{B}_d$ beams in any appreciable way. The width
difference for the $B_s$ system does not share the same fate, and this introduces an interesting difference between the two different systems of
neutral $B$ mesons. Recent data for the lifetime differences are \cite{HFAG}:
\eqa{
&&\Delta\Gamma_d/\Gamma_d=0.009\pm 0.037\pm 0.018 \label{DeltaGammad_exp}\\
&&\Delta\Gamma_s/\Gamma_s=0.104^{+0.076}_{-0.084} \label{DeltaGammas_exp} }

\end{itemize}

\begin{figure}
\begin{center}
\psfrag{xd}{\footnotesize\hspace{-0.2cm}$t/\tau_{\sss B_d}$} \psfrag{xs}{\footnotesize\hspace{0.4cm}$t/\tau_{\sss B_s}$} \psfrag{yd}{\footnotesize
\hspace{-1.2cm}$\mathcal{P}(B_d^0(t)=B_d^0)\ (\%)$} \psfrag{ys}{\footnotesize \hspace{-1.2cm}$\mathcal{P}(B_s^0(t)=B_s^0)\ (\%)$}
\includegraphics[width=6.5cm]{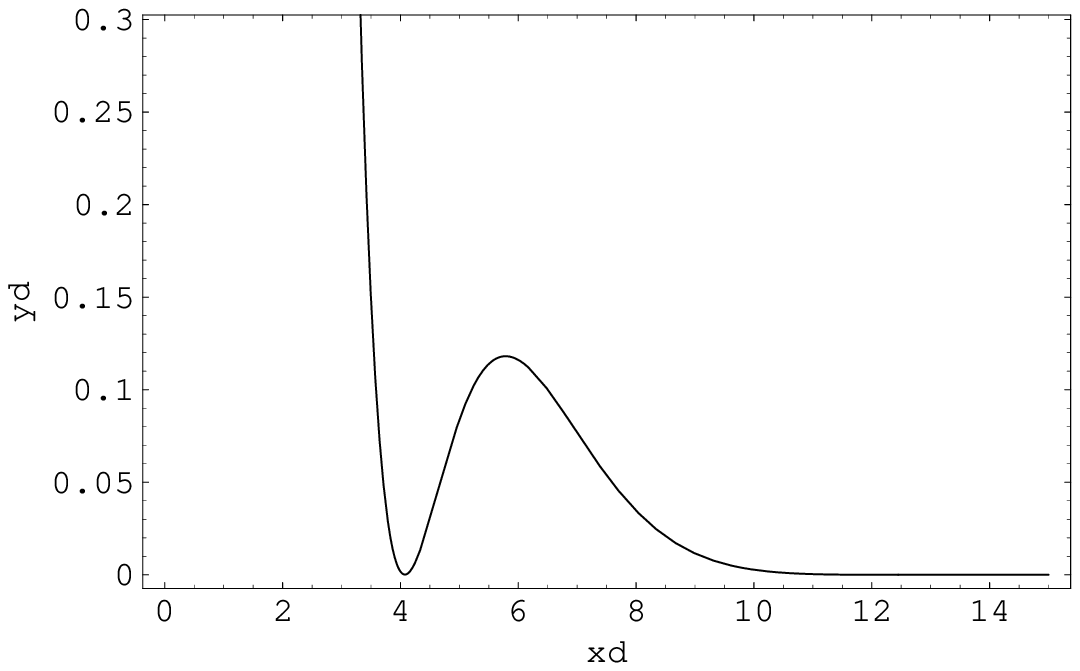}
\hspace{1cm}
\includegraphics[width=6.5cm]{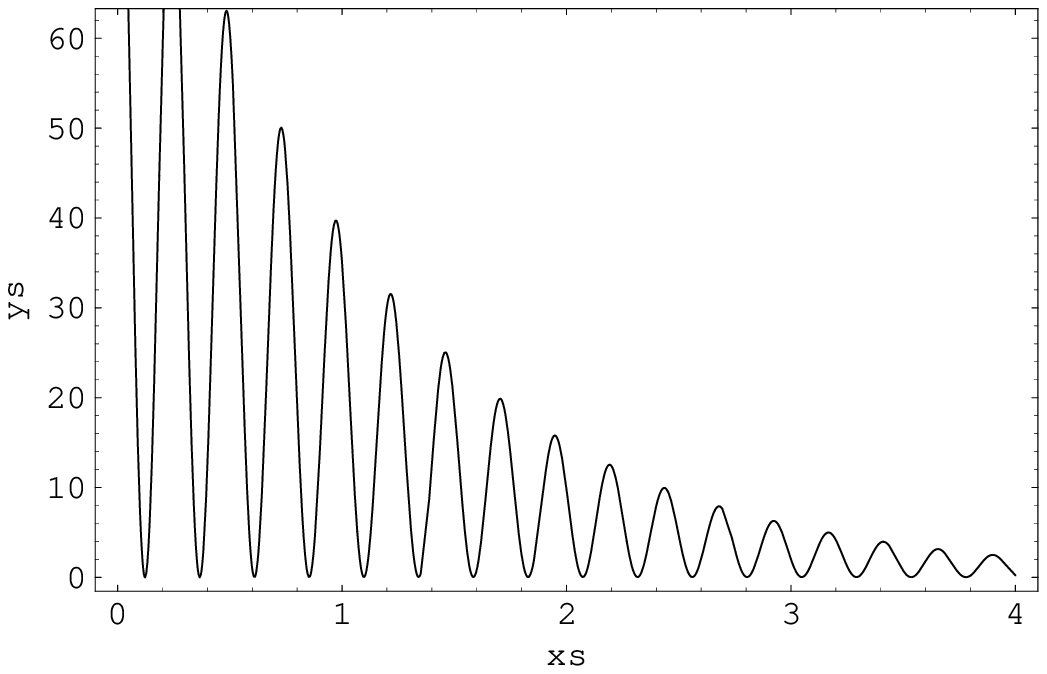}
\end{center}
\vspace{-0.5cm} \caption{\small Ocillation probabilities for $B_d$ and $B_s$ mesons. The time axis is given in units of the meson lifetimes. Clearly
the oscillation of the $B_s$ is much faster than the oscillation of the $B_d$ (note that they are not drawn to the same scale).} \label{figOsc}
\end{figure}

\begin{figure}
\begin{center}
\includegraphics[width=11cm,height=7cm]{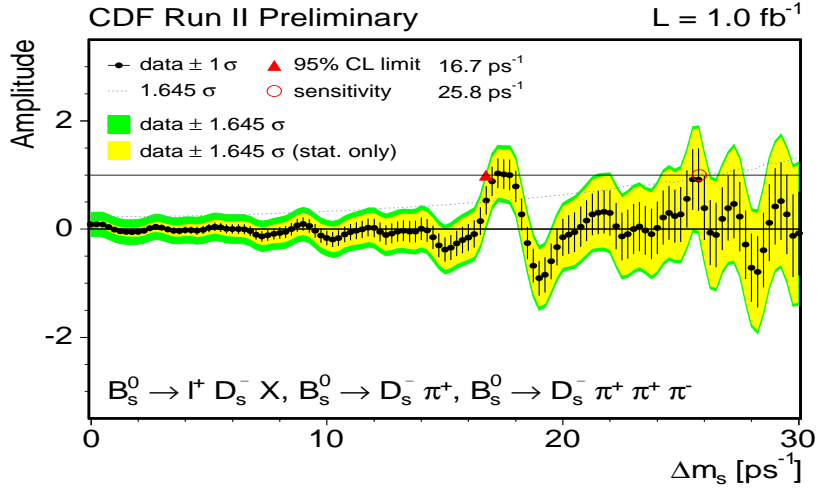}
\end{center}
\vspace{-0.7cm} \caption{\small Amplitude of $B_s^0-\bar{B}_s^0$ oscillations as a function of the frequency $\Delta M_s$. These data proved that
$B_s$ mesons oscillate, and measured their mass difference.} \label{OscExp}
\end{figure}

After the exposition of Section \ref{Mixing}, it is easy to understand why it has taken much longer to measure $\Delta M_s$, which was just measured
last year, than to measure $\Delta M_d$, which was measured in the 80's. Of course one of the reasons is that the production of $B_d$ mesons was
possible very early, and the $B$-factories operating at the $\Upsilon (4S)$ resonance have produced millions of $B_d-\bar{B}_d$ pairs, while the
detailed study of $B_s$ mesons has only been possible recently at Tevatron. The other reason is that while the $B_d$ meson oscillates relatively
slowly, the oscillations of the $B_s$ meson are very fast, oscillating on average 25 times before decay. Indeed, looking at eq. (\ref{P(t)}), it can
be seen that the function $g_+(t)$ measures the frequency at which a $P^0$ or a $\bar{P}^0$ meson turns back into itself. From eq. (\ref{g(t)}), the
probability of finding a $P^0$ at time $t$ from an original $P^0$ is given by
\eq{\mathcal{P}(P^0(t)=P^0)=|g_+(t)|^2\simeq \frac{1}{2}e^{-\bar{\Gamma}t}(1+\cos{\Delta M t})}
where $\bar{\Gamma}$ is the average lifetime and the approximation is valid if $\Delta\Gamma \ll \Gamma$. Then, $\Delta M$ is the frequency of the
oscillations, so from (\ref{DeltaMd_exp}) and (\ref{DeltaMs_exp}) we see that $B_s$ mesons oscillate about 35 times faster than $B_d$ mesons. Figure
\ref{figOsc} shows the oscillation probabilities of both neutral mesons. Figure \ref{OscExp} shows the experimental data revealing the oscillation
frequency $\Delta M_s$.

\subsection{CP Asymmetries for $B$ decays}

The small CP violation in mixing pointed out by (\ref{rd_exp}) and (\ref{rs_exp}) tells that $r$ is then a pure phase to an excellent approximation
($r\simeq e^{-i\phi_{\sss M}}$, see (\ref{phi_M,phi_D})). The time dependent CP asymmetry for $B_{d,s}$ decays into CP eigenstates adopts now a quite
simple form
\eq{\mathcal{A}_{\rm CP}(t)=\frac{\mathcal{A}_{\rm CP}^{dir}\cos{(\Delta M t)}+\mathcal{A}_{\rm CP}^{mix}\sin{(\Delta M t)}} {\cosh{(\Delta \Gamma
t/2)}-\mathcal{A}_{\Delta\Gamma}\sinh{(\Delta \Gamma t/2)}}\label{AcpB2} }
which defines two quantities, the \emph{direct} and \emph{mixing induced} CP asymmetries, of capital importance in this thesis:
\eqa{
\mathcal{A}_{\rm CP}^{dir}\!\!&\equiv&\!\!\frac{1-\m{\lambda_f}^2}{1+\m{\lambda_f}^2}\label{Adir}\\[2pt]
\mathcal{A}_{\rm CP}^{mix}\!\!&\equiv&\!\!-\,\frac{2{\rm Im}\lambda_f}{1+\m{\lambda_f}^2}\label{Amix} }
and also
\eq{\mathcal{A}_{\Delta\Gamma}\equiv -\,\frac{2{\rm Re}\lambda_f}{1+\m{\lambda_f}^2}}
These three quantities are not independent, but they are related through
\eq{\m{\mathcal{A}_{\rm CP}^{dir}}^2+\m{\mathcal{A}_{\rm CP}^{mix}}^2+\m{\mathcal{A}_{\Delta\Gamma}}^2=1}

Let us take a look at the denominator in (\ref{AcpB2}). When measuring the direct and mixing induced CP asymmetries, the oscillation period is the
relevant time, $\tau_{\rm osc}\sim \Delta M^{-1}$. During an oscillation period the denominator in (\ref{AcpB2}) is
\eq{\cosh{(\Delta\Gamma\,\tau_{\rm osc}/2)}-\mathcal{A}_{\Delta\Gamma}\sinh{(\Delta\Gamma\,\tau_{\rm osc}/2)} \sim 1+\mathcal{O}{\textstyle \left(
\frac{\Delta\Gamma}{\Delta M} \right)}}
And $\Delta\Gamma/\Delta M$ is very small for both $B_{d,s}$ systems. The time dependent CP asymmetry for $B$ decays into final CP eigenstates can
then be written in the following form
\eq{\mathcal{A}_{\rm CP}(t)=\mathcal{A}_{\rm CP}^{dir}\cos{(\Delta M t)}+\mathcal{A}_{\rm CP}^{mix}\sin{(\Delta M t)} \label{AcpB} }
In the case in which the relevant time is the lifetime, this is also basically true for $B_d$ mesons, due to eq. (\ref{DeltaGammad_exp}). However,
the possibility of a large width difference for the $B_s$ system (e.g. (\ref{DeltaGammas_exp})), might make possible the measurement of the
asymmetries $\mathcal{A}_{\Delta\Gamma}$. This can be important for phenomenology, an example of which will be given in Chapter \ref{SymFac2}.

\subsection{The $B_{d,s}-\bar{B}_{d,s}$ mixing angles and the $\Delta M_{d,s}$ mass differences}

\label{mixingangles}

The experimental facts that $\Delta M_{d,s}\!\gg\! \Delta \Gamma_{d,s}$ and $|r|_{d,s}\simeq 1$, have implications on the value of
$\Gamma_{12}^{d,s}/M_{12}^{d,s}$. In fact, from eqs. (\ref{B}) and (\ref{r}) one can solve for the real and imaginary parts of
$\Gamma_{12}^{d,s}/M_{12}^{d,s}$ in terms of $|r|_{d,s}^4$ and $\Delta\Gamma_{d,s}/\Delta M_{d,s}$. In the limit $|r|_{d,s}\to 1$ and
$\Delta\Gamma_{d,s}/\Delta M_{d,s}\to 0$, there is a trivial solution, $|\Gamma_{12}^{d,s}/M_{12}^{d,s}|\to 0$. However, this is not the only
solution. The set of two nonlinear equations always gives two solutions. The first is the one close to zero, and the second is the one that makes
$|\Gamma_{12}^{d,s}/M_{12}^{d,s}|$ a huge number, which for real data is of order $\sim 10^3$. This second solution is not realistic; for example in
the SM, $\ |\Gamma_{12}/M_{12}|\sim m_b^2/m_t^2\sim 10^{-3}$. Therefore, data implies that
\eq{\left| \frac{\Gamma_{12}^{d,s}}{M_{12}^{d,s}} \right|\ll 1}

When this result is plugged in eqs. (\ref{B}) and (\ref{r}), one gets
\eq{B\simeq |M_{12}|\ ,\quad r\simeq \sqrt{\frac{M_{12}^*}{M_{12}}}\ ,}
such that
\eqa{
\Delta M_{d,s}\!\!&\simeq& 2\m{M_{12}^{d,s}} \label{Massdifference}\\
\phi_{d,s}&\simeq&\arg{(M_{12}^{d,s}) \label{Mixingangle}} }
This means that in order to compute the mixing angles and the mass differences, it is enough to compute the mixing parameters $M_{12}^{d,s}$.
Moreover, the expression (\ref{Mixingangle}) is sometimes taken as the definition of the mixing angle. One should keep in mind, however, that the
mixing angle, as defined in this chapter, is not a physical quantity because it is sensitive to unphysical phase redefinitions. Therefore, it is
important to work with a consistent convention for the weak phases everywhere; then the fact of assigning a numerical value to the mixing angle
becomes sensible.

In some particular cases, ona can extract the mixing angle in a very clean way from a mixing induced CP asymmetry. Consider, for example, a decay
into a CP eigenstate which is dominated by a \emph{single} amplitude, which means that the dominant parts of the amplitude have all the same weak
phase. Then, working in a convention in which the global weak phase is zero, one has
\eq{\lambda_f=e^{-i\phi_{\sss M}}\frac{\bar{A}_f}{A_f}=\eta_f e^{-i\phi_{\sss M}}\frac{\bar{A}_{\bar{f}}}{A_f}=\eta_f e^{-i\phi_{\sss M}}\ ,}
and hence $\ \mathcal{A}_{\rm CP}^{mix}=-{\rm Im}\lambda_f=\eta_f\sin{\phi_{\sss M}}$. The prominent example is the case of $B_d\to J/\psi K_s$
\cite{Bigi:1981qs}. Since this decay is dominated by a single amplitude, then $\bar{A}_{\overline{J/\psi K_s}}/A_{J/\psi K_s}\simeq \eta_{J/\psi
K_s}=-1$, and therefore,
\eq{-\mathcal{A}_{\rm CP}^{mix}(B_d\to J/\psi K_s)\simeq \sin{\phi_d}\ .}
The neglected amplitude is both CKM and $\alpha_s(m_b)$ suppressed with respect to the dominant amplitude, so the corrections to this equation are
below the percent (or even the per mil) level \cite{Boos:2004xp,Ciuchini:2005mg,Li:2006vq}.

\section{CP Violation in the Standard Model}

In this section we present the SM mechanism for CP violation; a simple and beautiful explanation formulated by M.~Kobayashi and T.~Maskawa in 1972
\cite{Kobayashi:1973fv}, only 5 years after the experimental discovery of CP violation. This formulation required also the existence of a third
family of quarks (which had not yet been discovered), so together with the positive evidence of CP violation, this was a genuine \emph{prediction} of
the existence of the bottom and the top quarks.

Apart from the strong-CP issue --which we will not address here, and in any instance is a negligible effect--, the CP violation in the SM comes from
the electroweak sector, and in particular from the Yukawa couplings. These couplings are described by the Glashow-Weinberg-Salam theory of EW
interactions \cite{Glashow:1961tr,Weinberg:1967tq,Salam}.

The GWS theory is a gauge theory based on the $SU(2)_L\otimes U(1)_Y$ group. The `L' in $SU(2)_L$ indicates that the representations of the fermion
fields depend on the chirality: Left-handed fermions are doublets and right-handed fermions are singlets. (This is how P violation is introduced in
the SM.) Then, the kinetic terms in the lagrangian for the gauge and fermion fields are
\eqa{ \mathcal{L}_{K,F}&=&\bar{\psi}_{k_L}\,(i\pa + g \A^a T^a + g' \Bs \ Y_{k_L})\,\psi_{k_L} + \bar{\psi}_{k_R}\,(i\pa + g' \Bs \
Y_{k_R})\,\psi_{k_R}
\label{KF}\\
\mathcal{L}_{K,G}&=&-\frac{1}{4}(F_A^{a,\,\mu\nu})^2 - \frac{1}{4}(F_B^{\mu\nu})^2 }
where $A_\mu$, $B_\mu$ are the gauge fields, $T^a$ are the generators of $SU(2)$ in the fundamental representation, and a sum is understood over the
left-handed and right-handed field species $k_{L,R}$. $F_{\mu\nu}$ are the usual field-strength tensors and $Y_k$ are the hypercharges of the fermion
fields. Both kinetic terms are invariant under the full gauge group.

At this point a phenomenological problem arise. First, the chiral character of the gauge group spoils the gauge invariance of the fermion mass terms,
so unbroken gauge symmetry forbids massive fermions in this theory. Second, gauge invariance forbids also mass terms for gauge bosons. The problem is
that the observed fermions and gauge bosons that we want to describe are actually massive. The solution within the GWS theory is to postulate that
the gauge symmetry is spontaneously broken.

The minimal realization of the spontaneous symmetry breaking (SSB) in the GWS theory is achieved by means of a single scalar $SU(2)$ doublet. This
scalar field $\phi$, is assumed to acquire a vacuum expectation value $\av{\phi}=v$, breaking spontaneously three independent linear combinations of
generators and leaving unbroken the fourth combination. The unbroken combination of generators gives rise to the corresponding combination of the
gauge fields that constitute the physical (massless) photon. The three broken combinations give rise to the appearance of three goldstone bosons,
according to the Nambu-Goldstone theorem. These three degrees of freedom become, in an appropriate gauge, longitudinal polarization modes for the
other three independent combinations of gauge bosons, which then acquire a mass. These three combinations can be chosen such that they have diagonal
couplings with the photon (a definite electric charge). These are the physical $W^{\pm}$ and $Z$ bosons. The fourth remaining degree of freedom from
the original complex scalar doublet $\phi$ is a physical real scalar field $h$, the higgs boson. In this way, the kinetic term for the scalar field
$\phi$ can be written as
\eq{\mathcal{L}_{K,S}=(D^\mu\phi)^\dagger (D_\mu\phi)=\frac{1}{2}(\partial_\mu h)^2+\left[m_W^2W_\mu^+W^{-\,\mu}+\frac{1}{2}m_Z^2Z_\mu Z^\mu \right]
\left( 1+\frac{h}{v} \right)^2\ ,}
with $\,m_W\equiv vg/2\,$ and $\,m_Z\equiv v\sqrt{g^2+g'^2}/2\,$. This mechanism predicts a very definite relation between the masses of the $W$ and
$Z$ bosons that was verified precisely at LEP. This solves the issue of the massive gauge bosons.

The kinetic term for the fermions (\ref{KF}) can then be expanded in terms of the physical gauge fields, and the charged and neutral weak currents
identified. The electromagnetic currents then relate directly hypercharges and electric charges, and the hypercharge assignation can be made. The
rest of the scalar lagrangian consists of an ``opposite sign'' $\phi$ mass term, $\mu\phi^\dagger\phi$, and an interaction term, $\lambda \phi^4$. In
terms of physical fields these terms lead to interactions among gauge and higgs bossons, and to a higgs boson mass term. The parameters $\mu$ and
$\lambda$ combine into $v$ and $m_H$, the higgs mass.

The issue of the massive fermions is still unsolved, and CP violation has not appeared yet in the theory. Here is where the things get interesting.
For our purpose we focus on the quark fields $Q^i_L=(U^i_L,D^i_L)\ $, $U^i_R$ and $D^i_R$. Having introduced the scalar doublet, gauge invariance
allows (and hence requires) the following couplings between the fermions and the scalar,
\eqa{ \mathcal{L}_Y&=&-\lambda_d^{ij}\,\overline{Q^i_L}\cdot\phi\ D_R^j-
\lambda_u^{ij}\,\epsilon^{ab}\,\overline{Q_{L}^i}_a\cdot\phi_b^\dagger\ U_R^j+{\rm h.c}\nn\\
&=&-\frac{1}{\sqrt{2}}\lambda_d^{ij}\,\bar{D}_L^iD_R^j\,(v+h)-\frac{1}{\sqrt{2}}\lambda_u^{ij}\,\bar{U}_L^iU_R^j\,(v+h)+{\rm h.c} \label{LY} }
where $a$, $b$ are $SU(2)$ indices. The parameters $\lambda^{ij}$ are arbitrary complex matrices in flavor space. This piece of the SM lagrangian
provides masses for the quarks, quark-higgs interactions, and flavor mixing. Also, because the couplings can be complex, introduces the possibility
of CP violation.

Now, to identify the masses, we must rotate the quark fields to get a diagonal mass matrix. The squared matrices
$\lambda_u\!\cdot\!\lambda_u^\dagger$ and $\lambda_u^\dagger\!\cdot\!\lambda_u$ are hermitian, so they are diagonalizable with real and positive
eigenvalues: $\lambda_u\!\cdot\!\lambda_u^\dagger=V_uD_u^2V_u^\dagger$, $\lambda_u^\dagger\!\cdot\!\lambda_u=W_uD_u^2W_u^\dagger$, with $D_u$
diagonal. The same is valid for $\lambda_d$. So $\lambda_u=V_uD_uW_u^\dagger$ and $\lambda_d=V_dD_dW_d^\dagger$. Then we rotate the fields such that
\eqa{
U_L^i=V_u^{ij}u_L^j &,& U_R^i=W_u^{ij}u_R^j \nn\\
D_L^i=V_d^{ij}d_L^j &,& D_R^i=W_d^{ij}d_R^j }
In terms of the rotated fields $u_{L,R}$ and $d_{L,R}$, the Yukawa lagrangian (\ref{LY}) becomes
\eq{\mathcal{L}_Y=-m_d^i\bar{d}_L^id_R^i\left( 1+\frac{h}{v} \right)-m_u^i\bar{u}_L^iu_R^i\left( 1+\frac{h}{v} \right)}
where $\ m_q^i\equiv v D_q^{ii}/\sqrt{2}\ $ are the masses of the quarks.

This redefinition of the fields will only modify the fermionic part of the lagrangian $\mathcal{L}_{K,F}$, that is, the fermionic currents. In fact,
it only modifies the charged currents, because the neutral currents are changed by things like $V^\dagger V=1$. The charged currents become
\eq{J_W^{+\,\mu}=\frac{1}{\sqrt{2}}\bar{U}_L^i\gamma^\mu D_L^i=\frac{1}{\sqrt{2}}\bar{u}_L^i\gamma^\mu V_{\sss\rm CKM}^{ij} d_L^j}
which defines the \emph{CKM matrix}, $V_{\sss\rm CKM}^{ij}\equiv (V_u^\dagger V_d)^{ij}$. We see that in this mass eigenbasis for the quark fields,
neutral currents are still diagonal, but the CKM matrix generates flavor-changing charged currents.

The CKM matrix can in principle be any square complex matrix in flavor space, since there is absolutely no constraint on the values of the Yukawa
couplings $\lambda^{ij}$. The only constraint is unitarity, since $V_{\sss\rm CKM}V_{\sss\rm CKM}^\dagger=(V_u^\dagger V_d)(V_d^\dagger V_u)=1$. We
are ready now to see how CP violation arises in this picture.

Consider first the case of 2 fermion families. The most general $2\times 2$ unitary matrix can be parameterized by one angle and three phases,
\eq{ V_{\sss\rm CKM}=\left(
\begin{array}{cc}
\cos{\theta_{\sss C}} e^{i\alpha} & \sin{\theta_{\sss C}} e^{i\beta} \\
-\sin{\theta_{\sss C}} e^{i(\alpha+\gamma)} & \cos{\theta_{\sss C}} e^{i(\beta+\gamma)} \\
\end{array}
\right) }
However, we may rephase the three of the four quark fields to remove the three phases. This means that for 2 fermion families, the CKM matrix can
always be chosen to be \emph{real},
\eq{ V_{\sss\rm CKM}=\left(
\begin{array}{cc}
\cos{\theta_{\sss C}} & \sin{\theta_{\sss C}} \\
-\sin{\theta_{\sss C}} & \cos{\theta_{\sss C}} \\
\end{array}
\right) }
This is the Cabibbo matrix \cite{Cabibbo:1963yz}, and $\theta_{\sss C}$ is the Cabibbo angle. The conclusion is that with only 2 fermion families,
the SM \emph{cannot} account for CP violation. This was the argument that led to the prediction of the third family of quarks, once CP violation was
observed. For three generations of quarks, field rephasing cannot eliminate all the complex phases in the $3\times 3$ CKM matrix, but one complex
fase is left. Therefore, the SM with 3 families predicts CP violation provided this phase is not zero. Another conclusion is that CP can only be
violated in processes that involve the three families simultaneously, so in the SM, CP violation is generally a \emph{loop} effect.

\begin{figure}
\begin{center}
\includegraphics[height=4cm,width=8cm]{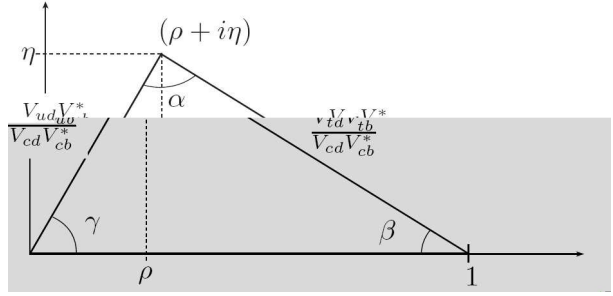}
\end{center}
\vspace{-0.5cm} \caption{\small The unitarity triangle, and the definitions of the angles.} \label{UTriangle}
\end{figure}

For numerical analyses, the most convenient parametrization of the CKM matrix is the \emph{standard parametrization} \cite{Chau:1984fp}. However, a
parametrization that makes its structure more clear is the \emph{Wolfenstein parametrization} \cite{Wolfenstein:1983yz},
\eq{ V_{\sss\rm CKM}= \left( \begin{array}{ccc} V_{ud} & V_{us} & V_{ub}\\V_{cd} & V_{cs} & V_{cb}\\V_{td} & V_{ts} & V_{tb} \end{array} \right)=
\left( \begin{array}{ccc}
1-\frac{1}{2}\lambda^2      &       \lambda                &              A\lambda^3(\rho-i\eta)   \\
-\lambda                    &     1-\frac{1}{2}\lambda^2   &                A\lambda^2             \\
A\lambda^3(1-\rho-i\eta)    &          -A\lambda^2         &                  1
\end{array} \right) + \mathcal{O}(\lambda^4)
}
where $\lambda\equiv \sin{\theta_{\sss C}}\simeq 0.22$ is treated as an expansion parameter. In this parametrization, the complex CP violating phase
is
\eq{\gamma\equiv \arg\left( -\frac{V_{ud}V_{ub}^*}{V_{cd}V_{cb}^*} \right)=-\arg{(V_{ub})}+\mathcal{O}(\lambda^4)
=\arg{(\rho+i\eta)}+\mathcal{O}(\lambda^4)\ ,}
which up to order $\lambda^4$ is localized in $V_{ub}=|V_{ub}|e^{-i\gamma}$ and $V_{td}=|V_{td}|e^{-i\beta}$, and the rest of the entries are real.
Here, $\beta$ is
\eq{\beta\equiv -\arg\left( -\frac{V_{td}V_{tb}^*}{V_{cd}V_{cb}^*} \right)=-\arg{(V_{td})}+\mathcal{O}(\lambda^4)
=\arg{(1-\rho+i\eta)}+\mathcal{O}(\lambda^4)\ ,}
and it is zero whenever $\gamma$ is zero. Therefore, in the SM, CP is violated if and only if $\gamma\ne 0$. Experimentally it is found that
$\gamma\sim 60^\circ-70^\circ$. A convenient graphical picture is provided by the \emph{Unitarity Triangle}, which arises from the unitarity of the
CKM matrix, in particular from the unitarity relation $V_{ud}V_{ub}^*+V_{cd}V_{cb}^*+V_{td}V_{tb}^*=0$ (see Fig.\ref{UTriangle}). CP violation can
then be quantified by the area of this triangle. SM fits provide already a very consistent and constrained Unitarity Triangle, as we saw already in
Fig.\ref{UTfit}.

\begin{figure}
\psfrag{b}{$b$} \psfrag{bb}{$\bar{b}$} \psfrag{q}{$q$} \psfrag{qb}{$\bar{q}$} \psfrag{t}{$t$} \psfrag{B}{$B_q$} \psfrag{Bb}{$\bar{B}_q$}
\psfrag{Vb}{$V_{tb}^*$}\psfrag{Vq}{$V_{tq}$}
\begin{center}
\hspace{1cm} \includegraphics[height=3.5cm,width=13cm]{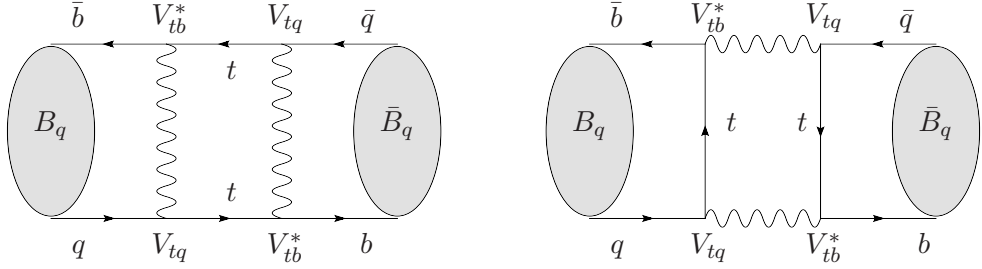}
\end{center}
\vspace{-0.5cm} \caption{\small Dominant SM contributions to the $B_q$-$\bar{B}_q$ mixing parameter $M_{12}^{q*}$. The CKM structure has been made
explicit.} \label{SMMixingBoxes}
\end{figure}

Let's now take a look at the mixing angles in the SM for $B_d$ and $B_s$ mesons. We saw in the previous section that for $B_q$ mesons the given
definition for the mixing angle is practically equivalent to $\phi_{\sss M}=\arg{(M_{12})}$. We also saw in Section \ref{effhamil} that $M_{12}$
contains contributions from $H_W^{\sss \Delta F=2}$ and also from transitions with intermediate on-shell states, at second order in $H_W^{\sss \Delta
F=1}$ (\emph{c.f.} eq.~\ref{M's y Gamma's}). The contribution from intermediate on-shell states is mainly non-perturbative. It is an important
contribution, for example, in kaon mixing. Fortunately, in the Standard Model the mixing of $B_q$ mesons is dominated by the perturbative $\Delta
F=2$ box diagrams with a top quark in the loop (see Fig. \ref{SMMixingBoxes}).

The CKM structure of these contributions is very simple. Both contributions are proportional to the product $(V_{tq}V_{tb}^*)^2$: $M_{12}^{d*}\propto
(V_{td}V_{tb}^*)^2$ and $M_{12}^{s*}\propto (V_{ts}V_{tb}^*)^2$. So it's easy to see that the $B_d$-$\bar{B}_d$ mixing angle in the SM is given by
\eq{\phi_d^{\sss SM}=2\beta + \mathcal{O}(\lambda^4)\ .}
The mixing angle $\phi_s^{\sss SM}$ is zero is at this level of approximation, which means that enters in terms which are suppressed by at least
$\lambda^4$. However, the phase itself is $\mathcal{O}(\lambda^2)$. The angle $\beta_s$ is defined as
\eq{\beta_s\equiv -\arg{\left( -\frac{V_{ts}V_{tb}^*}{V_{cs}V_{cb}^*} \right)}\ .}
Up to $\mathcal{O}(\lambda^6)$, only $V_{ts}$ acquires a phase, so $V_{ts}=-|V_{ts}|e^{-i\beta_s}$. In fact $\beta_s\simeq -\lambda^2\eta$. Then, the
$B_s$-$\bar{B}_s$ mixing angle in the SM is given by
\eq{\phi_s^{\sss SM}=2\beta_s + \mathcal{O}(\lambda^6)\ .}
This is the way in which the mixing angles are related to the angles of the unitarity triangles in the Standard Model.

\section*{Appendix: CP-averaged branching ratio for 2-Body decays}
\addcontentsline{toc}{section}{Appendix: CP-averaged branching ratio for 2-Body decays}

\fancyhead[LO]{\it Appendix: CP-averaged branching ratio for 2-Body decays}

\begin{figure}[h]
\begin{center}
\includegraphics[height=3.5cm]{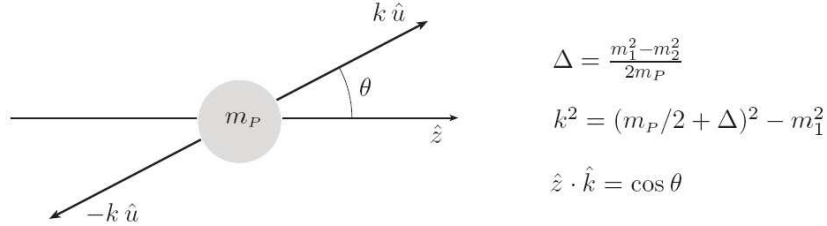}
\end{center}
\caption{\small Kinematic description in the CM frame of the decay of a particle $P$ into two final particles. Conservation of 4-momenta imposed by
the delta function in (\ref{dGamma}) implies the final state given by (\ref{p1,p2}), with $\Delta$ and $k\hat{u}$ given in this figure. The axis
$\hat{z}$ is chosen arbitrarily.} \label{DecayFigure}
\end{figure}

The differential decay rate $d\Gamma$ for the decay of a particle $P$ in the center of mass (CM) frame into a final state $f$ is (see, for example,
\cite{Peskin:1995ev})
\eq{d\Gamma(P\to f)=\frac{1}{2m_{\sss P}}\Big( \prod_i \frac{d^3p_i}{(2\pi)^3}\frac{1}{2 E_i} \Big) \mo{A(P\to f)}^2(2\pi)^4\delta(p_{\sss
P}-p_f)\label{dGamma}}
where $i$ runs over the set of particles in the final state with 4-momenta $p_i$ in the CM frame. $A(P\to f)$ is the amplitude of the process, and
$m_{\sss P}$ is the mass of the initial particle $P$ (the initial energy in the CM). For the case of 2 particles in the final state, the 4-momenta of
the final particles are
\eqa{
&&p_1=(m_{\sss P}/2+\Delta\,,k\hat{u})\nonumber\\
&&p_2=(m_{\sss P}/2-\Delta\,,-k\hat{u})\label{p1,p2} }
as described in Fig. \ref{DecayFigure}.

We can use the 4-delta function to integrate over the $\vec{p}_2$ 3-momenta and the modulus $\m{\vec{p}_1}$ in (\ref{dGamma}). This leads to the
following general formula:
\eq{\left( \frac{d\Gamma}{d\Omega} \right)_{\!\rm CM}\!=\frac{k}{32\pi^2 m_{\sss P}^2}\m{A(P\to f)}^2 \label{dGamma/dOmega}}
If the initial particle $P$ has spin 0, the process is spherically symmetric in the CM frame and the amplitude $A(P\to f)$ cannot have an angular
dependence\footnote{In the case in which $P$ has spin, but the polarization state is unknown, we must average over the spin states and we recover
spherical symmetry.}. In such a case the amplitude factorizes and the integrated decay rate can be written as
\eq{\Gamma(P\to f)=f_{\sss \textrm{PS}}\,\m{A(P\to f)}^2\label{Gamma=fps*A}}
where $f_{\sss \textrm{PS}}$ is called the \emph{phase space factor}. This is just the angular integral which depends only on kinematic variables:
\eq{f_{\sss \textrm{PS}}=\frac{1}{\sigma_s}\frac{k}{8\pi m_{\sss P}^2}}
Here $\sigma_s$ is a symmetry factor, which is $\sigma_s=1$ if the two final particles are distinguishable and $\sigma_s=2$ if they are identical.
The meaning of this factor is that the final state with $\theta$ and that with $\theta+\pi$ are indistinguishable, and must be counted only once.

In the case of 2 particles with the same mass in the final state, we have $m_1\!=\!m_2\!\equiv\! m_f$, $\Delta=0$, and \mbox{$k^2=(m_{\sss
P}^2-4m_f^2)/4$}, and the phase space factor is given by
\eq{f_{\sss \textrm{PS}}=\frac{1}{\sigma_s}\frac{1}{16\pi m_{\sss P}^2}\sqrt{m_{\sss P}^2-4m_f^2}\label{fps}}

The \emph{branching ratio} $BR(P\to f)$ is defined as the partial decay rate $\Gamma(P\to f)$ divided by the total decay rate $\Gamma(P\to {\rm
all})$. This quantity is naturally normalized to one, since the sum of the decay rates of \emph{all} the exclusive decay modes of $P$ must equal
$\Gamma(P\to {\rm all})$. The total decay rate is the inverse of the lifetime $\tau_{\sss \textrm{P}}$. In ``untagged'' neutral meson decays, there
is no information of whether the decaying particle is a $P^0$ or a $\bar{P}^0$. In such a case the measured branching ratio is an average between the
decays $P^0\to f$ and $\bar{P}^0\to f$,
\eq{\overline{BR}(P^0\to f)=\frac{\Gamma(P^0\to f)+\Gamma(\bar{P}^0\to f)}{2\,\Gamma(P^0\to {\rm all})}= \tau_{\sss \textrm{P}}f_{\sss
\textrm{PS}}\,\av{\m{A_f}^2}\label{CPavBR}}
where $\av{\m{A_f}^2}\equiv (\m{A_f}^2+\m{\bar{A}_f}^2)/2=(\m{A_f}^2+\m{\bar{A}_{\bar{f}}}^2)/2$, the last equality being true for $f$ a CP
eigenstate.  When the final state is a CP eigenstate then this is called a \emph{CP-averaged branching ratio}. In this thesis we will only deal with
CP-averaged branching ratios, so we will call them just $BR$, without the bar $\overline{BR}$. Moreover, we define the $g_{\sss \textrm{PS}}$ factor
as
\eq{g_{\sss \textrm{PS}}\equiv \tau_{\sss \textrm{P}}f_{\sss \textrm{PS}} \label{gps}}
Some numerical values for these factors, that will be used later, are
\eqa{
&&g_{\sss \textrm{PS}}(B_d^0\to \pi^+\pi^-)\ =\ 8.8 \times 10^9\,{\rm GeV}^{-2}\nn\\
&&g_{\sss \textrm{PS}}(B_d^0\to K^{\sss 0}\bar{K}^{\sss 0})\ =\ 8.6 \times 10^9\,{\rm GeV}^{-2}\nn\\
&&g_{\sss \textrm{PS}}(B_s^0\to K^+K^-)\ =\ 8.1 \times 10^9\,{\rm GeV}^{-2}\nn\\
&&g_{\sss \textrm{PS}}(B_s^0\to K^{\sss 0}\bar{K}^{\sss 0})\ =\ 8.1 \times 10^9\,{\rm GeV}^{-2} }
for input values of masses and lifetimes in \cite{Yao:2006px} and using the conversion factor
\eq{1\,{\rm sec}=1.519\times 10^{24}\,{\rm GeV}^{-1}}
For $B$ decays into two light mesons, the masses of the final mesons are sometimes neglected in front of $m_B$. If this is done, there are only two
$g_{\sss \textrm{PS}}$ factors: $g_{\sss \textrm{PS}}(B_d)=8.8\cdot 10^9\,{\rm GeV}^{-2}$ and $g_{\sss \textrm{PS}}(B_s)=8.2\cdot 10^9\,{\rm
GeV}^{-2}$, which often simplifies the numerics.

%


\part{Applications}

\fancyhead[LO]{\it \rightmark}

\chapter{\textit{Measuring} New Physics in $B$ Decays}

\label{Bdecays}

\n Up to now, the theoretical analyses of CP violation in B decays have been mostly focused on two complementary subjects. First, many different
methods have been proposed for extracting in a clean way the CP violating angles $\alpha$, $\beta$ and $\gamma$, within the SM (see e.g.
\cite{Yao:2006px,Harrison:1998yr}). We have now many ways of checking the compatibility of different measurements within the SM and detect the
presence of NP. Second, many studies have been made of observables within specific scenarios of NP in the \emph{forward} direction; that is: what
values for these observables does this specific model predict? These second type of studies are important because they provide a direct bridge
between new data and the reduction of the allowed regions in the NP parameter spaces\footnote{We use the \emph{plural} in ``parameter spaces''
because each NP scenario has its own parameters, and thus its own parameter space.}. In an extreme situation, they may even rule out some models (see
for example \cite{hep-ph/9910211}).

However, the identification of the NP, once this is found, requires NP studies in the \emph{inverse} direction; that is: given this non-standard
experimental data, what is the NP that describes it? This ``inverse problem'' is rather complex because of two reasons. First, given a specific
model, a signature in observable space is generally associated to many points in parameter space (``degeneracies''). In the most unfavorable
situation this can be a huge continuously connected region, meaning that this signature in observable space will give absolutely no clue on the basic
structure of the underlying theory (for a study of the inverse problem in supersymmetry see \cite{ArkaniHamed:2005px}). Second, the most important
information that must be extracted from NP signals is related to general characteristics of the NP. Indeed, solving the inverse problem in
supersymmetry is of no help if nature turns out \emph{not} to be supersymmetric. But model independent analyses of NP are generally unapproachable
because of the large number of parameters involved (see the discussion in Section \ref{BSMHeff}), so at the end one has to come up with arguments
that allow to reduce the number of model independent NP parameters.

A first step in this direction was taken in Ref. \cite{hep-ph/0404130}, where it was shown that an argument concerning new physics strong phases
allows to reduce the number of NP parameters to a manageable level, and measure them. The knowledge of these model independent NP parameters allows
then to identify partially the new physics, and to establish some of its generic features.

In this chapter we present an approach, within this framework, to \emph{measure} the NP parameters that appear in $B_s\to K^+K^-$ amplitudes. This
approach requires experimental information on branching ratios and CP asymmetries in $B_d\to\pi^+\pi^-$ and $B_s\to K^+K^-$, and is based on a SU(3)
analysis of these modes and the (motivated) assumption that any NP contributing to $\bar{b}\to\bar{d}$ penguin amplitudes is negligible at the
considered level of approximation.

\section{Parametrization of the SM Amplitudes}

The SM amplitude of a generic two body non-leptonic B decay ($B_q\to M_1M_2$) is commonly parameterized in terms of its CKM structure. We used this
fact already in the discussion of strong and weak phases in Section \ref{StrongAndWeakPhases}, where it was justified by invoking the form of the SM
effective Hamiltonian as written in Eq. (\ref{HeffSM}). However, we have never proved here that the Wilson coefficients in Eq. (\ref{HeffSM}) are CKM
independent; in fact, strictly speaking, they are not.

In this chapter we abandon temporarily the effective Hamiltonian description of the amplitudes, in favor of a diagrammatic approach that makes this
CKM issue more transparent.

\begin{figure}
\begin{center}
\includegraphics[width=3.3cm,height=3cm]{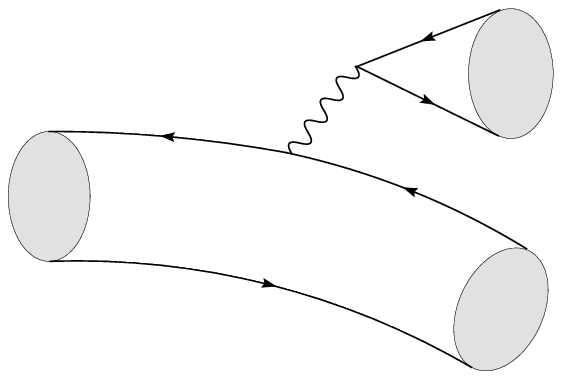}
\hspace{0.3cm}
\includegraphics[width=3.3cm,height=3cm]{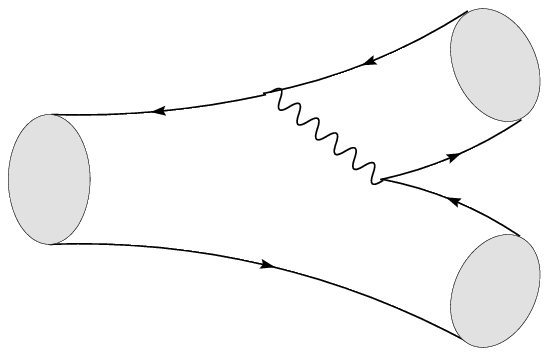}
\hspace{0.3cm}
\includegraphics[width=3.3cm,height=3cm]{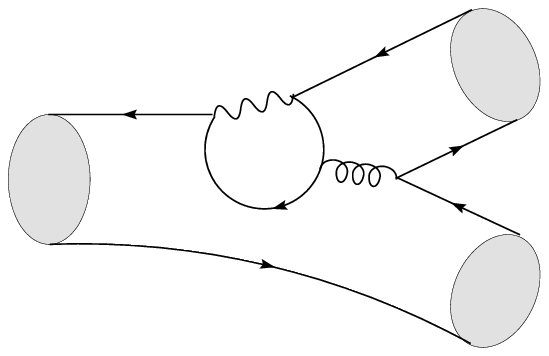}
\hspace{0.3cm}
\includegraphics[width=3.3cm,height=3cm]{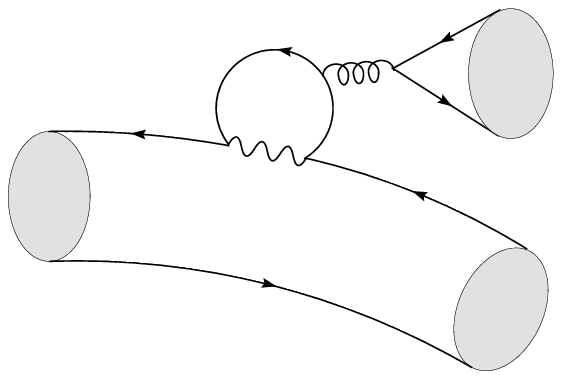}
\Text(-419,37)[lb]{\footnotesize $B_q$} \Text(-342,13)[lb]{\footnotesize $M_{\sss 1}$} \Text(-341,67)[lb]{\footnotesize $M_{\sss 2}$}
\Text(-400,47)[lb]{\footnotesize $\bar{b}$} \Text(-375,24)[lb]{\footnotesize $q$} \Text(-345,40)[lb]{\footnotesize $\bar{U}$}
\Text(-356,54)[lb]{\footnotesize $U'$} \Text(-357,79)[lb]{\footnotesize $\bar{D}$} \Text(-385,-5)[lb]{\footnotesize $(a)$}
\Text(-309,37)[lb]{\footnotesize $B_q$} \Text(-233,13)[lb]{\footnotesize $M_{\sss 1}$} \Text(-234,67)[lb]{\footnotesize $M_{\sss 2}$}
\Text(-290,47)[lb]{\footnotesize $\bar{b}$} \Text(-265,24)[lb]{\footnotesize $q$} \Text(-240,27)[lb]{\footnotesize $\bar{D}$}
\Text(-240,50)[lb]{\footnotesize $U'$} \Text(-250,75)[lb]{\footnotesize $\bar{U}$} \Text(-275,-5)[lb]{\footnotesize $(b)$}
\Text(-200,37)[lb]{\footnotesize $B_q$} \Text(-124,13)[lb]{\footnotesize $M_{\sss 1}$} \Text(-125,67)[lb]{\footnotesize $M_{\sss 2}$}
\Text(-180,47)[lb]{\footnotesize $\bar{b}$} \Text(-155,23)[lb]{\footnotesize $q$} \Text(-132,27)[lb]{\footnotesize $\bar{Q}$}
\Text(-132,50)[lb]{\footnotesize $Q$} \Text(-142,75)[lb]{\footnotesize $\bar{D}$} \Text(-167,45)[lb]{\tiny $\bar{u},\!\bar{c},\!\bar{t}$}
\Text(-165,-5)[lb]{\footnotesize $(c)$}
\Text(-92,37)[lb]{\footnotesize $B_q$} \Text(-15,13)[lb]{\footnotesize $M_{\sss 1}$} \Text(-14,67)[lb]{\footnotesize $M_{\sss 2}$}
\Text(-72,45)[lb]{\footnotesize $\bar{b}$} \Text(-45,24)[lb]{\footnotesize $q$} \Text(-15,40)[lb]{\footnotesize $\bar{D}$}
\Text(-26,54)[lb]{\footnotesize $Q$} \Text(-27,79)[lb]{\footnotesize $\bar{Q}$} \Text(-56,63)[lb]{\tiny $\bar{u},\!\bar{c},\!\bar{t}$}
\Text(-55,-5)[lb]{\footnotesize $(d)$}
\end{center}
\caption{\small Leading contributions to $B_q\to M_1M_2$ which do not involve the spectator quark ($\bar{b}\to\bar{D}q\bar{q}$). The quark labels
stand for $q,Q=\{u,d,s,c\}$, $U,U'=\{u,c\}\ $ and $D=\{d,s\}$. The color allowed (a) and color suppressed (b) tree amplitudes are proportional to
$V_{\sss U'\!\!D}V^*_{\sss U\!b}$. The color suppressed and color allowed penguin amplitudes (c) and (d) are proportional to $V_{\sss
\ell\!D}V^*_{\sss \ell\!b}$, where $\ell=\{u,c,t\}$ is the quark inside the loop. Note that not all of the diagrams contribute to a given decay.}
\label{BtoM1M2}
\end{figure}

We will restrict ourselves to a general situation in which the parametrization of Eq. (\ref{CKMdeco}) holds: decays which contain contributions from
tree or penguin diagrams where the two up-type quarks produced in the decay (if any) are both either up or charm quarks. The leading contributions,
which contain a single $W$ propagator, are then all proportional to either one of the CKM products $V_{uD}V^*_{ub}$, $V_{cD}V^*_{cb}\ $ or $\
V_{tD}V^*_{tb}$, where $D$ is the down-type quark in the final state which is not $q$ (see Fig. \ref{BtoM1M2}).

Because in a given decay the CKM elements always enter by pairs, it is customary to define the following product of CKM elements, $\lambda_\ell^{\sss
(D)}\equiv V_{\sss \ell\!D}^*V_{\sss \ell\!b}$. Moreover, the unitarity of the CKM matrix implies the triangle relation
\eq{\lambda_u^{\sss (D)}+\lambda_c^{\sss (D)}+\lambda_t^{\sss (D)}=0 \label{unit} }
which allows to eliminate the CKM factor $\lambda_t^{\sss (D)}$ from the amplitude. The amplitude is then a sum of two terms, proportional to
$\lambda_u^{\sss (D)}$ and $\lambda_c^{\sss (D)}$ respectively. We define the ``tree'' ($T_{\sss\!M_1\!M_2}^q$) and the ``penguin''
($P_{\sss\!M_1\!M_2}^q$) amplitudes as the coefficients of the two terms with different CKM structures. Thus the amplitude can be written as
\eq{A(B_q\to M_1M_2)=\lambda_u^{\sss (D)*}\,T_{\sss\!M_1\!M_2}^q+\lambda_c^{\sss (D)*}P_{\sss\!M_1\!M_2}^q \label{A(TP)}}

Besides the topologies shown in Fig. \ref{BtoM1M2}, there are also leading contributions which involve the spectator quark, mainly exchange,
annihilation and penguin annihilation diagrams \cite{Gronau:1995hm}. However, it has been argued that these can be neglected at leading order, since
they imply the fairly unlikely situation in which the two quarks inside the B meson annihilate each other or interact weakly before the decay of the
$b$ quark. This is called the \emph{spectator quark hypothesis}, and as any other hypothesis it should be handled with care (for example, these
contributions could be Cabibbo enhanced). Unlike the penguin annihilation topology, the exchange and annihilation contributions do not in general
obey the CKM structure discussed above, and do not fit in the scheme. However, for charmless decays and decays into CP eigenstates they do, and they
can be included in the description trivially. Since we will be focusing our discussion in two-body decays of neutral B mesons ($q=d,s$) into CP
eigenstates, this issue will be of no concern.

\begin{figure}
\begin{center}
\includegraphics[width=3.3cm,height=3cm]{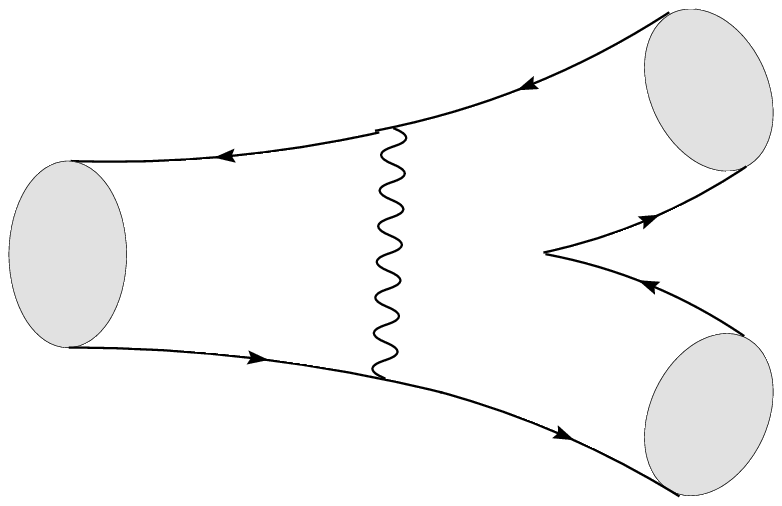}
\hspace{0.3cm}
\includegraphics[width=3.3cm,height=3cm]{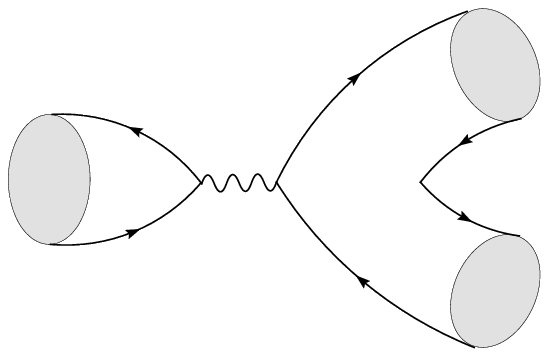}
\hspace{0.3cm}
\includegraphics[width=3.3cm,height=3cm]{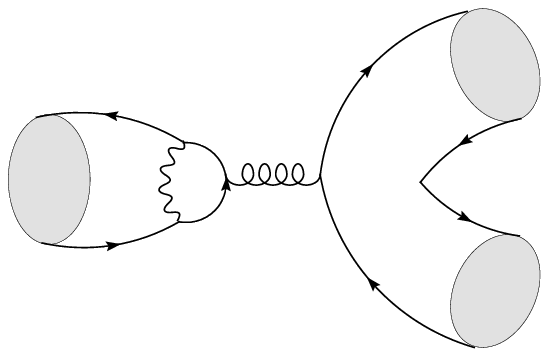}
\Text(-310,37)[lb]{\footnotesize $B_q^0$} \Text(-233,13)[lb]{\footnotesize $M_{\sss 1}$} \Text(-233,67)[lb]{\footnotesize $M_{\sss 2}$}
\Text(-290,48)[lb]{\footnotesize $\bar{b}$} \Text(-290,28)[lb]{\footnotesize $q$} \Text(-250,17)[lb]{\footnotesize $U'$}
\Text(-241,27)[lb]{\footnotesize $\bar{Q}$} \Text(-240,50)[lb]{\footnotesize $Q$} \Text(-250,75)[lb]{\footnotesize $\bar{U}$}
\Text(-275,-15)[lb]{\footnotesize $(a)$}
\Text(-201,37)[lb]{\footnotesize $B_q^+$} \Text(-124,13)[lb]{\footnotesize $M_{\sss 1}$} \Text(-124,67)[lb]{\footnotesize $M_{\sss 2}$}
\Text(-184,58)[lb]{\footnotesize $\bar{b}$} \Text(-184,21)[lb]{\footnotesize $q$} \Text(-132,25)[lb]{\footnotesize $Q$}
\Text(-132,50)[lb]{\footnotesize $\bar{Q}$} \Text(-145,75)[lb]{\footnotesize $U$} \Text(-145,4)[lb]{\footnotesize $\bar{D}$}
\Text(-170,-15)[lb]{\footnotesize $(b)$}
\Text(-91,37)[lb]{\footnotesize $B_q^0$} \Text(-15,13)[lb]{\footnotesize $M_{\sss 1}$} \Text(-15,67)[lb]{\footnotesize $M_{\sss 2}$}
\Text(-75,60)[lb]{\footnotesize $\bar{b}$} \Text(-75,19)[lb]{\footnotesize $q$} \Text(-23,25)[lb]{\footnotesize $Q$} \Text(-23,50)[lb]{\footnotesize
$\bar{Q}$} \Text(-36,75)[lb]{\footnotesize $Q'$} \Text(-36,3)[lb]{\footnotesize $\bar{Q}'$} \Text(-60,27)[lb]{\tiny $u,\!c,\!t$}
\Text(-55,-15)[lb]{\footnotesize $(c)$}
\end{center}
\caption{\small Exchange (a), Annihilation (b) and Penguin Annihilation (c) contributions. (a) and (c) only exist for $q=\{d,s\}$ and (b) only exists
for $q=\{u,c\}$. The penguin annihilation contribution always fits in the structure (\ref{A(TP)}). The exchange and annihilation topologies only have
such CKM structure for $U=U'$ and $q=U$ respectively. For final CP eigenstates this structure is automatically satisfied (there is no annihilation
contribution in this case).} \label{NonFactDiag}
\end{figure}

To be specific, the tree and penguin contributions, in terms of the diagrams in Fig. \ref{BtoM1M2}, are given by
\eqa{
T_{\sss\!M_1\!M_2}^q&=&T_u+C_u+(P_u-P_t)+(P^C_u-P^C_t)\nonumber\\
P_{\sss\!M_1\!M_2}^q&=&T_c+C_c+(P_c-P_t)+(P^C_c-P^C_t) }
where $\lambda_U^{\sss (D)*}T_U$,  $\lambda_U^{\sss (D)*}C_U$, $\lambda_\ell^{\sss (D)*}P_\ell$ and $\lambda_\ell^{\sss (D)*}P^C_\ell$ are
respectively the amplitudes (a), (b), (d) and (c) in Fig. \ref{BtoM1M2}.

The CKM products $\lambda_U^{\sss (D)}$ have the following numerical values \cite{Yao:2006px}:
\eqa{
&\lambda_u^{(d)}=0.0038e^{i\gamma}=\m{\lambda_u^{(d)}}e^{-i\gamma}  \qquad & \lambda_c^{(d)}=-0.0094=-\m{\lambda_c^{(d)}} \nonumber\\
&\lambda_u^{(s)}=0.00088e^{i\gamma}=\m{\lambda_u^{(s)}}e^{-i\gamma} \qquad & \lambda_c^{(s)}=0.04=\m{\lambda_c^{(s)}} }
where $\gamma$ is the CKM weak phase. Note the negative sign of $\lambda_c^{(d)}$. We denote the relative (strong) phase between
$T_{\sss\!M_1\!M_2}^q$ and $P_{\sss\!M_1\!M_2}^q$ by $\,\theta_{\sss\!M_1\!M_2}^q$:
\eq{\theta_{\sss\!M_1\!M_2}^q\equiv arg\left( \frac{P_{\sss\!M_1\!M_2}^q}{T_{\sss\!M_1\!M_2}^q} \right)}
Now we can give a general useful parameterization of the amplitude in (\ref{A(TP)}),
\eq{A(B_q\to M_1M_2)=\m{\lambda_u^{\sss (D)}}\m{T}e^{i\delta_T}\left(e^{i\gamma}\pm \mo{\frac{\lambda_c^{\sss (D)}}{\lambda_u^{\sss (D)}}}
\mo{\frac{P}{T}} e^{i\theta}\right)=\m{\lambda_u^{\sss (D)}}\m{T}e^{i\delta_T}\left(e^{i\gamma}\pm d e^{i\theta}\right)\label{A(dtheta)}}
the plus (minus) sign being for $D=s$ ($D=d$). We have dropped the labels $q$ and $M_1M_2$ for simplicity, and it will be done so systematically
whenever there is no ambiguity. We have also defined \cite{Fleischer:1999pa}
\eq{d=d_{\sss\!M_1\!M_2}^q\equiv \mo{\frac{\lambda_c^{\sss (D)}}{\lambda_u^{\sss (D)}}}\mo{\frac{P_{\sss\!M_1\!M_2}^q}
{T_{\sss\!M_1\!M_2}^q}}\label{d}}
and $\delta_T$ is the strong phase associated to $T_{\sss\!M_1\!M_2}^q$. Note that when (\ref{A(dtheta)}) is the full amplitude (that is, when there
are no NP contributions), the phase $\delta_T$ is not physical and does not appear in the observables. The amplitude for the CP-conjugate process
$\bar{B}_q\to \bar{M}_1\bar{M}_2$ shall be obtained from (\ref{A(dtheta)}) by changing the sign of the weak phase $\gamma$.

\section{Measuring the SM parameters}

\label{MeasuringSM}

Assuming that no new contributions come into play, and that the full amplitude is described within SM as in (\ref{A(dtheta)}), it is possible to
\emph{measure} the SM hadronic parameters $T_{\sss\!M_1\!M_2}^q$, $P_{\sss\!M_1\!M_2}^q$ and $\theta_{\sss\!M_1\!M_2}^q$. There are three potential
experimental measurements that can be carried out in neutral B decays that allow us to extract this information. These are the \emph{CP-averaged
branching ratio} (see the Appendix to Chapter \ref{CPV}), the \emph{direct CP asymmetry} and the \emph{mixing induced CP asymmetry} (see section
\ref{sectionCPB}). By virtue of the formulae (\ref{CPavBR}), (\ref{Adir}) and (\ref{Amix}) one can relate these observables to the theoretical
parameters in (\ref{A(dtheta)}):
\eq{ BR=g_{\sss \textrm{PS}}\m{\lambda_u^{\sss (D)}}^2\m{T}^2\left( 1 + d^2 \pm 2 d \cos{\gamma} \cos{\theta} \right)\label{BR(dtheta)} } \eq{
\mathcal{A}_{\rm CP}^{dir}=\pm\frac{2 d \sin{\gamma} \sin{\theta}}{ 1 + d^2 \pm 2 d \cos{\gamma} \cos{\theta}}\label{Adir(dtheta)} } \eq{
\mathcal{A}_{\rm CP}^{mix}=\frac{\sin{(2\gamma+\phi_{\sss M})}\pm 2 d \sin{(\gamma+\phi_{\sss M})} \cos{\theta}+ d^2 \sin{\phi_{\sss M}}} { 1 + d^2
\pm 2 d \cos{\gamma} \cos{\theta}}\label{Amix(dtheta)} }
where, again, the plus and minus signs correspond to $D=s$ and $D=d$, respectively.

The hadronic parameters $\m{T}$, $d$ and $\theta$ are specific of each particular decay. The CKM angle $\gamma$ and the mixing angle $\phi_{\sss M}$
are, on the contrary, universal theoretical parameters, and can be obtained from a multitude of different decays. Therefore, we are left with three
hadronic parameters ($\m{T}$, $d$ and $\theta$) to be obtained from three experimental values ($BR$, $\mathcal{A}_{\rm CP}^{dir}$ and
$\mathcal{A}_{\rm CP}^{mix}$)
 by means of equations (\ref{BR(dtheta)})-(\ref{Amix(dtheta)}).

\begin{figure}
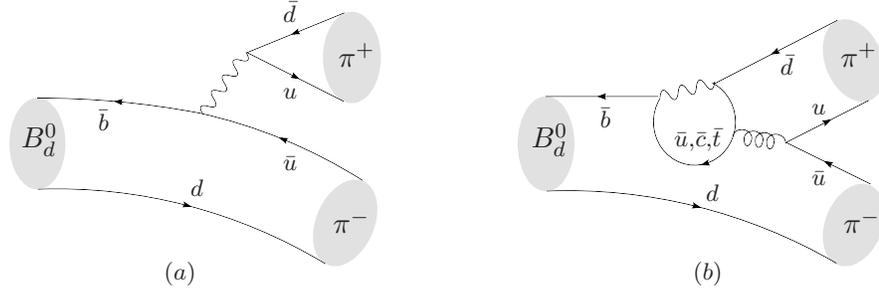

\begin{center}
\includegraphics[width=5cm,height=3.5cm]{TreeBsK+K-.eps}
\hspace{1.5cm}
\includegraphics[width=5cm,height=3.5cm]{PenguinBsK+K-.eps}
\Text(-329,44)[lb]{$B_d^0$} \Text(-211,15)[lb]{$\pi^-$} \Text(-209,78)[lb]{$\pi^+$} \Text(-300,54)[lb]{\footnotesize $\bar{b}$}
\Text(-265,29)[lb]{\footnotesize $d$} \Text(-230,95)[lb]{\footnotesize $\bar{d}$} \Text(-230,66)[lb]{\footnotesize $u$}
\Text(-230,39)[lb]{\footnotesize $\bar{u}$} \Text(-275,-5)[lb]{\footnotesize $(a)$}
\Text(-136,44)[lb]{$B_d^0$} \Text(-18,15)[lb]{$\pi^-$} \Text(-19,78)[lb]{$\pi^+$} \Text(-110,55)[lb]{\footnotesize $\bar{b}$}
\Text(-70,27)[lb]{\footnotesize $d$} \Text(-30,33)[lb]{\footnotesize $\bar{u}$} \Text(-30,60)[lb]{\footnotesize $u$} \Text(-42,75)[lb]{\footnotesize
$\bar{d}$} \Text(-75,-5)[lb]{\footnotesize $(b)$} \Text(-82,46)[lb]{\footnotesize $\bar{u},\!\bar{c},\!\bar{t}$}

\end{center}
\caption{\small Leading diagrams contributing to the decay $B_d^0 \to \pi^+ \pi^-$, (a) tree and (b) penguin contributions.} \label{BtoPiPi}
\end{figure}

As a matter of illustration and for later usage, consider the decay $B_d^0 \to \pi^+ \pi^-$. At the quark level (see Fig. \ref{BtoPiPi}) it
corresponds to a $\bar{b}\to \bar{d}$ transition ($D=d$), so we must choose the minus sign in (\ref{BR(dtheta)})-(\ref{Amix(dtheta)}). For the
branching ratio and the CP asymmetries we take an average of the Babar and Belle data \cite{Aubert:2005av,Chao:2003ue,Abe:2005dz},
\eqa{
BR(B_d^0\rightarrow \pi^+\pi^-)_{\rm exp}&=& (5.0\pm 0.4)\times 10^{-6} \label{BRpipiExp}\\
\mathcal{A}_{\rm CP}^{dir}(B_d^0\rightarrow \pi^+\pi^-)_{\rm exp}&=& -0.33\pm 0.11         \\
\mathcal{A}_{\rm CP}^{mix}(B_d^0\rightarrow \pi^+\pi^-)_{\rm exp}&=& 0.49\pm 0.12 }

\begin{figure}
\begin{center}
\includegraphics[width=12cm,height=5cm]{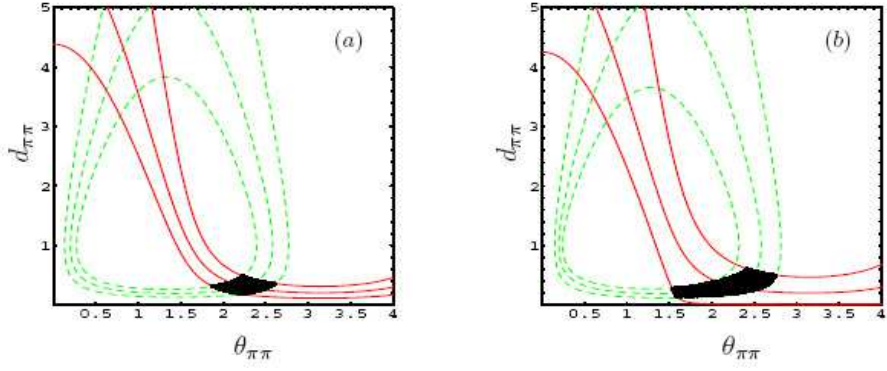}
\end{center}
\vspace{-0.3cm} \caption{\small Allowed regions in the $d-\theta$ plane for $B_d^0 \to \pi^+ \pi^-$, (a) taking $\gamma=61^\circ$ and (b) varying
$\gamma$ in the range $\gamma=(61^{+7}_{-5})^\circ$. The dashed lines show the central value and $\pm 1\sigma$ deviations for $\mathcal{A}_{\rm
CP}^{dir}(B_d^0 \to \pi^+ \pi^-)$, and the solid lines the same for $\mathcal{A}_{\rm CP}^{mix}(B_d^0 \to \pi^+ \pi^-)$.} \label{plotdtheta}
\end{figure}

Because $\mathcal{A}_{\rm CP}^{dir}$ and $\mathcal{A}_{\rm CP}^{mix}$ do not depend on the magnitude of the tree (c.f. (\ref{Adir(dtheta)}),
(\ref{Amix(dtheta)})), the calculation of the hadronic parameters can be divided in two steps. First we vary the asymmetries inside their
experimental range to find the allowed region in the $d-\theta$ plane. This is illustrated in Fig. \ref{plotdtheta}. Second, we vary the branching
ratio in its range together with the allowed values of $d$ and $\theta$, to find the range for $\m{T}$ according to (\ref{BR(dtheta)}). The results
are shown in Table \ref{tabledtheta}.

\begin{table}[b]
\begin{center}
\begin{tabular}{|c|c|c|c|c|}
\hline
&&&&\\[-12pt]
                     &   $d_{\pi\pi}$ &  $\theta_{\pi\pi}$ (deg) & $\mo{T_{\pi\pi}}(10^{-6}{\rm GeV})$  &  $\mo{P_{\pi\pi}/T_{\pi\pi}}$ \\[1pt]
\hline
&&&&\\[-12pt]
$\begin{array}{c}
\gamma=61^\circ\\
\end{array}$         &  $(0.18,0.49)$ &     $(107,150)$    &            $(4.87,6.11)$             &     $(0.07,0.20)$      \\[1pt]
\hline
&&&&\\[-12pt]
$\begin{array}{c}
\gamma=(61^{+7}_{-5})^\circ\\
\end{array}$         &  $(0.14,0.61)$ &     $(88,158)$     &             $(4.60,6.45)$            &     $(0.06,0.25)$      \\[1pt]
\hline
\end{tabular}
\end{center}
\caption{\small Allowed ranges for the hadronic parameters in the decay $B_d^0 \to \pi^+ \pi^-$, calculated from the experimental values for the
branching ratio and the two CP asymmetries. The values are taken from an average of Babar and Belle data. The sensibility on the error of $\gamma$ is
shown. The penguin-to-tree ratio is obtained from $d$ and (\ref{d}).} \label{tabledtheta}
\end{table}

Note that two different solutions in the $d-\theta$ plane are allowed by a single value of the pair $(\mathcal{A}_{\rm CP}^{dir}, \mathcal{A}_{\rm
CP}^{mix})$ (see the two separate regions in Figs. \ref{plotdtheta}). This is due to the non-linearity of the equations (\ref{Adir(dtheta)}) and
(\ref{Amix(dtheta)}). However, it should be clear that only one solution corresponds to a given pair $(T,P)$, so only one of the two regions in the
$d-\theta$ can be physical. QCDF predicts a value for $d_{\pi\pi}$ around $0.3$ \cite{hep-ph/0308039}, so the second solution ($d_{\pi\pi}>3$) is
likely to be the non-physical one and it's ruled out. This is also confirmed independently by an analysis concerning the channel $B_d\to
\pi^{\mp}K^{\pm}$ \cite{hep-ph/0410407}.

\section{Parameterization of the New Physics Amplitudes}

Besides the SM contributions, a given decay might also receive contributions from physics beyond the SM. Without a specific model for the NP, they
cannot be calculated. However, it is possible to make a model independent parameterization of these contributions under certain general circumstances
\cite{hep-ph/0404130}.

In the context of the low energy effective Hamiltonian, the new contributions will be of two types. First, there will be new contributions to the
Wilson Coefficients of the SM operators. Second, new operators will appear which are not generated by the SM. To this end we write the effective
Hamiltonian as a sum of the SM and NP contributions:
\eq{\mathcal{H}_{\rm eff}=\mathcal{H}_{\sss \rm SM}+\mathcal{H}_{\sss \rm NP}\label{HSM+HNP}}
Because the nature of the NP is unknown, $\mathcal{H}_{\sss \rm NP}$ contains all possible operators compatible with a given process. As it will
become clear below, we are mainly interested on the NP contributions to $\bar{b}\to \bar{s}q\bar{q}$ transitions. There are 20 dimension-six
operators that contribute potentially to such decays, and the general NP effective Hamiltonian in (\ref{HSM+HNP}) can be written as
\eqa{ &\mathcal{H}_{\sss \rm NP}^q={\displaystyle \sum_{\sss A,B=L,R}\frac{G_F}{\sqrt{2}}}& \Big[\,f^{\sss AB}_{q,1}\,\bar{s}_\alpha \gamma_A
b_\beta\,\bar{q}_\beta \gamma_B q_\alpha
+f^{\sss AB}_{q,2}\,\bar{s} \gamma_A b\, \bar{q} \gamma_B q  \nonumber \\
&&+g^{\sss AB}_{q,1}\,\bar{s}_\alpha \gamma^\mu\gamma_A b_\beta\,\bar{q}_\beta \gamma_\mu\gamma_B q_\alpha
+g^{\sss AB}_{q,2}\,\bar{s} \gamma^\mu\gamma_A b\, \bar{q} \gamma^\mu\gamma_B q \nonumber \\
&&+h^{\sss AB}_{q,1}\,\bar{s}_\alpha \sigma^{\mu\nu}\gamma_A b_\beta\,\bar{q}_\beta \sigma_{\mu\nu}\gamma_B q_\alpha +h^{\sss AB}_{q,2}\,\bar{s}
\sigma^{\mu\nu}\gamma_A b\, \bar{q} \sigma_{\mu\nu}\gamma_B q\,\Big] \label{HeffNP}}
where $\gamma_{\sss R,L}\equiv \frac{1}{2}(1\pm \gamma_5)$. The coefficients in (\ref{HeffNP}) will in general contain new (model dependent) weak
phases and the matrix elements of these operators will introduce new (process dependent) strong phases. The strong phases, as mentioned in section
\ref{StrongAndWeakPhases}, come mainly from rescattering. The important point is that the NP strong phases must come from rescattering of NP
operators. Let's discuss this in more detail.

In the SM, the strong phases may be large and must be taken into account because they arise from rescattering of tree operators, and although the
process of rescattering is size-consuming (the rescattered penguin amplitudes are about 5-10\% of the tree amplitudes), the tree amplitudes are big
and the rescattered penguins can be as large as the SM penguin amplitudes. This situation holds even when there are no tree operators that contribute
directly to the decay, since these can contribute through rescattering. In such cases, the rescattered penguin can be of the same order of the
largest contributions.

The NP operators, however, are expected to give \emph{direct} contributions not larger than the SM penguin contributions. This means that the
rescattering from such NP operators will be at most 5-10\% as big as the SM penguins, and at the level of other neglected diagrams. We can therefore
argue that these small contributions can be neglected, and since these are the main sources of new strong phases, the strong phases can be
approximately set to zero in the NP amplitude. Thus, without strong phases, the NP amplitude can be parameterized in terms of a single effective weak
phase:
\eq{\bra{f}\mathcal{H}_{\sss \rm NP}^q\ket{B}\equiv \mathcal{A}^q e^{i\Phi_q} \label{ANP}}
Note that there is a potential loop-hole in this argument. We can be sure about the size of the largest NP contributions to the decay in
consideration, and thus be confident in neglecting the strong phases coming from the rescattering of these diagrams. However, the possibility remains
of the existence of NP diagrams which do not contribute directly but through rescattering to the given decay. If they were much larger than the SM
penguins that do contribute, they could introduce sizeable strong phases in the NP amplitude. Other cases which get around this reasoning are when
the NP is light, or if there is a significant enhancement of certain matrix elements.

\section{Hints of New Physics in B decays}

The question arises of whether we should expect sizeable contributions from physics beyond the SM, or if these, even if they exist, are too small to
be observed at present. A second question would be what transitions are likely to be appreciably affected by NP. There are several indications that
may help to answer these questions. For example, the CP asymmetry in $\bar{b} \to \bar{s} q{\bar q}$ modes ($q=u,d,s$) is found to differ from that
in ${\bar b} \to {\bar c} c {\bar s}$ decays by 2.6$\sigma$ (they are expected to be approximately equal in the SM) \cite{HFAG,hep-ph/0507126}. In
addition, some $B\to\pi K$ measurements disagree with SM expectations \cite{hep-ph/0605094,hep-ph/0609006}, although the so-called $B \to \pi K$
puzzle \cite{hep-ph/0309012,hep-ph/0312259,hep-ph/0402112,hep-ph/0410407,hep-ph/0512032} has been reduced
\cite{hep-ph/0507156,hep-ph/0512253,hep-ph/0503077,hep-ph/0609128}. One also sees a discrepancy with the SM in triple-product asymmetries in $B \to
\phi K^*$ \cite{hep-ph/0303159,hep-ex/0408017,hep-ex/0505067}, and in the polarization measurements of $B \to \phi K^*$
\cite{hep-ex/0307026,hep-ex/0503013} and $B\to \rho K^*$ \cite{Yao:2006px,hep-ex/0505039,hep-ph/0508149}. These discrepancies are (almost) all not
yet statistically significant, being in the 1--2$\sigma$ range. However, if these hints are taken together, the statistical significance increases.
Furthermore, they are intriguing since they all point to New Physics in $\bar{b}\to\bar{s}$ transitions. For this reason it is interesting to
consider the effect of NP on $B$ decays dominated by the quark-level $\bar{b}\to\bar{s}$ transition, and assume that no NP affects
$\bar{b}\to\bar{d}$ transitions.

\section{Measuring the New Physics parameters through Flavor Symmetries}

\label{MeasuringNP}

For the sake of clarity we begin this section summarizing the basic ideas discussed so far.

The amplitudes for B decays that at the quark level are governed by the transition $\bar{b}\to\bar{D}q\bar{q}$, can be written as
\eqa{
&&A(\bar{b}\to\bar{d}q\bar{q})=\m{\lambda_u^{\sss (d)}}\m{T} \left(e^{i\gamma}- d e^{i\theta}\right) \label{A(b->d)} \\
&&A(\bar{b}\to\bar{s}q\bar{q})=\m{\lambda_u^{\sss (s)}}\m{T'}e^{i\delta_T}\left(e^{i\gamma}+ d' e^{i\theta'}\right) \label{A(b->s)} +\mathcal{A}^q
e^{i\Phi_q} }
These parameterizations rely on several conditions:

\begin{itemize}

\item SM diagrams which do not obey the CKM structure in (\ref{A(TP)}) can be neglected with respect to the penguin amplitudes.

\item Potential NP contributions to $\bar{b}\to\bar{d}$ decays can be neglected with respect to the SM penguin amplitudes.

\item There are no NP operators large enough to produce a sizeable contribution to the $\bar{b}\to\bar{s}$ decay through rescattering.

\end{itemize}

In section \ref{MeasuringSM} we showed how to measure the SM hadronic parameters in (\ref{A(dtheta)}) from measurements on the branching ratio and
the CP-asymmetries. The method was applied to obtain the hadronic parameters of the decay $B_d^0 \to \pi^+ \pi^-$, that according to the previous
conditions it is described entirely within SM as in (\ref{A(b->d)}). The same could be done, in principle, with a $\bar{b}\to\bar{s}$ decay. Taking
the amplitude in (\ref{A(b->s)}), the branching ratio and the asymmetries can be expressed in terms of the SM hadronic parameters \emph{and} the NP
parameters. The expressions are much more complicated in this case:
\eq{BR=BR^{\sss \rm SM}\cdot(1+\mathcal{B}^{\sss \rm NP})\label{BRNP}}
\eq{\mathcal{A}_{\rm CP}^{dir}=\frac{A_{dir}^{\sss \rm SM}+\mathcal{D}^{\sss \rm NP}}{1+\mathcal{B}^{\sss \rm NP}}\label{AdirNP}}
\eq{\mathcal{A}_{\rm CP}^{mix}=\frac{A_{mix}^{\sss \rm SM}\cdot \cos{\delta\phi_s^{\sss \rm NP}}+\mathcal{M}^{\sss \rm NP}}{1+\mathcal{B}^{\sss \rm
NP}} \label{AmixNP}}
where the SM functions $BR^{\sss \rm SM}$, $A_{dir}^{\sss \rm SM}$ and $A_{mix}^{\sss \rm SM}$ are those in (\ref{BR(dtheta)})-(\ref{Amix(dtheta)})
taking the plus sign and $D=s$, and the NP functions are given by
\eq{\mathcal{B}^{\sss \rm NP}=\frac{1}{\Delta}\big\{ z^2 +2z[\cos{(\Phi_q-\gamma)}\cos{\delta_T} + d \cos{\Phi_q}\cos{(\theta+\delta_T)} ] \big\}}
\eq{\mathcal{D}^{\sss \rm NP}=\frac{1}{\Delta}\big\{2z[\sin{(\Phi_q-\gamma)}\sin{\delta_T} + d \sin{\Phi_q}\sin{(\theta+\delta_T)} ] \big\}}
\eqa{ & \mathcal{M}^{\sss \rm NP}=&\hspace{-0.3cm}\frac{1}{\Delta} \big\{ z^2\sin{(2\Phi_q+\phi_s)}+2z[\sin{(\Phi_q+\phi_s+\gamma)}\cos{\delta_T}
+ d\cos{(\theta+\delta_T)}\sin{(\Phi_q+\phi_s)}] \nonumber \\
&&\hspace{-0.5cm}+\,[\cos{(2\gamma+\phi_s^{\sss \rm SM})}+2d\cos{(\theta)}\cos{(\gamma+\phi_s^{\sss \rm SM})} + d^2\cos{\phi_s^{\sss \rm
SM}}]\sin{\delta\phi_s^{\sss \rm NP}}\big\} \label{MNP} }
The $B^0_s-\bar{B}^0_s$ mixing angle $\phi_s$ can also suffer from the same NP contributions. We take this into account by adding a new
contribution\footnote{Note that $\phi_{\sss M}$ in (\ref{Amix(dtheta)}) should be understood in $A_{mix}^{\sss \rm SM}$ as $\phi_{\sss
M}=\phi_s^{\sss \rm SM}$}, $\phi_s=\phi_s^{\sss \rm SM}+\delta\phi_s^{\sss \rm NP}$. In order to simplify the expressions we have defined
\eq{\Delta=1+2d\cos{\gamma}\cos{\theta}+d^2\ ,\quad z=\mathcal{A}^q/\m{\lambda_u^{\sss (D)}T}\ .}
Note that in the limit $\mathcal{A}^q,\,\delta\phi_s^{\sss \rm NP}\to 0$ the NP functions $\mathcal{B}^{\sss \rm NP}$, $\mathcal{D}^{\sss \rm NP}$
and $\mathcal{M}^{\sss \rm NP}$ go to zero and the observables in (\ref{BRNP})-(\ref{AmixNP}) reduce to the SM functions.

The CKM angle $\gamma$ is considered an input parameter and it is obtained from the SM fit \cite{UTfit,CKMfitter}, $\gamma=(61^{+7}_{-5})^\circ$. The
mixing angle $\phi_s$ (including its NP contribution) can also be measured independently (see Chapter \ref{SymFac2}). However, three measured
observables are not enough in this case, since we have six parameters to determine, $\m{T}$, $d$, $\theta$, $\delta_T$, $\mathcal{A}^q$ and $\Phi_q$.

Fortunately, flavor symmetries allow us in certain situations to obtain the parameters $\m{T}$, $d$ and $\theta$ from a different decay, related to
the first by a symmetry such as U-spin. If this can be done, then $\delta_T$, $\mathcal{A}^q$ and $\Phi_q$ can be obtained from
(\ref{BRNP})-(\ref{AmixNP}). This strategy has been used in the literature for the SU(3)-related decays $B_d^0 \to \pi^+ \pi^-$ and $B_s^0 \to K^+
K^-$ \cite{Fleischer:1999pa,hep-ph/0404009,hep-ph/0410011}, although in pursue of slightly different goals.

\begin{figure}
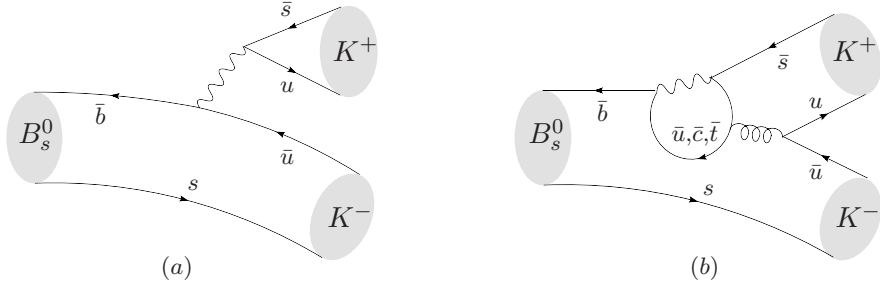

\begin{center}
\includegraphics[width=5cm,height=3.5cm]{TreeBsK+K-.eps}
\hspace{1.5cm}
\includegraphics[width=5cm,height=3.5cm]{PenguinBsK+K-.eps}
\Text(-329,44)[lb]{$B_s^0$} \Text(-212,15)[lb]{$K^-$} \Text(-210,78)[lb]{$K^+$} \Text(-300,54)[lb]{\footnotesize $\bar{b}$}
\Text(-265,29)[lb]{\footnotesize $s$} \Text(-230,95)[lb]{\footnotesize $\bar{s}$} \Text(-230,66)[lb]{\footnotesize $u$}
\Text(-230,39)[lb]{\footnotesize $\bar{u}$} \Text(-275,-5)[lb]{\footnotesize $(a)$}
\Text(-136,44)[lb]{$B_s^0$} \Text(-20,15)[lb]{$K^-$} \Text(-21,78)[lb]{$K^+$} \Text(-110,55)[lb]{\footnotesize $\bar{b}$}
\Text(-70,27)[lb]{\footnotesize $s$} \Text(-30,33)[lb]{\footnotesize $\bar{u}$} \Text(-30,60)[lb]{\footnotesize $u$} \Text(-42,75)[lb]{\footnotesize
$\bar{s}$} \Text(-75,-5)[lb]{\footnotesize $(b)$} \Text(-82,46)[lb]{\footnotesize $\bar{u},\!\bar{c},\!\bar{t}$}

\end{center}
\caption{\small Leading contributions to the decay $B_s^0 \to K^+ K^-$, (a) tree and (b) penguin contributions. These diagrams are related directly
to those of $B_d^0 \to \pi^+ \pi^-$ (see Fig. \ref{BtoPiPi}) by U-spin (mainly a permutation $d\leftrightarrow s$).} \label{BtoKK}
\end{figure}

In \cite{Fleischer:1999pa} (see also \cite{hep-ph/9906274,hep-ph/0204101}), the relationship between these decays was used to extract the CKM angles
$\beta$ and $\gamma$ in the SM. Assuming no NP in $B_s^0 \to K^+ K^-$, and obtaining the mixing phase $\phi_s^{\sss \rm SM}$ independently, $\beta$
(or $\phi_d^{\sss \rm SM}=2\beta$) and $\gamma$ can be obtained from $\mathcal{A}_{\rm CP}^{dir}(B_s^0 \to \pi^+ \pi^-)$ and $\mathcal{A}_{\rm
CP}^{mix}(B_s^0 \to \pi^+ \pi^-)$ once the parameters $d$ and $\theta$ have been obtained from $\mathcal{A}_{\rm CP}^{dir}(B_s^0 \to K^+ K^-)$ and
$\mathcal{A}_{\rm CP}^{mix}(B_s^0 \to K^+ K^-)$. There is, however, two-fold and four-fold ambiguities in the determination of $\gamma$ and $\beta$,
respectively.

In \cite{hep-ph/0404009}, the branching ratio for $B_s^0 \to K^+ K^-$ was computed from (\ref{BR(dtheta)}), also within the framework of SM, once the
hadronic parameters $d$, $\theta$ and $\m{T}$ were obtained from $B_d^0 \to \pi^+ \pi^-$. The prediction was compared with the experimental value by
CDF, looking for a clear discrepancy that would signal NP. Due to the present large experimental errors no discrepancy can be drawn, but once this
errors get squeezed, this method will certainly reveal if there is need for a nonzero value of $\mathcal{A}_u$.

The first time the NP amplitude for $B_s^0 \to K^+ K^-$ was included in the scheme was in \cite{hep-ph/0410011}. Although the lack of experimental
data on the CP asymmetries for $B_s^0 \to K^+ K^-$ makes it impossible to apply the method to extract the NP parameters, predictions can be computed
for the $B_s^0 \to K^+ K^-$ observables once a NP model has been used to compute the NP amplitude. This was done in refs.
\cite{hep-ph/0511295,hep-ph/0610109} and will be explained in Chapter \ref{SUSYcontributions}.

Up to this point, let's explain the method in somewhat more detail. The decays $B_d^0 \to \pi^+ \pi^-$ and $B_s^0 \to K^+ K^-$ are described at
leading order by the diagrams in Figs. \ref{BtoPiPi} and \ref{BtoKK} respectively. In the U-spin limit the contributions are equal except for the CKM
elements. This means that $T_{\sss KK}=T_{\pi\pi}$ and $P_{\sss KK}=P_{\pi\pi}$ hold up $SU(3)$ breaking effects. The $SU(3)$ breaking is accounted
for in the deviations from unity of the ratios
\eq{r_{\sss T}\equiv \frac{T_{\sss KK}}{T_{\pi\pi}}\ ,\quad r_{\sss PT}\equiv \frac{P_{\sss KK}/T_{\sss KK}}{P_{\pi\pi}/T_{\pi\pi}}\equiv \xi
e^{i\Delta\theta}.\label{UspinBreaking}}
An estimate of this U-spin breaking within QCDF can be found in \cite{Beneke:2003zw}. The SU(3) breaking in $r_{\sss T}$ is expected to be much
larger than that in $r_{\sss PT}$ since both factorizable and non-factorizable terms are present in the former, while factorizable contributions
cancel in the later. A more ambitious calculation of the ratio $\m{r_{\sss T}}$ using QCD sum rules \cite{Khodjamirian:2003xk} gives
\eq{\m{r_{\sss T}}=1.52^{+0.18}_{-0.14}\label{rT}}
A fair estimation for the SU(3) breaking parameter $r_{\sss PT}$ is still lacking. In section \ref{glimpse} we will make a quick comment on this
issue. For the moment it will be assumed that this is not more than $\pm 20\%$ in magnitude,
\eq{\xi=1.0\pm 0.2}
and neglect the U-spin breaking in its phase ($\Delta\theta=0$), whose impact has been shown to be very small \cite{hep-ph/0204101}.

\begin{figure}
\begin{center}
\includegraphics[height=5cm,width=14cm]{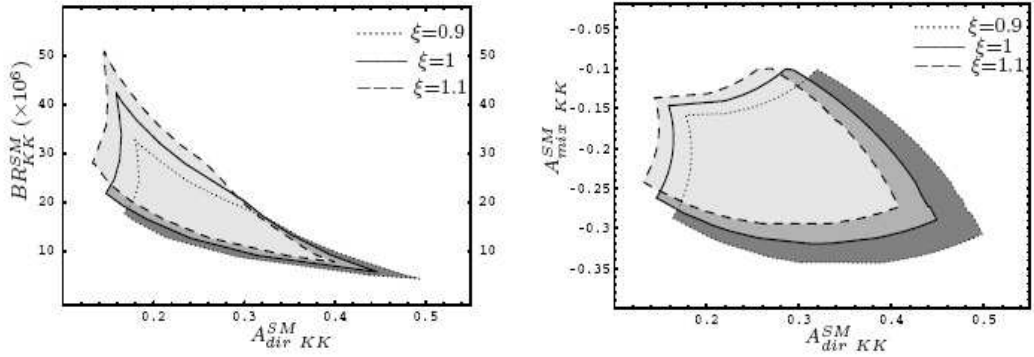}
\end{center}
\vspace{-.5cm} \caption{\small Correlations between the observables $A_{dir\ \sss KK}^{\sss SM}-BR^{\sss SM}_{\sss KK}$ and $A_{dir\ \sss KK}^{\sss
SM}-A_{mix\ \sss KK}^{\sss SM}$, for $\gamma=61^\circ$ and $\xi=1$, $\xi=0.9$ and $\xi=1.1$.} \label{CorrSM}
\end{figure}

With these values for the U-spin breaking parameters and the results for $\m{T_{\pi\pi}}$ and $\m{P_{\pi\pi}/T_{\pi\pi}}$ in Table \ref{tabledtheta}
it is possible to calculate the hadronic parameters for $B_s^0 \to K^+ K^-$ using (\ref{UspinBreaking}). These are shown in Table \ref{tableHadrBKK}.
Once the hadronic parameters have been calculated, we can give SM predictions for the observables\footnote{Regarding $BR_{\sss KK}^{\sss SM}$, it is
sometimes more useful to present the ratio of branching ratios of $B_s^0\to K^+K^-$ and $B_d^0\to \pi^+\pi^-$: $R_{d}^{s} \equiv BR(B_s^0 \to K^+
K^-)/BR(B_d^0 \to \pi^+ \pi^-)$ \cite{hep-ph/0410011}.} \cite{hep-ph/0511295,hep-ph/0610109}. These predictions are given in Table
\ref{tableSMpredsBKK} (see also \cite{hep-ph/0410407} for comparison), and the correlations between them are illustrated in Figure \ref{CorrSM}, for
different values of the U-spin breaking parameter $r_{\sss PT}$.

\begin{table}
\begin{center}
\begin{tabular}{|c|c|c|}
\hline
&&\\[-10pt]
 & $\mo{T_{\sss KK}}(10^{-6}{\rm GeV})$ & $\mo{P_{\sss KK}/T_{\sss KK}}$  \\[2pt]
\hline $\begin{array}{c}
\gamma=61^\circ\\
\xi=1 \end{array}$
& $(8.57,10.75)$ & $(0.07,0.20)$  \\
\hline $\begin{array}{c} \gamma=61^\circ\\ \xi=1\pm 0.2
\end{array}$
& $(8.57,10.75)$ & $(0.06,0.24)$  \\
\hline $\begin{array}{c} \gamma=(61^{+7}_{-5})^\circ\\ \xi=1
\end{array}$
& $(8.10,11.35)$ & $(0.06,0.25)$  \\
\hline
\end{tabular}
\end{center}
\caption{\small Allowed ranges for the hadronic parameters in the decay $B_s^0 \to K^+ K^-$. These are obtained from Table \ref{tabledtheta} by
U-spin correspondence as explained in the text. We take the central value in (\ref{rT}) for $\m{r_T}$. The impact of the uncertainty in the U-spin
breaking parameter $\xi$ and CKM-angle $\gamma$ is shown.} \label{tableHadrBKK}
\end{table}

\begin{table}
\begin{center}
\begin{tabular}{|c|c|c|c|c|}
\hline
&&&&\\[-10pt]
 & $BR_{\sss KK}^{\sss SM}\ (\times 10^6)$ & $R_{d}^{s \; SM}$ &
$A_{dir\ \sss KK}^{\sss SM}$ & $A_{mix\ \sss KK}^{\sss SM}$ \\[2pt]
\hline $\begin{array}{c}
\gamma=61^\circ\\
\xi=1 \end{array}$
& $(6.4,42.6)$ & $(1.2,9.3)$ & $(0.15,0.45)$ & $(-0.32,-0.10)$ \\
\hline $\begin{array}{c} \gamma=61^\circ\\ \xi=1\pm 0.2
\end{array}$
& $(4.2,61.9)$ & $(0.8,13.5)$ & $(0.12,0.56)$ & $(-0.38,-0.09)$ \\
\hline $\begin{array}{c} \gamma=(61^{+7}_{-5})^\circ\\ \xi=1
\end{array}$
& $(5.0,60.7)$ & $(0.9,13.2)$ & $(0.08,0.58)$ & $(-0.34,+0.08)$ \\
\hline
\end{tabular}
\end{center}
\caption{\small SM predictions for the branching ratio and CP-asymmetries of $B_s^0 \to K^+ K^-$.} \label{tableSMpredsBKK}
\end{table}

These results can be used as a consistency check when comparing with experiment, or can be used together with experimental data to extract the NP
parameters $\mathcal{A}^u$, $\Phi_u$ and $\delta_T$. Up to date the only available piece of data on $B_s^0 \to K^+ K^-$ is the value of given by CDF
\cite{hep-ph/0607021},
\eq{BR(B_s\to K^+K^-)_{\rm exp}=(24.4\pm 1.4\pm 4.6)\cdot 10^{-6}}
which fits fairly good in the SM prediction, and implies a value for $\mathcal{B}^{\sss \rm NP}$ compatible with zero,
\eq{\mathcal{B}^{\sss \rm NP}\in (-0.54,3.56)}
but with large errors that still allow for the presence of NP.

\section{A glimpse further}

\label{glimpse}

The method described above to \emph{measure} the NP parameters would rule out automatically all those NP models that fit the conditions listed in
Section \ref{MeasuringNP} but do not reproduce the measured values for these parameters. To that end one should \emph{compute} the contributions
coming from each particular model to be tested, which can be quite a tedious job. There are, however, several generic features of the NP flavor
structure that could be identified at once by the power of flavor symmetries. \emph{All} NP models which do not share these features would be ruled
out.

Consider the decay $B_s^0\to K^0 \bar{K}^0$. Only the diagrams (c) in Fig. \ref{BtoM1M2} contribute to this decay, so only the penguin
$P^s_{K^0\!\bar{K}^0}$ can be related to $P^s_{K^+\!K^-}$ by U-spin, and not the tree $T^s_{K^0\!\bar{K}^0}$ (which does not contain the ``pure''
tree (a)). However, because $\m{\lambda_u^{(s)}}\ll \m{\lambda_c^{(s)}}$, the amplitude for this process can be approximately written as
\eq{A(B_s^0\to K^0 \bar{K}^0)=\lambda_c^{(s)*} P^s_{K^0\!\bar{K}^0} + \mathcal{A}^d e^{i\Phi_d}}
Then measurements on $B_s^0\to K^0 \bar{K}^0$ allow to extract the NP parameters $\mathcal{A}^d$ and $\Phi_d$.

Consider the case in which the NP is isospin conserving. This means that whatever this NP is, it should predict $\mathcal{A}^d=\mathcal{A}^u$ and
$\Phi_d=\Phi_u$. Now, if the measured values for the NP parameters do not fit this rule,
then all isospin conserving NP models can be discarded.\\

The main advantage of using experimental measurements on a particular decay together with flavor symmetries to extract information in other channels,
is that the hadronic effects are taken into account in a model independent way. The price we pay is the large uncertainties coming from the SU(3)
breaking effects  (up to 30\%), which are still not under control.

A different approach is to use a particular theoretical framework to calculate the matrix elements, but the inclusion of hadronic effects always
leads to model dependence. It would be a step forward if both approaches could be combined to reach more reliable predictions.

An attempt in this direction has been made in \cite{hep-ph/0603239} using QCDF. According to the discussion at the end of Section \ref{QCDF}, and as
will be explained in detail in Chapter \ref{SymFac1}, the quantity
\eq{\Delta\ \equiv\ T^d_{K^0\!\bar{K}^0}-P^d_{K^0\!\bar{K}^0} \label{delta}}
for the decay $B_d^0\to K^0 \bar{K}^0$, is free from infrared divergencies when computed using QCDF. From (\ref{delta}) and the measured values for
the branching ratio and the direct CP-asymmetry for this decay, one can obtain its hadronic parameters $T^d_{K^0\!\bar{K}^0}$ and
$P^d_{K^0\!\bar{K}^0}$. QCDF predicts a relationship between these and the hadronic parameters for $B_s^0\to K^+ K^-$, which is now free of most of
the model dependence, and thus much more reliable than the pure QCDF computation.

As an example, one can obtain $d^s_{K^+\!K^-}$ in this way, and together with $d^d_{\pi^+\pi^-}$ from Table \ref{tabledtheta} obtain a value for the
U-spin breaking parameter $\xi$. This is described in detail in Chapter \ref{SymFac1}. The result as given in ref. \cite{hep-ph/0603239} is
\eq{\xi=0.81\pm 0.35 \label{xifromQCDF}}
This value can then be used in the analysis of Section \ref{MeasuringNP} instead of the poorly justified one taken above. However, one should
consider carefully up to what point these two approaches are independent when using the value in eq.~(\ref{xifromQCDF}) for $\xi$.

%


\chapter{Exploring $B\to KK$ Decays with Flavor Symmetries and QCDF}

\label{SymFac1}

Along the past chapters of this thesis we have seen continuously the difficulty and the phenomenological importance of the computation of the
hadronic parameters (``tree'' and ``penguin'') in two body non-leptonic $B$ decays. We have seen that this type of decays contain a huge amount of
information that can be used to understand and test the SM and to look and measure NP; but we have also seen that in order to do that the effects of
strong interactions have to be controlled in some way.

Moreover, the increasing precision that is being achieved experimentally, and the optimistic prospects for the future coming from the successes of
the CDF collaboration, the starting of LHC, and the possibility of a Super-B, urges the development of theoretical tools improving on the treatment
of hadronic uncertainties.

We have already seen several ways in which hadronic parameters in two-body non-leptonic $B$ decays can be extracted. First, in
Section~\ref{MeasuringSM} we have seen how these can be extracted directly from data. The argument is that three hadronic parameters can be extracted
from three independent observables, which for neutral $B$ decays are the branching ratio and the two components of the time dependent CP asymmetry.
One just has to invert the equations (\ref{BR(dtheta)})-(\ref{Amix(dtheta)}) and solve for the hadronic parameters. For the case of $B_d\to
\pi^+\pi^-$, the results were given in Table~\ref{tabledtheta}. One should be careful, though, when using this procedure in NP analyses, since this
is only correct when the decay at hand is not affected by the NP. In this thesis we bypass this potential problem by assuming that any NP affecting
$\bar{b}\to\bar{d}$ transitions is negligible at the considered level of accuracy.

A second approach that may be used to extract the hadronic parameters is QCDF as outlined in Sections~\ref{QCDFth} and \ref{QCDF}. The amplitude of a
$B$ decay in QCDF is given by the matrix elements in eq.~(\ref{ampQCDF}). Therefore, the hadronic parameters are given by $T_{\sss
M_1M_2}=\bra{M_1M_2}\mathcal{T}_A^u+\mathcal{T}_B^u\ket{B}$ and $P_{\sss M_1M_2}=\bra{M_1M_2}\mathcal{T}_A^c+\mathcal{T}_B^c\ket{B}$ (see Section
\ref{QCDF}). For the $B_d\to \pi^+\pi^-$ example, the hadronic parameters predicted by QCDF are shown in Table~\ref{HPQCDF} \cite{hep-ph/0308039}.
The corresponding QCDF prediction for the $B_d\to \pi^+\pi^-$ branching ratio is
\eq{BR(B_d\to \pi^+\pi^-)_{\rm \sss QCDF}= 8.9^{+4.0}_{-3.4}\,^{+3.6}_{-3.0}\,^{+0.6}_{-1.0}\,^{+1.2}_{-0.8} \times 10^{-6}}
which gives a central value which is high compared to the experimental number (\ref{BRpipiExp}), but with tremendous uncertainties. Clearly this is
presently a problematic decay for the QCDF approach.

\begin{table}
\begin{center}
\begin{tabular}{|c||c|ccc|}
\hline
\rule[-0.2cm]{0pt}{0.7cm}                                        &          Theory          &    S2    &   S3   &    S4  \\
\hline\hline
\rule[-0.3cm]{0pt}{0.8cm}%
$|T_{\pi\pi}|\ (10^{-6}{\rm GeV})$  & $7.7^{+0.04}_{-0.06}$    &   6.3    &   7.8  &   6.8  \\
\hline
\rule[-0.3cm]{0pt}{0.8cm}$|P_{\pi\pi}/T_{\pi\pi}|$               & $0.32^{+0.16}_{-0.09}$   &   0.49   &   0.37 &   0.48 \\
\hline
\end{tabular}
\end{center}
\vspace{-0.3cm} \caption{\small Hadronic parameters for $B_d\to \pi^+\pi^-$ from QCDF. S2, S3 and S4 correspond to different scenarios for the inputs
as described in ref.~\cite{hep-ph/0308039}.} \label{HPQCDF}
\end{table}

An alternative way to extract the hadronic parameters with the aim of giving predictions for the observables, is the flavor symmetry analysis
described in Section~\ref{MeasuringNP}. The process consists in establishing $SU(3)$ relations between the hadronic parameters such that the
knowledge of some hadronic parameters (that are extracted from data) allows to know the other hadronic parameters on which we are interested. Then,
after a sensible inclusion of $SU(3)$ breaking effects, one can calculate observables. The analysis in Chapter~\ref{Bdecays} relating $B_d\to
\pi^+\pi^-$ and $B_s\to K^+K^-$ gave the predictions plotted in Fig.~\ref{CorrSM}.

\begin{figure}
\begin{center}
\psfrag{aa}{\tiny $\xi=0.8$} \psfrag{bb}{\tiny $\xi=1.0$} \psfrag{cc}{\tiny $\xi=1.2$} \psfrag{dd}{\tiny QCDF} \psfrag{ee}{\tiny CDF}
\psfrag{BR}{\small\hspace{-1cm} $\stackrel{BR(B_s^0\to K^+K^-)\times 10^{6}}{}$}
\psfrag{Adir}{\hspace{-1cm}\begin{minipage}[l]{7cm}\vspace{0.5cm}{\tiny $A_{dir}(B_s^0\to K^+K^-)$}\end{minipage}}
\includegraphics[height=6cm,width=10cm]{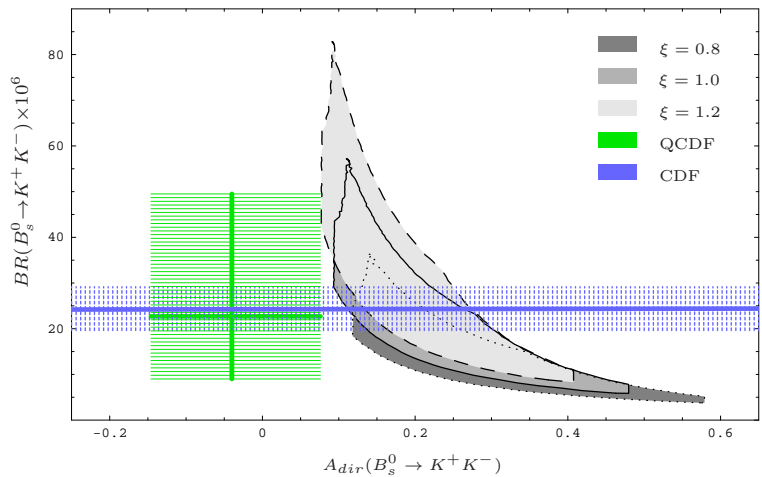}
\end{center}
\vspace{-0.5cm} \caption{\small Comparison between the predictions for $B_s\to K^+K^-$ from QCD factorization and flavor symmetry. The green box
shows the QCDF prediction \cite{hep-ph/0308039} (errors summed in quadrature), the gray regions show the predictions using $B_d\to \pi^+\pi^-$ and
flavor symmetry for different amounts of $SU(3)$ breaking (c.f. Chapter~\ref{Bdecays}), and the blue horizontal line is the experimental value for
the branching ratio \cite{hep-ex/0607021}.} \label{CompBsKK}
\end{figure}

A comparative plot between the predictions for $B_s\to K^+K^-$ from QCDF and flavor symmetry is shown in Fig.~\ref{CompBsKK}. As can be seen, the
theoretical predictions have uncertainties much above the experimental precision. Moreover, this decay will probably be measured with better accuracy
in Tevatron shortly, including the direct CP asymmetry. In LHCb both observables will be measured with very high accuracy. Therefore, it is mandatory
to improve substantially the theoretical predictions if the experimental efforts want to be exploited. But, how can this be done? To answer this
question it is necessary to understand what are the advantages and limitations of each of these approaches.

The flavor-symmetry approach has several positive features. It is a model-independent approach based on symmetries of the QCD lagrangian alone, as
described in Section~\ref{symmetries}. It also includes all the hadronic long-distance effects naturally, since they are contained inside the
hadronic parameters that are extracted from data. However, the major drawback is the inclusion of $SU(3)$ breaking corrections, which are obviously
completely unaccessible within the framework. These symmetry breaking effects depend on the dynamics and computing them requires the use of
non-perturbative methods. Alternatively, one might just assume that these effects are smaller than a certain percentage, and allow for arbitrary
breaking of this size. However, even in the cases in which this can be done model-independently, the associated uncertainties that are generated
preclude any precision in the predictions. Moreover, the predictions are also affected by large uncertainties related to the fact that the used data
has often large experimental errors. Therefore, the flavor-symmetry approach requires of two ingredients: more accurate experimental data, and more
reliable estimates of $SU(3)$ breaking.

On the other hand, QCD factorization is a purely theoretical framework based on QCD and the heavy-quark expansion. In principle its predictions are
unlimited up to a particular order in $\alpha_s$ and $\Lambda_{\rm\sss QCD}/m_b$. This means that it can predict all the hadronic parameters (not
only those for which a flavor-symmetry relation can be found) provided the non-perturbative input of form factors, decay constants and LCDAs is
given. However, as discussed at the end of Section~\ref{QCDF}, there are some difficulties with non-factorizable $1/m_b$ corrections that are
numerically relevant because they are chirally enhanced. When including these corrections, the infrared divergencies must be subtracted in a
model-dependent way, and the predictive power is partially lost. Moreover, as was already commented in Section~\ref{QCDF}, in QCDF the strong phases
are perturbatively calculable, and therefore appear first at order $\alpha_s$, which means that they are in general predicted to be small.
Phenomenologically, however, there are certain decays (for example $B\to \pi\pi$) in which the strong phases could be large, and not accountable by a
$\mathcal{O}(\alpha_s)$ correction. The QCD factorization approach should then be provided with a prescription to control the phenomenologically
relevant power-suppressed infrared divergencies, and with a mechanism that generates large strong phases whenever they are there.

In this chapter we present an approach that improves (at a phenomenological level) on the weak points mentioned above. This approach was proposed in
\cite{hep-ph/0603239} and since then it has been applied in \cite{hep-ph/0610109,hep-ph/0611280,arXiv:0705.0477,arXiv:0707.2046}.

\section{The theoretical input: $\Delta$}

\label{SectionDelta}

We consider now the following quantity defined as the difference between tree and penguin contributions,
\eq{\Delta\equiv T-P\ . \label{defDelta}}
This quantity is a hadronic, process-dependent, intrinsically non-perturbative object, and thus difficult to compute theoretically. Such hadronic
quantities are usually either extracted from data or computed using some factorization-based approach, according to the long discussion given above.
In the latter case, $\Delta$ could suffer from the aforementioned phenomenological problems of factorization.

However, for a certain class of decays, $T$ and $P$ share the same long-distance dynamics: the difference comes from the ($u$ or $c$) quark running
in the loop, which is dominated by short-distance physics~\cite{hep-ph/0603239}. Indeed, in such decays, $\Delta=T-P$ is not affected by the
breakdown of factorization that affects annihilation and hard-spectator contributions, and it can be computed in a well-controlled way leading to
safer predictions and smaller uncertainties.

Consider first the effect of the $A$-operators in eq.~(\ref{Aops}). The contribution to $\Delta$ will come from $\mathcal{T}_A^u-\mathcal{T}_A^c$.
The current-current pieces will be just $\sim a_1(\cdots)+a_2(\cdots)$ since they only appear in $\mathcal{T}_A^u$. They will be proportional to the
hard spectator functions $H_1$ and $H_2$ and they will be affected by the already familiar IR divergencies. However, the rest of the pieces will be
$\sim (a_i^u-a_i^c)$. So looking at eq.~(\ref{aQCDF}) we see that they will just receive contributions from the penguin contributions, proportional
to the functions $(P_i^u-P_i^c)$. These functions are well behaved, and dominated by short-distance physics. At order $\alpha_s$, and neglecting
electromagnetic corrections, penguin contractions are only present for $i=4,6$, so the only contribution to $\Delta$ (besides the $a_1$ and $a_2$
part) will be proportional to $\ (P_4^u-P_4^c)\pm r_\chi (P_6^u-P_6^c)$. Specifically, (see ref.~\cite{hep-ph/0308039}) we have
\eqa{
P_4^u(M_2)-P_4^c(M_2) &=& -\frac{C_F\alpha_s}{4\pi N_c}\,C_1 \Big[ G_{\sss M_2}(0)-G_{\sss M_2}(m_c^2/m_b^2) \Big]\ ,\nn\\
P_6^u(M_2)-P_6^c(M_2) &=& -\frac{C_F\alpha_s}{4\pi N_c}\,C_1 \Big[ \hat{G}_{\sss M_2}(0)-\hat{G}_{\sss M_2}(m_c^2/m_b^2) \Big]\ , }
where $G(x)$ and $\hat{G}(x)$ are certain penguin functions. The $B$-operators in Eq.~(\ref{Bops}) contain also the pieces proportional to $b_1$ and
$b_2$ that enter $\mathcal{T}_B^u$ and not $\mathcal{T}_B^c$, and these pieces will introduce the IR divergencies from weak annihilation. The rest of
the pieces, however, as can be seen looking at the specific form of the coefficients $b_i$ \cite{hep-ph/0308039}, cancel completely in the difference
$b_i^u-b_i^c$.

So now consider penguin-mediated decays that do not receive contributions from $a_{1,2}$ or $b_{1,2}$. For these decays, the quantity $\Delta$ is at
$\mathcal{O}(\alpha_s)$ free from dangerous IR divergencies in QCDF. This makes it a quantity that can be computed with acceptable accuracy and can
be used as a reliable theoretical input. In general, $\Delta$ is given by
\eq{\Delta_{\sss M_1M_2}=A_{\sss M_1M_2} \frac{C_F \alpha_s}{4\pi N}C_1 \big[ \bar{G}_{\sss M_2}(m_c^2/m_b^2)-\bar{G}_{\sss M_2}(0) \big]\ ,}
where $\bar{G}_{\sss M_2}\equiv G_{\sss M_2}\pm r_\chi^{\sss M_2}\hat{G}_{\sss M_2}$, with the plus or the minus sign depending on whether $M_1$ is a
pseudoscalar or a vector meson.

In our analysis of $B\to KK$ modes, we are mainly interested in $\Delta^{d}_{\sss KK}$ corresponding to $B_d\to K^0\bar{K}^0$. A careful computation
gives
\eq{|\Delta^{d}_{\sss KK}|=(2.96\pm 0.97)\times 10^{-7}\ . \label{ValueDeltaKK}}
Only the absolute value is physical, because an unphysical global phase in the amplitude would rotate simultaneously $P$ and $T$, rotating $\Delta$
accordingly. In particular, in the formulae for the hadronic parameters given in the following section, $\Delta$ can be chosen to be \emph{real},
introducing only a global phase in the amplitude. From now on we will always take it as a real quantity. The values of the inputs used in the
numerical evaluation of $\Delta^d_{\sss KK}$, and the impact of each one of them on the error, are given in tables \ref{inputsDeltaKK} and
\ref{errorsDeltaKK}.

\begin{table}
\begin{center}
\begin{tabular}{cccccc}
\hline\hline
\rule[-0.15cm]{0pt}{0.6cm}        $\mu$     &   $\alpha_1^K$ & $\alpha_2^K$ & $F^{B\to K}_0$  & $m_s(2\,{\rm GeV})$ & $m_c/m_b$        \\
\hline
\rule[-0.15cm]{0pt}{0.55cm}  $(m_b/2,2m_b)$ &  $0.2\pm 0.2$  & $0.1\pm 0.3$ & $0.34\pm 0.05$  &  $103\pm 20$ MeV    &   $0.30\pm0.06$  \\
\hline\hline
\end{tabular}
\end{center}
\vspace{-0.5cm} \caption{\small Inputs used in the computation of $\Delta^d_{\sss KK}$.} \label{inputsDeltaKK}
\end{table}

\begin{table}
\begin{center}
\begin{tabular}{c|cccccc}
\hline\hline
\rule[-0.15cm]{0pt}{0.6cm}                         &  $\mu$   &   $\alpha_1^K$    & $\alpha_2^K$ & $F^{B\to K}_0$ & $m_s(2\,{\rm GeV})$ & $m_c/m_b$ \\
\hline
\rule[-0.15cm]{0pt}{0.55cm}$|\Delta^{d}_{\sss KK}|$ & $19.7\%$ & $4.1\%$ & $0.9\%$ &    $15.8\%$  &      $5.7\%$   &           $53.7\%$              \\
\hline\hline
\end{tabular}
\end{center}
\vspace{-0.5cm} \caption{\small Relative impact of the input uncertainties on the error of $\Delta^d_{\sss KK}$.} \label{errorsDeltaKK}
\end{table}

\section{Tree and Penguin contributions}

\label{TandP}

\begin{figure}
\begin{center}
\includegraphics[height=5cm,width=8cm]{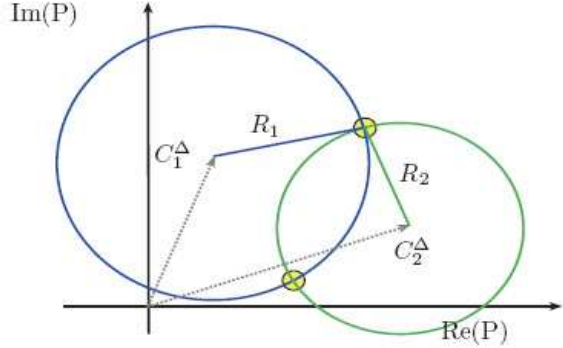}
\end{center}
\vspace{-0.5cm} \caption{\small Eqs.~(\ref{eqscircles}) can be written as $R_1=|P-C_1^{\Delta}|$ and $R_2=|P-C_2^{\Delta}|$, where the $R_i$ depend
only on data and CKM elements, and the $C_i^{\Delta}$ depend only on $\Delta$ and CKM elements. These are the equations for two circles with centers
in $C_i^{\Delta}$ and radii $R_i$ in the complex $P$ plane. Their crossing points are the solutions for $P$. The condition for the existence of
solutions is $|R_1-R_2|\le |C_1^{\Delta}-C_2^{\Delta}|\le |R_1+R_2|$, which translates into Eqs.~(\ref{cons}).} \label{circles}
\end{figure}

\begin{figure}
\begin{center}
\psfrag{A}{\hspace{-1.5cm}$\Adir(B_d\to K^{0}\bar{K}^{0})$} \psfrag{BR}{\hspace{-1.5cm}$BR(B_d\to K^{0}\bar{K}^{0})\times 10^6$}
\includegraphics[width=8cm]{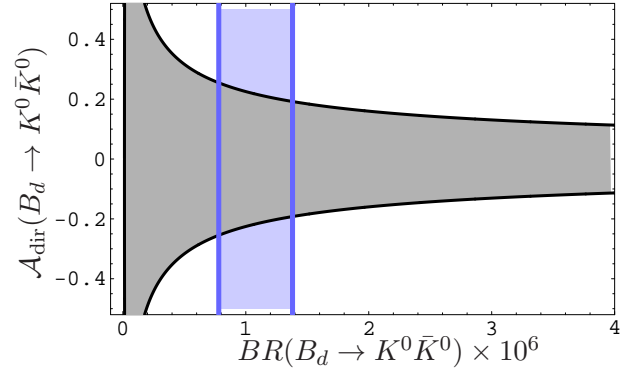}
\end{center}
\caption{\small Allowed values for the direct CP asymmetry of $B_d\to K^{0}\bar{K}^{0}$ in terms of its branching ratio, according to the value of
$\Delta_{KK}^d$. The vertical blue band corresponds to the experimental branching ratio, which sets an upper bound on $|\Adir|$ slightly above
$0.2$.} \label{BRAdir}
\end{figure}

We now describe how to obtain the hadronic parameters $T$ and $P$ from $\Delta$ and experimental data. The derivation will be completely general, and
is valid for any decay, independently of whether the computation of $\Delta$ is theoretically clean or not. We begin writing the two self-conjugated
amplitudes in terms of tree and penguin contributions,
\eq{A=\lambda_u^{(D)*} T +\lambda_c^{(D)*} P\ ,\quad \bar{A}=\lambda_u^{(D)} T +\lambda_c^{(D)} P\ .}
Now we put $T=P-\Delta$ and we square the amplitudes,
\eq{
\begin{array}{rcl}
|A|^2 & = & |\lambda_c^{(D)*}+\lambda_u^{(D)*}|^2\left| P + \frac{\lambda_u^{(D)*}}{\lambda_c^{(D)*}+\lambda_u^{(D)*}} \Delta \right|^2\ ,\\
&&\\
|\bar{A}|^2 & = & |\lambda_c^{(D)}+\lambda_u^{(D)}|^2\left| P + \frac{\lambda_u^{(D)}}{\lambda_c^{(D)}+\lambda_u^{(D)}} \Delta \right|^2\ .
\end{array}}
But the squared amplitudes are directly related to observables,
\eqa{|A|^2&=&BR(1+\Adir)/g_{PS}\ ,\nonumber\\
|\bar{A}|^2&=&BR(1-\Adir)/g_{PS}\ ,}
where $g_{PS}$ is the usual phase-space factor (c.f. Eq.~(\ref{gps})). So two equations can be written that relate observables with $P$ and $\Delta$,
\eqa{ \frac{BR(1+\Adir)/g_{PS}}{|\lambda_c^{(D)*}+\lambda_u^{(D)*}|^2}=
\left| P + \frac{\lambda_u^{(D)*}}{\lambda_c^{(D)*}+\lambda_u^{(D)*}} \Delta \right|^2\ ,\nonumber\\
\frac{BR(1-\Adir)/g_{PS}}{|\lambda_c^{(D)}+\lambda_u^{(D)}|^2}=\left| P + \frac{\lambda_u^{(D)}}{\lambda_c^{(D)}+\lambda_u^{(D)}} \Delta \right|^2\ .
\label{eqscircles} }
These are the equations for two circles in the complex $P$ plane, whose solutions are the two points of intersection. This will result in a two-fold
ambiguity in the determination of $P$ and $T$ (see Fig.~\ref{circles}). Before writing down the analytical solutions, notice that in order for
solutions to exist, the separation between the centers of these circles must be smaller than the sum of the radii but bigger than the difference.
This translates into a consistency condition between $BR$, $\Adir$ and $\Delta$:
\eqa{ |\Adir|  &\le&  \sqrt{\frac{\mathcal{R}_D^2\Delta^2}{2\widetilde{BR}}\Big(2-\frac{\mathcal{R}_D^2\Delta^2}{2\widetilde{BR}} \Big)}
\approx \frac{\mathcal{R}_D\Delta}{\sqrt{\widetilde{BR}}}\ ,\nn\\[5pt]
\widetilde{BR} &\ge& \frac{\mathcal{R}_D^2\Delta^2}{4}\ , \label{cons}}
where $\widetilde{BR}\equiv BR/g_{PS}\ $, and $\mathcal{R}_D$ is a specific combination of CKM factors, $\mathcal{R}_D=2|c_0^{\sss
(D)}|/\sqrt{c_2^{\sss (D)}}$ (see below). The approximation for the upper bound on $|\Adir|$ holds up to very small corrections in the usual
situation $\Delta\lesssim \mathcal{O}(10^{-7})$ and $BR\sim \mathcal{O}(10^{-6})$. The first condition in eq.~(\ref{cons}) turns out to be highly
nontrivial. For example, Fig.~\ref{BRAdir} shows the allowed values for the direct CP asymmetry of $B_d\to K^{0}\bar{K}^{0}$ in terms of its
branching ratio. It can be seen that for the present data of the branching ratio, the direct CP asymmetry is predicted to be less than about $\sim
20\%$.

The hadronic quantities $P$ and $T$ are then given by
\eqa{
Im[P] & = & \frac{\widetilde{BR}\,\Adir}{2 c_0^{(D)}\Delta}\ ,\nonumber\\
&&\nonumber\\
Re[P] & = & -c_1^{(D)}\,\Delta \pm
\sqrt{-Im[P]^2-\left(\frac{c_0^{(D)}\Delta}{c_2^{(D)}}\right)^2+\frac{\widetilde{BR}}{c_2^{(D)}}}\ ,\nonumber\\
&&\nonumber\\
T&=&P+\Delta\ , \label{eqTP}}
where the coefficients $c_i^{(D)}$ are again some specific combinations of CKM elements (see Table \ref{coeffs}).

\begin{table}
\begin{center}
\begin{tabular}{cccc}
\hline\hline
&&&\\[-10pt]
$c_0^{(d)}$ & $c_1^{(d)}$ & $c_2^{(d)}$ &  $\mathcal{R}_d$\\[2pt]
\hline
&&&\\[-12pt]
$\ -3.15\cdot 10^{-5}\ $  &  $\ -0.034\ $  &  $\ 6.93\cdot 10^{-5}\ $  &  $7.58\cdot 10^{-3}$  \\
\hline\hline
&&&\\[-10pt]
$c_0^{(s)}$    &    $c_1^{(s)}$   &   $c_2^{(s)}$  &  $\mathcal{R}_s$\\
\hline
&&&\\[-12pt]
$\ 3.11\cdot 10^{-5}\ $  &  $\ 0.011\ $ &  $\ 1.63\cdot 10^{-3}\ $ &  $1.54\cdot 10^{-3}$ \\
\hline\hline
\end{tabular}
\end{center}
\caption{\small Numerical values for the coefficients $c_i^{(D)}$ and $\mathcal{R}_D$ for $\gamma=62^\circ$.} \label{coeffs}
\end{table}

Equations (\ref{eqTP}) allow to extract the hadronic parameters $T$ and $P$ from experimental data on $BR$ and $\Adir$, information on sides of the
unitarity triangle and the weak phase $\gamma$, and the theoretical value for $\Delta$. This method is also powerful because if no experimental
information is available for $\Adir$, one can just vary $\Adir$ over its allowed range in eq.(\ref{cons}). So in fact $T$ and $P$ can be extracted
from $BR$, $\Delta$ and CKM elements. This is just the case for $B_d\to K^{0}\bar{K}^{0}$: while there is a preliminary experimental number for the
direct CP asymmetry, it is still too uncertain. Therefore, in our analysis we will assume that $\Adir(B_d\to K^{0}\bar{K}^{0})$ is unknown, and we
will give results for different values of this asymmetry between $-0.2$ and $0.2$, according to Fig.~\ref{BRAdir}. We will also see that this range
can be still made smaller by looking at data on $B_d\to \pi^+\pi^-$, but this will not be taken into account in our analysis.

Finally, we would like to mention that this procedure is also useful to extract hadronic parameters from charged $B$ decays, were the CP observables
are just two: branching ratio and CP asymmetry. An example is the decay $B^+\to \pi^+\phi$, for which the very small value $\Delta_{\pi^+\phi}\sim
10^{-8}\ $ indicates a very small CP asymmetry, as long as the branching ratio is not too small (at present there is only an upper bound).

\section{Flavor symmetries and QCDF}

\label{Flavor&QCDF}

\begin{figure}
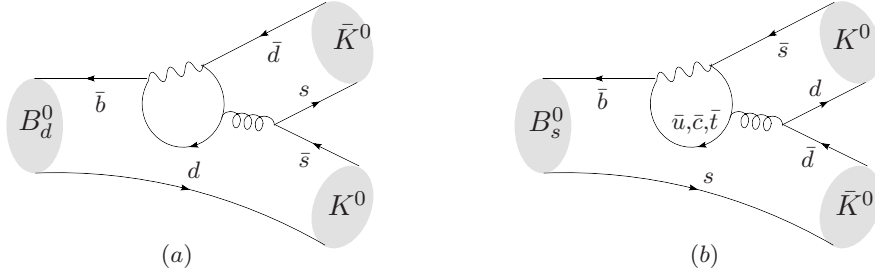

\begin{center}
\includegraphics[width=5cm,height=3.5cm]{PenguinBsK+K-.eps}
\hspace{1.5cm}
\includegraphics[width=5cm,height=3.5cm]{PenguinBsK+K-.eps}
\Text(-329,44)[lb]{$B_d^0$} \Text(-212,15)[lb]{$K^0$} \Text(-210,78)[lb]{$\bar{K}^0$} \Text(-300,54)[lb]{\footnotesize $\bar{b}$}
\Text(-265,29)[lb]{\footnotesize $d$} \Text(-235,73)[lb]{\footnotesize $\bar{d}$} \Text(-223,60)[lb]{\footnotesize $s$}
\Text(-223,33)[lb]{\footnotesize $\bar{s}$} \Text(-275,-5)[lb]{\footnotesize $(a)$}
\Text(-136,44)[lb]{$B_s^0$} \Text(-20,15)[lb]{$\bar{K}^0$} \Text(-21,78)[lb]{$K^0$} \Text(-110,55)[lb]{\footnotesize $\bar{b}$}
\Text(-70,27)[lb]{\footnotesize $s$} \Text(-33,33)[lb]{\footnotesize $\bar{d}$} \Text(-30,60)[lb]{\footnotesize $d$} \Text(-42,75)[lb]{\footnotesize
$\bar{s}$} \Text(-75,-5)[lb]{\footnotesize $(b)$} \Text(-82,46)[lb]{\footnotesize $\bar{u},\!\bar{c},\!\bar{t}$}

\end{center}
\caption{\small Diagrams contributing to (a) $B_d\to K^0\bar{K}^0$ and (b) $B_s\to K^0\bar{K}^0$ related through a $U$-spin transformation.}
\label{fig:uspink0}
\end{figure}

Using $U$-spin symmetry, we can relate the two penguin-mediated decays $\bar{B}_d \to K_0\bar{K}_0$ and $\bar{B}_s \to K_0\bar{K}_0$, as exemplified
in fig.~\ref{fig:uspink0}. Let us stress that we work with the operators of the effective Hamiltonian: internal loops have already been integrated
out to yield four-quark operators, so that the internal loop of the $u$-penguin is not affected by $U$-spin rotations. $U$-spin breaking should be
much smaller here than usual: it does not affect final-state interaction since both decays involve the same outgoing state, and it shows up mainly in
power-suppressed effects. This is confirmed by QCDF:
\eqa{ P^{s0}&=&f\,P^{d0} \Big[1+(A^d_{\sss KK}/P^{d0})\Big\{\delta\alpha_4^c +\delta\beta_3^c+2\delta\beta_4^c \Big\}\Big]
\ \equiv\ f\,P^{d0}\,(1+\delta_{P^{s0}})\,,\nn\\[7pt]
T^{s0}&=&f\,T^{d0}\Big[1+(A^d_{\sss KK}/T^{d0})\Big\{\delta\alpha_4^u+\delta\beta_3^u+2\delta\beta_4^u \Big\}\Big] \ \equiv\
f\,T^{d0}\,(1+\delta_{T^{s0}})\,, \label{Ps0Ts0}}
where we define the $U$-spin breaking differences $\delta\alpha_i^p\equiv\bar\alpha_i^p-\alpha_i^p$ (id. for $\beta$). The superscripts identify the
channel and the bar denotes quantities for decays with a spectator $s$ quark. Apart from the factorizable ratio~:
\eq{f=\frac{A_{\sss KK}^s}{A_{\sss KK}^d}=\frac{M_{B_s}^2 F_0^{\bar{B}_s\to K}(0)}{M_{B_d}^2 F_0^{\bar{B}_d\to K}(0)}=0.94\pm 0.20}
which should be computed on the lattice, $U$-spin breaking arises through $1/m_b$-suppressed contributions in which most long-distance contributions
have cancelled out.

First, the hard-spectator scattering ($\delta\alpha$) probes the difference between $B_d$- and $B_s$-distribution amplitudes:
\eqa{ \delta\alpha_4^p&=&\alpha_s C_F C_3 \pi/N_c^2 \times \delta\lambda_B \times
[\langle \bar{x} \rangle_{\sss K}^2 + r_\chi^{\sss K} \langle x \rangle_{\sss K} X_H^{\sss K}]\ ,\nn\\
\delta\lambda_B&=&B_{\sss KK}^s M_{B_s}/(A_{\sss KK}^s \lambda_{B_s})-B_{\sss KK}^d M_{B_d}/(A_{\sss KK}^d \lambda_{B_d})\ , }
$B_{\sss KK}^q=f_{B_q} f_K^2 {G_F}/{\sqrt{2}}$, $\langle \bar{x} \rangle_K$ and $M_{B_q}/\lambda_{B_q}$ are first and first inverse moments of $K$
and $B_q$ distribution amplitudes~\cite{hep-ph/0308039}, respectively. $\delta\lambda_B$ is expected small, since the dynamics of the heavy-light
meson in the limit $m_b\to\infty$ should vary little from $B_d$ and $B_s$. Second, the annihilation contributions ($\delta\beta$) contain a $U$-spin
breaking part when the gluon is emitted from the light quark in the $B_{d,s}$-meson (this effect from $A_1^i$ and $A_2^i$ defined
in~\cite{hep-ph/0308039} is neglected in the QCDF model for annihilation terms).

Taking the hadronic parameters in~\cite{hep-ph/0308039}, we obtain
\eq{|\delta_{P^{s0}}| \leq 0.03\ ,\quad |\delta_{T^{s0}}| \leq 0.03\ . \label{deltass0}}
These relations yield also the constraint $\Delta^s_{\sss KK} = f \Delta^d_{\sss KK}$  up to $1/m_b$-suppressed corrections, relating observables in
both decays.

Relations exist between $\bar{B}_d \to K_0\bar{K}_0$ and $\bar{B}_s \to K^+ K^-$ as well. A combination of $U$-spin and isospin rotations leads from
the penguin contribution in $\bar{B}_d \to K_0\bar{K}_0$ to that in $\bar{B}_s \to K_0\bar{K}_0$, then to $\bar{B}_s \to K^+ K^-$, up to electroweak
corrections (it corresponds to fig.~\ref{fig:uspink0} up to replacing $d\to u$ in the right-hand diagram). On the other hand, there are no such
relations between tree contributions, since $\bar{B}_s \to K^+ K^-$ contains tree contributions which have no counterpart in the penguin-mediated
decay $\bar{B}_d \to K_0\bar{K}_0$. This is seen in QCDF as well:
\eqa{ P^{s\pm}&\hspace{-0.3cm}=\hspace{-0.3cm}& f\,P^{d0}\Big[1 + ({A_{\sss
KK}^d}/{P^{d0}})\Big\{\frac{3}{2}(\delta\alpha_4^c+\delta\beta_3^c+2\delta\beta_4^c\Big\}\Big]
\,\equiv\,f\,P^{d0}\,(1+\delta_{P^{s\pm}})\ , \label{PspmTspm}\\
T^{s\pm}&\hspace{-0.3cm}=\hspace{-0.3cm}& f\,T^{d0}+A^s_{\sss KK}{\bar\alpha_1} \Big[1+\frac{1}{\bar\alpha_1}\Big\{\bar\beta_{1}+\delta\alpha_4^u +
\delta\beta_3^u+2\delta\beta_4^u \Big\}\Big] \,\equiv\,f\,T^{d0}+A^s_{\sss KK}{\bar\alpha_1}\,(1+\delta_{T^{s\pm}})\ .\nn }
From QCDF, we obtain the following bounds:
\eq{|\delta_{P^{s\pm}}| \leq 0.02\ ,\quad |\delta_{T^{s\pm}}| \leq 0.04\ . \label{deltasspm}}
The latter bound shows that the flavor-symmetry breaking corrections are smaller than $\ T^{d0}/(A^d_{\sss KK} \bar\alpha_1)=O(10\%)$. Fortunately,
$T^{s\pm}$ is strongly CKM suppressed in $B_s\to K^+ K^-$ so that the uncertainty on its QCDF determination will affect the branching ratio and
CP-asymmetries only marginally.

Finally, these relations between $B_d$ and $B_s$ hadronic parameters are affected by electroweak penguins, small in the SM but potentially enhanced
by NP effects.

\section{SM predictions for $B_s\to KK$ and SU(3) breaking}

In this section we present the results for the branching ratios and CP asymmetries of $B_s\to K^0 \bar{K}^0$ and $B_s\to K^+ K^-$. The knowledge of
the $B_s\to K^+ K^-$ hadronic parameters, which require data from $B_d\to K^0 \bar{K}^0$ (and not $B_d\to \pi^+ \pi^-$ as in Chapter \ref{Bdecays})
allows us to give also predictions for the $SU(3)$ breaking parameters $|r_T|$ and $\xi$ used in Chapter \ref{Bdecays} relating $B_s\to K^+ K^-$ to
$B_d\to \pi^+ \pi^-$.

Let us outline briefly the basic steps in the process.

\begin{itemize}

\item The $B_d\to K^0 \bar{K}^0$ hadronic parameters $P^{d0}$ and $T^{d0}$ are obtained from Eqs.~(\ref{eqTP}), with the value
of $\Delta^d_{\sss KK}$ given in Eq.~(\ref{ValueDeltaKK}) and the experimental value for the $B_d\to K^0 \bar{K}^0$ branching ratio
\cite{hep-ex/0608036}
\eq{BR(B_d\to K^0 \bar{K}^0)_{\rm exp}=(1.08\pm 0.30)\cdot 10^{-6}\ .}
Since the $B_d\to K^0 \bar{K}^0$ direct CP asymmetry ($A^{d0}_{\rm dir}$) is still uncertain, we vary its value between $-0.2$ and $0.2$ according to
the discussion in Section \ref{TandP} (see Fig.~\ref{BRAdir}).

\item The hadronic parameters $P^{s0}$, $T^{s0}$, $P^{s\pm}$ and $T^{s\pm}$ for $B_s\to K^0 \bar{K}^0$ and $B_s\to K^+ K^-$
are obtained as described in Section \ref{Flavor&QCDF}, in particular from Eqs.~(\ref{Ps0Ts0}), (\ref{PspmTspm}), (\ref{deltass0}) and
(\ref{deltasspm}). In order to account for well-behaved short-distance $1/m_b$-suppressed corrections not yet included, we stretch the $SU(3)$
breaking bounds to more conservative ranges:
\eq{|\delta_{P^{s0}}|,\,|\delta_{T^{s0}}|,\,|\delta_{P^{s\pm}}|,\,|\delta_{T^{s\pm}}|\le 0.05\ .}
This $\delta$'s are complex quantities and their phases are varied from $0$ to $2\pi$.

\item The observables are then obtained from the ranges of the hadronic parameters.

\end{itemize}

There is a further comment before presenting the results. We already mentioned in Section \ref{TandP} that the extraction of the hadronic parameters
results in a two-fold ambiguity (see Fig.~\ref{circles}). We lift this ambiguity on the basis that the rejected predictions would be in conflict with
the $B_d\to \pi^+ \pi^-$ analysis made in Chapter \ref{Bdecays}: First, this solution requires a way too large $U$-spin breaking in $\arg (P/T)$.
Second, a prediction of the rejected solution is that $A_{\rm dir}^{s\pm}<0$, which is clearly in contradiction with the $B_d\to \pi^+ \pi^-$
analysis, that predicts it positive (see Fig.~\ref{CorrSM}).

\begin{table}
\begin{center}
\vspace{2.05cm} \hspace{0.1cm}\colorbox{yellow}{\begin{minipage}[t]{11.7cm} \vspace{1.5cm}\hspace{0cm} \end{minipage}}
\vspace{-3.55cm}\\
\small
\begin{tabular}{|l||c|c|c|}
\hline
\rule[-0.15cm]{0pt}{0.55cm}  &  $BR^{s0}\,\times 10^6$ &  $A_{dir}^{s0}\,\times 10^2$  &  $A_{mix}^{s0}\,\times 10^2$ \\
\hline
\hline \rule[-0.15cm]{0pt}{0.55cm} $A_{dir}^{d0}=-0.2$ & $ 20.8\pm 6.7 \pm 4.6 \pm 1.6$  & $  0.9 \pm 0.3 \pm 0.1$  & $ -0.2 \pm 0.8 \pm 0.1$   \\
\hline \rule[-0.15cm]{0pt}{0.55cm} $A_{dir}^{d0}=-0.1$ & $ 20.6\pm 7.5 \pm 4.6 \pm 1.6$  & $  0.4 \pm 0.3 \pm 0.0$  & $ -0.7 \pm 0.8 \pm 0.1$   \\
\hline \rule[-0.15cm]{0pt}{0.55cm} $A_{dir}^{d0}=0$    & $ 20.5\pm 7.5 \pm 4.5 \pm 1.5$  & $  0.0 \pm 0.3 \pm 0.0$  & $ -0.8 \pm 0.6 \pm 0.1$   \\
\hline \rule[-0.15cm]{0pt}{0.55cm} $A_{dir}^{d0}=0.1$  & $ 20.6\pm 7.6 \pm 4.5 \pm 1.6$  & $ -0.4 \pm 0.3 \pm 0.0$  & $ -0.7 \pm 0.8 \pm 0.1$   \\
\hline \rule[-0.15cm]{0pt}{0.55cm} $A_{dir}^{d0}=0.2$  & $ 20.8\pm 7.0 \pm 4.6 \pm 1.8$  & $ -0.9 \pm 0.3 \pm 0.1$  & $ -0.2 \pm 0.7 \pm 0.2$   \\
\hline
\end{tabular}
\end{center}
\caption{\small Observables for $B_s\to K^0 \bar{K}^0$ as functions of the direct asymmetry $A_{\rm dir}(B_d\to K^0 \bar{K}^0)$ within the SM. The
errors for the branching ratios are split in: (1) Uncertainties of all inputs except $f$ and $\gamma$, (2) Uncertainty of $f$, (3) Uncertainty of
$\gamma$. For the asymmetries, the second error is due to $\gamma$. The preferred values for $A_{\rm dir}^{d0}$ are highlighted (see the text).}
\label{tableBsK0K0}
\end{table}

In Table \ref{tableBsK0K0} we present the results for the $B_s\to K^0 \bar{K}^0$ observables. An average over all the values for $A_{\rm dir}^{d0}$
gives a single prediction for the branching ratio,
\eq{BR(B_s\to K^0 \bar{K}^0)=(21\pm 8\pm 5\pm 2)\times 10^{-6}\ ,}
and the following ranges for the CP asymmetries,
\eqa{
-0.011 \le& A_{\rm dir}^{s0} &\le 0.011\ ,\\
-0.015 \le& A_{\rm mix}^{s0} &\le 0.005\ . }

In Table \ref{tableBsK+K-} we show the results for the $B_s\to K^+K^-$ observables. Again, a global average gives the following prediction for the
branching ratio,
\eq{BR(B_s\to K^+ K^-)=(24\pm 9\pm 5\pm 2)\times 10^{-6}\ ,\label{BRBK+K-th}}
and the following ranges for the CP asymmetries,
\eqa{
-0.22 \le& A_{\rm dir}^{s\pm} &\le 0.49\ ,\\
-0.55 \le& A_{\rm mix}^{s\pm} &\le 0.40\ . }

The information from the $B_d\to \pi^+ \pi^-$ analysis can also be used to chose some preferred values for $A_{\rm dir}^{d0}$ between those
considered in Tables \ref{tableBsK0K0} and \ref{tableBsK+K-}. Indeed, as can be seen from Table \ref{tableBsK+K-}, there is a sign anticorrelation
between $A_{\rm dir}^{d0}$ and $A_{\rm mix}^{s\pm}$. The analysis in Chapter \ref{Bdecays} clearly predicts negative values for $A_{\rm mix}^{s\pm}$,
which correspond, roughly, to positive values of $A_{\rm dir}^{d0}$. All these correlations between all these $B\to KK$ and $B\to \pi\pi$ modes turn
out to be extremely powerful. In the tables we have highlighted the preferred results (corresponding to $A_{\rm dir}^{d0}\gtrsim 0$). With these
restrictions the global averages given above get tightened, for example,
\eq{BR(B_s\to K^+ K^-)\to (20\pm 7\pm 4\pm 1)\times 10^{-6}\ ,\label{BRBK+K-thImp}}

\begin{table}
\begin{center}
\vspace{2.05cm} \hspace{0.1cm}\colorbox{yellow}{\begin{minipage}[t]{12.2cm} \vspace{1.55cm}\hspace{0cm} \end{minipage}}
\vspace{-3.55cm}\\
\small
\begin{tabular}{|l||c|c|c|}
\hline
\rule[-0.15cm]{0pt}{0.55cm}  &  $BR^{s\pm}\,\times 10^6$ &  $A_{dir}^{s\pm}\,\times 10^2$  &  $A_{mix}^{s\pm}\,\times 10^2$ \\
\hline
\hline \rule[-0.15cm]{0pt}{0.55cm} $A_{dir}^{d0}=-0.2$ & $ 24.9\pm 9.3 \pm 5.0 \pm 2.3$  & $ 20.1 \pm 15.7 \pm 2.6$  & $  26.2 \pm 11.5 \pm 1.9$ \\
\hline \rule[-0.15cm]{0pt}{0.55cm} $A_{dir}^{d0}=-0.1$ & $ 22.1\pm 8.5 \pm 4.7 \pm 1.7$  & $ 33.9 \pm 13.0 \pm 3.3$  & $   8.2 \pm 15.1 \pm 0.2$ \\
\hline \rule[-0.15cm]{0pt}{0.55cm} $A_{dir}^{d0}=0$    & $ 20.1\pm 7.3 \pm 4.4 \pm 1.2$  & $ 35.3 \pm 12.8 \pm 2.8$  & $  -9.1 \pm 15.3 \pm 1.1$ \\
\hline \rule[-0.15cm]{0pt}{0.55cm} $A_{dir}^{d0}=0.1$  & $ 18.4\pm 6.8 \pm 4.2 \pm 0.9$  & $ 27.3 \pm 27.9 \pm 2.1$  & $ -26.4 \pm 18.4 \pm 1.8$ \\
\hline \rule[-0.15cm]{0pt}{0.55cm} $A_{dir}^{d0}=0.2$  & $ 17.5\pm 6.5 \pm 4.1 \pm 0.9$  & $  0.5 \pm 33.6 \pm 1.4$  & $ -38.9 \pm 14.8 \pm 2.8$ \\
\hline
\end{tabular}
\end{center}
\caption{\small Observables for $B_s\to K^+K^-$ as functions of the direct asymmetry $A_{\rm dir}(B_d\to K^0 \bar{K}^0)$ within the SM. The errors
mean the same as in Table \ref{tableBsK0K0}.The preferred values for $A_{\rm dir}^{d0}$ are highlighted (see the text).} \label{tableBsK+K-}
\end{table}

The prediction for the branching ratio in Eq.~(\ref{BRBK+K-th}) or (\ref{BRBK+K-thImp}) can be already compared with the experimental value
\cite{hep-ph/0607021},
\eq{BR(B_s\to K^+ K^-)_{\rm exp}=(24.4\pm 1.4\pm 4.6)\times 10^{-6}\ ,\label{BRBK+K-exp}}
and fits nicely within the errors. Notice that the experimental errors are quite small. It is interesting to note that before this experimental
number was released, the current experimental value was $\sim 40\times 10^{-6}$. In \cite{hep-ph/0603239}, the discrepancy between the theoretical
results derived here and that high experimental number for the branching ratio led to the statement that \cite{hep-ph/0603239} \emph{``\dots the data
suggest a departure from the SM, to be further checked experimentally''}, so these results were genuine \emph{pre}dictions, and indeed, the `further
experimental check' fell right on top of the theoretical number.

\begin{figure}
\begin{center}
\psfrag{xi}{} \psfrag{xi}{} \psfrag{xi}{} \psfrag{=}{}\psfrag{0.8}{\hspace{-0.3cm}\tiny $\xi=0.8$}\psfrag{1.0}{\hspace{-0.4cm} \tiny $\xi=1.0$}
\psfrag{1.2}{\hspace{-0.3cm}\tiny $\xi=1.2$} \psfrag{QCDFx}{\hspace{0.1cm}\tiny QCDF} \psfrag{CDF}{\hspace{0.1cm}\tiny CDF}
\psfrag{DMVx}{\hspace{0.1cm}\tiny DMV} \psfrag{BR}{\small\hspace{-1cm}$\stackrel{BR(B_s^0\to K^+K^-)\times10^{6}}{}$}
\psfrag{Adir}{\hspace{-1cm}\begin{minipage}[l]{7cm}\vspace{0.5cm}{\tiny $A_{dir}(B_s^0\to K^+K^-)$}\end{minipage}}
\includegraphics[height=6cm,width=10cm]{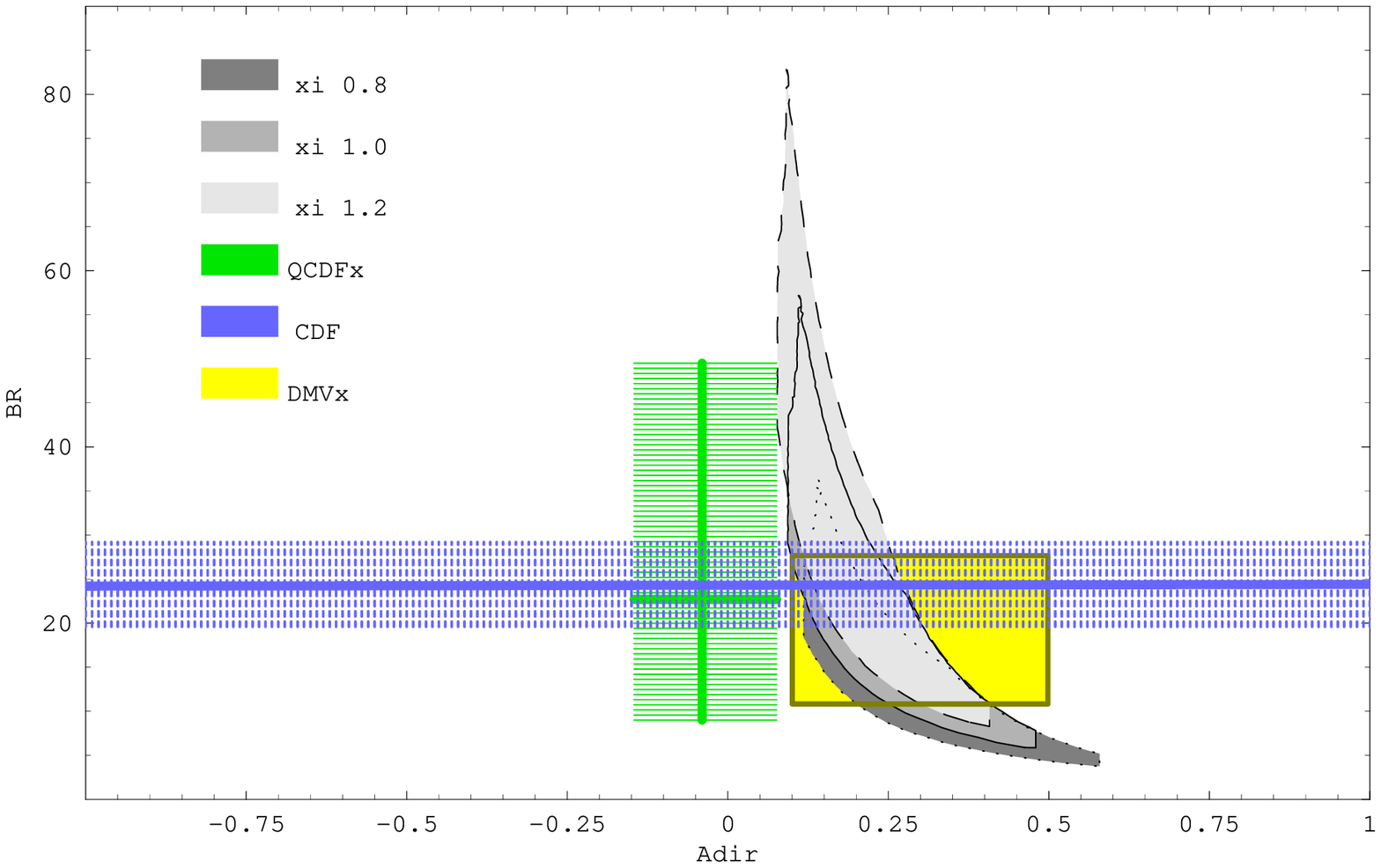}\hspace{0.5cm}
\raisebox{0.35cm}{\includegraphics[height=5.78cm,width=2cm]{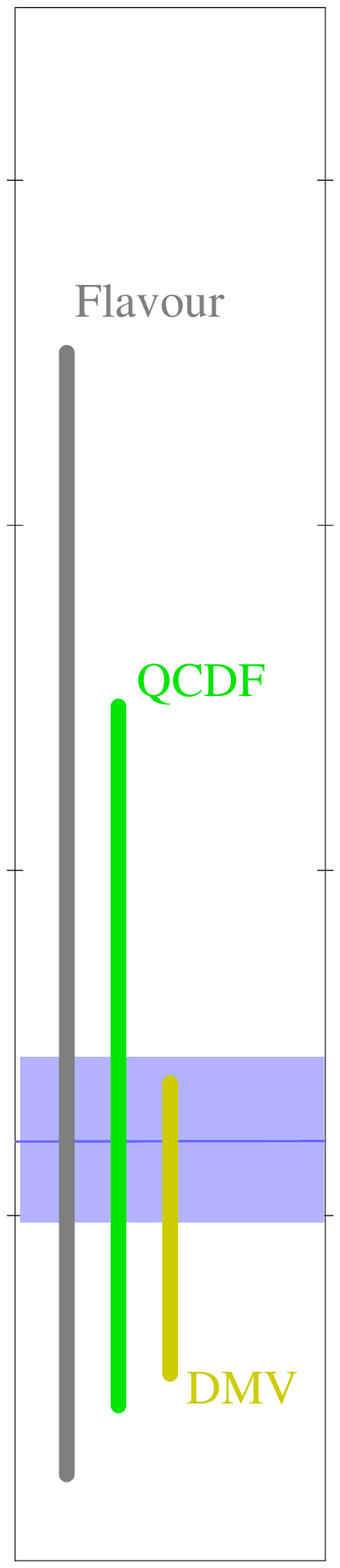}}
\end{center}
\vspace{-0.2cm} \caption{\small Comparison between the determinations of the $B_s\to K^+K^-$ branching ratio and direct CP asymmetry from flavor,
QCDF and the method described in this chapter (we refer to it as \emph{DMV}). This plot is an extension of Fig.~\ref{CompBsKK}, and shows how the new
predictions improve considerably the situation posed at the beginning of this chapter.} \label{CompBsKK2}
\end{figure}

The predictions for the $B_s\to K^+K^-$ observables can be compared to the other determinations (c.f. Fig.~\ref{CompBsKK}). This is shown in
Fig.~\ref{CompBsKK2}, where it can be easily seen that the improvement is substantial, specially in the prediction of the branching ratio.

Finally, we comment on how the results derived for the hadronic parameters can be used to predict the $SU(3)$ breaking parameters $|r_T|$ and $\xi$
connecting $B_d\to \pi^+\pi^-$ to $B_s\to K^+K^-$, introduced in Chapter \ref{Bdecays}. The point is that, once the $B_d\to \pi^+\pi^-$ hadronic
parameters $T_{\pi\pi}^{d\pm}$ and $P_{\pi\pi}^{d\pm}$ are extracted from data as in Chapter \ref{Bdecays}, and the $B_s\to K^+K^-$ hadronic
parameters $T^{s\pm}$ and $P^{s\pm}$ are obtained from $B_d\to K^0 \bar{K}^0$ as described here, the $SU(3)$ breaking factors that connect them can
be obtained.

To that end it is more useful to write the $B_d\to \pi^+\pi^-$ hadronic parameters obtained in Section \ref{MeasuringSM}, not as ranges, but as
numbers with errors:
\eqa{
|T^{d\pm}_{\pi\pi}|&=&(5.48 \pm 0.42)\times 10^{-6}\ ,\nn\\
|P^{d\pm}_{\pi\pi}/T^{d\pm}_{\pi\pi}|&=&0.13 \pm 0.05\ . }
Then Eqs.~(\ref{UspinBreaking}) give the values for the $SU(3)$ breaking parameters. In Table \ref{tableUspin} we show the derived $B_s\to K^+K^-$
hadronic parameters and the corresponding predictions for the $SU(3)$ breaking parameters.

\begin{table}[!h]
\begin{center}
\vspace{2.00cm} \hspace{0.1cm}\colorbox{yellow}{\begin{minipage}[t]{13cm} \vspace{1.50cm}\hspace{0cm} \end{minipage}}
\vspace{-3.50cm}\\
\small
\begin{tabular}{|l|c|c|c||c|c|}
\hline\rule[-0.15cm]{0pt}{0.55cm}
 &   $\vert T^{s\pm} \vert \times 10^6$  &  $\vert P^{s\pm}/T^{s\pm} \vert$  &  $\arg{(P^{s\pm}/T^{s\pm})}$  &  $|r_T|$ & $\xi$\\
\hline \rule[-0.15cm]{0pt}{0.55cm}$A_{dir}^{d0}=-0.2$  & $13.0\pm 2.4$ & $0.10\pm 0.04$ & $(38 \pm 28)^\circ$ & $2.4\pm 0.5$ & $0.74\pm 0.40$  \\
\hline \rule[-0.15cm]{0pt}{0.55cm}$A_{dir}^{d0}=-0.1$  & $12.3\pm 2.6$ & $0.10\pm 0.04$ & $(77 \pm 27)^\circ$ & $2.2\pm 0.5$ & $0.78\pm 0.41$  \\
\hline \rule[-0.15cm]{0pt}{0.55cm}$A_{dir}^{d0}=0$     & $11.7\pm 2.6$ & $0.11\pm 0.04$ & $(105\pm 25)^\circ$ & $2.1\pm 0.5$ & $0.82\pm 0.46$  \\
\hline \rule[-0.15cm]{0pt}{0.55cm}$A_{dir}^{d0}=0.1$   & $11.2\pm 2.6$ & $0.11\pm 0.05$ & $(134\pm 38)^\circ$ & $2.0\pm 0.5$ & $0.86\pm 0.51$  \\
\hline \rule[-0.15cm]{0pt}{0.55cm}$A_{dir}^{d0}=0.2$   & $10.8\pm 2.4$ & $0.12\pm 0.06$ & $(179\pm 30)^\circ$ & $2.0\pm 0.5$ & $0.90\pm 0.56$  \\
\hline
\end{tabular}
\end{center}
\vspace{-0.3cm} \caption{\small Hadronic parameters for $B_s\to K^+K^-$ and $U$-spin breaking parameters $|r_T|=|T^{s^\pm}/T_{\pi\pi}^{d^\pm}|$ and
$\xi=|P^{s^\pm}/T^{s^\pm}|/|P_{\pi\pi}^{d^\pm}/T_{\pi\pi}^{d^\pm}|$, that relate $B_s\to K^+K^-$ and $B_d\to \pi^+\pi^-$.} \label{tableUspin}
\end{table}

%


\chapter{Penguin-Mediated $B\to VV$ Decays and the $B_s-\bar{B}_s$ Mixing Angle}

\label{SymFac2}

\def\lg{^{\rm long}}
\def\Amix{\mathcal{A}_{\rm mix}}
\def\Adir{\mathcal{A}_{\rm dir}}

The method presented in the previous chapter is powerful for many reasons. One of the reasons is that the theoretical input ($\Delta$) is quite
clean, and it is the minimal information that one needs to obtain the hadronic parameters $T$ and $P$ without any information on the mixing-induced
CP asymmetry of neutral $B$ decays. This property was used in Chapter \ref{SymFac1} because there is still no experimental information on $A_{\rm
mix}(B_d\to K^0\bar{K}^0)$. However, because this asymmetry is the one that contains the information on the mixing angle (c.f. eq.~(\ref{Amix})),
this method can be used to extract the mixing angle from data, without the necessity of neglecting a part of the amplitude (see Section
\ref{mixingangles}), or doing a $SU(3)$ analysis with some other modes, which probably requires the input of other mixing angles.

In this chapter we use the method developed in Chapter \ref{SymFac1} to provide three alternative strategies to extract the $B_s-\bar{B}_s$ mixing
angle $\phi_s$ from non-leptonic B decays into pairs of vector mesons. In particular we focus on $B_{d,s}\to K^{*0} \bar{K}^{*0},\phi
\bar{K}^{*0},\phi\phi$ decays, for which the extraction of $\Delta$ is clean, according to the discussion in Section \ref{SectionDelta}.

However, the theoretical study of $B$ decays with vector-vector final states is complicated. The main reason is the fact that transverse amplitudes
do not factorize at leading order. In fact, naively one expects the transverse amplitudes to be suppressed in the heavy quark limit with respect to
the longitudinal one. This works quite well, for example, for $B\to \rho\rho$, where the longitudinal polarization fraction is $\sim 95\%$. However,
it does not work for some penguin mediated decays such as $B\to \phi K^{*0}$, where the measured polarization fraction is $\sim 50\%$. This has been
called the \emph{polarization puzzle}. It is therefore important to estimate properly these contributions when used; but because theoretically this
is a challenge, we will concentrate on longitudinal observables alone.

\section{Longitudinal observables in $B\to VV$ modes} \label{sec:longobs}

The amplitude for a $B$ meson decaying into 2 vector mesons can be written as
\eqa{
A(B\to V_1V_2)&=&\left[ \frac{4m_1m_2}{m_B^4}(\epsilon_1^*\cdot p_B)(\epsilon_2^*\cdot p_B) \right] A_0 \nonumber\\
&+&\left[\frac{1}{2}(\epsilon_1^*\cdot\epsilon_2^*) -\frac{(p_B\cdot\epsilon_1^*)(p_B\cdot\epsilon_2^*)}{m_B^2}
-\frac{i\epsilon_{\mu\nu\rho\sigma}\epsilon_1^{*\mu}\epsilon_2^{*\nu}p_1^\rho p_2^\sigma}{2 p_1\cdot p_2}\right] A_+\nonumber\\
&+&\left[\frac{1}{2}(\epsilon_1^*\cdot\epsilon_2^*) -\frac{(p_B\cdot\epsilon_1^*)(p_B\cdot\epsilon_2^*)}{m_B^2}
+\frac{i\epsilon_{\mu\nu\rho\sigma}\epsilon_1^{*\mu}\epsilon_2^{*\nu}p_1^\rho p_2^\sigma}{2 p_1\cdot p_2}\right] A_- }
where $A_{0,+,-}$ correspond to the amplitudes for longitudinal and transversely polarized final vector mesons. It is also customary to use the basis
$A_{0,\|,\bot}$, where $A_{\|,\bot}=(A_+\pm A_-)\sqrt{2}$.

The vector mesons in the final state decay typically into pairs of pseudoscalar particles. A full angular analysis of vector-vector modes provides
the following set of observables: three polarization fractions $f_0$, $f_\bot$ and $f_\|$ (only two of them are independent) and their CP-conjugate
counterparts $\bar{f}_{0,\bot,\|}$, two phases $\phi_{\bot,\|}$ (again, together with $\bar{\phi}_{\bot,\|}$), a total CP-averaged branching ratio
$BR$, and a total direct CP-asymmetry $\Adir$. The polarization fractions are defined as
\eq{ f_{0,\bot,\|}\equiv\frac{|A_{0,\bot,\|}|^2}{|A_0|^2+|A_\bot|^2+|A_\||^2}  \quad \quad {\bar
f}_{0,\bot,\|}\equiv\frac{|\bar{A}_{0,\bot,\|}|^2}{|\bar{A_0}|^2+|\bar{A_\bot}|^2+|\bar{A_\|}|^2} }
A full angular analysis is available for $B_d\to\phi K^{*0}$ from BaBar and Belle \cite{hep-ex/0408017,hep-ex/0503013}, and the same type of analysis
is expected for $B_d\to K^{*0} \bar{K}^{*0}$.

We will focus in this chapter on observables for the longitudinal polarization ($BR\lg$, $\Adir\lg$, $\Amix\lg$ and $\mathcal{A}_{\Delta\Gamma}\lg$),
where only $A_0$ occurs. These observables, free from the positive and negative helicity components, can be predicted with a much better accuracy.
Indeed the negative-helicity (positive-helicity) component of the amplitude is $1/m_b$-suppressed ($1/m_b^2$-suppressed) because of the nature of the
interactions involved (left-handed weak interaction, helicity-conserving strong interaction at high energies)~\cite{hep-ph/0405134,hep-ph/0612290}.
This suppression makes longitudinal observables better behaved and easier to compute than transverse ones.

Some decay channels exhibit the $1/m_b$-suppression of transverse amplitudes in a very striking way : the longitudinal polarization is very close to
1, e.g. $f_L\simeq 97\%$ for $B\to \rho^+\rho^-$. In such cases, the full observables (where $A_0$ is replaced by the sum $A=A_0+A_-+A_+$) coincide
with the longitudinal ones to a high degree of accuracy. On the other hand,  for penguin dominated $\Delta S=1$ decays, $f_L$ can be as low as $\sim
50\%$, so that the transverse amplitudes (or $\pm$ helicity amplitudes) contribute significantly to the full observables. Therefore, one must
determine whether purely longitudinal observables can be extracted from experimental measurements.

\begin{figure}
\begin{center}
\includegraphics[height=4cm,width=9cm]{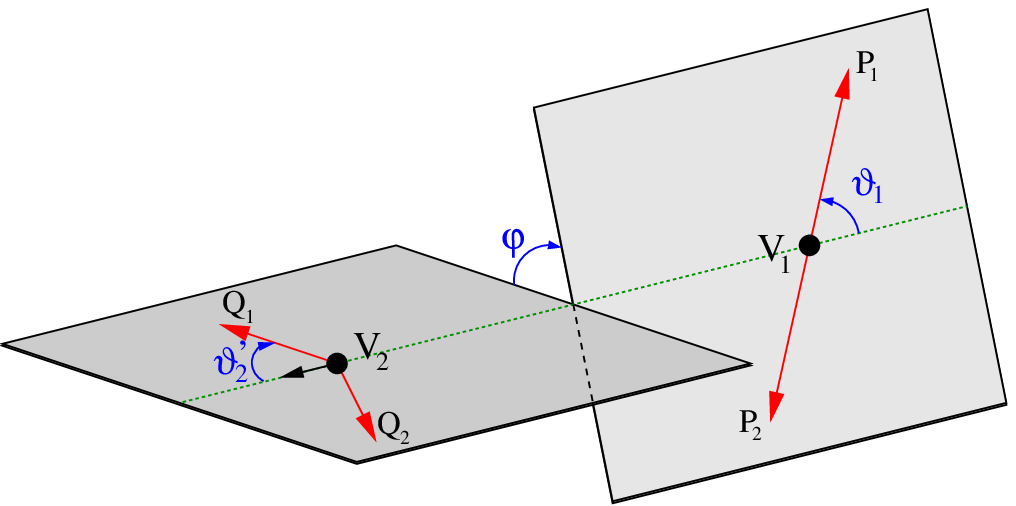}
\end{center}
\caption{\small Definition of the various angles in the decay $B\to V_1(\to P_1P_2)V_2(\to Q_1Q_2)$.} \label{BVVangles}
\end{figure}

We start from the normalized partial decay rate of $B\to V_1V_2$, where the two vector mesons go subsequently into pairs of pseudoscalar mesons. It
can be written~\cite{Kayser:1990ww}
\eqa{ \frac{d^3\Gamma}{\Gamma d\cos\theta_1 d\cos\theta_2 d\phi}&=&
 \frac{9}{8\pi}\frac{1}{|A_0|^2+|A_\| |^2+|A_\bot |^2} \nn\\[6pt]
&& \hspace{-3cm} \times\, \Bigg[|A_0|^2 \cos^2\theta_1 \cos^2\theta_2+|A_\| |^2 \frac{1}{2}\sin^2\theta_1 \sin^2\theta_2 \cos^2\phi \nn\\
&& \hspace{-3cm} +\,|A_\bot |^2 \frac{1}{2}\sin^2\theta_1 \sin^2\theta_2 \sin^2\phi+{\rm Re}[A_0^* A_\| ]
\frac{1}{2\sqrt{2}}\sin 2\theta_1 \sin 2\theta_2 \cos\phi \nn\\
&& \hspace{-3cm} +\,{\rm Im}[A_0^* A_\bot ] \frac{-1}{2\sqrt{2}}\sin 2\theta_1 \sin 2\theta_2 \sin\phi +{\rm Im}[A_\|^* A_\bot ]
\frac{-1}{2}\sin^2\theta_1 \sin^2\theta_2\sin 2\phi \Bigg] }
where $(\theta_1,\theta_2,\phi)$ are angles introduced to describe the kinematics of the decay $B\to V_1 V_2$ followed by $V_1 \to P_1 P_2$ and $V_2
\to Q_1 Q_2$. $\theta_1$ is the angle of one of the $V_1$ decay products in the rest frame of $V_1$ relative to the motion of $V_1$ in the rest frame
of the $B$-meson (same for $\theta_2$ with $V_2$). $\phi$ is the angle between the two planes formed by the decay products of $V_1$ and $V_2$
respectively (see Fig.~\ref{BVVangles} for a representation of the angles).

There are different ways to perform the angular integrations in order to extract the purely longitudinal component from the differential decay rate.
A first option consists in computing moments of $\cos\theta_1$ (or equivalently $\cos\theta_2$) :
\begin{equation}
\Gamma\lg
  \equiv \int \frac{d^3\Gamma}{d\cos\theta_1 d\cos\theta_2 d\phi}
      \left(\frac{5}{2}\cos^2\theta_1-\frac{1}{2}\right)
    d\cos\theta_1 d\cos\theta_2 d\phi = g_{PS} |A_0|^2/\tau_B
\end{equation}
where $g_{PS}$ is the product of phase-space and lifetime factors
\begin{equation} \label{gps}
g_{PS}=\frac{\tau_B}{16\pi M_B^3}\sqrt{[M_B^2-(m_1+m_2)^2][M_B^2-(m_1-m_2)^2]}
\end{equation}
A second possibility amounts to performing asymmetric integrations over one angle~\cite{hep-ph/0504178}
\begin{equation}
\Gamma\lg
  \equiv
 \int_{-1}^{1}  d\cos\theta_1 \int_T d\cos\theta_2
 \int_{0}^{2\pi} d\phi
 \frac{d^3\Gamma}{d\cos\theta_1 d\cos\theta_2 d\phi}
  = g_{PS} |A_0|^2/\tau_B
\end{equation}
with
\begin{equation}
\int_T d\cos\theta_2=\left(\frac{11}{9}\int_0^{\pi/3}-\frac{5}{9}\int_{\pi/3}^{2 \pi/3}+\frac{11}{9}\int_{2 \pi/3}^{\pi} \right) (-\sin\theta_2)
d\theta_2
\end{equation}
In the same way we can obtain the CP-conjugate $\Gamma\lg({\bar B_q^0} \to \bar{f})$ from the corresponding CP-conjugate distribution, leading to the
CP-averaged branching ratio of the longitudinal component
\begin{equation}
BR\lg =
 \frac{\tau_B}{2} \left(
 {\Gamma\lg(B^0_q\to f)
              +\Gamma\lg (\bar{B}^0_q\to \bar{f})} \right)
 = g_{PS} \frac{|A_0|^2+|\bar{A}_0|^2}{2}
\label{BR}
\end{equation}
where $\bar{A}_0$ is the CP-conjugate amplitude of $A_0$.

If we include the dependence on time in the above expressions, $B$-$\bar{B}$ mixing modifies the expressions. We will focus on CP-eigenstates $f_{\rm
\sss CP}$ in the final state $K^{*0} \bar{K}^{*0}$ and $\phi\phi$, as well as $\phi K^{*0}$ with a subsequent decay of $K^{*0}$ into a CP-eigenstate
($K_s \pi^0$ or $K_L \pi^0$).

The time evolution of these observables is obtained by considering the time dependence of $A_0(t)$~\cite{hep-ph/0002243}. Inserting this time
dependence one arrives at the usual expression for the longitudinal component of the time-dependent CP-asymmetry:
\eq{ \mathcal{A}^{\rm long}_{\rm CP}(t) \equiv \frac{\Gamma\lg (B^0_q(t)\to f_{\rm \sss CP})-\Gamma\lg (\bar{B}^0_q(t)\to f_{\rm \sss CP})}{\Gamma\lg
(B^0_q(t)\to f_{\rm \sss CP})+\Gamma\lg (\bar{B}^0_q(t)\to f_{\rm \sss CP})} =\frac{\Adir\lg\cos{(\Delta Mt)}+\Amix\lg\sin{(\Delta Mt)}}
{\cosh{(\Delta\Gamma t/2)}-\mathcal{A}\lg_{\Delta\Gamma}\sinh{(\Delta\Gamma t/2)}} }
where the direct and mixing-induced CP asymmetries are defined by:
\eq{ \Adir\lg\equiv \frac{|A_0|^2-|\bar{A}_0|^2}{|A_0|^2+|\bar{A}_0|^2},\quad \Amix\lg\equiv -2  \frac{{\rm
Im}(e^{-i\phi_M}A_0^{*}\bar{A_0})}{|A_0|^2+|\bar{A_0}|^2} \label{CPAs}}
together with the asymmetry related to the width difference :
\eq{ \mathcal{A}\lg_{\Delta\Gamma}\equiv -2 \frac{{\rm Re}(e^{-i\phi_M}A_0^{*}\bar{A_0})}{|A_0|^2+|\bar{A_0}|^2} \label{CPADelta} }
$\phi_M$ is the mixing angle and $\Delta \Gamma=\Gamma^H-\Gamma^L$. $\eta_f$ is the CP eigenvalue of the final state $f$ ($\pm 1$):
$\eta_{K^{*0}K^{*0}}=\eta_{\phi\phi}=1$, whereas $\eta_{K^{*0} \phi}=1$ if $K^{*0}$ decays into $K_s\pi^0$ and $-1$ if it decays into  $K_L\pi^0$. In
the latter case, the contribution from the strong process $K^{*0}\to K\pi$ is the same for both $B$ and $\bar{B}$ decays and it cancels in the
time-dependent CP-asymmetry Eq.(\ref{CPAs}), which depends only on the amplitudes $A_0$ and $\bar{A}_0$.

Finally, if the direct CP-asymmetries of all three helicity components are negligible, the longitudinal branching ratio can be estimated very easily
from: $BR\lg=BR^{\rm total} f_0$.

\section{Inputs} \label{sec:pt}

In this section we compute the values of the quantities $\Delta$ for the decays $B_{d,s}\to K^{*0} \bar{K}^{*0}$, $B_{d,s}\to \phi K^{*0}$ and
$B_s\to \phi \phi$. Because the final states can appear in three different polarization states, a different $\Delta$ should be associated to each
polarization. However, since we are only interested in longitudinal polarizations, we only give the results for this case.

For $B_{d,s}\to K^{*0} \bar{K}^{*0}$ we obtain,
\eqa{ |\Delta^d_{\sss K^*K^*}|&=&A_{\sss K^*K^*}^{d,0} \frac{C_F \alpha_s}{4\pi N_c}C_1\,|\bar{G}_{\sss K^*}(s_c)-\bar{G}_{\sss K^*}(0)|
=(1.85 \pm 0.79)\times 10^{-7}\ {\rm GeV}\qquad \label{delta0KK}\\
|\Delta^s_{\sss K^*K^*}|&=&A_{\sss K^*K^*}^{s,0} \frac{C_F \alpha_s}{4\pi N_c}C_1\,|\bar{G}_{\sss K^*}(s_c)-\bar{G}_{\sss K^*}(0)| =(1.62 \pm
0.69)\times 10^{-7}\ {\rm GeV}\qquad \label{delta0sKK} }
where $\bar{G}_V\equiv G_V-r_{\chi}^V \hat{G}_V$ are the usual penguin functions and $A_{V_1 V_2}^{q,0}$ are the naive factorization factors
combining decay constants and form factors (see Sections \ref{FFsDCsMDAs} and \ref{SectionDelta}). The numerical values of the used inputs are given
in Table~\ref{inputs}, which are taken from Ref.~\cite{hep-ph/0612290}. The contributions to each error from the various sources are detailed in
Table~\ref{errs}.

\begin{figure}
\begin{center}
\psfrag{BR}{\hspace{-1.5cm}$BR\lg(B_d\to K^{*0} \bar{K}^{*0})\times 10^6$} \psfrag{A}{\hspace{-1.5cm}$\Adir\lg(B_d\to K^{*0} \bar{K}^{*0})$}
\psfrag{-0.4}{\hspace{0.15cm}$-0.4$}\psfrag{-0.2}{\hspace{0.15cm}$-0.2$}\psfrag{0.2}{\hspace{0.15cm}$0.2$}\psfrag{0.4}{\hspace{0.15cm}$0.4$}
\psfrag{0}{$0$}\psfrag{2}{$2$}\psfrag{4}{$4$}\psfrag{6}{$6$}\psfrag{8}{$8$}
\includegraphics[width=10cm]{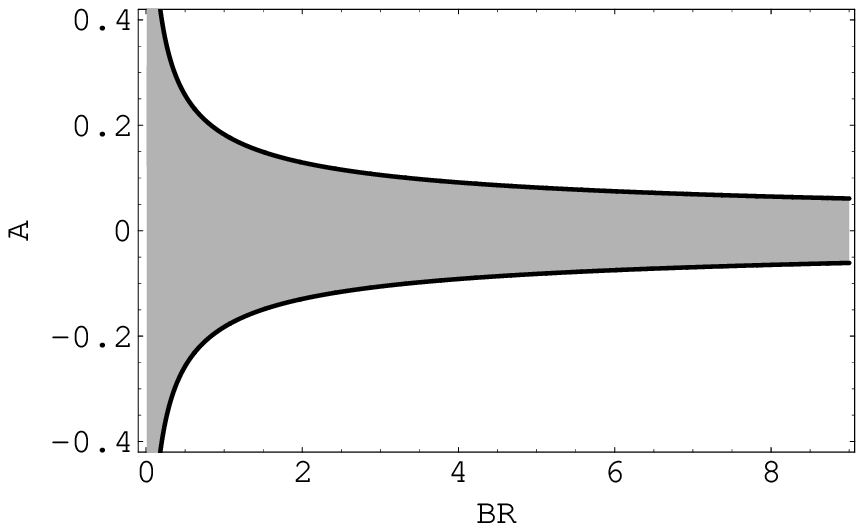}
\end{center}
\caption{\small Allowed region on the $BR\lg-\Adir\lg$ plane for $B_d\to K^{*0} \bar{K}^{*0}$, according to the value of $\Delta^{d}_{\sss K^*K^*}$.}
\label{BRvsAdirK*K*}
\end{figure}

The conditions in eqs.~(\ref{cons}) derived in Chapter \ref{SymFac1} can be applied to the $B_d\to K^{*0} \bar{K}^{*0}$ longitudinal branching ratio
and CP asymmetry given the value for $\Delta^{d}_{\sss K^*K^*}$ in eq.~(\ref{delta0KK}). The allowed region in the observable space is shown in
Fig.~\ref{BRvsAdirK*K*}.

For the $\Delta$, as well as for the other quantities computed in this chapter (as well as in Chapter \ref{SymFac1}), we quote as the central value
the value obtained from taking the central value of the inputs. To estimate the error, we vary one by one each of the inputs, compute the difference
with the central value, then add in quadrature the resulting uncertainties. The main sources of uncertainties are the scale of factorization $\mu$,
the mass of the charm quark $m_c$, and the form factor $A_0^{B\to K*}$.

\begin{table}
\begin{center}\small
\begin{tabular}{ccccc}
\hline\hline
\rule[-0.15cm]{0pt}{0.55cm}    $m_c(m_b)$    &        $f_B$        &      $f_{B_s}$      &   $\lambda_B,\lambda_{B_s}$   &   $\alpha_1^{\bot}(K^*)$  \\
\hline
\rule[-0.15cm]{0pt}{0.55cm}$\ 1.3\pm 0.2\ $  & $\ 0.21\pm 0.02\ $  &  $\ 0.24\pm 0.02\ $ &     $\ 0.35\pm 0.15\ $        &    $\ 0.06\pm 0.06\ $     \\
\hline\hline
\end{tabular}\\
\vspace{0.3cm}
\begin{tabular}{ccccc}
\hline\hline
\rule[-0.15cm]{0pt}{0.55cm}  $\alpha_2^{\bot}(K^*)$  &         $f_{K^*}$        &  $f_{K^*}^\bot(2{\rm GeV})$  &   $A_0^{B\to K^*}$    &   $A_0^{B_s\to K^*}$  \\
\hline
\rule[-0.15cm]{0pt}{0.55cm}    $\ 0.1\pm 0.2\ $      &   $\ 0.218\pm 0.004\ $   &    $\ 0.175\pm 0.025\ $      &  $\ 0.39\pm 0.06\ $   &   $\ 0.33\pm 0.05\ $  \\
\hline\hline
\end{tabular}\\
\vspace{0.3cm}
\begin{tabular}{ccccc}
\hline\hline
\rule[-0.15cm]{0pt}{0.55cm}      $f_\phi$         &  $f_\phi^\bot(2{\rm GeV})$  &     $A_0^{B_s\to \phi}$      &  $\alpha_2^{\bot}(\phi)$  \\
\hline
\rule[-0.15cm]{0pt}{0.55cm}$\ 0.221\pm 0.003\ $   &     $\ 0.175\pm 0.025\ $    &  $\ 0.38^{+0.10}_{-0.02}\ $  &   $\ 0.0\pm 0.3\ $        \\
\hline\hline
\end{tabular}
\end{center}
\caption{\small Input parameters required in QCD factorisation to compute the
  quantities $\Delta$'s and $\delta$'s described in the text. The masses and decay constants are given in GeV.}
\label{inputs}
\end{table}

\begin{table}
\begin{center}\small
\begin{tabular}{lccccccccc}
\hline\hline
\rule[-0.15cm]{0pt}{0.55cm}                     &   $m_c$  &  $A_0^{B\to K^*}$ &  $f_{K^*}$  & $\mu$  & $\alpha_1(K^*)$ & $\alpha_2(K^*)$ & $\alpha_1^\bot(K^*)$ & $\alpha_2^\bot(K^*)$  \\
\hline\hline
\rule[-0.15cm]{0pt}{0.55cm}$\Delta^d_{K^*K^*}$  &  37.3\%  &       13.2\%      &    0.2\%    & 44.2\% &        0.1\%    &        4.6\%    &           0.1\%      &     0.3\%         \\
\hline
\rule[-0.15cm]{0pt}{0.55cm}$\Delta^s_{K^*K^*}$  &  37.5\%  &       12.9\%      &    0.2\%    & 44.4\% &        0.1\%    &        4.7\%    &           0.1\%      &     0.3\%         \\
\hline\hline\\
\end{tabular}
\begin{tabular}{lcccccc}
\hline\hline
\rule[-0.15cm]{0pt}{0.55cm}                        &   $m_c$  &  $A_0^{B\to K^*}$  &  $f_{K^*}$ &  $f_{K^*}^\bot(2{\rm GeV})$ & $\mu$  & $\alpha_1(K^*)$  \\
\hline\hline
\rule[-0.15cm]{0pt}{0.55cm}$\Delta^d_{\phi K^*}$   &  44.2\%  &        2.0\%       &    ---     &           ---               & 52.3\% &        ---       \\
\hline
\rule[-0.15cm]{0pt}{0.55cm}$\Delta^s_{\phi K^*}$   &  35.0\%  &        ---         &    0.1\%   &            0.7\%            & 58.2\% &        0.7\%     \\
\hline
\rule[-0.15cm]{0pt}{0.55cm}$\Delta^s_{\phi \phi}$  &  44.1\%  &        ---         &    ---     &           ---               & 52.3\% &        ---       \\
\hline\hline\\
\end{tabular}
\begin{tabular}{lcccccccccccc}
\hline\hline
\rule[-0.15cm]{0pt}{0.55cm}                        & $\alpha_2(K^*)$ & $\alpha_2^\bot(K^*)$  & $A_0^{B\to \phi}$ & $f_{\phi}^\bot(2{\rm GeV})$ & $\alpha_2(\phi)$ & $\alpha_2^\bot(\phi)$\\
\hline\hline
\rule[-0.15cm]{0pt}{0.55cm}$\Delta^d_{\phi K^*}$   &        ---      &      ---              &          ---       &           0.4\%            &       0.7\%      &       0.3\%          \\
\hline
\rule[-0.15cm]{0pt}{0.55cm}$\Delta^s_{\phi K^*}$   &        0.1\%    &     0.1\%             &          5.0\%     &           0\%              &       0\%        &       0\%            \\
\hline
\rule[-0.15cm]{0pt}{0.55cm}$\Delta^s_{\phi \phi}$  &        ---      &       ---             &          2.1\%     &           0.4\%            &       0.7\%      &       0.3\%          \\
\hline\hline
\end{tabular}
\end{center}
\caption{\small Relative contributions from the inputs to the errors in $\Delta$ for the various decays. } \label{errs}
\end{table}

In a similar way, we can compute the corresponding longitudinal $\Delta$ for the decay modes $B_{d,s}\to \phi\bar{K}^{*0}$ and $B_s\to \phi\phi$:
\eqa{ |\Delta^d_{\phi K^*}|&=&A_{K^*\phi}^{d,0} \frac{C_F \alpha_s}{4\pi N_c}C_1\,|\bar{G}_{\phi}(s_c)-\bar{G}_{\phi}(0)|
=(1.02 \pm 1.11)\times 10^{-7}\ {\rm GeV}\quad \label{delta0phiK}\\
|\Delta^s_{\phi K^*}|&=&A_{\phi K^*}^{s,0} \frac{C_F \alpha_s}{4\pi N_c}C_1\,|\bar{G}_{\phi}(s_c)-\bar{G}_{\phi}(0)|
=(1.16 \pm 1.05)\times 10^{-7}\ {\rm GeV}\quad \label{delta0sphiK}\\
|\Delta^s_{\phi\phi}|&=&A_{\phi\phi}^{s,0} \frac{C_F \alpha_s}{4\pi N_c}C_1\,|\bar{G}_{\phi}(s_c)-\bar{G}_{\phi}(0)| =(2.06 \pm 2.24)\times 10^{-7}\
{\rm GeV}\quad \label{delta0phiphi} }

In the following Sections we show how to apply the results of this section and Chapter \ref{SymFac1} to the longitudinal contribution of
penguin-dominated $B\to VV$ modes. We will see that they can be used to extract the $B_s-\bar{B_s}$ mixing angle and some longitudinal observables
like branching ratios and time-dependent CP asymmetries within the Standard Model. In particular, we outline three different strategies to determine
the $B_s-\bar{B_s}$ mixing angle (in the SM and beyond). Indeed, concerning New Physics we will see that under the assumption of no significant New
Physics affecting the amplitude, while Strategy II can detect the presence of New Physics by comparing the obtained $\phi_s$ with $\phi_s^{SM}=2
\beta_s$, Strategy I and III can not only detect New Physics but allow also for the extraction of $\phi_s$ even in the presence of New Physics in the
mixing.

\section{First strategy to extract $\phi_s$: Bounding $T/P$}
\label{sec:bounding}

The $b\to s$ penguin-dominated decays like $B_s\to K^{*0} \bar{K}^{*0}$ are in principle clean modes to extract the mixing angle $\phi_s$. In this
section and those following, $\phi_s$ refers to the same mixing angle that will be measured, for instance, in the mixing induced CP asymmetry of $B_s
\to \psi \phi$ including possible New Physics contributions in the mixing. When focusing only on SM we will use the notation $\phi_s=2\beta_s$.

In an expansion in powers of $\lambda_u^{(s)}/\lambda_c^{(s)}$,  the amplitude for the decay $B_s\to K^{*0} \bar{K}^{*0}$  is given by:
\eq{\label{pr}\Amix\lg(B_s\to K^{*0} \bar{K}^{*0})\simeq \sin{\phi_s}+2\left| \frac{\lambda_u^{(s)}}{\lambda_c^{(s)}} \right| {\rm Re}\left(
\frac{T^s_{K^*K^*}}{P^s_{K^*K^*}} \right) \sin{\gamma}\cos{\phi_s}+\cdots}
In order to determine the accuracy of this relation, we must assess the size of the CKM-suppressed hadronic contribution $T$. Notice that this
relation is valid even in presence of New Physics in the mixing. In the SM, one can derive from the Wolfenstein parametrisation that eq.~(\ref{pr})
is of order $\lambda^2$ (with $\lambda=V_{us}$), and both pieces shown on the r.h.s of eq.({\ref{pr}}) are of this same order. However, despite the
smallness of the ratio $|\lambda_u^{(s)}/\lambda_c^{(s)}|=0.044$, a significant value of the hadronic ratio ${\rm Re}(T/P)$  could spoil the
potentially safe extraction of $\sin{\phi_s}$. The deviation from  $\sin \phi_s$ is:
\eq{\label{deltass}\Delta S(B_s\to K^{*0} \bar{K}^{*0})\equiv 2\left| \frac{\lambda_u^{(s)}}{\lambda_c^{(s)}} \right| {\rm Re}\left(
\frac{T^s_{K^*K^*}}{P^s_{K^*K^*}} \right) \sin{\gamma}\cos{\phi_s}}
We want to set bounds on ${\rm Re}(T/P)$, which can be related to the inputs:
\eq{ {\rm Re}\left( \frac{T}{P} \right)={\rm Re}\left( \frac{P+\Delta}{P} \right)=1+{\rm Re}\left( \frac{\Delta}{P} \right)= 1+\frac{{\rm
Re}(P)\,\Delta}{{\rm Re}(P)^2+{\rm Im}(P)^2} \label{ReP/T}}
Eqs.~(\ref{eqTP}) show that the maximum of ${\rm Re}(T/P)$ is reached for $\Adir\lg=0$ together with the positive branch for ${\rm Re}(P)$. The
following bound is obtained
\eq{ {\rm Re}\left( \frac{T}{P} \right)\le\,1+\left(-c_1^{(s)}+\sqrt{-(c_0^{(s)}/c_2^{(s)})^2+(1/c_2^{(s)})\,\widetilde{BR}/\Delta^2}\right)^{-1}
\label{formula} }
where the lower bound for $BR\lg$ and the upper bound for $\Delta$ must be used. In a similar way, the minimum of ${\rm Re}(T/P)$ occurs for
$\mathcal{A}_{dir}\lg=0$, for the negative branch of Eq.~(\ref{eqTP}) for the solution of ${\rm Re}(P)$
\eq{ {\rm Re}\left( \frac{T}{P} \right)\ge\,1+\left(-c_1^{(s)}-\sqrt{-(c_0^{(s)}/c_2^{(s)})^2+(1/c_2^{(s)})\,\widetilde{BR}/\Delta^2}\right)^{-1}
\label{formula2} }
where the lower bound for $BR\lg$ and the upper bound for $\Delta$ must be used once again. As a conclusion, we obtain a range for ${\rm Re}(T/P)$
from two inputs: the branching ratio $BR\lg(B_s\to K^{*0} \bar{K}^{*0})$ and $\Delta^s_{K^*K^*}$, given in Eq.(\ref{delta0sKK}).

\begin{figure}
\begin{center}
\psfrag{S}{\small \hspace{-1cm}$\Delta S(B_s\to K^{*0} \bar{K}^{*0})$} \psfrag{BR}{\small \hspace{-1cm}$BR\lg(B_s\to K^{*0} \bar{K}^{*0})\times
10^6$} \psfrag{10}{\small $10$}\psfrag{20}{\small $20$}\psfrag{30}{\small $30$}\psfrag{40}{\small $40$}\psfrag{50}{\small $50$} \psfrag{0.03}{\small
$0.03$}\psfrag{0.035}{\small $0.035$}\psfrag{0.04}{\small $0.04$} \psfrag{0.045}{\small $0.045$}\psfrag{0.05}{\small $0.05$} \psfrag{0.055}{\small
$0.055$}\psfrag{0.06}{\small $0.06$}
\includegraphics[height=6cm]{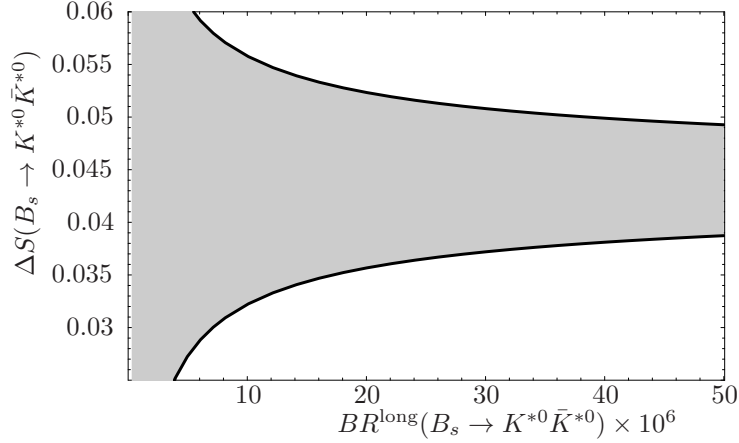}
\end{center}
\caption{\small Absolute bounds on $\Delta S(B_s\to K^{*0} \bar{K}^{*0})$ as a function of the longitudinal branching ratio $BR\lg(B_s\to K^{*0}
\bar{K}^{*0})$.} \label{DeltaS}
\end{figure}

Using Eq.~(\ref{deltass}), these upper and lower bounds on ${\rm Re}(T/P)$ are converted into a bound on the pollution $\Delta S(B_s\to
K^{*0}\bar{K}^{*0})$. The latter is plotted as a function of the longitudinal $BR\lg(B_s\to K^{*0} \bar{K}^{*0})$ in Fig.\ref{DeltaS}.

Once a measurement of  $\Amix\lg(B_s\to K^{*0} \bar{K}^{*0})$ is available, upper and lower bounds for
 $\phi_s$ are easily obtained.
For instance, if we take as a lower bound for the branching ratio $BR\lg(B_s\to K^{*0} \bar{K}^{*0})\gtrsim 5\times 10^{-6}$, Fig.~\ref{DeltaS} gives
$0.03<\Delta S(B_s\to K^{*0} \bar{K}^{*0})<0.06$.  In the case of a moderately large branching ratio $BR\lg(B_s\to K^{*0} \bar{K}^{*0})\sim
(30-40)\,\times 10^{-6}$, the bounds get sharper, with $0.04<\Delta S(B_s\to K^{*0} \bar{K}^{*0})<0.05$ and
\eq{\big(\Amix\lg(B_s\to K^{*0} \bar{K}^{*0})-0.05\big)\ < \sin{\phi_s} <\ \big(\Amix\lg(B_s\to K^{*0} \bar{K}^{*0})-0.04\big)}

The same strategy can be applied to $B_s \to \phi K^{*0}$ and $B_s\to\phi\phi$ decays
\begin{itemize}
\item Take the experimental value for the longitudinal branching ratio $BR\lg$
  (once available), and the theoretical value for
$\Delta$ from Eqs.~(\ref{delta0sphiK}) or (\ref{delta0phiphi}).
\item Apply Eqs.~(\ref{formula}) and (\ref{formula2}) to constrain the range of ${\rm Re}(T/P)$.
\item Derive the allowed range for $\Delta S$ according to the equivalent of (\ref{deltass})
\item From the measured value of $\Amix\lg$, determine $\phi_s$ from
\eq{\big(\Amix\lg-\Delta S_{max}\big)\ < \sin{\phi_s} <\ \big(\Amix\lg-\Delta S_{min}\big)}
\end{itemize}
A weak mixing angle $\phi_s$ different from $\phi_s^{SM}$ would signal the presence of New Physics.

Interestingly, if the longitudinal direct CP asymmetry becomes available and happens to be inconsistent with zero, the bounds for ${\rm Re}(T/P)$ in
Eq.~(\ref{formula}) and (\ref{formula2}) can be tightened. Eq.~(\ref{ReP/T}) can be exploited to derive expressions similar to Eq.~(\ref{formula})
and (\ref{formula2}) with a non-vanishing $\Adir\lg$, leading to stronger bounds on ${\rm Re}(T/P)$ and consequently on $\sin \phi_s$.

\section{Second strategy :  Measuring CP asymmetries and BR}
\label{sec:measuring}

In this section, we show how we can extract  mixing angles and related CKM phases in a clean way from experimental data, the length of two sides of
the unitarity triangle and the theoretical quantity $\Delta$. The only theoretical requirement is that the decay must allow for a safe way of
computing  $\Delta$. The approach is general in the same sense as in the previous section, since it can be applied to any B decay into two
pseudoscalars or vectors. But it yields different results for the four groups of decays:
\begin{enumerate}
\item $B_d$ decay through a $b\to d$ process, e.g. $B_{d}\to K^{*0} \bar{K}^{*0}$
\item $B_s$ decay through a $b\to s$ process, e.g. $B_{s}\to K^{*0} \bar{K}^{*0}$
\item $B_d$ decay through a $b\to s$ process, e.g. $B_{d}\to \phi
  \bar{K}^{*0}$ (with a subsequent decay into a CP eigenstate)
\item $B_s$ decay through a $b\to d$ process, e.g. $B_{s}\to \phi
  \bar{K}^{*0}$
(with a subsequent decay into a CP eigenstate)
\end{enumerate}
As far as weak interactions are concerned, the difference between $B_d$ and $B_s$ decays consists in the mixing angle, whereas $b\to d$ and $b\to s$
processes differ through the CKM elements $\lambda_{u,c}^{(D)}$, where $D=d$ or $s$.

In the case of a $B_d$ meson decaying through a $b\to D$ process $(D=d,s)$, we can extract the angles $\alpha$ \cite{hep-ph/0611280} and $\beta$ from
the identities:
\eqa{ \sin^2{\alpha}&=&\frac{\widetilde{BR}}{2|\lambda_u^{(D)}|^2|\Delta|^2}\left( 1-
\sqrt{1-{(\Adir)}^2-{(\Amix)}^2} \right)\\
\sin^2{\beta}&=&\frac{\widetilde{BR}}{2|\lambda_c^{(D)}|^2|\Delta|^2}\left( 1-\sqrt{1- {(\Adir)}^2-{(\Amix)}^2} \right) }
In the case of a $B_s$ meson decaying through a $b\to D$ process $(D=d,s)$, we can extract the angles $\beta_s$ \cite{hep-ph/0701116} and $\gamma$,
assuming no New Physics in the decay, from the following expressions:
\eqa{ \label{sbetas} \sin^2{\beta_s}&=&\frac{\widetilde{BR}}{2|\lambda_c^{(D)}|^2|\Delta|^2}\left( 1-
\sqrt{1-{(\Adir)}^2-{(\Amix)}^2} \right)\\
\sin^2{\left(\beta_s+\gamma\right)}&=& \frac{\widetilde{BR}}{2|\lambda_u^{(D)}|^2|\Delta|^2}\left( 1- \sqrt{1-{(\Adir)}^2-{(\Amix)}^2} \right) }
If the obtained $\beta_s$ differs from its SM value, this would signal the presence of New Physics. Notice that this strategy is obtained by
combining the definition of $\Delta$ with the unitarity of the CKM matrix, so it is designed to work only in the context of the SM. Consequently the
previous expressions should be understood as a way of testing the SM. This is an important difference with Strategies I and III where one can obtain
a value for the weak mixing phase also in the presence of New Physics in the mixing (but not in the decay).

While the previous equations are quite general (they can be used for $B\to PP$ decays), it is understood that $BR$ and $A_{\rm dir,mix}$ refer to the
longitudinal branching ratio and longitudinal CP-asymmetries, respectively, when they are applied to $B\to VV$ decays.

Eq.~(\ref{sbetas}) provides a new way to perform a consistency test for the SM value of  $|\sin \beta_s |$ from the measurements of $\Amix\lg(B_s\to
K^{*0} \bar{K}^{*0})$, $\Adir\lg(B_s\to K^{*0} \bar{K}^{*0})$ and $BR\lg(B_s\to K^{*0} \bar{K}^{*0})$. The same strategy can be applied to $B_s\to
\phi \bar{K}^{*0}$ and $B_s\to \phi\phi$ using the corresponding sum rules. This sum rule offers several advantages : it is independent of CKM
angles, and all the hadronic input is concentrated on a single well-controlled quantity $\Delta$.

Note that all these equations depend actually on the corresponding branching ratio and $\mathcal{A}_{\Delta \Gamma}\lg$. The asymmetry
$\mathcal{A}_{\Delta \Gamma}\lg$ is indeed related to the direct and mixing-induced CP-asymmetries through the equality
${(\Adir\lg)}^2+{(\Amix\lg)}^2+{(\mathcal{A}_{\Delta \Gamma}\lg)}^2 = 1$. It was already noticed in \cite{hep-ph/0204101} in the context of $B_s \to
K^+ K^-$ and in \cite{hep-ph/9804253} in the context of $B \to J/\psi K^*, D^{\ast +}_s \bar D^{\ast}$ decays that it is possible to extract
$\mathcal{A}_{\Delta \Gamma}\lg$ directly from the ``untagged'' rate:
\begin{equation}
\Gamma\lg(B_s(t)\to VV)+\Gamma\lg(\overline{B_s}(t)\to VV) \propto R_{\rm H}e^{-\Gamma_{\rm H}^{(s)}t}+ R_{\rm L}e^{-\Gamma_{\rm L}^{(s)}t}
\end{equation}
If the time dependence of both exponentials can be separated, one obtains
\begin{equation}
\mathcal{A}_{\Delta\Gamma}\lg(B_s\to VV)= \frac{R_{\rm H}-R_{\rm L}}{R_{\rm H}+R_{\rm L}},
\end{equation}
The branching ratio and $\mathcal{A}_{\Delta\Gamma}\lg$ are thus the only required observables to extract $\beta_s$ through this method, which offers
the advantage of concentrating in $\Delta$ all the hadronic input needed to bound the tree-to-penguin ratio.

\section{Third strategy : Relating $B_s\to K^{*0} \bar{K}^{*0}$ and $B_d\to K^{*0} \bar{K}^{*0}$}
\label{sec:flavour}

Once an angular analysis of $B_d\to K^{*0} \bar{K}^{*0}$ is performed, it is possible to extract the CP-averaged branching ratio corresponding to the
longitudinal helicity final state. Eqs.~(\ref{eqTP}) can be used to extract the hadronic parameters, if one assumes that no New Physics contributes
in an appreciable way. If flavor symmetries are sufficiently accurate for this particular process, this estimate can be converted into a fairly
precise determination of hadronic parameters for the $b\to s$ channel $B_s\to K^{*0} \bar{K}^{*0}$. We have seen in Chapter \ref{SymFac1} that for
$B_{d,s}\to KK$ modes a $U$-spin analysis combined with QCD factorization leads to tight constraints on the ratio of the tree contributions to both
decay modes, as well as that for the penguins. In this section we show how to relate $B_d\to K^{*0} \bar{K}^{*0}$ and $B_s\to K^{*0} \bar{K}^{*0}$
decay modes following the same approach.

\begin{figure}
\begin{center}
\psfrag{ReP}{\small \hspace{-1cm}${\rm Re}(P_{\sss K^*K^*}^{d})\ (\times 10^{-6})$} \psfrag{ImP}{\small \hspace{-1cm}${\rm Im}(P_{\sss K^*K^*}^{d})\
(\times 10^{-6})$}
\includegraphics[height=6cm]{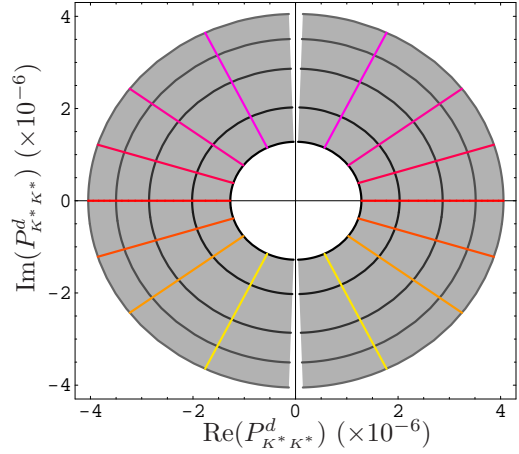}
\end{center}
\vspace{-0.5cm} \caption{\small Values for $P_{\sss K^*K^*}^{d}$ corresponding to a real $\Delta_{K^*K^*}$, as a function of the longitudinal
observables $BR^0(B_d\to K^{*0} \bar{K}^{*0})$ and $Adir^0(B_d\to K^{*0} \bar{K}^{*0})$. The contours from black to gray correspond to branching
ratios from $10^{-6}$ to $10^{-5}$ in steps of $2.5\times 10^{-6}$. The contours from red to yellow correspond to values of the CP asymmetry from $0$
to its maximum allowed value in steps of 30\% of its maximum  value. The contours from red to pink are the same for negative CP asymmetry.}
\label{figP}
\end{figure}

We define the parameters $\delta_{\sss K^*K^*}^P$ and $\delta_{\sss K^*K^*}^T$ as
\begin{equation}
\begin{array}{rclcrcl}
P_{K^*K^*}^{s}&=& f\,P_{K^*K^*}^{d}(1+\delta_{\sss K^*K^*}^P)\ ,&\quad &
T_{K^*K^*}^{s}&=& f\,T_{K^*K^*}^{d}(1+\delta_{\sss K^*K^*}^T)\\
\end{array}
\label{TsPs}
\end{equation}
where the factor $f$ is given by
\eq{\frac{m_{B_s}^2A_0^{B_s\to K^*}}{m_B^2A_0^{B\to K^*}}=0.88\pm 0.19}

We compute $|\delta_{\sss K^*K^*}^{P,T}|$ using QCDF. These parameters are affected by the model dependent treatment of annihilation and
spectator-scattering contributions, so the results should be considered
 as an estimate. A significant part of long-distance dynamics is common
to both decays, and we find the following upper bounds
\begin{equation}
|\delta_{\sss K^*K^*}^P|\le 0.12\ , \qquad |\delta_{\sss K^*K^*}^T|\le 0.15
\end{equation}
where the largest contribution comes from the lower value of $\lambda_B$.

We could in principle apply the same strategy to $B_{d,s} \to \phi K^{*0}$, but the corresponding $\delta$'s are much larger. Indeed, the computation
leads to corrections up to $\delta_{\sss \phi K^*}\sim 50\%$. This shows that $U$-spin symmetry cannot be expected to hold at a high accuracy for any
pair of flavour-related processes. $K^{(*)}K^{(*)}$ offer a much more interesting potential than other final states such as $\phi K^{*0}$. Moreover,
we cannot perform a similar analysis for $\phi\phi$ since $B_d\to\phi\phi$ is a pure weak-annihilation process, contrary to $B_s\to\phi\phi$ mediated
through penguins. Therefore we focus on the  precise $B_s \to K^{*0} {\bar K^{*0}}$ modes in the remaining part of this section. Notice that the
large hadronic uncertainties affecting $B_s \to \phi\phi$ and $B_s \to \phi K^{*0}$ have no impact when we use these modes in the strategies
described in Sections \ref{sec:bounding} and \ref{sec:measuring}, since we exploited a quantity $\Delta$ where they cancel out.

\begin{figure}
\begin{center}
\psfrag{BRd}{\hspace{-1.5cm}$BR\lg(B_d\to K^{*0} \bar{K}^{*0})\times 10^6$} \psfrag{BRs}{\hspace{-2.2cm}$BR\lg(B_s\to K^{*0} \bar{K}^{*0})\times
10^6$}
\includegraphics[height=6cm]{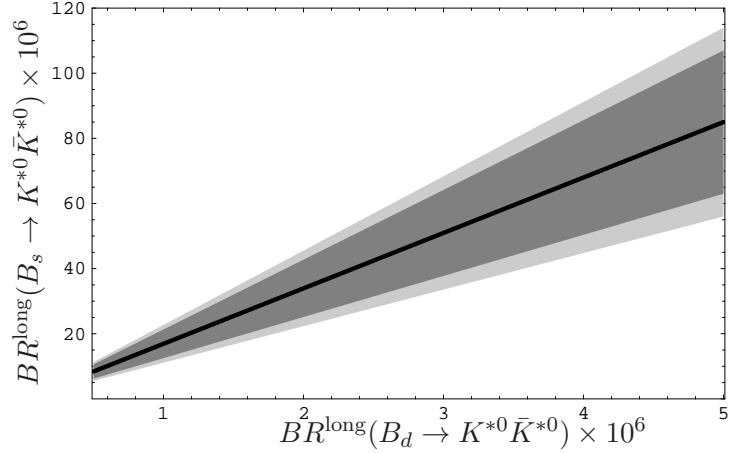}
\end{center}
\caption{\small Longitudinal branching ratio for $B_s\to K^{*0} \bar{K}^{*0}$ in terms of the longitudinal $B_d\to K^{*0} \bar{K}^{*0}$ branching
ratio. The light-shaded area corresponds to the uncertainty on the ratio of form factors $f$, whereas the dark-shaded area comes from varying the
various hadronic inputs.} \label{BstoKK}
\end{figure}

Once the hadronic parameters $P_{K^*K^*}^{s}$ and $T_{K^*K^*}^{s}$ have been obtained from Eq.(\ref{TsPs}), one can give predictions for the $B_s\to
K^{*0} \bar{K}^{*0}$ observables. Note that the branching ratio $BR\lg(B_d\to K^{*0} \bar{K}^{*0})$ is an experimental input in this analysis, and
this piece of information is not  available yet. The result for the branching ratio of $B_s\to K^{*0} \bar{K}^{*0}$ is given in terms of the $B_d\to
K^{*0} \bar{K}^{*0}$ branching ratio in Fig.\ref{BstoKK}. Once the branching ratio of $B_d\to K^{*0} \bar{K}^{*0}$ is measured one can use this plot
to find the SM prediction for $BR\lg(B_s\to K^{*0} \bar{K}^{*0})$.

The ratio of branching ratios $BR\lg(B_s\to K^{*0}\bar{K}^{*0})/BR\lg(B_d\to K^{*0}\bar{K}^{*0})$ and the asymmetries turn out to be quite
insensitive to the exact value of  $BR\lg(B_d\to K^{*0}\bar{K}^{*0})$ as long as $BR\lg(B_d\to K^{*0}\bar{K}^{*0})\gtrsim 5 \times 10^{-7}$. The
numerical values are summarised in Table~\ref{Tableresults}.

\begin{table}
\begin{center}
\begin{tabular}{|c|}
\hline $\begin{array}{c}
\\
\quad\displaystyle \left(\frac{BR\lg(B_s\to K^{*0}\bar{K}^{*0})}{BR\lg(B_d\to K^{*0}\bar{
K}^{*0})}\right)_{\sss SM}=17\pm 6 \quad\\
\\
\end{array}$\\
\hline $\begin{array}{c}
\\
\Adir\lg(B_s\to K^{*0}\bar{K}^{*0})_{\sss SM}=0.000\pm 0.014\\
\\
\end{array}$\\
\hline $\begin{array}{c}
\\
\Amix\lg(B_s\to K^{*0}\bar{K}^{*0})_{\sss SM}=0.004\pm 0.018\\
\\
\end{array}$\\
\hline
\end{tabular}
\end{center}
\caption{\small Results for the longitudinal observables related to $B_s\to K^{*0}\bar{K}^{*0}$ according to Sec.~\ref{sec:flavour}. These are
predictions for the SM contributions under the standard assumption of no New Physics in $b\to d$ transition. We used $\phi_s^{SM}=2 \beta_s=-2^\circ$
for $\Amix\lg$, and we assumed $BR\lg(B_d\to K^{*0}\bar{K}^{*0})\gtrsim 5 \times 10^{-7}$. The quoted uncertainty includes the errors associated to
all input parameters including the variation of $\gamma$ inside the range $56^\circ \leq \gamma\ \leq 68^\circ$ \cite{UTfit,CKMfitter}.}
\label{Tableresults}
\end{table}

Under the standard assumption that  New Physics contribution to $b \to d$ penguins is negligible, and since the experimental input comes entirely
from $B_d\to K^{*0}\bar{K}^{*0}$ (a $b \to d$ penguin), the results given in Table \ref{Tableresults} are SM predictions. In  presence of New Physics
in $b\to s$ penguins the full prediction can be obtained by adding to the SM piece extra contributions to the amplitude and weak mixing angle as
explained in \cite{hep-ph/0410011,hep-ph/0404130,hep-ph/0406192}.

\begin{figure}
\begin{center}
\psfrag{phis}{\hspace{-0.5cm}$\phi_s$ (Degree)} \psfrag{Amix}{\hspace{-1cm}$\Amix\lg(B_s\to K^{*0}\bar{K}^{*0})$}
\includegraphics[height=6cm]{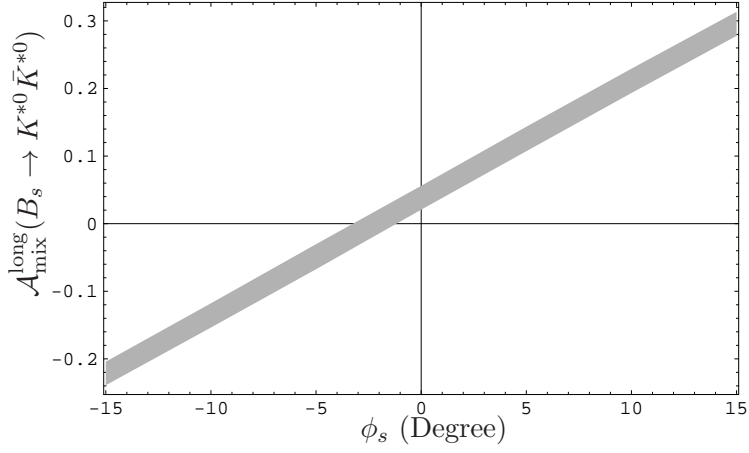}
\end{center}
\caption{\small Relation between $\Amix\lg(B_s\to K^{*0}\bar{K}^{*0})$ and the $B_s-\bar{B}_s$ mixing angle $\phi_s$. We assumed $BR\lg(B_d\to
K^{*0}\bar{K}^{*0})\gtrsim 5 \times 10^{-7}$ and $\gamma=62^\circ$. A measurement of this asymmetry leads to a prediction for $\phi_s$, which
includes hadronic pollution and SU(3) breaking effects, according to Sec.~\ref{sec:flavour}.} \label{Amix-phis}
\end{figure}

One may also use this as a strategy to extract the mixing angle $\phi_s$. If one assumes no New Physics in the decay $B_s\to K^{*0}\bar{K}^{*0}$,
this method relates directly $\Amix\lg(B_s\to K^{*0}\bar{K}^{*0})$ and $\phi_s$. Fig.\ref{Amix-phis} shows $\Amix\lg(B_s\to K^{*0}\bar{K}^{*0})$ vs.
$\phi_s$. Once this asymmetry is measured, this plot can be used as a way to extract $\phi_s$, and this result can be compared to the one found in
tree decays such as $B\to DK$. A disagreement would point out New Physics. Moreover, it is possible to distinguish whether New Physics affects the
decay or the mixing itself, by confronting $BR\lg(B_s\to K^{*0}\bar{K}^{*0})$ and $\Adir\lg(B_s\to K^{*0}\bar{K}^{*0})$ with the SM predictions given
in Table~\ref{Tableresults}. If the predictions for the branching ratio and the direct CP asymmetry agree with experiment, but the $\phi_s$ extracted
from $\Amix\lg(B_s\to K^{*0}\bar{K}^{*0})$ differs from $\phi_s^{SM}$, this will be a clear indication of New Physics in $B_s-\bar{B}_s$ mixing. An
interesting comparison will be allowed between the value for $\phi_s$ obtained here and the measurement of $\phi_s$ from the mixing induced
CP-asymmetry of $B_s\to D K$ decay \cite{hep-ph/0310081,hep-ph/0304027}.

\section{Discussion} \label{sec:discussion}

The increasing list of measured non-leptonic two-body $B_d$- and $B_s$-decays provides many tests of the CKM mechanism of CP violation in the
Standard Model. Of particular interest is the determination of angles through time-dependent CP-asymmetries. For instance $\phi_s$, related to
$B_s-\bar{B}_s$ mixing, should be constrained : it is tiny in the Standard Model, and can be measured through many penguin-dominated decays. However,
for such determination to be valid, one must assess the size of the various hadronic quantities involved as precisely as possible.

\begin{table}
\begin{center}\small
\begin{tabular}{|c||c|c|c|}
\hline
& Strategy 1 & Strategy 2 & Strategy 3\\
\hline\hline \hspace{0.0cm}Inputs\hspace{0.0cm} & \footnotesize $\begin{array}{c}
\\
BR\lg(B_s\to K^{*0} \bar{K}^{*0})\\
\Amix\lg(B_s\to K^{*0} \bar{K}^{*0})\\
\Delta^s_{K^*K^*},\ \gamma\\
\\
\end{array}$& \footnotesize
$\begin{array}{c}
BR\lg(B_s\to K^{*0} \bar{K}^{*0})\\
\Adir\lg(B_s\to K^{*0} \bar{K}^{*0})\\
\Amix\lg(B_s\to K^{*0} \bar{K}^{*0})\\
\Delta^s_{K^*K^*}
\end{array}$& \footnotesize
$\begin{array}{c}
BR\lg(B_d\to K^{*0} \bar{K}^{*0})\\
\Amix\lg(B_s\to K^{*0} \bar{K}^{*0})\\
\Delta^d_{K^*K^*},\ \delta_T,\ \delta_P,\ \gamma
\end{array}$\\
\hline Outputs & $\phi_s$& $|\sin{\beta_s}|$, $\gamma$& \footnotesize $\begin{array}{c}
\\
BR\lg(B_s\to K^{*0} \bar{K}^{*0})_{\sss SM}\\
\Adir\lg(B_s\to K^{*0} \bar{K}^{*0})_{\sss SM}\\
\phi_s\\
\\
\end{array}$\\
\hline\hline Advantages &
\begin{minipage}[c]{3cm}
\vspace{0.2cm} Applies also to $B_s\to\phi K^{*0}$ and $B_s\to\phi \phi$ \vspace{0.2cm}
\end{minipage}&
\begin{minipage}[c]{3cm}
\vspace{0.2cm} Applies also to $B_s\to\phi K^{*0}$ and $B_s\to\phi \phi$ \vspace{0.2cm}
\end{minipage}&
\begin{minipage}[c]{3cm}
\vspace{0.2cm} It can be easily generalized to include New Physics in the decay and mixing. \vspace{0.2cm}
\end{minipage}\\
\hline Limitations &
\begin{minipage}[c]{3cm}
\vspace{0.2cm} It assumes no New Physics in $b \to s$ decay. \vspace{0.2cm}
\end{minipage}&
\begin{minipage}[c]{3cm}
\vspace{0.2cm} It assumes no New Physics in $b \to s$ decay. \vspace{0.2cm}
\end{minipage}&
\begin{minipage}[c]{3cm}
\vspace{0.2cm} Does not apply to $B_s\to\phi K^{*0}$ or $B_s\to\phi \phi$ because $\delta_{T,P}$ are big. \vspace{0.2cm}
\end{minipage}\\
\hline
\end{tabular}
\end{center}
\caption{\small Comparison between the three strategies for $B_s\to K^{*0} \bar{K}^{*0}$.} \label{comparative}
\end{table}

In this chapter we have applied ideas presented in Chapter \ref{SymFac1} for $B_{d,s}\to KK$ to vector-vector modes mediated through penguins :
$B_{d,s}\to K^{*0} \bar{K}^{*0}$, $\phi\phi$ and $\phi K^{*0}$ (with the condition that $K^{*0}$ decays into a definite-CP eigenstate). In order to
combine flavor symmetries with QCD factorization, we have restricted our analysis to longitudinal observables, which are under better theoretical
control. These observables have been related to the angular analysis performed experimentally in Sec.~\ref{sec:longobs}. Penguin-mediated modes offer
the very interesting feature that the difference between tree and penguin contributions $\Delta=T-P$ should be dominated by short-distance physics.
It can be computed fairly accurately using QCD factorization, and it can be used to determine tree and penguin contributions from observables as
explained in Chapter \ref{SymFac1}. This theoretical piece of information is used to relate CP-asymmetries of $B_{d,s}\to K^{*0} \bar{K}^{*0}$,
$\phi\phi$ and $\phi K^{*0}$ to CKM angles according to different strategies. For illustration, we have focused on $B_{d,s}\to K^{*0} \bar{K}^{*0}$,
where all three strategies apply.

In Sec.~\ref{sec:bounding}, we have proposed to use $\Delta=T-P$ to put stringent bounds on the pollution due to hadronic uncertainties. Indeed, even
though the ratio $|\lambda_u^{(s)}/\lambda_c^{(s)}|=0.044$ is small, a large value of the hadronic quantity ${\rm Re}(T/P)$ could spoil the naively
safe extraction of $\sin{\phi_s}$ from the mixed asymmetry of $B_s\to K^{*0}\bar{K}^{*0}$. This strategy to control the pollution can be applied to
all penguin-mediated processes of interest here. A similar analysis for $\sin{2\beta}$ from $B_d\to \phi K_S$ can be found in \cite{arXiv:0707.2046}.

In Sec.~\ref{sec:measuring}, we have suggested a second approach, using $\Amix\lg(B_s\to K^{*0} \bar{K}^{*0})$, $\Adir\lg(B_s\to K^{*0}
\bar{K}^{*0})$ and $BR\lg(B_s\to K^{*0} \bar{K}^{*0})$ to extract $|{\rm sin} \beta_s|$. In principle, one can also use an alternative set of
experimental quantities : the branching ratio together with a direct measurement of the longitudinal untagged rate. The sum rule needed for the
$B_s\to K^{*0} \bar{K}^{*0}$ is independent of the CKM angle $\gamma$ and the input on hadronic dynamics is limited to a single well-controlled
quantity: $\Delta^{s}_{K^*K^*}$. This strategy can also be applied to extract $\beta_s$ from $B_s\to \phi \bar{K}^{*0}$ and $B_s \to \phi\phi$ using
the corresponding sum rule.

In Sec.~\ref{sec:flavour}, we proposed a last method to determine  $\phi_s$, by relying on the prediction of the mixing induced CP-asymmetry
$\Amix\lg(B_s\to K^{*0} \bar{K}^{*0})$ as a function of the $BR\lg(B_d \to K^{*0} \bar{K}^{*0})$ and the theoretical input $\Delta^d_{K^*K^*}$. In
this strategy, tree pollution is controlled using the hadronic information from flavour symmetry and QCD factorisation. The outcome of our analysis
is presented in Fig.~\ref{Amix-phis}. This strategy requires data on $B_d \to K^{*0} \bar{K}^{*0}$ and on the mixing-induced CP-asymmetry
$\Amix\lg(B_s\to K^{*0} \bar{K}^{*0})$. The input from $B_s$ decay is therefore minimal : $\Amix\lg(B_s\to K^{*0} \bar{K}^{*0})$, while all other
inputs can be obtained from $B$-factories.

A comparison among the three different strategies discussed in this chapter is given in Table \ref{comparative}, where the needed inputs are
enumerated as well as the predicted observables and the range of validity.

If both hadronic machines and super-$B$ factories \cite{arXiv:0709.0451} running at $\Upsilon(5S)$ provide enough information on $B_s$-decays, it
will be interesting to compare the determination from $\phi_s$ following those methods, which rely on penguin-mediated decays, with the value
obtained from tree processes like $B_s \to DK$. Differences between the values obtained through these two procedures would provide a clear hint of
physics beyond the Standard Model. In such a situation, the different methods presented in this letter would yield very useful cross-checks for the
penguin-dominated vector modes.

%


\chapter{Supersymmetric Contributions to $B_s^{\sss 0}\to K^+K^-$ and $B_s^{\scriptscriptstyle 0}\to K^{\sss 0}\bar{K}^{\sss 0}$}

\label{SUSYcontributions}

\label{SUSYContributions}

\n In Section \ref{MeasuringNP} we saw that the NP parameters for the decay $B_s^{\scriptscriptstyle 0}\to K^+K^-$ can be extracted from the measured
values for the branching ratio and the CP-asymmetries, once the SM hadronic parameters are known. These hadronic parameters can be obtained from
other decays related by flavor symmetries to $B_s^{\scriptscriptstyle 0}\to K^+K^-$, such as $B_d^{\scriptscriptstyle 0}\to \pi^+\pi^-$ or
$B_d^{\scriptscriptstyle 0}\to K^0 \bar{K}^0$, provided that these are not affected by NP. In Chapter \ref{Bdecays} we obtained the
$B_s^{\scriptscriptstyle 0}\to K^+K^-$ SM hadronic parameters from $B_d^{\scriptscriptstyle 0}\to \pi^+\pi^-$ using flavor symmetry, and in Chapter
\ref{SymFac1} we obtained the SM hadronic parameters of $B_s^{\scriptscriptstyle 0}\to K^+K^-$ and $B_s^{\scriptscriptstyle 0}\to K^0\bar{K}^0$ from
$B_d^{\scriptscriptstyle 0}\to K^0 \bar{K}^0$ using an interesting combination of flavor symmetry and QCDF.

Experimental values for the CP-asymmetries in $B_s^{\scriptscriptstyle 0}\to K^+K^-$ are still not available, so the method of Chapter \ref{Bdecays}
cannot be applied yet to measure the NP parameters $\mathcal{A}^u$ and $\Phi_u$. However, it is always interesting to study which existing NP models
can generate sizeable contributions to these parameters, since once $\mathcal{A}^u$ and $\Phi_u$ are measured, these NP models will be the most
sensitive to this new data. If they generate too big NP parameters, their parameter space will be drastically constrained. If the NP parameters are
indeed measured to be large, then these models will prevail over more conservative ones.

On the other hand, being able to introduce hadronic effects in a fairly reliable way allows to give full predictions for observables in the presence
of NP. Of course, no extension of the SM will be given full support until the SM predictions (and in particular those predictions derived in Chapters
\ref{Bdecays} and \ref{SymFac1}) are found inconsistent with experiment. However, due to the place of honor that supersymmetry holds within the NP
models in the present, it is worth to study in detail how can it modify the predictions for these observables. For this reason, in this chapter we
compute the $B_s^{\scriptscriptstyle 0}\to K^+K^-$ and $B_s^{\scriptscriptstyle 0}\to K^0\bar{K}^0$ observables in a generic SUSY model.

Naively, one would guess that all NP contributions to $\mathcal{A}^u$ and $\Phi_u$ are suppressed by $M_W^2/M_{\sss NP}^2$, where $M_{\sss NP} \sim
1$ TeV, and are therefore small. However, SUSY contributions involving squark-gluino loops can be important since they involve the strong coupling
constant $\alpha_s$. This contributions are thus proportional to $\alpha_s/M_{\sss NP}^2$ and can compete with the SM contributions which are of
order $\alpha/M_W^2$, since $(\alpha_s/\alpha)(M_W^2/M_{\sss NP}^2) \sim 1$. Indeed, these are the dominant effects, and are the only ones which are
considered below.  As we will see, they can account for an important enhancement of the CP asymmetries.

The NP amplitude for a $B\to f$ decay is given generically by (see (\ref{ANP}))
\eq{\mathcal{A}^q e^{i\Phi_q}=\bra{f}\mathcal{H}_{\rm eff}^{\sss \rm NP}\ket{B}=\frac{G_F}{\sqrt{2}}\sum_i C_i(\mu)\av{O_i(\mu)} \label{B->f NP}}

The procedure to follow is the one motivated in Chapter \ref{RenormGroup}:

\begin{itemize}

\item We consider all operators generated at the heavy scale, taken to
be $M_W$. We compute the SUSY contributions to the coefficients of these operators.

\item Using the renormalization group, we run the operator
coefficients down to the factorization scale $\mu$. Operator mixing takes place in this step.

\item We compute the matrix elements of the various operators at
$m_b$. This allows us to calculate $\mathcal{A}^u$, $\Phi_u$, $\mathcal{A}^d$ and $\Phi_d$.

\end{itemize}
Once we have computed the values of the NP parameters, we use the SM hadronic parameters calculated in Chapter \ref{SymFac1} together with equations
(\ref{BRNP})-(\ref{MNP}) to give full predictions for the observables.

\section{The Effective Hamiltonian in the presence of NP}

We begin by listing all the operators which are generated by the New Physics at the heavy scale. In general, NP can induce many new operators with
Dirac structures that cannot fit in the SM operator basis presented in Section \ref{SMHeff}. Therefore it is mandatory to generalize the set of
effective operators beyond the one generated by the SM. Computing the contributions in which we are interested, we will come up with a set of
operators whose basis we choose as \cite{hep-ph/9909297}:
\eqa{
&O_1^q=(\bar{b}_\alpha s_\alpha)_{V-A}(\bar{q}_\beta q_\beta)_{V+A}
\qquad\quad  & O_2^q=(\bar{b}_\alpha s_\beta)_{V-A}(\bar{q}_\beta q_\alpha)_{V+A}\nn\\
&O_3^q=(\bar{b}_\alpha s_\alpha)_{V-A}(\bar{q}_\beta q_\beta)_{V-A}
\qquad\quad  & O_4^q=(\bar{b}_\alpha s_\beta)_{V-A}(\bar{q}_\beta q_\alpha)_{V-A}\nn\\
&O_5^q=(\bar{b}_\alpha q_\alpha)_{V-A}(\bar{q}_\beta s_\beta)_{V+A}
\qquad\quad & O_6^q=(\bar{b}_\alpha q_\beta)_{V-A}(\bar{q}_\beta s_\alpha)_{V+A}\nn\\
&Q_{8g} = \frac{g_s}{8 \pi^2} m_b {\bar b} \sigma_{\mu \nu} (1-\gamma_5)G^{\mu \nu} s \quad\  &
\label{NPops} }
and the corresponding effective Hamiltonian considered here is the following
\eq{ \mathcal{H}_{\rm eff}^{\sss \rm NP}=\frac{G_F}{\sqrt{2}} \left[ \sum_{i,q=u,d} \left( c_i^q(\mu) \, O_i^q + \tilde{c}_i^q(\mu) \, \tilde{O}_i^q
\right) + C_{8g}(\mu) \, Q_{8g} + {\tilde C}_{8g}(\mu) \, {\tilde Q}_{8g} \right] ~, \label{HSUSY} }
Despite the fact that, at the quark level, $B_s^{\scriptscriptstyle 0}\to K^+K^-$ is $\bar{b} \to \bar{s} u{\bar u}$, $d$-quark operators must be
included above since they mix with the $u$-quark operators upon renormalization to $m_b$. The same is true for $B_s^{\scriptscriptstyle 0}\to
K^0\bar{K}^0$ and the $u$-quark operators. The operators $\tilde{O}_i^q$ and ${\tilde Q}_{8g}$ are obtained from $O_i^q$ and $Q_{8g}$ by chirality
flipping, i.e, $\scriptstyle V-A\ \leftrightarrow\ V+A$.

The above list of operators includes new-physics contributions to electroweak-penguin operators. As we will see, these effects can be significant.
This shows that, although the SM electroweak-penguin contributions to $B_s^{\scriptscriptstyle 0}\to K^+K^-$ are negligible, the same does not hold
for the NP.

\section{Wilson Coefficients}

\label{WilsonCoefficients}

\begin{figure}
\psfrag{bb}{$b_\alpha$} \psfrag{sb}{$s_\gamma$} \psfrag{Tar}{$T_{\alpha\rho}^a$} \psfrag{Tsg}{$T_{\sigma\gamma}^a$} \psfrag{Trb}{$T_{\rho\beta}^b$}
\psfrag{Tds}{$T_{\delta\sigma}^b$} \psfrag{Trg}{$T_{\rho\gamma}^b$} \psfrag{Tsb}{$T_{\sigma\beta}^a$} \psfrag{Tbs}{$T_{\sigma\beta}^b$}
\psfrag{Tsd}{$T_{\sigma\delta}^b$} \psfrag{Tsg2}{$T_{\sigma\gamma}^b$} \psfrag{Tds2}{$T_{\delta\sigma}^a$} \psfrag{Trg2}{$T_{\rho\gamma}^b$}
\psfrag{Tsd2}{$T_{\delta\sigma}^a$} \psfrag{ga}{$\tilde{g}^a$}\psfrag{gb}{$\tilde{g}^b$}
\begin{center}
\psfrag{b}{$d_\delta$} \psfrag{s}{$d_\beta$} \psfrag{Dr}{$D_{i\rho}$} \psfrag{Ds}{$D_{j\sigma}$}
\includegraphics{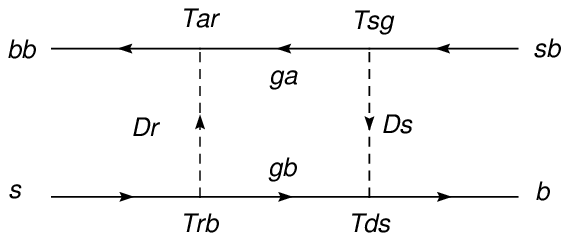}
\hspace{0.7cm}
\includegraphics{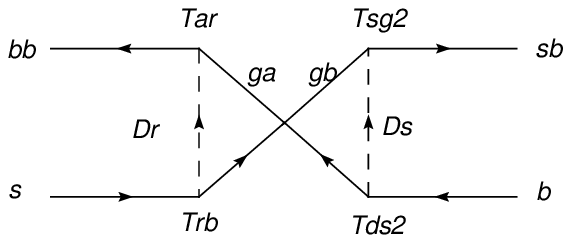}
\Text(-275,-15)[lb]{$(a)$} \Text(-87,-15)[lb]{$(b)$}
\end{center}
\vspace{0.1cm}
\begin{center}
\psfrag{b}{$q_\delta$} \psfrag{s}{$q_\beta$} \psfrag{Dr}{$D_{i\rho}$} \psfrag{Ds}{$Q_{j\sigma}$}
\includegraphics{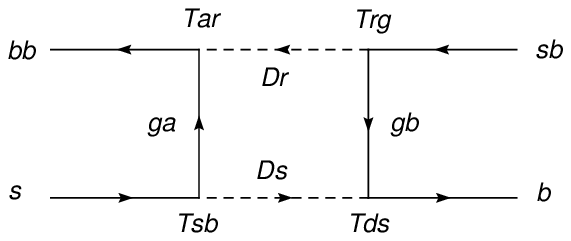}
\hspace{0.7cm}
\includegraphics{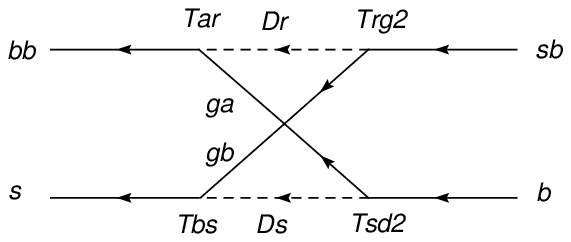}
\Text(-275,-17)[lb]{$(c)$} \Text(-87,-17)[lb]{$(d)$}
\end{center}
\caption{\small Gluino box contributions to $b\to sqq$.} \label{SUSYBoxes}
\end{figure}

\begin{figure}
\psfrag{bb}{$b_\alpha$} \psfrag{sb}{$s_\gamma$} \psfrag{b}{$q_\delta$} \psfrag{s}{$q_\beta$} \psfrag{Dr}{$D_{i\rho}$} \psfrag{Ds}{$D_{j\sigma}$}
\psfrag{Tar}{$T_{\alpha\rho}^a$} \psfrag{Tsg}{$T_{\sigma\gamma}^a$} \psfrag{Trs}{$T_{\rho\sigma}^b$} \psfrag{Tdb}{$T_{\delta\beta}^b$}
\psfrag{Tdb2}{$T_{\delta\beta}^c$} \psfrag{Trg}{$T_{\rho\gamma}^b$} \psfrag{Tsd}{$T_{\sigma\delta}^b$} \psfrag{Tsg2}{$T_{\sigma\gamma}^b$}
\psfrag{Tds2}{$T_{\delta\sigma}^a$} \psfrag{Trg2}{$T_{\rho\gamma}^b$} \psfrag{Tsd2}{$T_{\sigma\delta}^a$}
\psfrag{ga}{$\tilde{g}^a$}\psfrag{gb}{$\tilde{g}^b$}\psfrag{gc}{$g^c$}\psfrag{gb2}{$g^b$}
\begin{center}
\includegraphics[height=2.8cm]{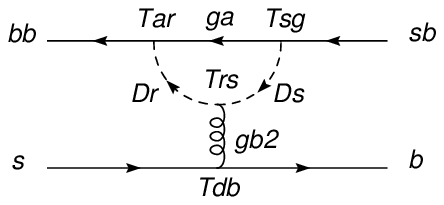}
\hspace{0.7cm}
\includegraphics[height=3.2cm]{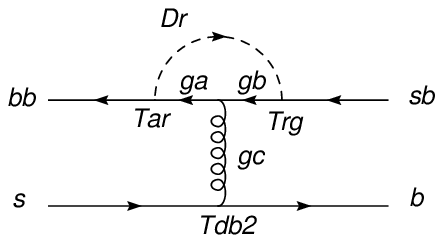}
\end{center}
\caption{\small Gluino-squark penguin contributions to $b\to sqq$.} \label{SUSYpengs}
\end{figure}

In order to calculate the relevant SUSY contributions to the Wilson coefficients in (\ref{HSUSY}) we must compute the corresponding diagrams in the
full theory. These, as stated above, are the contributions involving squark-gluino loops. Specifically, we have to deal with box and penguin diagrams
(see Figs. \ref{SUSYBoxes} and \ref{SUSYpengs}). The contributions from the box diagrams to the Wilson coefficients at the matching scale are
\cite{hep-ph/9909297}
\eqa{ c^u_{1,\rm box}&=&\frac{\alpha_s^2}{2\sqrt{2} G_F m_{\tilde{g}}^2} (\Gamma^{D_L}_{ib})^*\Gamma^{D_L}_{is}(\Gamma^{U_R}_{ju})^*\Gamma^{U_R}_{ju}
       \Big[ \frac{1}{18}F(x_{\tilde{d}_i\tilde{g}},x_{\tilde{u}_j\tilde{g}})-\frac{5}{18}G(x_{\tilde{d}_i\tilde{g}},x_{\tilde{u}_j\tilde{g}}) \Big]\nn\\
c^d_{1,\rm box}&=&\frac{\alpha_s^2}{2\sqrt{2} G_F m_{\tilde{g}}^2} (\Gamma^{D_L}_{ib})^*\Gamma^{D_L}_{is}(\Gamma^{D_R}_{jd})^*\Gamma^{D_R}_{jd}
       \Big[ \frac{1}{18}F(x_{\tilde{d}_i\tilde{g}},x_{\tilde{d}_j\tilde{g}})
       -\frac{5}{18}G(x_{\tilde{d}_i\tilde{g}},x_{\tilde{d}_j\tilde{g}}) \Big]\nn\\
c^u_{2,\rm box}&=&\frac{\alpha_s^2}{2\sqrt{2} G_F m_{\tilde{g}}^2} (\Gamma^{D_L}_{ib})^*\Gamma^{D_L}_{is}(\Gamma^{U_R}_{ju})^*\Gamma^{U_R}_{ju}
       \Big[ \frac{7}{6}F(x_{\tilde{d}_i\tilde{g}},x_{\tilde{u}_j\tilde{g}})
       +\frac{1}{6}G(x_{\tilde{d}_i\tilde{g}},x_{\tilde{u}_j\tilde{g}}) \Big]\nn\\
c^d_{2,\rm box}&=&\frac{\alpha_s^2}{2\sqrt{2} G_F m_{\tilde{g}}^2} (\Gamma^{D_L}_{ib})^*\Gamma^{D_L}_{is}(\Gamma^{D_R}_{jd})^*\Gamma^{D_R}_{jd}
       \Big[ \frac{7}{6}F(x_{\tilde{d}_i\tilde{g}},x_{\tilde{d}_j\tilde{g}})
       +\frac{1}{6}G(x_{\tilde{d}_i\tilde{g}},x_{\tilde{d}_j\tilde{g}}) \Big]\nn\\
c^u_{3,\rm box}&=&\frac{\alpha_s^2}{2\sqrt{2} G_F m_{\tilde{g}}^2} (\Gamma^{D_L}_{ib})^*\Gamma^{D_L}_{is}(\Gamma^{U_L}_{ju})^*\Gamma^{U_L}_{ju}
       \Big[ -\frac{5}{9}F(x_{\tilde{d}_i\tilde{g}},x_{\tilde{u}_j\tilde{g}})
       +\frac{1}{36}G(x_{\tilde{d}_i\tilde{g}},x_{\tilde{u}_j\tilde{g}}) \Big]\nn\\
c^d_{3,\rm box}&=&\frac{\alpha_s^2}{2\sqrt{2} G_F m_{\tilde{g}}^2} \Big\{(\Gamma^{D_L}_{ib})^*\Gamma^{D_L}_{is}(\Gamma^{D_L}_{jd})^*\Gamma^{D_L}_{jd}
       \Big[ -\frac{5}{9}F(x_{\tilde{d}_i\tilde{g}},x_{\tilde{d}_j\tilde{g}})
       +\frac{1}{36}G(x_{\tilde{d}_i\tilde{g}},x_{\tilde{d}_j\tilde{g}}) \Big]\nn\\
    && \qquad\qquad\quad+\,(\Gamma^{D_L}_{ib})^*\Gamma^{D_L}_{js}(\Gamma^{D_L}_{jd})^*\Gamma^{D_L}_{id}
    \Big[ \frac{1}{3}F(x_{\tilde{d}_i\tilde{g}},x_{\tilde{d}_j\tilde{g}})
    -\frac{7}{12}G(x_{\tilde{d}_i\tilde{g}},x_{\tilde{d}_j\tilde{g}}) \Big]\Big\}\nn
} \eqa{ c^u_{4,\rm box}&=&\frac{\alpha_s^2}{2\sqrt{2} G_F m_{\tilde{g}}^2}
(\Gamma^{D_L}_{ib})^*\Gamma^{D_L}_{is}(\Gamma^{U_L}_{ju})^*\Gamma^{U_L}_{ju}
       \Big[ \frac{1}{3}F(x_{\tilde{d}_i\tilde{g}},x_{\tilde{u}_j\tilde{g}})
       +\frac{7}{12}G(x_{\tilde{d}_i\tilde{g}},x_{\tilde{u}_j\tilde{g}}) \Big]\nn\\
c^d_{4,\rm box}&=&\frac{\alpha_s^2}{2\sqrt{2} G_F m_{\tilde{g}}^2} \Big\{(\Gamma^{D_L}_{ib})^*\Gamma^{D_L}_{is}(\Gamma^{D_L}_{jd})^*\Gamma^{D_L}_{jd}
       \Big[ \frac{1}{3}F(x_{\tilde{d}_i\tilde{g}},x_{\tilde{d}_j\tilde{g}})
       +\frac{7}{12}G(x_{\tilde{d}_i\tilde{g}},x_{\tilde{d}_j\tilde{g}}) \Big]\nn\\
    && \qquad\qquad\quad+\,(\Gamma^{D_L}_{ib})^*\Gamma^{D_L}_{js}(\Gamma^{D_L}_{jd})^*\Gamma^{D_L}_{id}
    \Big[ -\frac{5}{9}F(x_{\tilde{d}_i\tilde{g}},x_{\tilde{d}_j\tilde{g}})
    +\frac{1}{36}G(x_{\tilde{d}_i\tilde{g}},x_{\tilde{d}_j\tilde{g}}) \Big]\Big\}\nn\\
c^d_{5,\rm box}&=&\frac{\alpha_s^2}{2\sqrt{2} G_F m_{\tilde{g}}^2} (\Gamma^{D_L}_{ib})^*\Gamma^{D_R}_{js}(\Gamma^{D_R}_{jd})^*\Gamma^{D_L}_{id}
       \Big[ \frac{1}{18}F(x_{\tilde{d}_i\tilde{g}},x_{\tilde{d}_j\tilde{g}})
       -\frac{5}{18}G(x_{\tilde{d}_i\tilde{g}},x_{\tilde{d}_j\tilde{g}}) \Big]\nn\\
c^d_{6,\rm box}&=&\frac{\alpha_s^2}{2\sqrt{2} G_F m_{\tilde{g}}^2} (\Gamma^{D_L}_{ib})^*\Gamma^{D_R}_{js}(\Gamma^{D_R}_{jd})^*\Gamma^{D_L}_{id}
       \Big[ \frac{7}{6}F(x_{\tilde{d}_i\tilde{g}},x_{\tilde{d}_j\tilde{g}})
       +\frac{1}{6}G(x_{\tilde{d}_i\tilde{g}},x_{\tilde{d}_j\tilde{g}}) \Big]\nn\\
c^u_{5,\rm box}&=&c^u_{6,\rm box}\ =\ 0 \label{cbox}}

The coefficients $c^u_{5,\rm box}$ and $c^u_{6,\rm box}$ are zero since their structure arises from the diagrams (a) and (b), but these diagrams
don't contribute to u-type coefficients, as mentioned above. The coefficients $\tilde{c}^q_{i,\rm box}$ are just like (\ref{cbox}) with the
replacement $L\leftrightarrow R$. Also $\ x_{i\tilde{g}}\equiv m_i^2/m_{\tilde{g}}^2$, and the loop functions are given by
\eqa{
F(x,y)&=&-\frac{x\ln{x}}{(x-y)(x-1)^2}-\frac{y\ln{y}}{(y-x)(y-1)^2}-\frac{1}{(x-1)(y-1)}\ ,\nn\\
G(x,y)&=&\frac{x^2\ln{x}}{(x-y)(x-1)^2}+\frac{y^2\ln{y}}{(y-x)(y-1)^2}+\frac{1}{(x-1)(y-1)\ .} \label{F,G}}

The contributions to the Wilson coefficients from the penguin diagrams are given by
\eq{c^q_{1,\rm peng}=c^q_{3,\rm peng}=-\frac{c^q_{2,\rm peng}}{3}=-\frac{c^q_{4,\rm peng}}{3} =\frac{\alpha_s^2}{2\sqrt{2} G_F m_{\tilde{g}}^2}
(\Gamma^{D_L}_{ib})^*\Gamma^{D_L}_{is} \Big[ \frac{1}{2}A(x_{\tilde{d}_i\tilde{g}})+\frac{2}{9}B(x_{\tilde{d}_i\tilde{g}}) \Big] \label{cpeng}}
where the loop functions in this case are
\eqa{
A(x)&=&\frac{1}{2(1-x)}+\frac{(1+2x)\ln{x}}{6(1-x)^2}\ ,\nn\\
B(x)&=&-\frac{11-7x+2x^2}{18(1-x)^3}-\frac{\ln{x}}{3(1-x)^4\ .} \label{A,B}}

Again, the coefficients of the operators with opposite chirality are obtained via the substitution $L\leftrightarrow R$.

It is possible to cast this Wilson coefficients into a more convenient form. To that end we will deal with the two $\,6\times 6\,$ squark mass
matrices $\tilde{M}^u$ and $\tilde{M}^d$,
\eq{\tilde{M}^{q\,2}=\left(
\begin{array}{cc}
\tilde{M}^{q\,2}_{LL} & \tilde{M}^{q\,2}_{LR} \\
\sss  & \sss \\
\tilde{M}^{q\,2}_{RL} & \tilde{M}^{q\,2}_{RR}
\end{array} \right)
}
These matrices are hermitian and so are $\tilde{M}^{q\,2}_{LL}$ and $\tilde{M}^{q\,2}_{RR}$, while
$\tilde{M}^{q\,2}_{LR}=\tilde{M}^{q\,2\,\dagger}_{RL}$. These are the mass matrices in the interaction basis for the squark fields,
$\{\tilde{u}_i^I,\tilde{d}_j^I\}$ and are diagonalized by the rotation matrices $\Gamma^D$ and $\Gamma^U$ that link this basis to the mass eigenbasis
$\{\tilde{u}_i,\tilde{d}_j\}$,
\eq{\tilde{d}^I=(\Gamma^D)^\dagger\,\tilde{d}}
and similar for up-type squarks.

It turns out that the entries of the left-right down-squark mass submatrix are suppressed (at least) by a factor $m_b/M_{\sss \rm SUSY}$
\cite{hep-ph/9909297}, and we neglect it. Generically, SUSY can induce large FCNC's. Experimental bounds -- from neutral meson mixings, $b\to
s\gamma$, etc.-- provide strong constraints on the SUSY parameter space. In particular, $B_d-\bar{B}_b$ and $K-\bar{K}$ mixing allow us to decouple
the down squark from the strange and bottom squarks. The same can be said for the up squark based on $D-\bar{D}$ mixing, although this doesn't
directly decouple the up from the top squarks \cite{hep-ph/9909297}.

These simplifications imply, first, that $\tilde{M}^{q\,2}_{LR}\simeq 0$. Second, the d-type rotation matrices can be parametrized by two mixing
angles $\theta_{R,L}$ and two \emph{weak} phases $\delta_{L,R}$,
\eq{\Gamma^D\simeq \left(
\begin{array}{ccc|ccc}
1   &              0                   &                0                 &&& \\
0   &        \cos{\theta_L}            &   \sin{\theta_L}e^{i\delta_L}    &&& \\
0   &  -\sin{\theta_L}e^{-i\delta_L}   &         \cos{\theta_L}           &&& \\
\hline
&&& 1   &              0                   &                0                  \\
&&& 0   &        \cos{\theta_R}            &   \sin{\theta_R}e^{i\delta_R}     \\
&&& 0   &  -\sin{\theta_R}e^{-i\delta_R}   &         \cos{\theta_R}
\end{array} \right)
}
The same approximation is made for $\Gamma^U$, although we just shall need the first and fourth rows and no angles arise from the u-type sector in
our computations. We choose the mixing angles to take values $\m{\theta_{L,R}}\le 45^\circ$, which defines the mass eigenstates $\tilde{s}$ and
$\tilde{b}$ as the ones more closely aligned to the $s$ and $b$ quarks respectively.

By substituting the simplified rotation matrices into (\ref{cbox}) and (\ref{cpeng}) we find
\eqa{ c_1^q&=&\frac{\alpha_s^2\sin{2\theta_L}e^{i\delta_L}}{4\sqrt{2}G_Fm_{\tilde{g}}^2} \left[ \frac{1}{18}F(x_{\sss \tilde{b}_L\tilde{g}},x_{\sss
\tilde{q}_R\tilde{g}})- \frac{5}{18}G(x_{\sss \tilde{b}_L\tilde{g}},x_{\sss \tilde{q}_R\tilde{g}})+
\frac{1}{2}A(x_{\sss \tilde{b}_L\tilde{g}})+\frac{2}{9}B(x_{\sss \tilde{b}_L\tilde{g}})\right]\nn\\
&&-(x_{\sss \tilde{b}_L\tilde{g}} \to x_{\sss \tilde{s}_L\tilde{g}})\ ,\nn\\
c_2^q&=&\frac{\alpha_s^2\sin{2\theta_L}e^{i\delta_L}}{4\sqrt{2}G_Fm_{\tilde{g}}^2} \left[ \frac{7}{6}F(x_{\sss \tilde{b}_L\tilde{g}},x_{\sss
\tilde{q}_R\tilde{g}})+ \frac{1}{6}G(x_{\sss \tilde{b}_L\tilde{g}},x_{\sss \tilde{q}_R\tilde{g}})-
\frac{3}{2}A(x_{\sss \tilde{b}_L\tilde{g}})-\frac{2}{3}B(x_{\sss \tilde{b}_L\tilde{g}})\right]\nn\\
&&-(x_{\sss \tilde{b}_L\tilde{g}} \to x_{\sss \tilde{s}_L\tilde{g}})\ ,\nn\\
c_3^q&=&\frac{\alpha_s^2\sin{2\theta_L}e^{i\delta_L}}{4\sqrt{2}G_Fm_{\tilde{g}}^2} \left[ -\frac{5}{9}F(x_{\sss \tilde{b}_L\tilde{g}},x_{\sss
\tilde{q}_L\tilde{g}})+ \frac{1}{36}G(x_{\sss \tilde{b}_L\tilde{g}},x_{\sss \tilde{q}_L\tilde{g}})+
\frac{1}{2}A(x_{\sss \tilde{b}_L\tilde{g}})+\frac{2}{9}B(x_{\sss \tilde{b}_L\tilde{g}})\right]\nn\\
&&-(x_{\sss \tilde{b}_L\tilde{g}} \to x_{\sss \tilde{s}_L\tilde{g}})\ ,\nn\\
c_4^q&=&\frac{\alpha_s^2\sin{2\theta_L}e^{i\delta_L}}{4\sqrt{2}G_Fm_{\tilde{g}}^2} \left[ \frac{1}{3}F(x_{\sss \tilde{b}_L\tilde{g}},x_{\sss
\tilde{q}_L\tilde{g}})+ \frac{7}{12}G(x_{\sss \tilde{b}_L\tilde{g}},x_{\sss \tilde{q}_L\tilde{g}})-
\frac{3}{2}A(x_{\sss \tilde{b}_L\tilde{g}})-\frac{2}{3}B(x_{\sss \tilde{b}_L\tilde{g}})\right]\nn\\
&&-(x_{\sss \tilde{b}_L\tilde{g}} \to x_{\sss \tilde{s}_L\tilde{g}})\ , \label{WC} }
with the loop functions $F,G,A,B$ given by (\ref{F,G}) and (\ref{A,B}). The expressions for the coefficients $\tilde{c}_i^q$ are obtained from those
in (\ref{WC}) via the exchange $L\leftrightarrow R$, and the coefficients $c_{5,6}^q$ and $\tilde{c}_{5,6}^q$ vanish. In addition
$x_{\tilde{q}_i\tilde{g}}\equiv m_{\tilde{q}_i}^2/m_{\tilde{g}}^2$, where $m_{\tilde{q}_i}$ ($q=d,u$) is the mass of the \textit{i}-th squark mass
eigenstate.

Using the same approximations, the SUSY contribution to the WC of the chromomagnetic operator is given by
\eq{ \lambda_t \frac{2 \alpha_s}{3 \pi} C^{\rm SUSY}_{8g}=\frac{8}{3} \frac{\alpha_s^2 \sin 2\theta_L e^{i \delta_L}}{4\sqrt{2} G_F m_{\tilde{g}}^2}
\left[f_8^{\rm SUSY}(x_{\sss \tilde{b}_L \tilde{g}})-(b_L \leftrightarrow s_L)\right]}
where the loop function is
\eq{f_8^{\rm SUSY}(x)= \frac{-11 + 51\,x - 21\,x^2 - 19\,x^3+6\,x\,\left(-1 + 9\,x \right) \,\log (x)}{72\, {\left(-1+x\right)}^4}}

\section{RG evolution of the Wilson Coefficients}

Once the Wilson coefficients have been calculated at the perturbative scale $M_W$, the next step is to compute their renormalization-group running,
including operator mixing, down to the factorization scale. This scale, as mentioned in Section \ref{WEH}, is taken to be $m_b$. The QCD evolution of
the Wilson coefficients is given by (\ref{Cevol}),
\eq{\vec{C}(\mu)=U_5(\mu,M_W)\vec{C}(M_W) \label{RG}}
where $U_5(\mu,M_W)$ is the evolution matrix, given at leading order by (\ref{C(mu)}). For our purposes it is sufficient to work at leading order in
$\alpha_s$\footnote{The inclusion of next-to-leading-order corrections in $\alpha_s$ in the anomalous dimension matrix has an impact of less than
10\%.} and neglecting electromagnetic corrections (zeroth order in $\alpha$).

The easiest way to perform the running is to translate our basis (\ref{NPops}) to the usual basis of 12 $\Delta B=1$ SM operators in eq.
(\ref{OpsSM}). The coefficients $C_i$ in (\ref{HeffSM}) are related to the NP $c_i^q$'s through
\eq{\begin{array}{ll}
c_1^u=-\lambda_t\,(C_5 + C_7)\qquad\quad & c_1^d=-\lambda_t\,(C_5 - \frac{1}{2}C_7)\\
c_2^u=-\lambda_t\,(C_6+C_8)\qquad & c_2^d=-\lambda_t\,(C_6 -\frac{1}{2}C_8)\\
c_3^u=-\lambda_t\,(C_3+C_9)\qquad & c_3^d=-\lambda_t\,(C_3 -\frac{1}{2}C_9)\\
c_4^u=-\lambda_t\,(C_4 + C_{10})\qquad & c_4^d=-\lambda_t\,(C_4-\frac{1}{2}C_{10}) \label{cs vs Cs}
\end{array}}
Note that the $c_{5,6}^{u,d}$ are zero at the $M_W$ scale in our case. We take them to be zero also at the $m_b$ scale since the electroweak
combination $c_{5,6}^u(m_b)-c_{5,6}^d(m_b)$ is at LO a function only of $c_{5,6}^{u,d}(M_W)$, and the QCD combination $(c_{5,6}^{u}(m_b)+2
c_{5,6}^d(m_b))$ is mostly dominated by the same combination at $M_W$, taking into account that all NP penguin Wilson coefficients are of similar
size.

The running is then performed over the operators $C_i$ by means of eq. (\ref{C(mu)}) and the anomalous dimension matrix in Table \ref{tableADM}.

\section{Hadronic matrix elements for $B_s^{\scriptscriptstyle 0}\to K^+K^-$ \& $B_s^{\scriptscriptstyle 0}\to K^0\bar{K}^0$}

The final step in the program is to compute the hadronic matrix elements of the operators in Eq.~(\ref{NPops}) for $B_s^{\scriptscriptstyle 0}\to
K^+K^-$ and $B_s^{\scriptscriptstyle 0}\to K^0\bar{K}^0$. These are calculated using the naive factorization approach as explained in Section
\ref{factorization}, plus the contribution from the chromomagnetic operator.

We consider first the contribution from $O_1^u$, $O_2^u$, $O_3^u$ and $O_4^u$ to $B_s^{\scriptscriptstyle 0}\to K^+K^-$. From eq. (\ref{GFformula})
we have
\eqa{ A(B_s^{\scriptscriptstyle 0}\to K^+K^-)_{1,2,3,4}&=&\sum_{p=u,c}\lambda_p^{\sss (s)}
(a_4+r_\chi^{\sss K} a_6+a_{10}+r_\chi^{\sss K} a_8)A_{\sss KK}\nn\\
&&\hspace{-4cm}=-\lambda_t^{\sss (s)}A_{\sss KK}\left[ \left( C_4+\frac{C_3}{3} \right) + r_\chi^{\sss K} \left( C_6+\frac{C_5}{3} \right)
+ \left( C_{10}+\frac{C_9}{3} \right) + r_\chi^{\sss K} \left( C_8+\frac{C_7}{3} \right) \right]\nn\\[7pt]
&&\hspace{-4cm}=A_{\sss KK}\left[ \left( c_4^u+\frac{c_3^u}{3} \right)+ r_\chi^{\sss K} \left( c_2^u+\frac{c_1^u}{3} \right) \right]\ , }
where we have used the unitarity relation (\ref{unit}) and the relationship between the two different sets of Wilson coefficients (\ref{cs vs Cs}).
In the same way, for $B_s^{\scriptscriptstyle 0}\to K^0\bar{K}^0$, one finds
\eq{ A(B_s^{\scriptscriptstyle 0}\to K^0\bar{K}^0)_{1,2,3,4} =A_{\sss KK}\left[ \left( c_4^d+\frac{c_3^d}{3} \right)+ r_\chi^{\sss K} \left(
c_2^d+\frac{c_1^d}{3} \right) \right]\ . }
The contributions from the operators $\tilde{O}_1^q$, $\tilde{O}_2^q$, $\tilde{O}_3^q$ and $\tilde{O}_4^q$ are the same with the substitutions
$c_i^q\to -\tilde{c}_i^q$. Finally, the factorized matrix element $A_{\sss KK}$ (c.f. eq. (\ref{Afactors})) and the chiral factor $r_\chi^{\sss K}$
(c.f. eq. (\ref{rxi})) are given by
\eq{A_{\sss KK}=i\frac{G_F}{\sqrt{2}}m_B^2f_KF_0^{B\to K}\ ,\quad r_\chi^{\sss K}=\frac{2m_K^2}{m_b\,(m_u+m_s)}}

The contribution from the chromomagnetic operator can be obtained from the order $\alpha_s$ QCDF formulae as found, for instance, in
\cite{hep-ph/0308039}. In reference to the QCDF coefficients in eq. (\ref{aQCDF}), the coefficient $C_{8g}^{\rm eff}$ contributes only to the penguin
contractions $P_4^p(K)$ and $P_6^p(K)$. These only enter in $a_4^p(KK)$ and $a_6^p(KK)$, so at the end the contribution to the amplitude is
\eq{A(B_s^{\scriptscriptstyle 0}\to KK)_{8g}=\frac{C_F\alpha_s}{4\pi N_c} \left[ -2C_{8g}^{\rm eff}\int_0^1\frac{dx}{1-x}\Phi_K(x)-2C_{8g}^{\rm eff}
r_\chi^{\sss K} \right]\ .}

The integral over the kaon light-cone distribution amplitude can be expanded in terms of Gegenbauer polynomials as in eq. (\ref{GegenbauerExp}),
which gives ${\rm integral}=3+3\alpha_1^K+3\alpha_2^K$. However, because we are working in naive factorization, we may just take the asymptotic form
for the distribution amplitudes and set $\alpha_i^K\to 0$.

Then, the NP amplitudes for $B_s^{\scriptscriptstyle 0}\to K^+K^-$ and $B_s^{\scriptscriptstyle 0}\to K^0\bar{K}^0$ in naive factorization read
\eqa{
\mathcal{A}^ue^{i\Phi_u}&=&\bra{K^+K^-} \Heff^{\rm\sss NP}\ket{B_s^{\sss 0}}\nn\\[5pt]
&=&A_{\sss KK}\left[ \left( \bar{c}_4^u+\frac{\bar{c}_3^u}{3} \right)+ r_\chi^{\sss K} \left( \bar{c}_2^u+\frac{\bar{c}_1^u}{3} \right)
+\lambda_t \frac{2\alpha_s}{3\pi}\bar{C}_{8g}^{\rm eff}\left( 1+\frac{r_\chi^{\sss K}}{3} \right)\right]\ ,\nn\\[5pt]
\mathcal{A}^de^{i\Phi_d}&=&\bra{K^0\bar{K}^0} \Heff^{\rm\sss NP}\ket{B_s^{\sss 0}}\nn\\[5pt]
&=&A_{\sss KK}\left[ \left( \bar{c}_4^d+\frac{\bar{c}_3^d}{3} \right)+ r_\chi^{\sss K} \left( \bar{c}_2^d+\frac{\bar{c}_1^d}{3} \right)
+\lambda_t \frac{2\alpha_s}{3\pi}\bar{C}_{8g}^{\rm eff}\left( 1+\frac{r_\chi^{\sss K}}{3} \right)\right]\ ,\label{NPamplitudes}\\
\nn }
where the barred coefficients are $\bar{c}_i\equiv c_i-\tilde{c}_i$ and $\bar{C}_{8g}^{\rm eff}\equiv C_{8g}^{\rm eff}-\tilde{C}_{8g}^{\rm eff}$.

\section{NP contributions to $B_s^{\sss 0}-\bar{B}_s^{\sss 0}$ mixing}

As mentioned in Section \ref{MeasuringNP}, the NP will in general affect also the mixing, and not only the decay amplitude. As the NP correction to
this angle -- $\delta\phi_s^{NP}$-- is necessary in order to compute the mixing induced CP asymmetry (\ref{AmixNP}), we must consider the SUSY
contributions to $B_s^{\sss 0}-\bar{B}_s^{\sss 0}$ mixing, $M_{12}^{\rm \sss NP}$. Adding this new contribution, the mixing angle $\phi_s$ is given
by (\ref{Mixingangle})
\eqa{
\phi_s&\!\!=&\!\!\phi_s^{\rm \sss SM}+\delta\phi_s^{\rm \sss NP}=\arg{(M_{12}^{\rm \sss SM}\!+\!M_{12}^{\rm \sss NP})} \nn\\[5pt]
&\!\!=&\!\!\arg{\left[M_{12}^{\rm \sss SM}\left(1+\frac{M_{12}^{\rm \sss NP}}{M_{12}^{\rm \sss SM}}\right)\right]}= \arg{(M_{12}^{\rm \sss
SM})}+\arg{\left(1+\frac{M_{12}^{\rm \sss NP}}{M_{12}^{\rm \sss SM}}\right)}\quad }
so that the NP contribution to the mixing angle is
\eq{\delta\phi_s^{\rm \sss NP}=\arg{\left(1+\frac{M_{12}^{\rm \sss NP}}{M_{12}^{\rm \sss SM}}\right)}}

\begin{figure}
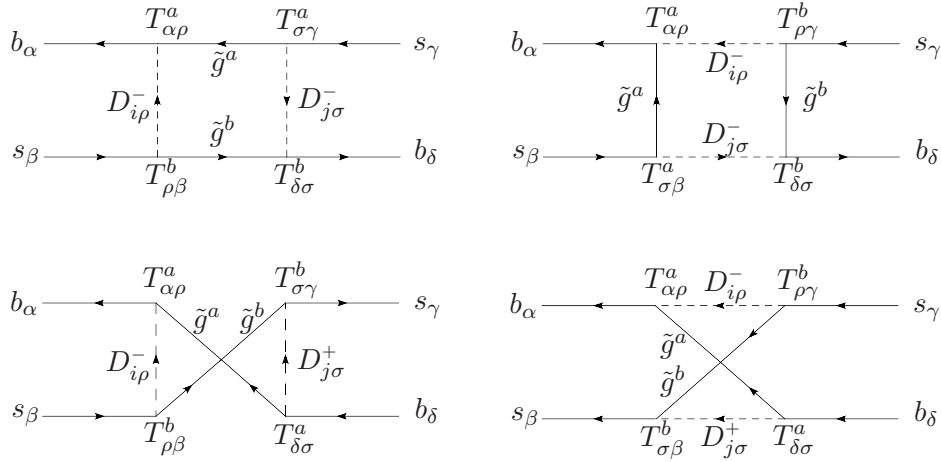

\psfrag{bb}{$b_\alpha$} \psfrag{b}{$b_\delta$} \psfrag{s}{$s_\beta$} \psfrag{sb}{$s_\gamma$} \psfrag{Tar}{$T_{\alpha\rho}^a$}
\psfrag{Tsg}{$T_{\sigma\gamma}^a$} \psfrag{Trb}{$T_{\rho\beta}^b$} \psfrag{Tds}{$T_{\delta\sigma}^b$} \psfrag{Trg}{$T_{\rho\gamma}^b$}
\psfrag{Tsb}{$T_{\sigma\beta}^a$} \psfrag{Tbs}{$T_{\sigma\beta}^b$} \psfrag{Tsd}{$T_{\sigma\delta}^b$} \psfrag{Tsg2}{$T_{\sigma\gamma}^b$}
\psfrag{Tds2}{$T_{\delta\sigma}^a$} \psfrag{Trg2}{$T_{\rho\gamma}^b$} \psfrag{Tsd2}{$T_{\delta\sigma}^a$} \psfrag{Dr}{$D^{\sss -}_{i\rho}$}
\psfrag{ga}{$\tilde{g}^a$}\psfrag{gb}{$\tilde{g}^b$}
\begin{center}
\psfrag{Ds}{$D^{\sss -}_{j\sigma}$}
\includegraphics{BoxMix1.eps}
\hspace{0.7cm}
\includegraphics{BoxMix2.eps}
\end{center}
\vspace{0.1cm}
\begin{center}
\psfrag{Ds}{$D^{\sss +}_{j\sigma}$}
\includegraphics{BoxMix3.eps}
\hspace{0.7cm}
\includegraphics{BoxMix4.eps}
\end{center}
\caption{\small Gluino box contributions to $B_s^{\sss 0}-\bar{B}_s^{\sss 0}$ mixing.} \label{SUSYBs-Bs}
\end{figure}

These contributions, in our case, come from the squark-gluino box diagrams in Figure \ref{SUSYBs-Bs}. These diagrams are calculated like the ones in
Figure \ref{SUSYBoxes} (see for instance \cite{Hagelin:1992tc}) and match to the following $\Delta B=2$ effective Hamiltonian,
\eq{\Heff^{\Delta B=2}=\sum_{i=1}^5 C_iO_i+\sum_{i=1}^3 \widetilde{C}_i\,\widetilde{O}_i\ ,}
where the operators are defined as
\eq{
\begin{array}{ll}
O_1 = (s_L\gamma_\mu b_L)(s_L\gamma^\mu b_L)\ ,& O_4 = (s_R b_L)(s_L b_R)\ ,\\[3pt]
O_2 = (s_R b_L)(s_R b_L)\ ,& O_5 = (s_R^\alpha b_L^\beta)(s_L^\beta b_R^\alpha)\ ,\\[3pt]
O_3 = (s_R^\alpha b_L^\beta)(s_R^\beta b_L^\alpha)\ ,& \widetilde{O}_i=O_i|_{L\leftrightarrow R}\ .\\
\end{array}
}
When only LL or RR mixing is present in our scenario, only $C_1$ or $\widetilde{C}_1$ are generated. When both LL and RR mixings coexist, the
operators $O_4$ and $O_5$ contribute. However, in our scenario the operators $O_2$ and $O_3$ are not generated at all. The Wilson coefficients at the
matching scale $M_W$ are given by

\eqa{ C_1^{\rm\sss SUSY}&\!\!=\!\!&\frac{\alpha_s^2\sin^2{2\theta_L}e^{2i\delta_L}}{4m_{\tilde{g}}^2} \left[ \frac{11}{36}\Big(G(x_{\sss
\tilde{b}_L\tilde{g}},x_{\sss \tilde{b}_L\tilde{g}})+G(x_{\sss \tilde{s}_L\tilde{g}},x_{\sss \tilde{s}_L\tilde{g}})
-2G(x_{\sss \tilde{b}_L\tilde{g}},x_{\sss \tilde{s}_L\tilde{g}})\Big)\right.\nn\\
&&\hspace{3cm}\left.-\frac{1}{9}\Big(F(x_{\sss \tilde{b}_L\tilde{g}},x_{\sss \tilde{b}_L\tilde{g}}) +F(x_{\sss \tilde{s}_L\tilde{g}},x_{\sss
\tilde{s}_L\tilde{g}})
-2F(x_{\sss \tilde{b}_L\tilde{g}},x_{\sss \tilde{s}_L\tilde{g}})\Big)\right]\ ,\nn\\[10pt]
\widetilde{C}_1^{\rm\sss SUSY}&\!\!=\!\!&\frac{\alpha_s^2\sin^2{2\theta_R}e^{2i\delta_R}}{4m_{\tilde{g}}^2} \left[ \frac{11}{36}\Big(G(x_{\sss
\tilde{b}_R\tilde{g}},x_{\sss \tilde{b}_R\tilde{g}})+G(x_{\sss \tilde{s}_R\tilde{g}},x_{\sss \tilde{s}_R\tilde{g}})
-2G(x_{\sss \tilde{b}_R\tilde{g}},x_{\sss \tilde{s}_R\tilde{g}})\Big)\right.\nn\\
&&\hspace{3cm}\left.-\frac{1}{9}\Big(F(x_{\sss \tilde{b}_R\tilde{g}},x_{\sss \tilde{b}_R\tilde{g}}) +F(x_{\sss \tilde{s}_R\tilde{g}},x_{\sss
\tilde{s}_R\tilde{g}})
-2F(x_{\sss \tilde{b}_R\tilde{g}},x_{\sss \tilde{s}_R\tilde{g}})\Big)\right]\ ,\nn\\[10pt]
C_4^{\rm\sss SUSY}&\!\!=\!\!&\frac{\alpha_s^2\sin{2\theta_L}\sin{2\theta_R}e^{i(\delta_L+\delta_R)}}{4m_{\tilde{g}}^2} \left[
-\frac{1}{3}\Big(G(x_{\sss \tilde{b}_R\tilde{g}},x_{\sss \tilde{b}_L\tilde{g}})+G(x_{\sss \tilde{s}_R\tilde{g}},x_{\sss \tilde{s}_L\tilde{g}})
-G(x_{\sss \tilde{b}_R\tilde{g}},x_{\sss \tilde{s}_L\tilde{g}})\right.\nn\\
&&\hspace{-1.5cm}\left.-G(x_{\sss \tilde{s}_R\tilde{g}},x_{\sss \tilde{b}_L\tilde{g}})\Big) -\frac{7}{3}\Big(F(x_{\sss \tilde{b}_R\tilde{g}},x_{\sss
\tilde{b}_L\tilde{g}})+F(x_{\sss \tilde{s}_R\tilde{g}},x_{\sss \tilde{s}_L\tilde{g}})
-F(x_{\sss \tilde{b}_R\tilde{g}},x_{\sss \tilde{s}_L\tilde{g}})-F(x_{\sss \tilde{s}_R\tilde{g}},x_{\sss \tilde{b}_L\tilde{g}})\Big)\right]\ ,\nn\\[10pt]
C_5^{\rm\sss SUSY}&\!\!=\!\!&\frac{\alpha_s^2\sin{2\theta_L}\sin{2\theta_R}e^{i(\delta_L+\delta_R)}}{4m_{\tilde{g}}^2} \left[
\frac{5}{9}\Big(G(x_{\sss \tilde{b}_R\tilde{g}},x_{\sss \tilde{b}_L\tilde{g}})+G(x_{\sss \tilde{s}_R\tilde{g}},x_{\sss \tilde{s}_L\tilde{g}})
-G(x_{\sss \tilde{b}_R\tilde{g}},x_{\sss \tilde{s}_L\tilde{g}})\right.\nn\\
&&\hspace{-1.5cm}\left.-G(x_{\sss \tilde{s}_R\tilde{g}},x_{\sss \tilde{b}_L\tilde{g}})\Big) -\frac{1}{9}\Big(F(x_{\sss \tilde{b}_R\tilde{g}},x_{\sss
\tilde{b}_L\tilde{g}})+F(x_{\sss \tilde{s}_R\tilde{g}},x_{\sss \tilde{s}_L\tilde{g}})
-F(x_{\sss \tilde{b}_R\tilde{g}},x_{\sss \tilde{s}_L\tilde{g}})-F(x_{\sss \tilde{s}_R\tilde{g}},x_{\sss \tilde{b}_L\tilde{g}})\Big)\right]\ ,\nn\\[10pt]
}
where the loop functions $F$ and $G$ are given in (\ref{F,G}). The Wilson coefficients at the scale $m_b$ are obtained applying the RG procedure. The
two loop running gives \cite{hep-ph/0112303}
\eq{
\begin{array}{ll}
C_1(m_b) = 0.848\,C_1(M_W)\ ,& C_4(m_b) = 2.395\,C_4(M_W)+0.485\,C_5(M_W)\ ,\\[3pt]
\widetilde{C}_1(m_b) = 0.848\,\widetilde{C}_1(M_W)\ ,& C_5(m_b) = 0.061\,C_4(M_W)+0.904\,C_5(M_W)\ .\\
\end{array}
}
The calculation of the matrix elements in terms of the B-parameters gives
\eqa{ M_{12}^s &=& M_{B_s} f_{B_s}^2 \Bigg[{1 \over 3} B_1(m_b) (C_1(m_b) + \widetilde{C}_1(m_b))
+ {1 \over 4} B_4(m_b) \left(M_{B_s} \over \mu_b + m_s\right)^2 C_4(m_b) \nn\\
&&\hspace{2cm} +{1 \over 12} B_5(m_b) \left(M_{B_s} \over m_b + m_s\right)^2 C_5(m_b) \Bigg] , }
where the B-parameters can be found in \cite{hep-lat/0110091} and their central values are
\eq{B_1(m_b)=0.86\,,\ B_4(m_b)=1.17\,,\ B_5(m_b)=1.94\,.}
In the SM, only $C_1^{\rm\sss SM}$ is non-zero, and it is given by
\eq{C_1^{\rm\sss SM}(M_W)=\frac{G_F^2 M_W^2}{4\pi^2}\lambda_t^{(s)\,2}S_0(x_t)\ ,}
where $\,x_t\equiv m_t^2/M_W^2\,$ and $S_0$ is \cite{hep-ph/9512380}
\eq{S_0(x)=\frac{x^4-12x^3+15x^2-4x+6x^3\ln{x}}{4(x-1)^3}\ .}
Finally we can write
\eqa{ \delta\phi_s^{\rm \sss NP}=\arg \Bigg(1 &+& \frac{C_1^{\rm\sss SUSY}(m_b)+\widetilde{C}_1^{\rm\sss SUSY}(m_b)}{C_1^{\rm\sss SM}(m_b)}
+\frac{3}{4}\frac{B_4(m_b)}{B_1(m_b)}\left(\frac{M_{B_s}}{m_b+m_s}\right)^2\frac{C_4^{\rm\sss SUSY}(m_b)}{C_1^{\rm\sss SM}(m_b)}\nn\\
&+&\frac{1}{4}\frac{B_5(m_b)}{B_1(m_b)}\left(\frac{M_{B_s}}{m_b+m_s}\right)^2\frac{C_5^{\rm\sss SUSY}(m_b)}{C_1^{\rm\sss SM}(m_b)} \Bigg)
\label{arg(M12)} }
We have now all the ingredients to calculate the NP amplitude and the $B_s^{\sss 0}\to KK$ observables in terms of the set of SUSY parameters.

\section{Additional Bounds on the SUSY parameter space}

Let us address now the issue of the input parameter space. To that end we begin summarizing the basic points discussed so far.

\begin{itemize}

\item The final aim is to see how the space of allowed values for the $B_s^0\to K^+K^-$ and $B_s^0\to K^0\bar{K}^0$ observables
within a general SUSY model is increased with respect to that of the SM alone.

\item The expressions for full observables in the presence of NP are given in (\ref{BRNP})-(\ref{MNP}) and (\ref{BR(dtheta)})-(\ref{Amix(dtheta)}).
These depend on the following set of parameters: $\m{T_{\sss KK}}$, $d_{\sss KK}$, $\theta_{\sss KK}$, $\gamma$, $\lambda_u^{(s)}$, $\phi_s^{\sss
SM}$, $\delta\phi_s^{\sss NP}$, $\mathcal{A}^u$ and $\Phi_u$.

\item The CKM parameters $\gamma$, $\lambda_u^{(s)}$ and $\phi_s^{\sss SM}=2\beta_s$ are measured independently.

\item The SM hadronic parameters $\m{T_{\sss KK}}$, $d_{\sss KK}$ and $\theta_{\sss KK}$ are obtained from
$B_d^0\to K^0\bar{K}^0$ as described in Chapter \ref{SymFac1}, paying especial attention to the correlations, since these parameters must be varied
together with the scanning of the NP parameter space.

\item The NP parameters $\mathcal{A}^u$, $\Phi_u$, $\mathcal{A}^d$ and $\Phi_d$ are the magnitudes and weak phases of the NP amplitudes.
This amplitudes have been calculated in this chapter within supersymmetry, as a function of the following SUSY parameters:

\begin{itemize}

\item The masses of the squarks and the gluino. We take $\ m_{\tilde{u}_L} =
m_{\tilde{g}} = m_{\tilde{d}_{L,R}} = m_{\tilde{b}_{L,R}} = 250 \, {\rm GeV}$ and $\ 250~{\rm GeV} \le m_{\tilde{u}_{R}},m_{\tilde{s}_{L,R}} \le
1000~{\rm GeV}$.

\item The mixing and weak angles. These are taken unconstrained: $-\pi/4 \le \theta_{L,R} \le \pi/4\,$ (see Section \ref{WilsonCoefficients})
and $-\pi \le \delta_{L,R}\le \pi$.

\end{itemize}

The Wilson coefficients are sensitive to the $\tilde{s}-\tilde{b}$ mass splitting. They vanish for $m_{\tilde{s}}=m_{\tilde{b}}$ and grow when the
splitting is large. We therefore expect these contributions to be most important for large values of $m_{\tilde{s}}$ (keeping $m_{\tilde{b}}$ fixed).
In the same way, NP effects in ${\bar b} \to {\bar d} q {\bar q}$ transitions depend on the difference $m_{\tilde{d}}-m_{\tilde{b}}$. By setting
$m_{\tilde{d}}=m_{\tilde{b}}$ we ensure that $\bar{b}\to\bar{d}$ decays get no such contributions, which is consistent with the discussion in
Sec.~\ref{WilsonCoefficients}. A difference between $\mathcal{A}^u e^{\Phi_u}$ and $\mathcal{A}^d e^{\Phi_d}$ is only possible in the presence of a
nonzero $\tilde{u}-\tilde{d}$ mass splitting. Without it there are no contributions to isospin-violating operators. However, this mass splitting must
be very small in the left-handed sector due to $SU(2)_L$ invariance.  We therefore set $m_{\tilde{u}_L}=m_{\tilde{d}_L}$, but allow for a significant
mass splitting in the right-handed sector.

\end{itemize}

A quick look at the form of the Wilson coefficients in (\ref{WC}) shows that the NP amplitude (\ref{NPamplitudes}) has the following structure,
\eq{\mathcal{A}^u e^{i\Phi_u}=(\mathcal{F}\sin{2\theta_L} e^{i\delta_L}-\widetilde{\mathcal{F}}\sin{2\theta_R} e^{i\delta_R})A}
where $\mathcal{F}$ and $\widetilde{\mathcal{F}}$ are functions of the masses of the squarks and gluino. They are related by a change
$L\leftrightarrow R$, and for $m_{\tilde{q}_L}=m_{\tilde{q}_R}$ they are equal ($\mathcal{F}=\widetilde{\mathcal{F}}$). Clearly, a way of maximizing
the NP amplitude is to set $\theta_R=\theta_L=\pi/4$ (maximal $\tilde{s}-\tilde{b}$ mixing). In such a situation we have,
\eq{\mathcal{A}^u=2\mathcal{F}\,\m{A}\m{\sin{(\delta_L-\Phi_u)}}}
which shows that if $\delta_{L,R}$ are unconstrained, then $\mathcal{A}^u$ and $\Phi_u$ are not correlated and can be maximized independently.

However, the range in the SUSY parameter space has to be constrained to fit to several existing experimental bounds. In the following we show what
these bounds are and how they constrain the various ranges taken for the input parameters.

\subsection{Bounds from $B\to \pi K$}

The same SUSY contributions to $B_s^{\scriptscriptstyle 0}\to KK$ will also affect $B\to \pi K$, since they both share the same quark-level decay of
the $b$ quark. In particular, there will be effects on the quantities $R_*$ and $A_{\rm CP}(\pi^+ \bar{K^0})$ \cite{hep-ph/0308039}, whose
definitions and measured values are \cite{HFAG}
\eqa{ R_* &\!\!\equiv& \!\!\frac{BR(B^+\rightarrow \pi^+K^0)+BR(B^-\rightarrow \pi^-\bar{K}^0)}
{2[BR(B^+\rightarrow \pi^0K^+)+BR(B^-\rightarrow \pi^0K^-)]}=1.00\pm 0.08 \label{R*}\\
\mathcal{A}_{\sss CP}(\pi^+K^0)&\!\!\equiv& \!\!\frac{BR(B^+\rightarrow \pi^+K^0)-BR(B^-\rightarrow \pi^-\bar{K}^0)} {BR(B^+\rightarrow
\pi^+K^0)+BR(B^-\rightarrow \pi^-\bar{K}^0)}=-0.020\pm 0.034\qquad }
In order to incorporate these two constraints we follow the approach in \cite{hep-ph/9909297} where QCDF is used, except for the strong phase related
to $A_{\rm CP}$ which we take as a free parameter.

\subsection{Bounds from $B\to X_s\gamma$}

The bounds from $B\to X_s\gamma$ are usually important constraints on any FCNC NP analysis, and this is no exception here. For the branching ratio of
$B\to X_s\gamma$ we take \cite{HFAG}
\eq{BR(B\to X_s\gamma)=(3.55\pm 0.26)\times 10^{-4}\label{ExpBRXsGamma}}
and we consider a 2-$\sigma$ range, allowing for various theoretical uncertainties. As we shall see, even a wide range of this sort has a
considerable impact in our analysis. This bounds are incorporated following \cite{hep-ph/9805303}.

\subsection{Bounds from $B_s^{\sss 0}-\bar{B}_s^{\sss 0}$ mixing}

Data from $B_s^{\sss 0}-\bar{B}_s^{\sss 0}$ mixing also provides a potential constraint in the SUSY parameter space. In particular, the recently
measured mass difference $\Delta M_s$ in eq.~(\ref{DeltaMs_exp}) \cite{Abulencia:2006ze}, together with the SM fit \cite{UTfit,CKMfitter} gives
\eq{\left(\frac{\Delta M_s}{\Delta M_s^{\rm \sss SM}}\right)_{\rm exp}=(0.81\pm 0.19) \label{Ms/MsSMexp}\ ,}
which constrains the SUSY parameter space through the modulus in (\ref{arg(M12)}), since from (\ref{Massdifference}) we have
\eq{\frac{\Delta M_s}{\Delta M_s^{\rm \sss SM}}=\frac{2\m{M_{12}^{\rm \sss SM}+M_{12}^{\rm \sss NP}}}{2\m{M_{12}^{\rm \sss SM}}}
=\mo{1+\frac{M_{12}^{\rm \sss NP}}{M_{12}^{\rm \sss SM}}}\ .\label{Ms/MsSM}}
This bound also turns out to be important.\\

There are other traditionally important constraints, like $B_s^{\sss 0}\to\mu^+\mu^-$, that are very sensitive to other SUSY parameters, mostly
$\tan\beta$ and $m_A$. However, for low $\tan\beta$ and values of $m_A$ above $200\,{\rm GeV}$, they have no effect on our region in the SUSY
parameter space.

Taking into account the various constraints, the contributions from LL and RR mixing have been analyzed. $\Delta M_s$ is the strongest constraint,
and it is the relevant one when considering only LL or RR mixing separately. In particular, it has a large impact on the phases $\Phi_u$ and
$\Phi_d$. In the case of $B_s^{\sss 0}\to K^+K^-$, LL mixing gives the largest contribution to the amplitude, more than twice that of RR. However, in
the case of $B_s^{\sss 0}\to K^0\bar{K}^0$ both contributions are similar in size.

When both LL and RR mixings are allowed simultaneously, the constraints on the SUSY parameter space are changed. In this case new operators for
$B_s^{\sss 0}-\bar{B}_s^{\sss 0}$ mixing are generated, so that the effect is not simply the combination of the two separate contributions (for
instance, see Ref.~\cite{hep-ph/0605182}). We find that (i) now $BR(B\to X_s\gamma)$ is also important, not only $\Delta M_s$, and (ii) the global
effect of the constraints is weaker.  The upshot is that there is a certain enhancement of the NP amplitudes when both LL and RR mixings are
combined. In particular, the weak phases $\Phi_u$ and $\Phi_d$ are not so strongly constrained as when either LL or RR mixing is taken to vanish.

\section{Results: NP amplitudes and Observables}

\begin{figure}
\begin{center}
\includegraphics[height=7cm]{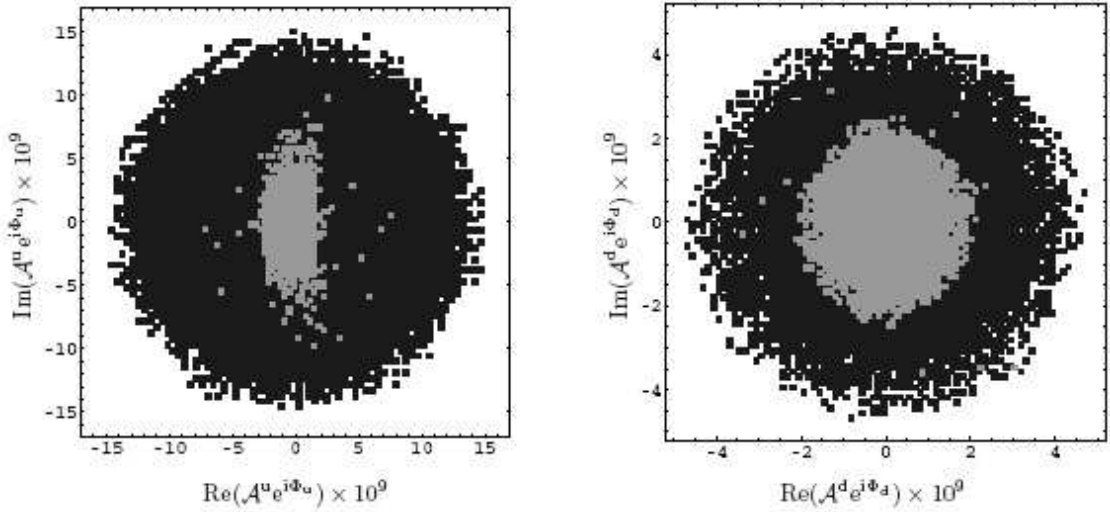}
\end{center}
\caption{\small SUSY contribution to the NP amplitudes $\mathcal{A}^u e^{i \Phi_u}$ (left) and $\mathcal{A}^d e^{i \Phi_d}$ (right) in the scenario
with simultaneous LL and RR mixings.  The dark regions correspond to the variation of the SUSY parameters over the considered parameter space. The
light regions satisfy the experimental bounds, including the recent measurement of $\Delta M_s$.} \label{plotANP}
\end{figure}

In Fig.~\ref{plotANP} we show the allowed ranges for $\mathcal{A}^u e^{i \Phi_u}$ and $\mathcal{A}^d e^{i \Phi_d}$ in the scenario with simultaneous
LL and RR mixings. The dark regions correspond to the values that these amplitudes take when varying the parameter space over the initial ranges. The
light regions show how these values are reduced by the existing experimental constraints mentioned above. There are two important remarks. First, we
see that the above constraints do indeed greatly reduce the allowed SUSY parameter space. Second, even so, the effect on $\mathcal{A}^u e^{i \Phi_u}$
and $\mathcal{A}^d e^{i \Phi_d}$ can be significant.

At this stage we can identify what are the effects of the various constraints in reducing the SUSY parameter space. The bound from $B \to X_s \gamma$
affects only the left-handed sector. In particular, for large $m_{\tilde{s}}$, the regions with $|\theta_L| \gtrsim 10^\circ$, $|\delta_L| \lesssim
60^\circ$ and $|\theta_L| \gtrsim 10^\circ$, $|\delta_L-\pi| \lesssim 60^\circ$ are excluded.  The bound from $\Delta M_s$ is much stronger: when
$m_{\tilde{s}}\gtrsim 400~{\rm GeV}$, any values of $|\theta_{L,R}|\gtrsim 5^\circ$ are excluded, as well as those regions in which
$\delta_L+\delta_R\approx -3\pi/2,-\pi/2,\pi/2,3\pi/2$.\footnote{For further details of the $\Delta M_s$ constraint on this parameter space, see
Ref.~\cite{hep-ph/0605182}.} After these bounds are imposed on the parameter space, the constraints from $B\to \pi K$ have very little effect on the
regions in Fig.~\ref{plotANP}.

Note that the allowed region for $\mathcal{A}^u$ is much larger than that for $\mathcal{A}^d$, by approximately a factor of 3. In the isospin limit,
these should be equal, so this factor of 3 is a measure of isospin breaking in this NP scenario. In particular, for $m_{\tilde{u}_R}={\rm 250~GeV}$
(zero $\tilde{u}_R$-$\tilde{d}_R$ mass splitting), the values of $\mathcal{A}^u$ reduce to those for $\mathcal{A}^d$.

\begin{figure}
\begin{center}
\includegraphics[width=14.5cm,height=6.8cm]{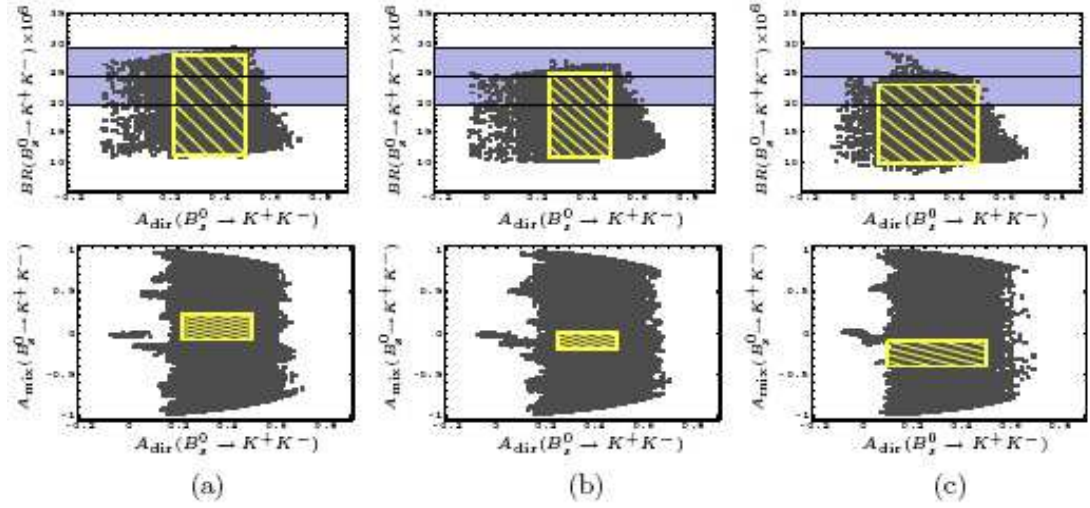}
\end{center}
\vspace{-0.5cm} \caption{\small Predictions, in the form of scatter plots, for the correlations between $BR(B_s^{\scriptscriptstyle 0}\to
K^+K^-)-A_{\rm dir}(B_s^{\scriptscriptstyle 0}\to K^+K^-)$ (up) and $A_{\rm mix}(B_s^{\scriptscriptstyle 0}\to K^+K^-)-A_{\rm
dir}(B_s^{\scriptscriptstyle 0}\to K^+K^-)$ (down) in the presence of SUSY, for (a) $A_{dir}^{d0}=-0.1$, (b) $A_{dir}^{d0}=0$ and (c)
$A_{dir}^{d0}=0.1$. The dashed rectangles correspond to the SM predictions. The horizontal band shows the experimental value for
$BR(B_s^{\scriptscriptstyle 0}\to K^+K^-)$ at  $1\sigma$.} \label{plotBK+K-}
\end{figure}

We now examine the effect of these contributions on the observables. By adding the SUSY contributions to the SM amplitudes, it is possible to compute
the branching ratio and the CP asymmetries in the presence of SUSY. Fig.~\ref{plotBK+K-} shows the allowed values for the $B_s^{\scriptscriptstyle
0}\to K^+K^-$ observables, for three different values of $A_{dir}^{d0}$, compared with the predictions of the SM and with the recent experimental
value for the $B_s^{\scriptscriptstyle 0}\to K^+K^-$ branching ratio reported by the CDF collaboration \cite{hep-ph/0607021}:
\eq{ BR(B_s^{\scriptscriptstyle 0}\to K^+K^-)_{\rm exp}=(24.4\pm 1.4 \pm 4.6) \times 10^{-6} ~. }
The agreement between the CDF measurement and the prediction of the SM in Ref.~\cite{hep-ph/0603239} erases any discrepancy between experiment and
the SM. This branching ratio will now be an important future constraint. The branching ratio within SUSY should not deviate much from the SM
prediction so as not to generate any disagreement with data. Indeed, Fig.~\ref{plotBK+K-} shows that the impact of SUSY on the branching ratio of
$B_s^{\scriptscriptstyle 0}\to K^+K^-$ is practically negligible. Interestingly, for positive values of $A_{dir}^{d0}$ (preferred region), the SM
predicts a smaller value for $BR(B_s^{\scriptscriptstyle 0}\to K^+K^-)$, but it is now compatible with the new data. Still, it is in this case that
SUSY shows a larger deviation in the correct direction.

A completely different picture arises for the CP asymmetries. The results for the direct CP asymmetry reveal that SUSY can have an impact. This is
not surprising: SUSY introduces a term in the total amplitude which is of the same order of magnitude as that of the SM and carries a weak phase that
is not constrained.  The mixing-induced CP asymmetry gets affected in a more dramatic way.  The interpretation is that the SUSY contribution to the
mixing angle $\phi_s$ can be large (in fact it can take all values between $-\pi$ and $\pi$), while in the SM it is tiny: $\phi_s^{SM}\simeq
-2^\circ$. Any experimental measurement falling inside the dark area in the plots, but outside the dashed rectangle, would not only signal NP but
clearly could be accommodated by supersymmetry.

\begin{figure}
\begin{center}
\includegraphics[width=14.5cm,height=6.8cm]{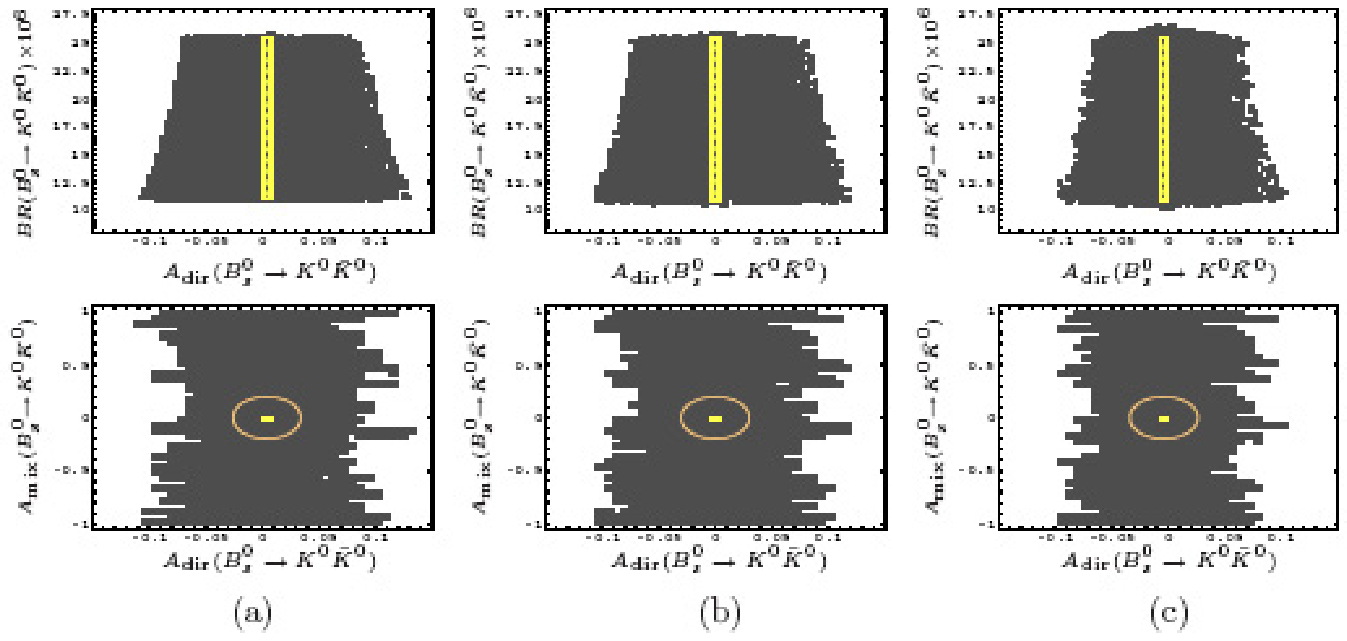}
\end{center}
\vspace{-0.5cm} \caption{\small Predictions, in the form of scatter plots, for the correlations between $BR(B_s^{\sss 0}\to K^0\bar{K}^0)-A_{\rm
dir}(B_s^{\sss 0}\to K^0\bar{K}^0)$ (up) and $A_{\rm mix}(B_s^{\sss 0}\to K^0\bar{K}^0)-A_{\rm dir}(B_s^{\sss 0}\to K^0\bar{K}^0)$ (down) in the
presence of SUSY, for (a) $A_{dir}^{d0}=-0.1$, (b) $A_{dir}^{d0}=0$ and (c) $A_{dir}^{d0}=0.1$. The dashed rectangles correspond to the SM
predictions. These are quite small in the three lower plots, so they are indicated by a circle.} \label{plotBK0K0}
\end{figure}

Fig.~\ref{plotBK0K0} shows the results for $B_s^{\scriptscriptstyle 0}\to K^0\bar{K}^0$. Although the branching ratio is little changed in the
presence of SUSY, the enhancement of the CP asymmetries due to the inclusion of the SUSY contributions is in this case even more important. The
reason is that, within the SM, the CP asymmetries are much smaller in $B_s^{\scriptscriptstyle 0}\to K^0\bar{K}^0$ than they are for
$B_s^{\scriptscriptstyle 0}\to K^+K^-$, because of the absence of the tree diagram. Thus the impact of SUSY is much greater. This is evident by
looking at the three lower plots in Fig.~\ref{plotBK0K0}, where the tiny rectangles corresponding to the SM predictions can hardly be observed. We
have drawn a circle around them to indicate their position.

These are a good illustration of the general scenario discussed in the introduction.  While the branching ratios in this case are relatively
insensitive to supersymmetry, the direct and mixing-induced CP asymmetries of these decays are greatly affected. Thus, these CP asymmetries are the
observables to focus on in order to observe NP, particularly SUSY, while the branching ratio of $B_s^{\scriptscriptstyle 0}\to K^+K^-$ can become an
important constraint on models beyond the SM other than SUSY.

\section{Summary of the results}

In this chapter we have considered the branching ratios and CP asymmetries for the decays $B_s^{\sss 0}\to K^+K^-$ and $B_s^{\sss 0}\to K^0\bar{K}^0$
in a supersymmetric (SUSY) model, focusing on the dominant gluino-squark contributions \cite{hep-ph/9909297}. The determination of the SM
contributions has been taken from the combination of $B_d^{\scriptscriptstyle 0}\to K^0\bar{K}^0$ and QCD factorization described in
Chapter~\ref{SymFac1} \cite{hep-ph/0603239}. In ref.~\cite{hep-ph/0511295} the SM contributions were taken from $B_d \to \pi^+\pi^-$, as was done in
Chapter~\ref{Bdecays} \cite{hep-ph/0410011}, but the uncertainties are much bigger. Here we have described the analysis of
ref.~\cite{hep-ph/0610109}. We have also included the constraints coming from $BR(B \to X_s \gamma)$, $B \to \pi K$ and $\Delta M_s$, and we find the
following results.
\begin{itemize}

\item The new-physics (NP) amplitudes are $\mathcal{A}^u e^{i \Phi_u}$ ($B_s^{\scriptscriptstyle 0}\to K^+K^-$) and
$\mathcal{A}^d e^{i\Phi_d}$ ($B_s^{\scriptscriptstyle 0}\to K^0\bar{K}^0$). We find that both can get significant contributions from SUSY. In the
isospin limit, these quantities are equal. However, our calculations show that, for the region of parameters considered, in SUSY there can be a
difference of up to a factor of 3 between the NP amplitudes. This indicates the possible level of isospin breaking in this type of theory. In
particular, in the SUSY model considered here, large isospin violation is possible when there is large mass splitting in
$\widetilde{u}_R$-$\widetilde{d}_R$.

\item The branching ratio $BR(B_s^{\sss 0}\to K^+K^-)$ is very little affected by SUSY. At most, the SM prediction can be increased by 15\% for
$A_{dir}^{d0}=0.1$. In fact, SUSY can somewhat improve the already good agreement between the SM prediction and the new precise CDF measurement
\cite{hep-ph/0607021}. The impact of SUSY on $BR(B_s^{\sss 0}\to K^0\bar{K}^0)$ is even smaller, reflecting the reduced allowed region for
$\mathcal{A}^d e^{i\Phi_d}$ as compared to $\mathcal{A}^u e^{i \Phi_u}$.

\item The situation is very different for the CP asymmetries; the size of the effect depends strongly on the decay and the type of asymmetry.
For $B_s^{\sss 0}\to K^+K^-$, the direct CP asymmetry within SUSY is in the range $-0.1\lesssim A_{dir}(B_s^{\sss 0}\to K^+K^-)^{SUSY} \lesssim 0.7$
for $-0.1 \leq A_{dir}^{d0}\leq 0.1$. Depending on the value of $A_{dir}^{d0}$, it may be possible to disentangle the SUSY contribution from that of
the SM. This is due to the competition between the tree and the NP amplitudes for each value of $A_{dir}^{d0}$. As for $A_{mix}(B_s^{\sss 0}\to
K^+K^-)$, its value can vary all the way from $-1$ to $+1$, signaling a large impact from SUSY.

\item Turning to $B_s^{\scriptscriptstyle 0}\to K^0\bar{K}^0$, the CP asymmetries are particularly
promising. This decay is dominated by the penguin amplitude in the SM, and so the direct CP asymmetry is strongly suppressed: it is predicted to be
at most of the order of 1\%. However, in the presence of SUSY, the direct CP asymmetry can be 10 times larger. The mixing-induced CP asymmetry is
also predicted to be very small in the SM. However, $A_{mix}(B_s^{\scriptscriptstyle 0}\to K^0\bar{K}^0)^{SUSY}$ covers the entire range, and so this
asymmetry can be large in the presence of SUSY.

\end{itemize}

\part{Conclusions}

\chapter{Concluding Remarks}

\label{Conclus}

Non-leptonic weak decays of B mesons are suitable for the combined study of different aspects in the phenomenology of particle physics. We have seen
their relevance in studies of CP violation, strong interactions, heavy quark physics, flavor physics and physics beyond the SM. All these topics are
interconnected and related to the search for new physics; a search that will most probably succeed in the next few years, but with consequences that
will surely require theoretical and experimental work for a much longer time.

This is a very interesting moment in the history of particle physics, in which we expect to finally break the TeV wall and go beyond a theory of
elementary particles formulated forty years ago. The outcome can be a completely new understanding of the world we live in, leaving aside the period
of speculation and uncertainty of the last decades. A joint effort of the whole particle physics community has brought us here, and the same effort
will be necessary to make this breakthrough possible. It is an impressive world-wide project that will have a historic result.

The direct search for new particles is a crucial part of this program. The main objective of the LHC, with its experiments ATLAS and CMS, is to find
the elusive higgs particle, the last requirement of the SM. However, a more important task will be the search for other completely new particles that
might help to understand all the puzzles that the SM leaves unanswered. In fact, from the theoretical point of view, the observation of new particles
might be much more valuable than the establishment of the higgs' existence.

Despite the importance of the direct searches, the correct identification of the new physics and its properties, and the understanding of its
theoretical consequences, requires also from indirect analyses. Some of the major open questions in particle physics that might find a partial answer
in the next future experiments are related to flavor physics. B physics has been the center of attention in the experiments of the last decade, with
a huge amount of data collected at the B factories and at Tevatron, and will keep a place of honor in the future, with the B physics experiment LHCb
at the LHC and a future Super B. The reason for this is that B physics is a powerful probe of physics beyond the SM in a completely complementary way
to direct searches. In fact, all the B physics experiments carried out so far have put stringent bounds on the nature of the new physics, so thus far
the indirect searches are on the head of the list. And whether or not the direct production of new particles takes place, B physics will still be an
important way to test the SM and look for new physics.

However, in order to look for tiny new physics effects, the SM predictions must be under strict control. In this thesis we have focused on
non-leptonic B decays. These type of decays are very difficult to study because of their sensitivity to long distance strong interaction effects. So
in order to look for new physics in these decays one must find a way to deal with hadronic phenomena leading to predictions of acceptable accuracy.
Indeed, the claim of a NP signal in non-leptonic B decays will have to rely on a solid inclusion of long distance QCD contributions. This also means
that the study of these decays might have a double prize: understanding the NP, and understanding QCD. We have already obtained benefits from both
sides.

Towards this end, we have presented analyses of two body non-leptonic B decays within two approaches. The first one is the flavor symmetry approach,
which takes advantage of the approximate $SU(3)$ symmetry of the QCD lagrangian to establish relations between hadronic quantities for different
decay modes, in such a way that predictions can be made without the need of direct computations of these hadronic quantities. One can then use data
on some sort to predict the values of other observables. In Chapter \ref{Bdecays} we have used this approach to extract the SM hadronic parameters in
$B_s\to K^+K^-$ decays from $B_d\to \pi^+\pi^-$ data. This has provided predictions for $B_s\to K^+K^-$ observables within the SM that can be
compared with experiment. A deviation between the theoretical predictions and the experimental data will be a signal of NP. A useful general
parametrization of the NP contributions to this decay has been given, together with a method to \emph{measure} the new physics parameters. Therefore,
if NP was found in this decay, the measurement of these NP parameters would provide interesting information on the nature of this NP, as for example
if the NP is isospin violating, and so on. The limitations of the $SU(3)$ analysis have been discussed in detail, in particular the problem of
estimating the symmetry breaking effects, that constitute the main source of uncertainties.

A second approach that has been discussed here is the QCD factorization method. This provides a tool to compute matrix elements from first
principles, based on a heavy quark expansion. Therefore this approach provides genuine QCD predictions. However, some of the phenomenological
difficulties of this approach are those non-factorizable power suppressed contributions that are numerically important for realistic $b$ quark
masses. In Chapter \ref{SymFac1} we have proposed a method to get around this difficulty. A combination of QCDF with flavor symmetry relations has
been used to study $B_s\to KK$ decays, giving predictions for the observables with uncertainties reduced by more than a $50\%$. This method has also
been used in Chapter \ref{SymFac2} for $B_s\to VV$ decays, and used to propose three strategies to measure the $B_s-\bar{B_s}$ mixing angle. This
mixing angle is predicted to be extremely small within the SM, and it is therefore an excellent place to look for NP. However, the claim of a NP
contribution to this mixing angle cannot be easily disentangled from a NP contribution to the decay amplitudes in these type of decays, and we have
discussed the ways in which these strategies could help to make this distinction.

This proposal is clearly an improvement on the flavor symmetry and the QCDF approaches, giving more reliable results with smaller uncertainties. In
fact, the results presented for $B_s\to K^+K^-$ were \emph{pre}dictions that were confirmed very nicely --maybe unfortunately-- by the CDF
collaboration shortly after. We would like to make some further comments concerning this method. The applicability of the approach has to be checked
individually for each mode: it can only be applied to those decays for which $\Delta$ receives no contributions from annihilation or hard spectator
scattering graphs, such as $B\to K^{(*)}K^{(*)}$, $B_d\to\phi K^{(*)}$, $B^+\to \pi^+\phi$, etc, as discussed in Section \ref{SectionDelta}. The
predictions derived in this way include most of the long distance physics, which is contained inside the experimental input. The used theoretical
input is minimal, and is the most reliable input that QCDF can offer, free from the troublesome IR-divergencies. Moreover, the theoretical error is
under control and is likely to be reduced in the near future due to, for example, the fast progress that is taking place in lattice simulations.

There is, however, an honest criticism due to the fact that some long distance effects that are controversial in QCDF, are also absent here. The
prominent one is the contribution from the charming penguins, which could easily account for a significant discrepancy between theory and experiment
that would not be due to NP.

Finally, in Chapter \ref{SUSYContributions} we have studied $B_s\to KK$ decays in supersymmetry. Using the SM analyses performed in the previous
chapters, and computing the most relevant contributions from SUSY to the decay amplitudes and to the mixing, we have studied the departure from SM
expectations that SUSY could provide in the branching ratios and CP asymmetries. The conclusion is that the enhancements are important, which means
that moderately precise experimental measurements on these observables, if consistent with the SM predictions, will provide useful constraints on the
SUSY parameter space, especially on the CP violating phases.

To be able foresee what is the most relevant direction that future work on these topics should follow requires a considerable scientific maturity,
which I lack. However, there are some general aspects concerning this issue that are of a general opinion.

On the QCD side, an all-order proof of factorization within QCDF would be appealing, but maybe not too useful for practical purposes. A further
development of the SCET formulation will probably put factorization theorems on a much more firm basis, specially if the issue of the long distance
charm loops finds a phenomenologically satisfactory solution. The final aim is to have a completely systematic way to compute QCD corrections to
matrix elements up to any desired accuracy (even if it means that the actual computation is extremely difficult and lengthy). Moreover, a similar
progress should be expected from the lattice community, which is certainly conceivable.

On the NP side, the progress should be faster, since immediate future events will have a drastic influence on any work in this direction. The SM fits
are almost completely systematized, and any tension between theory and experiment can be dragged to the origin with ease. Probably a similar
systematization should be built for the largest possible of phenomenologically relevant NP models that exist in the literature, so that the explosion
of new phenomena that might occur in the first five years at the LHC can be used to accept and discard models in the blink of an eye.

However, the true model of particle physics at the TeV scale might have nothing to do with any of the NP models that have been proposed up to now.
Therefore, model-independent studies of NP are necessary in order to be sure that we are not missing anything new, or old.


\begin{thebibliography}{99}

\addcontentsline{toc}{chapter}{Bibliography}



\bibitem{Neubert:1993mb}
  M.~Neubert,
  Phys.\ Rept.\  {\bf 245}, 259 (1994)
  [arXiv:hep-ph/9306320].

\bibitem{Bauer:2000ew}
  C.~W.~Bauer, S.~Fleming and M.~E.~Luke,
  Phys.\ Rev.\  D {\bf 63}, 014006 (2001)
  [arXiv:hep-ph/0005275].

\bibitem{Bauer:2000yr}
  C.~W.~Bauer, S.~Fleming, D.~Pirjol and I.~W.~Stewart,
  Phys.\ Rev.\  D {\bf 63}, 114020 (2001)
  [arXiv:hep-ph/0011336].

\bibitem{Bauer:2001yt}
  C.~W.~Bauer, D.~Pirjol and I.~W.~Stewart,
  Phys.\ Rev.\  D {\bf 65}, 054022 (2002)
  [arXiv:hep-ph/0109045].

\bibitem{Beneke:2002ph}
  M.~Beneke, A.~P.~Chapovsky, M.~Diehl and T.~Feldmann,
  Nucl.\ Phys.\  B {\bf 643}, 431 (2002)
  [arXiv:hep-ph/0206152].

\bibitem{Chay:2002vy}
  J.~Chay and C.~Kim,
  Phys.\ Rev.\  D {\bf 65}, 114016 (2002)
  [arXiv:hep-ph/0201197].

\bibitem{Hill:2002vw}
  R.~J.~Hill and M.~Neubert,
  Nucl.\ Phys.\  B {\bf 657}, 229 (2003)
  [arXiv:hep-ph/0211018].

\bibitem{hep-ph/9512380}
  G.~Buchalla, A.~J.~Buras and M.~E.~Lautenbacher,
  Rev.\ Mod.\ Phys.\  {\bf 68}, 1125 (1996)
  [arXiv:hep-ph/9512380].

\bibitem{Buras:1998raa}
  A.~J.~Buras,
  arXiv:hep-ph/9806471.

\bibitem{Buchalla:2005us}
  G.~Buchalla, G.~Hiller, Y.~Nir and G.~Raz,
  JHEP {\bf 0509}, 074 (2005)
  [arXiv:hep-ph/0503151].

\bibitem{Buchmuller:1985jz}
  W.~Buchmuller and D.~Wyler,
  Nucl.\ Phys.\  B {\bf 268}, 621 (1986).

\bibitem{Buras:2003jf}
  A.~J.~Buras,
  Acta Phys.\ Polon.\  B {\bf 34}, 5615 (2003)
  [arXiv:hep-ph/0310208].

\bibitem{Hall:1990ac}
  L.~J.~Hall and L.~Randall,
  Phys.\ Rev.\ Lett.\  {\bf 65}, 2939 (1990).

\bibitem{Neubert:1997uc}
  M.~Neubert and B.~Stech,
  Adv.\ Ser.\ Direct.\ High Energy Phys.\  {\bf 15}, 294 (1998)
  [arXiv:hep-ph/9705292].

\bibitem{Buras:1998us}
  A.~J.~Buras and L.~Silvestrini,
  Nucl.\ Phys.\  B {\bf 548}, 293 (1999)
  [arXiv:hep-ph/9806278].

\bibitem{Cheng:1994zx}
  H.~Y.~Cheng,
  Phys.\ Lett.\  B {\bf 335}, 428 (1994)
  [arXiv:hep-ph/9406262].

\bibitem{Cheng:1998uy}
  H.~Y.~Cheng and B.~Tseng,
  Phys.\ Rev.\  D {\bf 58}, 094005 (1998)
  [arXiv:hep-ph/9803457].

\bibitem{Ali:1997nh}
  A.~Ali and C.~Greub,
  Phys.\ Rev.\  D {\bf 57}, 2996 (1998)
  [arXiv:hep-ph/9707251].

\bibitem{Ali:1998eb}
  A.~Ali, G.~Kramer and C.~D.~Lu,
  Phys.\ Rev.\  D {\bf 58}, 094009 (1998)
  [arXiv:hep-ph/9804363].

\bibitem{Cheng:1999gs}
  H.~Y.~Cheng, H.~n.~Li and K.~C.~Yang,
  Phys.\ Rev.\  D {\bf 60}, 094005 (1999)
  [arXiv:hep-ph/9902239].

\bibitem{Beneke:1999br}
  M.~Beneke, G.~Buchalla, M.~Neubert and C.~T.~Sachrajda,
  Phys.\ Rev.\ Lett.\  {\bf 83}, 1914 (1999)
  [arXiv:hep-ph/9905312].

\bibitem{Beneke:2000ry}
  M.~Beneke, G.~Buchalla, M.~Neubert and C.~T.~Sachrajda,
  Nucl.\ Phys.\  B {\bf 591}, 313 (2000)
  [arXiv:hep-ph/0006124].

\bibitem{Beneke:2001ev}
  M.~Beneke, G.~Buchalla, M.~Neubert and C.~T.~Sachrajda,
  Nucl.\ Phys.\  B {\bf 606}, 245 (2001)
  [arXiv:hep-ph/0104110].

\bibitem{Du:2001ns}
  D.~s.~Du, D.~s.~Yang and G.~h.~Zhu,
  Phys.\ Rev.\  D {\bf 64}, 014036 (2001)
  [arXiv:hep-ph/0103211].

\bibitem{Du:2002up}
  D.~s.~Du, H.~j.~Gong, J.~f.~Sun, D.~s.~Yang and G.~h.~Zhu,
  Phys.\ Rev.\  D {\bf 65}, 094025 (2002)
  [Erratum-ibid.\  D {\bf 66}, 079904 (2002)]
  [arXiv:hep-ph/0201253].

\bibitem{hep-ph/0308039}
  M.~Beneke and M.~Neubert,
  Nucl.\ Phys.\  B {\bf 675}, 333 (2003)
  [arXiv:hep-ph/0308039].

\bibitem{hep-ph/0612290}
  M.~Beneke, J.~Rohrer and D.~Yang,
  Nucl.\ Phys.\  B {\bf 774}, 64 (2007)
  [arXiv:hep-ph/0612290].

\bibitem{Beneke:2000wa}
  M.~Beneke and T.~Feldmann,
  Nucl.\ Phys.\  B {\bf 592}, 3 (2001)
  [arXiv:hep-ph/0008255].

\bibitem{hep-ph/0405134}
  A.~L.~Kagan,
  Phys.\ Lett.\  B {\bf 601}, 151 (2004)
  [arXiv:hep-ph/0405134].

\bibitem{Braun:1988qv}
  V.~M.~Braun and I.~E.~Filyanov,
  Z.\ Phys.\  C {\bf 44}, 157 (1989)
  [Sov.\ J.\ Nucl.\ Phys.\  {\bf 50}, 511.1989\ YAFIA,50,818 (1989\ YAFIA,50,818-830.1989)].

\bibitem{Braun:1989iv}
  V.~M.~Braun and I.~E.~Filyanov,
  Sov.\ J.\ Nucl.\ Phys.\  {\bf 52}, 126 (1990)
  [Z.\ Phys.\  C {\bf 48}, 239 (1990\ YAFIA,52,199-213.1990)].

\bibitem{Bauer:2001cu}
  C.~W.~Bauer, D.~Pirjol and I.~W.~Stewart,
  Phys.\ Rev.\ Lett.\  {\bf 87}, 201806 (2001)
  [arXiv:hep-ph/0107002].

\bibitem{Bjorken:1988kk}
  J.~D.~Bjorken,
  Nucl.\ Phys.\ Proc.\ Suppl.\  {\bf 11}, 325 (1989).

\bibitem{Bander:1979px}
  M.~Bander, D.~Silverman and A.~Soni,
  Phys.\ Rev.\ Lett.\  {\bf 43}, 242 (1979).

\bibitem{hep-ph/0603239}
  S.~Descotes-Genon, J.~Matias and J.~Virto,
  Phys.\ Rev.\ Lett.\  {\bf 97}, 061801 (2006)
  [arXiv:hep-ph/0603239].

\bibitem{hep-ph/0610109}
  S.~Baek, D.~London, J.~Matias and J.~Virto,
  JHEP {\bf 0612}, 019 (2006)
  [arXiv:hep-ph/0610109].

\bibitem{hep-ph/0611280}
  A.~Datta, M.~Imbeault, D.~London and J.~Matias,
  Phys.\ Rev.\  D {\bf 75}, 093004 (2007)
  [arXiv:hep-ph/0611280].

\bibitem{arXiv:0705.0477}
  S.~Descotes-Genon, J.~Matias and J.~Virto,
  Phys.\ Rev.\  D {\bf 76}, 074005 (2007)
  [arXiv:0705.0477 [hep-ph]].

\bibitem{arXiv:0707.2046}
  J.~Virto,
  AIP Conf.\ Proc.\  {\bf 964}, 90 (2007)
  [arXiv:0707.2046 [hep-ph]].

\bibitem{Ciuchini:1997hb}
  M.~Ciuchini, E.~Franco, G.~Martinelli and L.~Silvestrini,
  Nucl.\ Phys.\  B {\bf 501}, 271 (1997)
  [arXiv:hep-ph/9703353].

\bibitem{Ciuchini:2001gv}
  M.~Ciuchini, E.~Franco, G.~Martinelli, M.~Pierini and L.~Silvestrini,
  Phys.\ Lett.\  B {\bf 515}, 33 (2001)
  [arXiv:hep-ph/0104126].

\bibitem{Bauer:2004tj}
  C.~W.~Bauer, D.~Pirjol, I.~Z.~Rothstein and I.~W.~Stewart,
  Phys.\ Rev.\  D {\bf 70}, 054015 (2004)
  [arXiv:hep-ph/0401188].

\bibitem{Beneke:2004bn}
  M.~Beneke, G.~Buchalla, M.~Neubert and C.~T.~Sachrajda,
  Phys.\ Rev.\  D {\bf 72}, 098501 (2005)
  [arXiv:hep-ph/0411171].

\bibitem{Jain:2007dy}
  A.~Jain, I.~Z.~Rothstein and I.~W.~Stewart,
  arXiv:0706.3399 [hep-ph].

\bibitem{Weinberg:1975ui}
  S.~Weinberg,
  Phys.\ Rev.\  D {\bf 11}, 3583 (1975).

\bibitem{Hooft:1976up}
  G.~'t Hooft,
  Phys.\ Rev.\ Lett.\  {\bf 37}, 8 (1976).

\bibitem{Khodjamirian:2003xk}
  A.~Khodjamirian, T.~Mannel and M.~Melcher,
  Phys.\ Rev.\  D {\bf 68}, 114007 (2003)
  [arXiv:hep-ph/0308297].

\bibitem{Gronau:1991dq}
  M.~Gronau,
  Phys.\ Lett.\  B {\bf 265}, 389 (1991).

\bibitem{Nir:1991cu}
  Y.~Nir and H.~R.~Quinn,
  Phys.\ Rev.\ Lett.\  {\bf 67}, 541 (1991).

\bibitem{Lipkin:1991st}
  H.~J.~Lipkin, Y.~Nir, H.~R.~Quinn and A.~Snyder,
  Phys.\ Rev.\  D {\bf 44}, 1454 (1991).

\bibitem{Swart:1963gc}
  J.~J.~de Swart,
  Rev.\ Mod.\ Phys.\  {\bf 35}, 916 (1963).

\bibitem{Grinstein:1996us}
  B.~Grinstein and R.~F.~Lebed,
  Phys.\ Rev.\  D {\bf 53}, 6344 (1996)
  [arXiv:hep-ph/9602218].

\bibitem{Gronau:1994rj}
  M.~Gronau, O.~F.~Hernandez, D.~London and J.~L.~Rosner,
  Phys.\ Rev.\  D {\bf 50}, 4529 (1994)
  [arXiv:hep-ph/9404283].

\bibitem{Gronau:1995hn}
  M.~Gronau, O.~F.~Hernandez, D.~London and J.~L.~Rosner,
  Phys.\ Rev.\  D {\bf 52}, 6374 (1995)
  [arXiv:hep-ph/9504327].

\bibitem{Gronau:2006eb}
  M.~Gronau, Y.~Grossman, G.~Raz and J.~L.~Rosner,
  Phys.\ Lett.\  B {\bf 635}, 207 (2006)
  [arXiv:hep-ph/0601129].

\bibitem{Escribano:2007mq}
  R.~Escribano, J.~Matias and J.~Virto,
  arXiv:0708.0119 [hep-ph].

\bibitem{hep-ph/9909297}
  Y.~Grossman, M.~Neubert and A.~L.~Kagan,
  JHEP {\bf 9910}, 029 (1999)
  [arXiv:hep-ph/9909297].

\bibitem{Neubert:1998re}
  M.~Neubert,
  JHEP {\bf 9902}, 014 (1999)
  [arXiv:hep-ph/9812396].

\bibitem{Savage:1989ub}
  M.~J.~Savage and M.~B.~Wise,
  Phys.\ Rev.\  D {\bf 39}, 3346 (1989)
  [Erratum-ibid.\  D {\bf 40}, 3127 (1989)].

\bibitem{Fleischer:1997um}
  R.~Fleischer and T.~Mannel,
  Phys.\ Rev.\  D {\bf 57}, 2752 (1998)
  [arXiv:hep-ph/9704423].

\bibitem{Buras:1997cv}
  A.~J.~Buras, R.~Fleischer and T.~Mannel,
  Nucl.\ Phys.\  B {\bf 533}, 3 (1998)
  [arXiv:hep-ph/9711262].

\bibitem{Grossman:1997gr}
  Y.~Grossman, G.~Isidori and M.~P.~Worah,
  Phys.\ Rev.\  D {\bf 58}, 057504 (1998)
  [arXiv:hep-ph/9708305].

\bibitem{Neubert:1998pt}
  M.~Neubert and J.~L.~Rosner,
  Phys.\ Lett.\  B {\bf 441}, 403 (1998)
  [arXiv:hep-ph/9808493].

\bibitem{Fleischer:1999pa}
  R.~Fleischer,
  Phys.\ Lett.\  B {\bf 459}, 306 (1999)
  [arXiv:hep-ph/9903456].

\bibitem{Grossman:2003qp}
  Y.~Grossman, Z.~Ligeti, Y.~Nir and H.~Quinn,
  Phys.\ Rev.\  D {\bf 68}, 015004 (2003)
  [arXiv:hep-ph/0303171].

\bibitem{Matias:2001ch}
  J.~Matias,
  Phys.\ Lett.\  B {\bf 520}, 131 (2001)
  [arXiv:hep-ph/0105103].

\bibitem{Wu:1957my}
  C.~S.~Wu, E.~Ambler, R.~W.~Hayward, D.~D.~Hoppes and R.~P.~Hudson,
  Phys.\ Rev.\  {\bf 105}, 1413 (1957).

\bibitem{Garwin:1957hc}
  R.~L.~Garwin, L.~M.~Lederman and M.~Weinrich,
  Phys.\ Rev.\  {\bf 105}, 1415 (1957).

\bibitem{GellMann:1955jx}
  M.~Gell-Mann and A.~Pais,
  Phys.\ Rev.\  {\bf 97}, 1387 (1955).

\bibitem{Christenson:1964fg}
  J.~H.~Christenson, J.~W.~Cronin, V.~L.~Fitch and R.~Turlay,
  Phys.\ Rev.\ Lett.\  {\bf 13}, 138 (1964).

\bibitem{Yao:2006px}
  W.~M.~Yao {\it et al.}  [Particle Data Group],
  J.\ Phys.\ G {\bf 33}, 1 (2006).

\bibitem{Kobayashi:1973fv}
  M.~Kobayashi and T.~Maskawa,
  Prog.\ Theor.\ Phys.\  {\bf 49}, 652 (1973).

\bibitem{UTfit}
  M.~Bona {\it et al.}  [UTfit Collaboration],
  JHEP {\bf 0610}, 081 (2006);\\
  \texttt{http://utfit.roma1.infn.it/}.

\bibitem{CKMfitter}
  J.~Charles {\it et al.}  [CKMfitter Group],
  Eur.\ Phys.\ J.\  C {\bf 41}, 1 (2005);\\
  \texttt{http://ckmfitter.in2p3.fr/}.

\bibitem{Sakharov:1967dj}
  A.~D.~Sakharov,
  Pisma Zh.\ Eksp.\ Teor.\ Fiz.\  {\bf 5}, 32 (1967)
  [JETP Lett.\  {\bf 5}, 24 (1967\ SOPUA,34,392-393.1991\ UFNAA,161,61-64.1991)].

\bibitem{Kmixing}
  V.~L.~Fitch, P.~A.~Piroué and R.~B.~Perkins,
  Nuovo Cim. {\bf 22} 1160 (1961).

\bibitem{Albajar:1986it}
  C.~Albajar {\it et al.}  [UA1 Collaboration],
  Phys.\ Lett.\  B {\bf 186}, 247 (1987)
  [Erratum-ibid.\  {\bf 197B}, 565 (1987)].

\bibitem{Albrecht:1987dr}
  H.~Albrecht {\it et al.}  [ARGUS COLLABORATION Collaboration],
  Phys.\ Lett.\  B {\bf 192}, 245 (1987).

\bibitem{Abulencia:2006mq}
  A.~Abulencia {\it et al.}  [CDF - Run II Collaboration],
  Phys.\ Rev.\ Lett.\  {\bf 97}, 062003 (2006)
  [AIP Conf.\ Proc.\  {\bf 870}, 116 (2006)]
  [arXiv:hep-ex/0606027].

\bibitem{Aubert:2007wf}
  B.~Aubert {\it et al.}  [BABAR Collaboration],
  Phys.\ Rev.\ Lett.\  {\bf 98}, 211802 (2007)
  [arXiv:hep-ex/0703020].

\bibitem{Weisskopf:1930au}
  V.~Weisskopf and E.~P.~Wigner,
  Z.\ Phys.\  {\bf 63}, 54 (1930).

\bibitem{Lee:1965hi}
  T.~D.~Lee and L.~Wolfenstein,
  Phys.\ Rev.\  {\bf 138}, B1490 (1965).

\bibitem{Bigi:2000yz}
  I.~I.~Y.~Bigi and A.~I.~Sanda,
  Camb.\ Monogr.\ Part.\ Phys.\ Nucl.\ Phys.\ Cosmol.\  {\bf 9}, 1 (2000).

\bibitem{Harrison:1998yr}
  P.~F.~.~Harrison and H.~R.~.~Quinn  [BABAR Collaboration],

\bibitem{HFAG}
  E.~Barberio {\it et al.}  [Heavy Flavor Averaging Group (HFAG) Collaboration],
  arXiv:0704.3575 [hep-ex];
  \texttt{http://www.slac.stanford.edu/xorg/hfag/}.

\bibitem{Abulencia:2006ze}
  A.~Abulencia {\it et al.}  [CDF Collaboration],
  Phys.\ Rev.\ Lett.\  {\bf 97}, 242003 (2006)
  [arXiv:hep-ex/0609040].

\bibitem{Bigi:1981qs}
  I.~I.~Y.~Bigi and A.~I.~Sanda,
  Nucl.\ Phys.\  B {\bf 193}, 85 (1981).

\bibitem{Boos:2004xp}
  H.~Boos, T.~Mannel and J.~Reuter,
  Phys.\ Rev.\  D {\bf 70}, 036006 (2004)
  [arXiv:hep-ph/0403085].

\bibitem{Ciuchini:2005mg}
  M.~Ciuchini, M.~Pierini and L.~Silvestrini,
  Phys.\ Rev.\ Lett.\  {\bf 95}, 221804 (2005)
  [arXiv:hep-ph/0507290].

\bibitem{Li:2006vq}
  H.~n.~Li and S.~Mishima,
  JHEP {\bf 0703}, 009 (2007)
  [arXiv:hep-ph/0610120].

\bibitem{Glashow:1961tr}
  S.~L.~Glashow,
  Nucl.\ Phys.\  {\bf 22}, 579 (1961).

\bibitem{Weinberg:1967tq}
  S.~Weinberg,
  Phys.\ Rev.\ Lett.\  {\bf 19}, 1264 (1967).

\bibitem{Salam}
  A.~Salam, in \textit{Elementary Particle Theory},
  ed. N.~Savartholm (Almquist and Wiksells, Stockholm, 1969), p.367.

\bibitem{Cabibbo:1963yz}
  N.~Cabibbo,
  Phys.\ Rev.\ Lett.\  {\bf 10}, 531 (1963).

\bibitem{Chau:1984fp}
  L.~L.~Chau and W.~Y.~Keung,
  Phys.\ Rev.\ Lett.\  {\bf 53}, 1802 (1984).

\bibitem{Wolfenstein:1983yz}
  L.~Wolfenstein,
  Phys.\ Rev.\ Lett.\  {\bf 51}, 1945 (1983).

\bibitem{Peskin:1995ev}
  M.~E.~Peskin and D.~V.~Schroeder,
{\it  Reading, USA: Addison-Wesley (1995) 842 p}

\bibitem{hep-ph/9910211}
  P.~Ball, J.~M.~Frere and J.~Matias,
  Nucl.\ Phys.\  B {\bf 572}, 3 (2000)
  [arXiv:hep-ph/9910211].

\bibitem{ArkaniHamed:2005px}
  N.~Arkani-Hamed, G.~L.~Kane, J.~Thaler and L.~T.~Wang,
  JHEP {\bf 0608}, 070 (2006)
  [arXiv:hep-ph/0512190].

\bibitem{hep-ph/0404130}
  A.~Datta and D.~London,
  Phys.\ Lett.\  B {\bf 595}, 453 (2004)
  [arXiv:hep-ph/0404130].

\bibitem{Gronau:1995hm}
  M.~Gronau, O.~F.~Hernandez, D.~London and J.~L.~Rosner,
  Phys.\ Rev.\  D {\bf 52}, 6356 (1995)
  [arXiv:hep-ph/9504326].


\bibitem{Aubert:2005av}
  B.~Aubert {\it et al.}  [BaBar Collaboration],
  Phys.\ Rev.\ Lett.\  {\bf 95}, 151803 (2005)
  [arXiv:hep-ex/0501071].

\bibitem{Chao:2003ue}
  Y.~Chao {\it et al.}  [Belle Collaboration],
  Phys.\ Rev.\  D {\bf 69}, 111102 (2004)
  [arXiv:hep-ex/0311061].

\bibitem{Abe:2005dz}
  K.~Abe {\it et al.}  [Belle Collaboration],
  Phys.\ Rev.\ Lett.\  {\bf 95}, 101801 (2005)
  [arXiv:hep-ex/0502035].

\bibitem{hep-ph/0410407}
  A.~J.~Buras, R.~Fleischer, S.~Recksiegel and F.~Schwab,
  Acta Phys.\ Polon.\  B {\bf 36}, 2015 (2005)
  [arXiv:hep-ph/0410407].

\bibitem{hep-ph/0507126}
  Y.~F.~Zhou,
  Eur.\ Phys.\ J.\  C {\bf 46}, 713 (2006)
  [arXiv:hep-ph/0507126].

\bibitem{hep-ph/0605094}
  S.~Baek,
  JHEP {\bf 0607}, 025 (2006)
  [arXiv:hep-ph/0605094].

\bibitem{hep-ph/0609006}
  Y.~L.~Wu, Y.~F.~Zhou and C.~Zhuang,
  Phys.\ Rev.\  D {\bf 74}, 094007 (2006)
  [arXiv:hep-ph/0609006].

\bibitem{hep-ph/0309012}
  A.~J.~Buras, R.~Fleischer, S.~Recksiegel and F.~Schwab,
  Eur.\ Phys.\ J.\  C {\bf 32}, 45 (2003)
  [arXiv:hep-ph/0309012].

\bibitem{hep-ph/0312259}
  A.~J.~Buras, R.~Fleischer, S.~Recksiegel and F.~Schwab,
  Phys.\ Rev.\ Lett.\  {\bf 92}, 101804 (2004)
  [arXiv:hep-ph/0312259].

\bibitem{hep-ph/0402112}
  A.~J.~Buras, R.~Fleischer, S.~Recksiegel and F.~Schwab,
  Nucl.\ Phys.\  B {\bf 697}, 133 (2004)
  [arXiv:hep-ph/0402112].

\bibitem{hep-ph/0512032}
  A.~J.~Buras, R.~Fleischer, S.~Recksiegel and F.~Schwab,
  Eur.\ Phys.\ J.\  C {\bf 45}, 701 (2006)
  [arXiv:hep-ph/0512032].

\bibitem{hep-ph/0507156}
  R.~Fleischer,
  Int.\ J.\ Mod.\ Phys.\  A {\bf 21}, 664 (2006)
  [arXiv:hep-ph/0507156].

\bibitem{hep-ph/0512253}
  R.~Fleischer,
  J.\ Phys.\ G {\bf 32}, R71 (2006)
  [arXiv:hep-ph/0512253].

\bibitem{hep-ph/0503077}
  Y.~L.~Wu and Y.~F.~Zhou,
  Phys.\ Rev.\  D {\bf 72}, 034037 (2005)
  [arXiv:hep-ph/0503077].

\bibitem{hep-ph/0609128}
  C.~W.~Chiang and Y.~F.~Zhou,
  JHEP {\bf 0612}, 027 (2006)
  [arXiv:hep-ph/0609128].

\bibitem{hep-ph/0303159}
  A.~Datta and D.~London,
  Int.\ J.\ Mod.\ Phys.\  A {\bf 19}, 2505 (2004)
  [arXiv:hep-ph/0303159].

\bibitem{hep-ex/0408017}
  B.~Aubert {\it et al.}  [BABAR Collaboration],
  Phys.\ Rev.\ Lett.\  {\bf 93}, 231804 (2004)
  [arXiv:hep-ex/0408017].

\bibitem{hep-ex/0505067}
  K.~Senyo  [BELLE Collaboration],
  arXiv:hep-ex/0505067.

\bibitem{hep-ex/0307026}
  B.~Aubert {\it et al.}  [BABAR Collaboration],
  Phys.\ Rev.\ Lett.\  {\bf 91}, 171802 (2003)
  [arXiv:hep-ex/0307026].

\bibitem{hep-ex/0503013}
  K.~F.~Chen {\it et al.}  [BELLE Collaboration],
  Phys.\ Rev.\ Lett.\  {\bf 94}, 221804 (2005)
  [arXiv:hep-ex/0503013].

\bibitem{hep-ex/0505039}
  J.~Zhang {\it et al.}  [BELLE Collaboration],
  arXiv:hep-ex/0505039.

\bibitem{hep-ph/0508149}
  S.~Baek, A.~Datta, P.~Hamel, O.~F.~Hernandez and D.~London,
  Phys.\ Rev.\  D {\bf 72}, 094008 (2005)
  [arXiv:hep-ph/0508149].

\bibitem{hep-ph/0404009}
  D.~London and J.~Matias,
  Phys.\ Rev.\  D {\bf 70}, 031502 (2004)
  [arXiv:hep-ph/0404009].

\bibitem{hep-ph/0410011}
  D.~London, J.~Matias and J.~Virto,
  Phys.\ Rev.\  D {\bf 71}, 014024 (2005)
  [arXiv:hep-ph/0410011].

\bibitem{hep-ph/9906274}
  R.~Fleischer and J.~Matias,
  Phys.\ Rev.\  D {\bf 61}, 074004 (2000)
  [arXiv:hep-ph/9906274].

\bibitem{hep-ph/0204101}
  R.~Fleischer and J.~Matias,
  Phys.\ Rev.\  D {\bf 66}, 054009 (2002)
  [arXiv:hep-ph/0204101].

\bibitem{hep-ph/0511295}
  S.~Baek, D.~London, J.~Matias and J.~Virto,
  JHEP {\bf 0602}, 027 (2006)
  [arXiv:hep-ph/0511295].

\bibitem{Beneke:2003zw}
  M.~Beneke,
{\it In the Proceedings of 2nd Workshop on the CKM Unitarity Triangle, Durham, England, 5-9 Apr 2003, pp FO001}
  [arXiv:hep-ph/0308040].

\bibitem{hep-ph/0607021}
  G.~Cynolter and E.~Lendvai,
  J.\ Phys.\ G {\bf 34}, 1711 (2007)
  [arXiv:hep-ph/0607021].

\bibitem{hep-ex/0607021}
  A.~Abulencia {\it et al.}  [CDF Collaboration],
  Phys.\ Rev.\ Lett.\  {\bf 97}, 211802 (2006)
  [arXiv:hep-ex/0607021].

\bibitem{hep-ex/0608036}
  B.~Aubert {\it et al.}  [BABAR Collaboration],
  Phys.\ Rev.\ Lett.\  {\bf 97}, 171805 (2006)
  [arXiv:hep-ex/0608036].

\bibitem{Kayser:1990ww}
  B.~Kayser, M.~Kuroda, R.~D.~Peccei and A.~I.~Sanda,
  Phys.\ Lett.\  B {\bf 237}, 508 (1990).

\bibitem{hep-ph/0504178}
  C.~Sharma and R.~Sinha,
  Phys.\ Rev.\  D {\bf 73}, 014016 (2006)
  [arXiv:hep-ph/0504178].

\bibitem{hep-ph/0002243}
  C.~W.~Chiang,
  Phys.\ Rev.\  D {\bf 62}, 014017 (2000)
  [arXiv:hep-ph/0002243].

\bibitem{hep-ph/0701116}
  J.~Matias,
  Acta Phys.\ Polon.\  B {\bf 38}, 2901 (2007)
  [arXiv:hep-ph/0701116].

\bibitem{hep-ph/9804253}
  A.~S.~Dighe, I.~Dunietz and R.~Fleischer,
  Eur.\ Phys.\ J.\  C {\bf 6}, 647 (1999)
  [arXiv:hep-ph/9804253].

\bibitem{hep-ph/0406192}
  A.~Datta, M.~Imbeault, D.~London, V.~Page, N.~Sinha and R.~Sinha,
  Phys.\ Rev.\  D {\bf 71}, 096002 (2005)
  [arXiv:hep-ph/0406192].

\bibitem{hep-ph/0310081}
  R.~Fleischer,
  Eur.\ Phys.\ J.\  C {\bf 33}, S268 (2004)
  [arXiv:hep-ph/0310081].

\bibitem{hep-ph/0304027}
  R.~Fleischer,
  Nucl.\ Phys.\  B {\bf 671}, 459 (2003)
  [arXiv:hep-ph/0304027].

\bibitem{arXiv:0709.0451}
  M.~Bona {\it et al.},
  arXiv:0709.0451 [hep-ex].

\bibitem{Hagelin:1992tc}
  J.~S.~Hagelin, S.~Kelley and T.~Tanaka,
  Nucl.\ Phys.\  B {\bf 415}, 293 (1994).

\bibitem{hep-ph/0112303}
  D.~Becirevic {\it et al.},
  Nucl.\ Phys.\  B {\bf 634}, 105 (2002)
  [arXiv:hep-ph/0112303].

\bibitem{hep-lat/0110091}
  D.~Becirevic, V.~Gimenez, G.~Martinelli, M.~Papinutto and J.~Reyes,
  JHEP {\bf 0204}, 025 (2002)
  [arXiv:hep-lat/0110091].

\bibitem{hep-ph/9805303}
  A.~L.~Kagan and M.~Neubert,
  Eur.\ Phys.\ J.\  C {\bf 7}, 5 (1999)
  [arXiv:hep-ph/9805303].

\bibitem{hep-ph/0605182}
  S.~Baek,
  JHEP {\bf 0609}, 077 (2006)
  [arXiv:hep-ph/0605182].




\end{thebibliography}
\end{document}